\documentclass[twoside,11pt]{report}
\usepackage{amssymb}
\usepackage{amsmath}
\usepackage{epsfig}
\usepackage[spanish, activeacute]{babel}
\usepackage{graphicx}

\usepackage{fancyhdr}
\def\secondpage{\clearpage\null\vfill
\pagestyle{empty}
\begin{minipage}[b]{0.9\textwidth}
\footnotesize\raggedright
\setlength{\parskip}{0.5\baselineskip}
Copyright \copyright 2004--\the\year\ Dr Edgar  \'{A}valos\par
\end{minipage}
\vspace*{2\baselineskip}
\cleardoublepage
\rfoot{\thepage}}

\makeatletter
\g@addto@macro{\maketitle}{\secondpage}
\makeatother

\input{tcilatex}

\begin{document}

\begin{titlepage}
    \centering
    \vfill 
    \includegraphics[width=\textwidth]{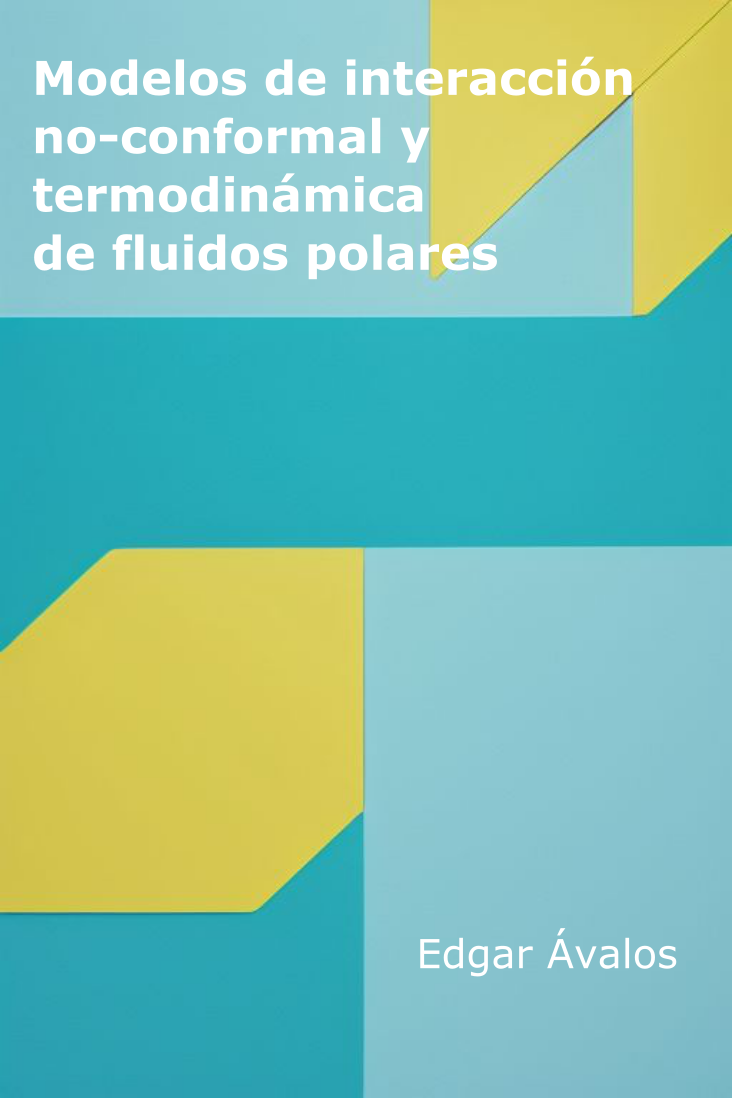} 
    \vfill
\end{titlepage}

\title{Modelos de interacci\'{o}n no-conformal y termodin\'{a}mica de
fluidos polares.}
\author{Edgar \'{A}valos}
\maketitle

\begin{abstract}
\indent En esta tesis desarrollamos un potencial modelo simple para mol\'{e}%
culas polares que representa de manera efectiva y precisa la termodin\'{a}%
mica de gases diluidos. Este potencial modela las interacciones dipolares
cuya parte no-polar sea esf\'{e}rica, como es el caso de las mol\'{e}culas
de Stockmayer (SM), o las diat\'{o}micas, como las mol\'{e}culas dipolares
de Lennard-Jones con dos centros (2CLJ). Las predicciones basadas en este
modelo para el segundo coeficiente virial para los fluidos de SM y 2CLJ
concuerdan muy bien con los recientes resultados num\'{e}ricos obtenidos por
C. Vega \textit{et. al} \cite{carloses}. Este modelo se usa para predecir la
temperatura cr\'{\i}tica de un fluido de SM con momento dipolar variable as%
\'{\i} como tambi\'{e}n sus propiedades en equilibrio l\'{\i}quido-vapor.
Finalmente aplicamos este modelo a mol\'{e}culas polares reales. Una caracter%
\'{\i}stica relevante de este trabajo es que relaciona la forma del perfil
del potencial con las propiedades termodin\'{a}micas.

\end{abstract}

\tableofcontents

\chapter{\textbf{Introducci\'{o}n}}

Los electrones en una mol\'{e}cula est\'{a}n en continuo movimiento y t\'{\i}%
picamente no est\'{a}n homog\'{e}neamente distribuidos. En general, existen m%
\'{u}ltiples desplazamientos de carga el\'{e}ctrica. Si estos
desplazamientos no se compensan entre s\'{\i}, se produce un desplazamiento
resultante conocido como momento dipolar. A las mol\'{e}culas que contienen
momentos dipolares se les llama polares y, por extensi\'{o}n, se aplica el
mismo adjetivo a las sustancias constituidas con mol\'{e}culas polares. Las
caracter\'{\i}sticas polares de una mol\'{e}cula dependen --adem\'{a}s del
tipo de enlaces y del n\'{u}mero de electrones-- de su geometr\'{\i}a. La
interacci\'{o}n entre las mol\'{e}culas polares se describe se hace a trav%
\'{e}s de la electrost\'{a}tica . Sin embargo, la interacci\'{o}n entre
puentes de hidr\'{o}geno --como caso especial de la interacci\'{o}n
dipolar--- no est\'{a} bien caracterizada.

Por otro lado, desde un punto de vista pr\'{a}ctico, existe un gran n\'{u}%
mero de sustancias polares en la industria, por ejemplo los alcoholes. Esta
variedad de sustancias requiere tener bien caracterizado su comportamiento a
nivel termodin\'{a}mico. Es decir, se requieren ecuaciones de estado capaces
de reproducir, dentro del error experimental, diversas propiedades termodin%
\'{a}micas.

Los primeros trabajos sobre la mec\'{a}nica estad\'{\i}stica de los fluidos
polares fueron revisados y compilados por Wertheim \cite{Wertheim 1979},
J.M. Deutch \cite{Deutch 1973} y P.L. Rossky \cite{Rossky 1985}; mientras
que los m\'{a}s recientes por Telo da Gama \textit{et al}.\cite{Telo da Gama
2000}.

Actualmente existen diversas l\'{\i}neas de trabajo en el \'{a}rea de los
fluidos polares. Entre estas podemos mencionar en primer lugar la investigaci%
\'{o}n de diferentes propiedades estructurales en una amplia variedad de
condiciones, tales como los ferrofluidos y los fluidos electro-reol\'{o}%
gicos \cite{van Leeuwen PRL 1993, Safran 2000}. Una segunda l\'{\i}nea de
investigaci\'{o}n est\'{a} concentrada en la predicci\'{o}n de propiedades
termodin\'{a}micas de modelos \textquotedblleft realistas\textquotedblright\
basada en modelos te\'{o}ricos de ecuaciones de estado. En esta tesis
seguimos esta l\'{\i}nea de trabajo. Nuestro prop\'{o}sito es obtener
modelos efectivos de la interacci\'{o}n entre mol\'{e}culas dipolares. La
interacci\'{o}n dipolar tiene dos caracter\'{\i}sticas especiales. La
primera, consiste en que su alcance es mayor que las interacciones no
polares y la segunda es que la interacci\'{o}n depende de la orientaci\'{o}n
entre las mol\'{e}culas.

Brevemente hacemos una retrospectiva de c\'{o}mo han abordado este problema
otros investigadores. Durante la d\'{e}cada de los a\~{n}os 60's --en el
siglo pasado-- S. Kielich \cite{Kielich 1961, Kielich 1962, Kielich 1963,
Kielich 1965} desarrolla a lo largo de sus numerosos trabajos, un formalismo
tensorial destinado a calcular una ecuaci\'{o}n de estado para gases
multipolares. En sus investigaciones Kielich representa la energ\'{\i}a de
interacci\'{o}n por pares como la suma de dos contribuciones. Una debida a
las fuerzas centrales --independientes de la orientaci\'{o}n-- y la otra
debida a las fuerzas con dependencia angular --a las que Kielich se refiere
como \textquotedblleft Fuerzas tensoriales\textquotedblright . El segundo
coeficiente virial del sistema polar se expresa como una perturbaci\'{o}n
que hace la contribuci\'{o}n tensorial a la debida \'{u}nicamente a las
fuerzas centrales. El tratamiento perturbativo queda limitado a momentos
dipolares peque\~{n}os.

D. J. Evans \cite{Evans: 1974, Evans: 1974b, Evans: 1974c} en 1974, para
hacer una estimaci\'{o}n del segundo coeficiente virial, $B(T)$, en vapor de
agua (una sustancia notablemente polar), usa un modelo que desarroll\'{o}
Kistenmacher --uno de sus colaboradores\footnote{%
Afortunadamente el Profesor Evans con gran amabilidad y diligencia nos envi%
\'{o} desde Canberra, Australia, por correo postal, una gran cantidad de
literatura que de otra forma habr\'{\i}a sido dif\'{\i}cil de conseguir,
pues mucho de este material apareci\'{o} en revistas locales y en memorias
de congresos realizados durante la d\'{e}cada de los 70's.}. Ese modelo
consiste en sumar una correcci\'{o}n por energ\'{\i}a de correlaci\'{o}n a
un potencial de Hartree-Fock. La correcci\'{o}n a distancias largas est\'{a}
dominada por la interacci\'{o}n debida a las fuerzas de dispersi\'{o}n, dada
en t\'{e}rminos de la distancia entre los \'{a}tomos de oxigeno, $\varphi
_{D}(r)=-C_{6}/r^{6}$. Evans propone un m\'{e}todo num\'{e}rico para
calcular $B(T)$ usando diferentes expresiones para $C_{6}$. En este trabajo
Evans compara sus resultados con distintas series de datos experimentales y
concluye que no es suficiente sumar --de manera sencilla-- los t\'{e}rminos
de energ\'{\i}a de dispersi\'{o}n.

En 1983 Massih y Mansoori en un ambicioso trabajo \cite{Massih: 1983}
extienden la teor\'{\i}a de la soluci\'{o}n conformal para incluir fluidos
polares. La teor\'{\i}a de la soluci\'{o}n conformal se refiere a sustancias
cuyos potenciales de interacci\'{o}n est\'{a}n relacionados entre s\'{\i} y
con uno de referencia, $U_{00}$, de acuerdo a $U_{aa}\left( r\right)
=f_{aa}U_{00}\left( r/g_{aa}\right) $. Donde $f_{aa}$ y $g_{aa}$ son \textit{%
factores de forma} que --en el caso de fluidos polares-- estos autores
encuentran que son funciones de la temperatura. Este trabajo tiene dos
puntos d\'{e}biles. En primer lugar s\'{o}lo trata con sustancias conformales%
\footnote{%
Se dice que dos sustancias son conformales si sus respectivos potenciales de
interacci\'{o}n difieren solo en dos factores de escala: uno de energ\'{\i}a
y uno de longitud.} y los fluidos polares suelen ser no conformales. En
segundo lugar los autores, en este trabajo, no reportaron comparaciones con
otros modelos o con experimentos.

Una d\'{e}cada despu\'{e}s, de nuevo Mansoori --esta vez en colaboraci\'{o}n
con C. A. Hwang-- propone un modelo basado en un n\'{u}cleo duro y una serie
de funciones de Yukawa \cite{Hwang 1994}. Estos autores incluyen la interacci%
\'{o}n dipolar (promediada angularmente)\ como un t\'{e}rmino adicional en
serie. Este t\'{e}rmino adicional contiene un par\'{a}metro, $z_{d}$, que
tiene cierto significado a nivel molecular, sin embargo no es independiente
del resto de los par\'{a}metros restantes --el de energ\'{\i}a $\epsilon $ y
el de tama\~{n}o del n\'{u}cleo duro $\sigma $--. Al final de este trabajo,
los autores concluyen que los resultados que obtuvieron para fluidos polares
no fueron satisfactorios. Despu\'{e}s de 10 a\~{n}os Mansoori sigue
considerando \'{u}til la investigaci\'{o}n de estos sistemas \cite{Mansoori
Mail}.

En el mismo a\~{n}o de 1994,\ L.A. Weber publica un trabajo sobre la predicci%
\'{o}n de $B(T)$ de diversas sustancias \cite{Weber:1994}. En este trabajo,
Weber aprovecha el bien conocido factor ac\'{e}ntrico de Pitzer \cite%
{Prausnitz 2001}, logrando reproducir una gran cantidad de propiedades
termodin\'{a}micas de diversas sustancias polares. El trabajo de Weber y
otros similares \cite{Lee 1998} basados en m\'{e}todos fundamentalmente emp%
\'{\i}ricos son de gran utilidad, tanto pr\'{a}ctica como de referencia. Sin
embargo, no establecen una conexi\'{o}n entre la termodin\'{a}mica y su
descripci\'{o}n en t\'{e}rminos de interacciones a nivel molecular.

Con excepci\'{o}n del \'{u}ltimo, todos los trabajos citados comparten un
tratamiento a base de potenciales modelo efectivos. Los potenciales
efectivos han sido ampliamente usados como modelos simplificados para
estimar propiedades termodin\'{a}micas por medio de ecuaciones integrales o
de simulaciones. Este tratamiento puede usarse --\textit{e.g.--} para
incorporar efectos de tres cuerpos \cite{Attard 1992, Madden 1999} dentro de
un potencial binario esf\'{e}rico o tambi\'{e}n para modelar el efecto del
solvente en una soluci\'{o}n \cite{McMillan 1945, Rowlinson 1984}.

Hay que subrayar que existen otros importantes tratamientos para abordar los
fluidos polares, como es el caso de las teor\'{\i}as de perturbaciones \cite%
{Stell 1972, Stell 1973, Stell 1974, Dave 1997, Dave 1998}, o funcionales de
la densidad \cite{Groh}.

Hace algunos a\~{n}os el Profesor F. del R\'{\i}o y su equipo de
colaboradores \cite{del Rio 1998, del Rio 1998 II} introdujeron una familia
de potenciales esf\'{e}ricos y no conformales ANC (Approximate
Non-Conformal) que representan con gran precisi\'{o}n las interacciones de
una gran cantidad de sustancias --modelo y reales-- en la fase vapor;
inclusive en regiones tan densas como la de la fase l\'{\i}quido \cite%
{Guzman 2001}. Esta familia de interacciones permite establecer una conexi%
\'{o}n simple, vers\'{a}til y poderosa entre las caracter\'{\i}sticas
moleculares y las propiedades termodin\'{a}micas de los fluidos.

La teor\'{\i}a de fluidos no conformales ANC extiende el principio de
estados correspondientes para los gases (PEC), permitiendo la estimaci\'{o}n
de propiedades termodin\'{a}micas de sustancias constituidas con mol\'{e}%
culas cuya interacci\'{o}n es no conformal \footnote{%
Este m\'{e}todo esta \'{\i}ntimamente relacionado con el `\textit{mapeo}' de
propiedades, cuya principal caracter\'{\i}stica consiste en escribir la
termodin\'{a}mica de un sistema en t\'{e}rminos de la ecuaci\'{o}n de estado
de un sistema equivalente usado como referencia\cite{Gil-Villegas 1996}.}.
La teor'ia ANC usa una familia de potenciales efectivos esf'ericos que
dependen de la posici\'{o}n y profundidad del m'inimo ($r_{\text{m}}$ y $%
\epsilon $) y de\ un par\'{a}metro $s$ --la suavidad-- que se identifica con
la forma del potencial. Esta familia de potenciales ha permitido reproducir
datos de $B(T)$ de una amplio conjunto de sustancias no conformales as\'{\i}
como de sus mezclas binarias \cite{Guzman 2001}.

En este trabajo de investigaci\'{o}n extendemos la teor\'{\i}a ANC para
incluir el tratamiento de sustancias polares. Desarrollaremos potenciales de
interacci\'{o}n molecular que describan adecuadamente la termodin\'{a}mica
de sustancias cuyas mol\'{e}culas constitutivas sean polares. Estos
potenciales deber\'{a}n ser capaces de reproducir diferentes propiedades
termodin\'{a}micas.

Las principales metas que logramos al final de esta investigaci\'{o}n son:

\begin{itemize}
\item Obtuvimos un potencial esf\'{e}rico que caracteriza diversos modelos
de fluidos polares.

\item La interacci\'{o}n resultante tiene la forma de un potencial ANC con
par\'{a}metros --de energ\'{\i}a $\epsilon _{\text{ef}}$, di\'{a}metro $r_{%
\text{mef}}$ y suavidad $s_{\text{ef}}$-- que dependen del momento dipolar
permanente de las mol\'{e}culas.

\item Estos par\'{a}metros --en contraste a los potenciales desarrollados
para mol\'{e}culas no polares-- resultan depender tambi\'{e}n de la
temperatura.

\item El potencial que obtuvimos conduce a expresiones anal\'{\i}ticas de
las propiedades termodin\'{a}micas.

\item Realizamos estimaciones cuantitativas de los efectos de las caracter%
\'{\i}sticas moleculares sobre esas propiedades.
\end{itemize}

El resto de este trabajo se aborda como sigue: en el cap\'{\i}tulo 2 se har%
\'{a} una revisi\'{o}n, de ninguna manera exhaustiva, de los fluidos polares
y de sus caracter\'{\i}sticas moleculares. Se describir\'{a}n algunas de sus
propiedades macrosc\'{o}picas.

El cap\'{\i}tulo 3 presenta las herramientas te\'{o}ricas para describir los
fluidos no conformales. Se hace una revisi\'{o}n del principio de estados
correspondientes y sus extensiones emp\'{\i}ricas. Se introduce la Teor\'{\i}%
a ANC y sus principales caracter\'{\i}sticas. Se hacen estudios de simulaci%
\'{o}n de diversos modelos dipolares no conformales.

El cap\'{\i}tulo 4 describe el potencial de Stockmayer generalizado (GSM)
--objeto medular en la construcci\'{o}n de los potenciales efectivos. Se
calculan formalmente los par\'{a}metros efectivos usando dos niveles de
descripci\'{o}n --con una y con dos suavidades. Se estudia la termodin\'{a}%
mica de los fluidos GSM.

En el cap\'{\i}tulo 5 se aplican los m\'{e}todos desarrollados a los modelos
moleculares no esf\'{e}ricos en los que el momento dipolar puede incluso
estar inclinado respecto al eje molecular.

En el cap\'{\i}tulo 6 se aplica el modelo de interacci\'{o}n propuesto a las
sustancias reales.

Finalmente, en el cap\'{\i}tulo 7 presentamos nuestras conclusiones y
perspectivas.

\chapter{\textbf{Sistemas polares}}

\section{Introducci\'{o}n}

En este cap\'{\i}tulo se describen las caracter\'{\i}sticas moleculares de
los fluidos polares. Se revisan las fuerzas intermoleculares, la descripci%
\'{o}n de las de las interacciones repulsivas por medio del principio de
exclusi\'{o}n de Pauli y la descripci\'{o}n de las fuerzas de van der Waals.
Se menciona el tipo de interacci\'{o}n polar en la estamos interesados en
esta tesis. Se describen las principales propiedades termodin\'{a}micas. Se
describe cualitativamente --a trav\'{e}s de la teor\'{\i}a de van der
Waals-- la relaci\'{o}n entre la interacci\'{o}n dipolar con las propiedades
termodin\'{a}micas. Se describen las contribuciones polar y covalente en la
interacci\'{o}n por puentes de hidrogeno.

\section{Caracter\'{\i}sticas moleculares de los fluidos polares}

Los momentos dipolares moleculares se originan por desplazamientos no
compensados de carga el\'{e}ctrica. Los \'{a}tomos al agruparse para formar
compuestos distribuyen sus electrones para formar una estructura molecular
estable. Dos factores importantes que influyen en la distribuci\'{o}n electr%
\'{o}nica son: la electronegatividad de los elementos participantes y la
geometr\'{\i}a molecular. Los \'{a}tomos con mayor electronegatividad
tienden a atraer electrones con mayor intensidad que los de menor
electronegatividad. La electronegatividad, en la tabla peri\'{o}dica,
aumenta a medida que nos acercamos al fl\'{u}or por la izquierda o desde
abajo. El fl\'{u}or es el elemento mas electronegativo. En una mol\'{e}cula
como el CCl$_{4}$, aunque hay diferencia entre las electronegatividades de
los \'{a}tomos que participan, la estructura molecular tiene simetr\'{\i}a
tetra\'{e}drica y por tanto el momento dipolar resultante es cero. Un caso
diferente es el CHCl$_{3}$. En este caso la diferencia de
electronegatividades no es la misma en cada enlace, de manera que uno de los
extremos de la mol\'{e}cula tiene una carga parcial positiva, mientras que
los extremos restantes tienen cargas parciales negativas. El efecto
resultante es una distribuci\'{o}n no compensada de carga y por tanto esta
mol\'{e}cula tiene un momento dipolar no nulo.

\section{Fuerzas intermoleculares}

Las fuerzas intermoleculares tienen fundamentalmente su origen en las
propiedades el\'{e}ctricas de las mol\'{e}culas y en el comportamiento cu%
\'{a}ntico de electrones y n\'{u}cleos at\'{o}micos. Las mol\'{e}culas
cuando est\'{a}n lejos tienden a acercarse entre s\'{\i}, pero si se acercan
demasiado tienden a repelerse. Esta competencia entre lo atractivo y lo
repulsivo, puede dar lugar a estados de equilibrio termodin\'{a}mico que
suelen estar descritos por ecuaciones de estado, como la de van der Waals 
\cite{Kihara} con $V^{\ast }=V/N$,

\begin{equation}
p=\frac{kT}{V^{\ast }-b}-\frac{a}{V^{\ast 2}}\text{ ,}
\label{van der Waals eos}
\end{equation}%
donde $p$, $V$, $N$, $k$ y $T$ son, respectivamente, la presi\'{o}n, el
volumen del sistema, el n\'{u}mero de part\'{\i}culas, la constante de
Boltzmann y la temperatura. En esta expresi\'{o}n $a$ y $b$ representan,
respectivamente, los efectos atractivos y repulsivos.

\section{Origen de las interacciones}

\subsection{Interacciones repulsivas}

Cuando las mol\'{e}culas se aproximan lo suficiente, la repulsi\'{o}n
electrost\'{a}tica entre electrones es significativa. Adicionalmente, el
principio de exclusi\'{o}n de Pauli prohibe que dos fermiones compartan el
mismo estado cu\'{a}ntico. No existe una representaci\'{o}n matem\'{a}tica
sencilla pare estas fuerzas, pero se pueden modelar mediante potenciales
efectivos de la forma $u_{\text{rep}}=A/r^{n}$, donde $n$ es un exponente
entero y $A$ es una constante a especificar.

\subsection{Interacciones de van der Waals}

Estas fuerzas (generalmente atractivas) tienen su origen en la distribuci%
\'{o}n instant\'{a}nea de la carga en \'{a}tomos y mol\'{e}culas debido las
posiciones de electrones y n\'{u}cleos. En una mol\'{e}cula, una vez que la
distribuci\'{o}n espacial, $\left\vert \Psi \right\vert ^{2}$, de su nube
electr\'{o}nica ha sido determinada resolviendo la ecuaci\'{o}n de Schr\"{o}%
dinger, las fuerzas intermoleculares pueden ser calculadas a partir de la
electrost\'{a}tica cl\'{a}sica (teorema de electroest\'{a}tico de Feynman) 
\cite{Masunov, Levine} \footnote{%
Los puentes de hidr\'{o}geno tradicionalmente han sido considerados como
interacciones dipolares. Sin embargo, actualmente existen evidencias de que
pueden incluir caracter\'{\i}sticas covalentes \cite{Isaacs 1999, Poater
2003}.}.

La energ\'{\i}a potencial, $u$, entre dos cargas el\'{e}ctricas $Q_{1}$ y $%
Q_{2}$ es $Q_{1}Q_{2}/r$, donde $r$ es la distancia entre ellas. Esto es
suficiente para deducir que la energ\'{\i}a entre un momento dipolar fijo, $%
\overrightarrow{\mu }$, y una carga individual, $Q$, separados por una
distancia $r$ es, $-\overrightarrow{\mu }\cdot \overrightarrow{E}=$ $-Q\mu
\cos \theta /r^{2}$, donde $\overrightarrow{E}$ es el campo el\'{e}ctrico
debido a la carga$\ $y $\theta $ es la orientaci\'{o}n relativa entre el
momento dipolar y la carga. Esta energ\'{\i}a depende de la orientaci\'{o}n.
En las condiciones en las que se encuentra un fluido real podemos suponer
con muy buena aproximaci\'{o}n que la temperatura es lo suficientemente alta
como para que la energ\'{\i}a de interacci\'{o}n entre los momentos
dipolares sea mucho menor que la energ\'{\i}a t\'{e}rmica $kT$ \footnote{%
Para citar un ejemplo concreto de una sustancia polar t\'{\i}pica, podemos
considerar al HCl , \textit{\'{a}cido clorh\'{\i}drico}. En condiciones
normales de presi\'{o}n, $P=P_{\text{atm}}=1$ atm y temperatura, $T=298$ K,
para una orientaci\'{o}n fija, la energ\'{\i}a t\'{e}rmica $kT$ es
aproximadamente $716$ veces mayor que la energ\'{\i}a de interacci\'{o}n
entre los dipolos $U_{dd}$. A\'{u}n considerando una temperatura mas baja, $%
T=180$ K, que corresponde al l\'{\i}mite inferior en el que se conocen datos
experimentales de $B(T)$, se tiene que $kT/U_{dd}\approx 260$.}. Esto
equivale a asumir que los momentos dipolares pueden rotar libremente y que
no existe una direcci\'{o}n preferencial de alineamiento. En tales casos la
energ\'{\i}a de interacci\'{o}n de un momento dipolar en libre rotaci\'{o}n
\ y que interact\'{u}a con una carga individual es $-Q^{2}\mu ^{2}/6kTr^{4}$%
. Tambi\'{e}n dos momentos dipolares que rotan libremente interact\'{u}an.
En este caso la energ\'{\i}a de interacci\'{o}n es $-\mu _{1}^{2}\mu
_{2}^{2}/3kTr^{6}$ y se conoce como \textit{energ\'{\i}a de Keesom}\footnote{%
En el siguiente cap\'{\i}tulo mostramos que en efecto, en un r\'{e}gimen de
temperaturas suficientemente altas, la interacci\'{o}n entre dos dipolos
permanentes var\'{\i}a como $1/r^{6}$.}. La energ\'{\i}a de interacci\'{o}n
entre momentos dipolares que rotan libremente deja entonces de depender de
su orientaci\'{o}n relativa, pero adquiere una dependencia de la temperatura%
\footnote{%
Debido a que la interacci\'{o}n dipolar es de largo alcance, esta aproximaci%
\'{o}n s\'{o}lo es buena cuando el momento dipolar no es demasiado grande.
El umbral de validez de esta aproximaci\'{o}n depende de las propiedades en
que estemos interesados estudiar y en las unidades de reducci\'{o}n. R. J.
Sadus realiz\'{o} estudios de simulaci\'{o}n de coexistencia de fases de un
fluido dipolar y compar\'{o} resultados cuando la contribuci\'{o}n dipolar
est\'{a} angularmente promediada y cuando se deja sin promediar \cite{Sadus
1996}.}. Para un momento dipolar de valor fijo, la energ\'{\i}a de Keesom se
intensifica a medida que la temperatura disminuye.

Cuando una carga se aproxima a una mol\'{e}cula que no tiene momento dipolar
permanente, las energ\'{\i}as que acabamos de considerar son cero, sin
embargo, hay evidencia de que existe una fuerza atractiva en esa situaci\'{o}%
n. Esto se debe a que el monopolo induce un desplazamiento de carga en la mol%
\'{e}cula no polar generando un momento dipolar inducido que interact\'{u}a
con el monopolo. La energ\'{\i}a de interacci\'{o}n en este caso es $%
-Q^{2}\alpha /2r^{4}$. Aqu\'{\i} $\alpha $ es la polarizabilidad, que se
define como $\mu _{\text{inducido}}=\alpha E$, donde $E$ es el campo el\'{e}%
ctrico producido por la carga. Tambi\'{e}n una mol\'{e}cula con momento
dipolar permanente interact\'{u}a con una mol\'{e}cula polarizable. Si el
momento dipolar puede rotar libremente la energ\'{\i}a es $-\mu \alpha
/r^{6} $. Esta se conoce como \textit{energ\'{\i}a de Debye} \footnote{%
N\'{o}tese que --a diferencia de la energ\'{\i}a de Keesom-- la de Debye no
incluye el factor de Bolzmann $1/kT$. En el caso de Keesom, ese factor da m%
\'{a}s peso a las orientaciones que tengan una energ\'{\i}a menor, \textit{%
i.e.}, m\'{a}s negativa. Sin embargo, en el caso de la interacci\'{o}n entre
un dipolo y una mol\'{e}cula polarizable, la temperatura no favorece
orientaciones privilegiadas.}.

Las energ\'{\i}as que acabamos de describir se pueden calcular usando
argumentos cl\'{a}sicos, sin embargo, ninguna explica satisfactoriamente,
por ejemplo, la interacci\'{o}n entre dos mol\'{e}culas en un gas de nitr%
\'{o}geno. Esta interacci\'{o}n s\'{o}lo puede entenderse usando mec\'{a}%
nica cu\'{a}ntica. En una mol\'{e}cula no polar la nube electr\'{o}nica fluct%
\'{u}a constantemente haciendo que se generen momentos dipolares inst\'{a}%
ntaneos. Estos momentos dipolares inst\'{a}ntaneos interact\'{u}an
atractivamente. La energ\'{\i}a entre dos mol\'{e}culas con energ\'{\i}a de
ionizaci\'{o}n\ $h\nu $ puede aproximarse por, 
\begin{equation}
u_{\text{London}}=-\frac{3}{4}\frac{h\nu \alpha ^{2}}{r^{6}}
\label{U London}
\end{equation}%
Esta se llama \textit{energ\'{\i}a de dispersi\'{o}n de London}\emph{\ }\cite%
{Israelachvili 1985, Bonaccurso}. En la expresi\'{o}n (\ref{U London}) $\nu $
es la frecuencia de absorci\'{o}n electr\'{o}nica. Al conjunto de las
fuerzas atractivas originadas por interacci\'{o}n momento dipolar-momento
dipolar, momento dipolar inducido-momento dipolar y entre momentos dipolares
instant\'{a}neos (Keesom, Debye y London), se le llama \textit{interacciones
de van der Waals}, estas tres decrecen como $1/r^{6}$ y usualmente la
dispersi\'{o}n de London es la fuerza dominante. En el Cuadro (\ref%
{interacciones}) (Ref. \cite{Israelachvili 1985}), se resumen las
interacciones que hemos analizado. En este Cuadro abreviamos momento dipolar
simplemente como dipolo. En esta tesis estudiaremos los efectos asociados a
la contribuci\'{o}n del momento dipolar permanente. 
\begin{table}[tbp] \centering%
\begin{tabular}{|l|l|l|}
\hline
Tipo de interacci\'{o}n & Objetos & Energ\'{\i}a de interacci\'{o}n \\ \hline
carga-carga & $Q_{1}$, $Q_{2}$ & $Q_{1}Q_{2}/r$ (Coulomb) \\ 
carga-dipolo & $\mu _{\text{fijo}}$, $Q$ & $-Q\mu \cos \theta /r^{2}$ \\ 
& $\mu _{\text{rotando}}$, $Q$ & $-Q^{2}\mu ^{2}/6kTr^{4}$ \\ 
dipolo-dipolo & $\mu _{1},\mu _{2}$ fijos & $-\mu _{1}\mu _{2}(2\cos \theta
_{1}\cos \theta _{2}-$ \\ 
&  & $\sin \theta _{1}\sin \theta _{2}\cos \phi )/r^{3}$ \\ 
& $\mu _{1},\mu _{2}$ rotando & $-\mu _{1}^{2}\mu _{2}^{2}/3kTr^{6}$ (Keesom)
\\ 
carga-no polar & $Q$, $\alpha $ & $-Q^{2}\alpha /2r^{4}$ \\ 
dipolo-no dipolar & $\mu _{\text{fijo}}$, $\alpha $ & $-Q^{2}\alpha (1+3\cos
^{2}\theta _{2})/2r^{6}$ \\ 
& $\mu _{\text{rotando}}$, $\alpha $ & $-\mu \alpha /r^{6}$ (Debye) \\ 
no polares & $\alpha $, $\alpha $ & $-3h\nu \alpha ^{2}/4r^{6}$ (London) \\ 
puente de hidr\'{o}geno &  & $\approx -1/r^{2}$ \\ \hline
\end{tabular}
\caption{Interacciones intermoleculares} \label{interacciones} 
\end{table}%

\section{Propiedades de los fluidos polares}

Para que una sustancia hierva las fuerzas intermoleculares deben romperse.
Las sustancias formadas por mol\'{e}culas polares son generalmente menos vol%
\'{a}tiles que aquellas sustancias no polares con peso molecular equivalente
debido a que el punto de ebullici\'{o}n depende del n\'{u}mero de enlaces y
de electrones. A mayor cantidad de electrones en una mol\'{e}cula mayor es
la distorsi\'{o}n en la nube electr\'{o}nica y por tanto mayor ser\'{a} el
momento dipolar inducido. Las fuerzas atractivas se incrementan en
consecuencia y esto representa, a nivel macrosc\'{o}pico, una mayor
resistencia a que se rompa la estructura por medios t\'{e}rmicos, esto es,
las sustancias polares usualmente tendr\'{a}n un punto de ebullici\'{o}n
relativamente grande.

Por otro lado, uno de los efectos macrosc\'{o}picos en un fluido polar en
que se pone en evidencia la superioridad de las fuerzas atractivas debidas a
la interacci\'{o}n polar, es una presi\'{o}n menor. Dicho en una forma muy
simple, el predominio de las fuerzas atractivas de origen polar entre las mol%
\'{e}culas en un sistema a volumen constante, tiende a mantener a las mol%
\'{e}culas, m\'{a}s cercanas unas de otras. En consecuencia, la presi\'{o}n
sobre las paredes contenedoras en los sistemas polares es menor que en los
no polares. De la misma forma, pero ahora en un sistema a presi\'{o}n
constante, las fuerzas atractivas facilitan un empaquetamiento molecular m%
\'{a}s comprimido que en el caso no polar.

El segundo coeficiente virial, $B(T)$, es una cantidad que cuantifica la
desviaci\'{o}n por efectos\ de las fuerzas de interacci\'{o}n a la presi\'{o}%
n de un gas ideal. En efecto si desarrollamos la ecuaci\'{o}n de estado de
van der Waals, Ec. (\ref{van der Waals eos}), como una serie de potencias de 
$1/V^{\ast }$ obtenemos, 
\begin{equation}
pV^{\ast }=kT\left[ 1+\left( b-\frac{a}{kT}\right) \frac{1}{V^{\ast }}%
+\cdots \right] \text{,}  \label{van del waals virial}
\end{equation}%
donde la cantidad $b-a/kT$ corresponde a\footnote{%
La Ec.(\ref{van del waals virial}) la escribimos en general en el siguiente
cap\'{\i}tulo como la Ec. (\ref{Ec Virial}).} $B(T)$ , $a$ y $b$ son
respectivamente cantidades proporcionales a las fuerzas repulsivas y
atractivas. En la secci\'{o}n anterior mostramos que la interacci\'{o}n
entre momentos dipolares permanentes es atractiva, de manera que, $a$\ es
proporcional a la magnitud del momento dipolar. A mayor intensidad del
momento dipolar, el valor de $a$ se vuelve progresivamente m\'{a}s grande.
En consecuencia --a temperaturas suficientemente bajas-- $B(T)$ se hace cada
vez m\'{a}s negativo y por tanto, la presi\'{o}n disminuye a medida que se
consideran momentos dipolares cada vez m\'{a}s intensos. En la Fig. (\ref%
{FIGBSMantes}) podemos ver un ejemplo de esta situaci\'{o}n para el caso de
un potencial de Stockmayer. Los puntos son datos de simulaci\'{o}n que
corresponden al valor de $B(T)$ de tres sistemas con momento dipolar
progresivamente mayor y la l\'{\i}neas asociadas a los puntos corresponden a
la predicci\'{o}n de la teor\'{\i}a ANC desarrollada en este trabajo.

La ecuaci\'{o}n de estado de van der Waals es la relaci\'{o}n m\'{a}s simple
que predice transici\'{o}nes de fase. Sin embargo exiten muchas m\'{a}s que
describen con mayor precisi\'{o}n otras propiedades.
 \begin{figure}[h]
        \begin{center}
        \includegraphics[width=.8\hsize]{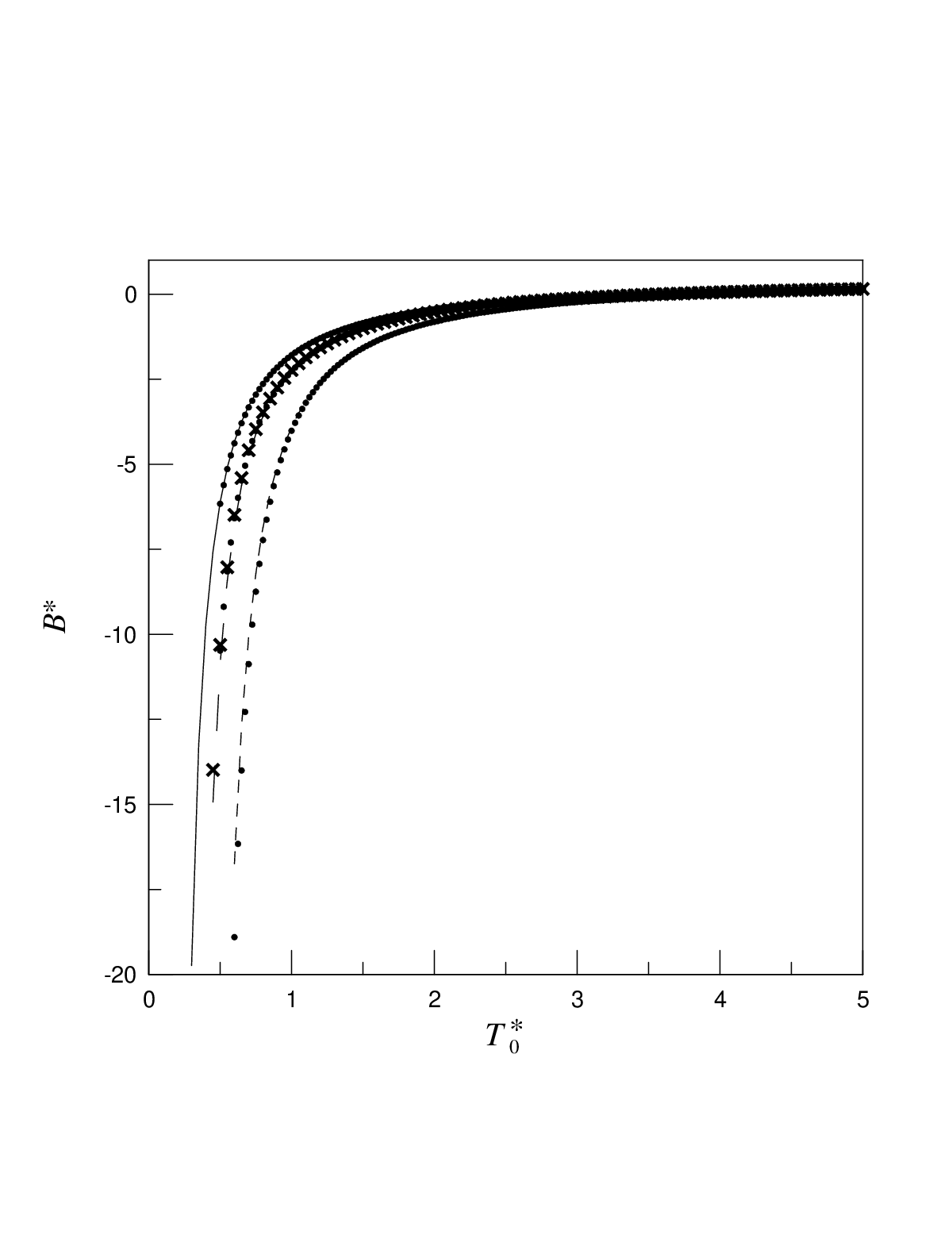}
        \end{center}
        \caption{%
Segundo coeficiente virial para el potencial de
Stockmayer calculado para distintos momentos dipolares. Los puntos
corresponden a los valores calculados por integraci\'{o}n num\'{e}rica 
\protect\cite{carloses} y las diferentes l\'{\i}neas representan la predicci%
\'{o}n de la teor\'{\i}a ANC desarrollada en este trabajo: $\protect\mu %
_{0}^{\ast }=$0.0 (l\'{\i}nea continua), $\protect\mu _{0}^{\ast }=$0.841
(segmentos largos) y $\protect\mu _{0}^{\ast }=$1.189 (segmentos cortos).
                     }%
       \label{FIGBSMantes}
\end{figure}

%

El punto cr\'{\i}tico de un fluido polar es mayor que el de uno no polar. En
un fluido no polar, justo por debajo del punto cr\'{\i}tico, dos estados con
diferentes vol\'{u}menes pueden coexistir a la misma temperatura. La inclusi%
\'{o}n del momento dipolar a\~{n}ade una contribuci\'{o}n atractiva
adicional que condensa la incipiente regi\'{o}n del gas transform\'{a}ndolo
todo en l\'{\i}quido. Se requiere de una cantidad adicional de energ\'{\i}a
cin\'{e}tica --\textit{i.e.}-- subir la temperatura cr\'{\i}tica para
asegurar la coexistencia de un fluido polar. Estos resultados se verifican
por experimentaci\'{o}n y por simulaci\'{o}n computacional.

La interacci\'{o}n entre mol\'{e}culas dipolares tiene tiene dos caracter%
\'{\i}sticas especiales. Primera, es una interacci\'{o}n de largo alcance y
la segunda, es porque que requiere una descripci\'{o}n de las posiciones
angulares relativas entre los momentos dipolares.

\section{Puentes de hidr\'{o}geno}

Los \textit{hidr\'{\i}dos}\emph{\ }son compuestos que muchos elementos
forman con el hidr\'{o}geno. Los puntos de ebullici\'{o}n de los hidr\'{\i}%
dos formados con elementos del Grupo IVA de la tabla peri\'{o}dica se
incrementan conforme descendemos en el grupo. El incremento del punto de
ebullici\'{o}n se explica debido a que las mol\'{e}culas son cada vez m\'{a}%
s grandes y por tanto las fuerzas de van der Waals hacen que la fase l\'{\i}%
quido sea estructuralmente m\'{a}s estable. Sin embargo con los hidr\'{\i}%
dos de los grupos VA, VIA y VIIA (encabezados respectivamente por N, O y F),
sucede algo diferente. A pesar de que en la mayor parte la tendencia es la
misma que la del grupo IVA, el punto de fusi\'{o}n del hidr\'{\i}do del
primer elemento de cada grupo es anormalmente alto. En los l\'{\i}quidos
formados por NH$_{3}$, H$_{2}$O y HF existen fuerzas intermoleculares de
atracci\'{o}n que requieren de m\'{a}s energ\'{\i}a para romperlas --\textit{%
los puentes de hidr\'{o}geno, }HB (hydrogen bond). El metanol es una mol\'{e}%
cula polar que puede formar puentes de hidr\'{o}geno entre sus mol\'{e}%
culas. Al igual que el metanol, el agua tambi\'{e}n es una mol\'{e}cula
polar. Esta sustancia tiene m\'{a}s \'{a}tomos por mol\'{e}cula disponibles
para formar puentes que el metanol, de tal manera que su estructura es m\'{a}%
s estable y requiere m\'{a}s energ\'{\i}a para hervir.

Los HB son fuerzas intermoleculares que operan de manera similar a la de una
interacci\'{o}n polar muy intensa. Los HB suelen presentarse entre las mol%
\'{e}culas polares covalentes que contienen H y algunos de los elementos
peque\~{n}os muy electronegativos como F, O, N, Cl y S. Desde un punto de
vista electrost\'{a}tico\footnote{%
En la literatura los HB frecuentemente han sido considerados como
interacciones de tipo polar. De manera que, seg\'{u}n el teorema electrost%
\'{a}tico de Feynman, su descripci\'{o}n ser\'{\i}a b\'{a}sicamente electrost%
\'{a}tica. Sin embargo, actualmente se sabe que los HB no tienen un origen
puramente electrost\'{a}tico \cite{Isaacs 1999, Poater 2003}.}, un HB t\'{\i}%
pico es un caso especial de una interacci\'{o}n momento dipolar-momento
dipolar muy fuerte que se forma entre un \'{a}tomo de hidr\'{o}geno con
carga parcial positiva, H$^{\delta +}$ --el cual est\'{a} ligado a un \'{a}%
tomo A en la mol\'{e}cula donadora-- y un \'{a}tomo con carga parcial
negativa en la mol\'{e}cula receptora, X$^{\delta -}$. En un puente de hidr%
\'{o}geno, A$-$H$^{\delta +}\cdots $X$^{\delta ^{-}}$, la $\delta ^{-}$
encima de X indica \textit{carga parcial negativa}. Esto significa que el
extremo X de la mol\'{e}cula es negativo con respecto al extremo H. La $%
\delta ^{+}$ sobre el H indica \textit{carga parcial positiva}, es decir,
que el extremo H de la mol\'{e}cula es positivo con respecto al extremo X.
El HB puede ocurrir entre mol\'{e}culas iguales, \textit{i.e.}, A$=$X.

En la literatura los HB frecuentemente han sido considerados como
interacciones de tipo polar. Sin embargo, actualmente se sabe que los HB no
tienen un origen puramente electrost\'{a}tico. Existen an\'{a}lisis te\'{o}%
ricos \cite{Poater 2003} y evidencias experimentales de que los HB tienen
--parcialmente-- un car\'{a}cter covalente. La mol\'{e}culas de hielo se
mantienen unidas a trav\'{e}s de HB. En 1999 Isaacs \textit{et al}. \cite%
{Isaacs 1999} encontraron indicios de que las funciones de onda de las mol%
\'{e}culas vecinas de hielo est\'{a}n traslapadas. Esto es un indicio de que
los HB no son del todo interacciones electrost\'{a}ticas.

La energ\'{\i}a de interacci\'{o}n asociada a los HB est\'{a} dentro del
intervalo de 2.5 a los 10 kcal$/$mol. En comparaci\'{o}n, la energ\'{\i}a t%
\'{\i}pica de las interacciones de van der Waals es de menor a 1 kcal$/$mol;
mientras que la energ\'{\i}a de los enlaces covalentes es del orden de 100
kcal$/$mol \cite{Israelachvili 1985}.

Los HB no tienen una representaci\'{o}n matem\'{a}tica bien conocida; sin
embargo, debido a que son fuerzas muy localizadas espacialmente, y anisotr%
\'{o}picas, se modelan usando sitios de interacci\'{o}n, es decir, se supone
una regi\'{o}n espacial muy localizada donde act\'{u}a la fuerza \cite{ENFE}.

\section{Resumen}

En este cap\'{\i}tulo describimos caracter\'{\i}sticas moleculares y las
propiedades termodin\'{a}micas de los fluidos polares. Introducimos la
electronegatividad como una importante propiedad que determina el origen de
los desplazamientos de carga el\'{e}ctrica que son responsables del momento
dipolar. Describimos las fuerzas intermoleculares repulsivas y de van der
Waals --Keesom, Debye y London. Las repulsivas, las describimos en t\'{e}%
rminos del principio de exclusi\'{o}n de Pauli y las de van der Waals, en t%
\'{e}rminos de interacciones electrost\'{a}ticas. Relacionamos --a trav\'{e}%
s de la ecuaci\'{o}n de estado de van der Waals-- y manera cualitativa, la
interacci\'{o}n dipolar con las propiedades termodin\'{a}micas de presi\'{o}%
n, segundo coeficiente virial y temperatura cr\'{\i}tica. Introdujimos el
puente de hidr\'{o}geno como una contribuci\'{o}n polar con caracter\'{\i}%
sticas especiales y mencionamos las limitaciones de su descripci\'{o}n
debidas a su car\'{a}cter parcialmente covalente. Hemos restringido esta
tesis a los efectos asociados a los momentos dipolares permanentes.

\chapter{\textbf{Teor\'{\i}a de fluidos no conformales}}

\section{Introducci\'{o}n}

En este cap\'{\i}tulo hacemos una breve revisi\'{o}n de los principales
resultados de la mec\'{a}nica estad\'{\i}stica en equilibrio. Introducimos
el potencial de interacci\'{o}n molecular y discutimos las distintas
aproximaciones necesarias para describirlo. Revisamos los modelos b\'{a}%
sicos: Lennard-Jones y Stockmayer. Describimos el principio de estados
correspondientes, las sustancias conformales y la relaci\'{o}n con el
potencial de interacci\'{o}n molecular. Recordamos las extensiones emp\'{\i}%
ricas para tratar sustancias no conformales e introducimos la Teor\'{\i}a
ANC como una extensi\'{o}n no emp\'{\i}rica, sino basada en las caracter%
\'{\i}sticas moleculares. Introducimos el potencial efectivo y sus
requisitos para la descripci\'{o}n de la termodin\'{a}mica. Desarrollamos
dos versiones de la Teor\'{\i}a ANC: con uno y con dos par\'{a}metros de
suavidad. Para una funci\'{o}n de Lennard-Jones, obtenemos los par\'{a}%
metros ANC correspondientes. Evidenciamos la necesidad de una extensi\'{o}n
de la teor\'{\i}a ANC para incluir el tratamiento de fluidos polares.
Calculamos el potencial efectivo entre momentos dipolares puntuales y su
segundo coeficiente virial. Estudiamos por simulaci\'{o}n, los efectos de
cambio de forma del potencial en la termodin\'{a}mica en equilibrio de los
fluidos polares. Para el potencial ANC estudiamos --tambi\'{e}n por simulaci%
\'{o}n-- la influencia del par\'{a}metro de suavidad en el equilibrio l\'{\i}%
quido-vapor.

\section{Resultados de la mec\'{a}nica estad\'{\i}stica}

La mec\'{a}nica estad\'{\i}stica para sistemas en equilibrio permite
calcular propiedades termodin\'{a}micas de sistemas formados por part\'{\i}%
culas cuando se conocen las energ\'{\i}as totales\ cin\'{e}tica, $K(\mathbf{p%
}^{N})$, y potencial, $U(\mathbf{q}^{N})$. B\'{a}sicamente existen dos t\'{e}%
cnicas para calcular las propiedades termodin\'{a}micas: \textit{funciones
de distribuci\'{o}n} y \textit{funciones de partici\'{o}n}. En este trabajo
usaremos la segunda; la cual establece que los valores de las variables
termodin\'{a}micas corresponden a valores promedio de variables microsc\'{o}%
picas. Para un sistema formado por $N$ part\'{\i}culas que ocupan un volumen 
$V$ en equilibrio con un ba\~{n}o t\'{e}rmico a la temperatura $T$, la relaci%
\'{o}n que liga la funci\'{o}n de partici\'{o}n con la energ\'{\i}a libre de
Helmholtz, $A$, es 
\begin{equation}
\beta A(N,V,T)=-\ln Z(N,V,T)  \label{energia helmholtz}
\end{equation}
donde $\beta =1/kT$ y $Z$, la funci\'{o}n de partici\'{o}n can\'{o}nica,
tiene la forma \cite{Hansen 1976, Anta tesis} 
\begin{equation}
Z(N,V,T)=\frac{1}{N!h^{3N}}\int \left\{ d\mathbf{p}_{N}\right\} \int \left\{
d\mathbf{q}_{N}\right\} e^{-\beta H(\left\{ \mathbf{p}_{N}\right\} ,\left\{ 
\mathbf{q}_{N}\right\} )}  \label{Z canonica}
\end{equation}
En donde $H$ es la energ\'{\i}a total del sistema representada como la suma
de sus energ\'{\i}as cin\'{e}tica y potencial, $H=K+U$, y $\left\{ \mathbf{q}%
_{N}\right\} $, $\left\{ \mathbf{p}_{N}\right\} $ son respectivamente, los
conjuntos de las coordenadas y los momentos de las $N$ part\'{\i}culas.

Para completar la descripci\'{o}n del sistema, recurrimos a las relaciones
entre las derivadas de $A$ y las variables termodin\'{a}micas; por ejemplo
la presi\'{o}n, $P$, se obtiene de sustituir la Ec. (\ref{energia helmholtz}%
) en la expresi\'{o}n siguiente: 
\begin{equation}
P=-\left( \frac{\partial A}{\partial V}\right) _{T}\text{.}  \label{presion}
\end{equation}
La funci\'{o}n de partici\'{o}n, Ec. (\ref{Z canonica}), la podemos escribir
como el producto de una parte ideal, $Z_{\text{id}}$ --independiente del
potencial-- y otra parte configuracional, $Q_{N}$ --que depende del
potencial, 
\begin{equation}
Z(N,V,T)=Z_{\text{id}}(N,V,T)Q_{N}(N,V,T)  \label{Z factorizada}
\end{equation}
donde 
\begin{eqnarray}
Z_{\text{id}} &=&\frac{V^{N}}{N!\lambda ^{3N}} \\
Q_{N} &=&\frac{1}{V^{N}}\int \left\{ d\mathbf{q}_{N}\right\} e^{-\beta
U(\left\{ \mathbf{q}_{N}\right\} )}
\end{eqnarray}
La cantidad $\lambda $ es la longitud de onda de de Broglie, $\lambda
=h/\left( 2\pi mkT\right) ^{1/2}$, de las mol\'{e}culas en una muestra de
fluido a temperatura $T$ y con sendas masas $m$ \footnote{%
La Ec. (\ref{Z canonica}) representa un l\'{\i}mite cl\'{a}sico que queda
bien justificado cuando se tratan sistemas en donde los efectos cu\'{a}%
nticos no son importantes. La longitud de onda de de Broglie, $\lambda
=h/\left( 2\pi mkT\right) ^{1/2}$, representa un valor por debajo del cual
los efectos cu\'{a}nticos dejan de ser relevantes; $\lambda $ es mayor
mientras m\'{a}s ligera es la part\'{\i}cula o menor sea la temperatura. Los
efectos de difracci\'{o}n \ de las ondas asociadas a part\'{\i}culas de
materia se aten\'{u}an a medida que el tama\~{n}o de las rendijas de difracci%
\'{o}n, $d$, supere a $\lambda $. En este caso $d$ corresponde a la separaci%
\'{o}n media de $N$ mol\'{e}culas contenidas en un volumen $V$; es decir, $%
d=\left( V/N\right) ^{1/3}$. De esta manera si $d/\lambda >>1$, tendremos un
r\'{e}gimen en donde es suficiente un tratamiento cl\'{a}sico.
\par
Para tener una idea del orden de magnitud de la raz\'{o}n $d/\lambda $,
hemos calculado esta cantidad para los siguientes sistemas: arg\'{o}n en
condiciones normales de presi\'{o}n, $P=P_{\text{atm}}=1$ atm y temperatura, 
$T=298$ K, $d/\lambda =148$; arg\'{o}n l\'{\i}quido ($P_{\text{atm}}$ y $%
T=87 $ K), $d/\lambda =53$; vapor de helio ($P_{\text{atm}}$ y $T=87$ K), $%
d/\lambda =44$ y finalmente helio l\'{\i}quido ($P_{\text{atm}}$ y $T=4$ K), 
$d/\lambda =$1.29.
\par
De esta manera, con excepci\'{o}n del helio e hidr\'{o}geno, todas las
sustancias en estado fluido aceptan un tratamiento cl\'{a}sico ya que la
distancia media entre sus part\'{\i}culas excede considerablemente a la
longitud de onda de de Broglie correspondiente.}.

En el caso l\'{\i}mite de un gas ideal en el que las part\'{\i}culas no
interact\'{u}an, $U=0$, la funci\'{o}n de partici\'{o}n se reduce a $Z_{%
\text{id}}$. Notemos que de la Ec. (\ref{presion}) obtenemos f\'{a}cilmente
la bien conocida ecuaci\'{o}n de estado para el gas ideal, $PV=NkT$. De esta
manera, una vez conocidas la energ\'{\i}a totales en un sistema dado, la mec%
\'{a}nica estad\'{\i}stica permite encontrar la ecuaci\'{o}n de estado.

\subsection{El potencial de interacci\'{o}n molecular}

Es necesario conocer la energ\'{\i}a total de interacci\'{o}n, $U$, para
poder evaluar la funci\'{o}n de partici\'{o}n configuracional, $Q_{N}$. Para
esto, b\'{a}sicamente tenemos dos caminos: el experimental y el te\'{o}rico.
Los diferentes m\'{e}todos experimentales los podemos separar en dos grandes
grupos; el primero consiste en estudiar las interacciones de mol\'{e}culas
en haces moleculares mediante experimentos de dispersi\'{o}n y la segunda v%
\'{\i}a experimental consiste en medir propiedades macrosc\'{o}picas
directamente relacionadas con $U$, e invertir despu\'{e}s los datos
experimentales para obtener el potencial.

El m\'{e}todo te\'{o}rico para obtener $U$ consiste en calcular directamente
las fuerzas intermoleculares a partir de la mec\'{a}nica cu\'{a}ntica. El
obtener resultados exactos v\'{\i}a la soluci\'{o}n de la ecuaci\'{o}n de
Schr$\ddot{o}$dinger suele ser complicado\footnote{%
En la actualidad existe un novedoso m\'{e}todo conocido como din\'{a}mica
molecular \textit{ab initio} en el que se generan las trayectorias usando
fuerzas que se obtienen \textquotedblleft sobre la marcha\textquotedblright
,\ realizando c\'{a}lculos de la estructura electr\'{o}nica \cite%
{CarParinello}. La distribuci\'{o}n de la densidad electr\'{o}nica --junto
con las coordenadas nucleares y las posibles restricciones externas--
definen un hamiltoniano cl\'{a}sico. Realizando promedios de funciones
apropiadas de este hamiltoniano, se pueden calcular los observables de inter%
\'{e}s, \textit{e.g.}, la funci\'{o}n de distribuci\'{o}n radial, y de aqu%
\'{\i}, las propiedades termodin\'{a}micas \cite{Tuckerman 2002}.}, por lo
que se recurre a algunas aproximaciones para conjeturar la forma funcional
de $U$.

La primera aproximaci\'{o}n --la de Born-Oppenheimer \cite{G. Gangopadhyay
2004} -- consiste en resolver el problema electr\'{o}nico para una
configuraci\'{o}n est\'{a}tica de los n\'{u}cleos. La raz\'{o}n de esta hip%
\'{o}tesis es que los protones son 1840 veces m\'{a}s pesados que los
electrones y esto los hace moverse m\'{a}s lento \footnote{%
Afortunadamente las correcciones necesarias debidas a la aproximaci\'{o}n de
Born-Oppenheimer son muy peque\~{n}as \cite{Coulson 1974}. \textit{e.g.},
para el H$_{2}^{+}$ (dos protones ligados con un electr\'{o}n) el error es
solamente de 720 J$/$mol (0.0075 eV), en tanto, que la energ\'{\i}a de
disociaci\'{o}n \cite{Herzberg 1969} es de 266 kJ$/$mol (2.77 eV). La relaci%
\'{o}n entre estos n\'{u}meros es aproximadamente de 370$:$1.}.

La segunda aproximaci\'{o}n consiste en ignorar los grados internos de
libertad de las part\'{\i}culas; esto es, considerarlas esf\'{e}ricas y r%
\'{\i}gidas. Al hacer esto evitamos considerar las energ\'{\i}as de rotaci%
\'{o}n y vibraci\'{o}n molecular que est\'{a}n lejos de ser consideradas en
esta tesis. De esta manera, la energ\'{\i}a $U$ depende s\'{o}lo de las
posiciones del centro de masa $\left\{ \mathbf{r}_{N}\right\} $y de las
orientaciones relativas $\left\{ \mathbf{\omega }_{N}\right\} $ de las mol%
\'{e}culas, es decir $\left\{ \mathbf{q}_{N}\right\} =\left\{ \mathbf{r}_{N},%
\mathbf{\omega }_{N}\right\} $

La \'{u}ltima aproximaci\'{o}n que necesitamos para estimar $U$ consiste en
truncar las contribuciones colectivas de muchos cuerpos a la energ\'{\i}a
potencial. Expliquemos este punto brevemente, expresamos la energ\'{\i}a
potencial del sistema como una serie creciente de contribuciones simples,
pares, triples y as\'{\i} sucesivamente 
\begin{equation}
U(1,2,...,N)=\sum_{i}u_{\text{1}}(i)+\frac{1}{2}\sum_{ij}u_{\text{2}}(ij)+%
\frac{1}{3!}\sum_{ijk}u_{\text{3}}(ijk)+\cdot \cdot \cdot
\label{U potencial}
\end{equation}
donde $u_{\text{1}}$ es el potencial de una part\'{\i}cula, $u_{2}$ el
potencial de interacci\'{o}n de dos part\'{\i}culas o \textit{potencial
binario}, $u_{\text{3}}$ el potencial triple o \textit{de tres cuerpos},
etc. Cada uno de estos t\'{e}rminos hemos de dividirlo por un factor de $n!$
a fin de no contar con la misma interacci\'{o}n m\'{a}s de una vez. El
potencial de una sola part\'{\i}cula responde a la existencia de un campo
externo tal como puede ser un campo el\'{e}ctrico o gravitacional, o la
presencia de una pared o interfase. El resto de los t\'{e}rminos surgen como
consecuencia de las interacciones que existen entre las part\'{\i}culas
debidas a las fuerzas intermoleculares. La aproximaci\'{o}n a la que nos
referimos aqu\'{\i} consiste en ignorar tanto los campos externos ($u_{\text{%
1}}=0$) como las contribuciones de tres y m\'{a}s cuerpos. El orden de
magnitud de la contribuci\'{o}n que estamos ignorando debida a las energ%
\'{\i}as entre tres cuerpos es aproximadamente de un 5\% a un 10\% para el
caso del arg\'{o}n en el punto triple \cite{Orlando tesis}.

Esta hip\'{o}tesis se hace debido a que la fuerza intermolecular corresponde
al negativo del gradiente de la energ\'{\i}a de interacci\'{o}n; de manera
que, la cantidad importante aqu\'{\i} es el cambio en la energ\'{\i}a\
respecto a las coordenadas.

El t\'{e}rmino sobreviviente --el debido al potencial binario-- lo podemos a
su vez subdividir en dos contribuciones. El negativo del gradiente de la
primera contribuci\'{o}n es positiva y corresponde a fuerzas repulsivas\ que
decaen r\'{a}pidamente con la separaci\'{o}n de las mol\'{e}culas y cuya
naturaleza la hemos ya descrito en el cap\'{\i}tulo anterior. La segunda
contribuci\'{o}n es atractiva y consiste en tres componentes denominadas
\textquotedblleft de van der Waals\textquotedblright , seg\'{u}n provengan
de la polarizaci\'{o}n permanente de las mol\'{e}culas\ (Keesom), de la
polarizaci\'{o}n inducida en una mol\'{e}cula por la polarizaci\'{o}n
permanente de otra (Debye) o de la interacci\'{o}n entre momentos dipolares
instant\'{a}neos (London). Estas componentes de van der Waals fueron ya
descritas en el cap\'{\i}tulo anterior.

\subsection{La forma del potencial de interacci\'{o}n}

\begin{figure}[h]
        \begin{center}
        \includegraphics[width=.8\hsize]{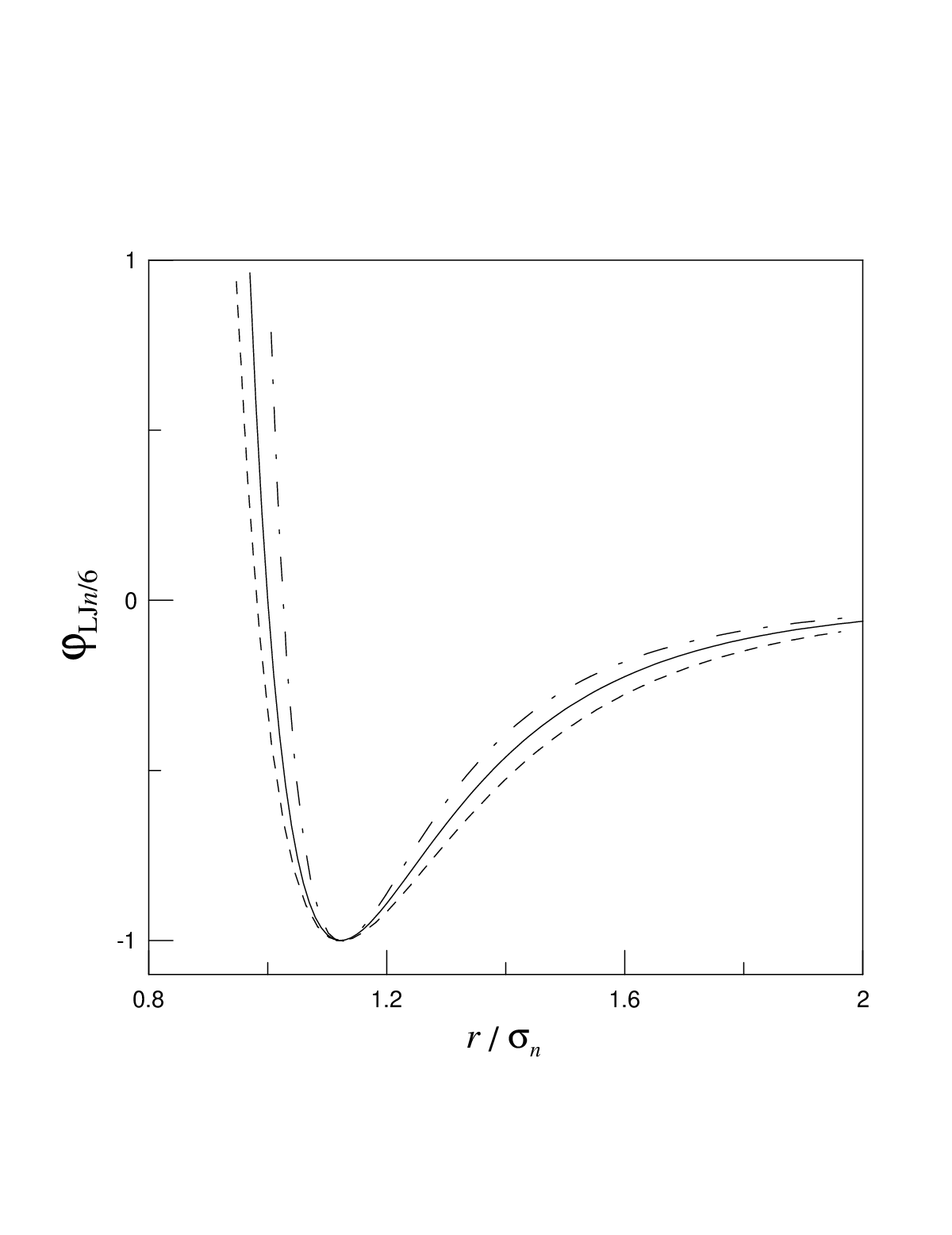}
        \end{center}
        \caption{%
Variaci\'{o}n de la \textit{suavidad} del potencial definido por la Ec.( 
\protect\ref{LJ n/6 reducido}). A medida que $n$ disminuye, la intensidad de
las fuerzas tanto atractivas como repulsivas disminuye; esto es, el perfil
del potencial se hace m\'{a}s suave. Aqu\'{\i} se muestran los casos $n=18$
(l\'{\i}nea de segmentos y puntos), $12$ (l\'{\i}nea continua) y $9$ (l\'{\i}%
nea de segmentos).
                     }%
       \label{FIGLJn6}
\end{figure}

Entre los modelos de interacci\'{o}n binarios m\'{a}s populares por su
facilidad de manejo est\'{a}n los del tipo Mie \cite{Israelachvili 1985}. En
particular el de Lennard-Jones n/6, $\varphi _{\text{LJ}n\text{/6}}$, 
\begin{equation}
\varphi _{\text{LJ}n\text{/6}}(r,\epsilon _{n})=\text{C}_{n}\epsilon _{n}%
\left[ \frac{1}{r^{n}}-\frac{1}{r^{6}}\right] \text{,}  \label{LJ n/6}
\end{equation}
El par\'{a}metro $n$ da cuenta de la intensidad de la contribuci\'{o}n
repulsiva total, $\epsilon _{n}$ es una unidad caracter\'{\i}stica de energ%
\'{\i}a y C$_{n}$ es una constante. A fin de comparar los perfiles
correspondientes a distintos valores de $n$, rescribimos la Ec. (\ref{LJ n/6}%
) en t\'{e}rminos de la unidad reducida de distancia, $z_{n}\equiv r/\sigma
_{n}$, donde $\sigma _{n}$ indica la posici\'{o}n en la que el potencial
cambia de signo. Elegimos $\epsilon _{n}$ como unidad de energ\'{\i}a. 
\begin{equation}
\left. \varphi _{\text{LJ}n\text{/6}}(z_{n})\right/ \epsilon _{n}=\text{C}%
_{n}\left[ \frac{1}{\left( z_{n}\right) ^{n}}-\frac{1}{\left( z_{n}\right)
^{6}}\right]  \label{LJ n/6 reducido}
\end{equation}
La constante C$_{n}$ es un factor que hace que la profundidad del pozo del
potencial tenga tama\~{n}o unitario, \textit{e.g.} para $n=12$, C$_{n}=4$ 
\footnote{%
Similarmente para $n=$18 y 10.5 , C$_{n}=3\sqrt{3}/2=$2.59808 y $%
(49/12)(7/4)^{1/3}=$4.92071, respectivamente.}. En la Fig. (\ref{FIGLJn6})
se muestra el potencial definido por la Ec. (\ref{LJ n/6 reducido}) para los
casos $n=18$ (l\'{\i}nea de segmentos y puntos), $12$ (l\'{\i}nea continua)
y $9$ (l\'{\i}nea de segmentos). Hemos normalizado los diferentes perfiles
para hacer coincidir sus m\'{\i}nimos en un mismo punto, $(r_{\text{m}%
_{n}},-1)$, siendo $r_{\text{m}_{n}}$ la posici\'{o}n del m\'{\i}nimo de
cada perfil. En particular, si $n=12$, recuperamos la forma m\'{a}s conocida
del potencial de Lennar-Jones; en el que la posici\'{o}n de su m\'{\i}nimo
es $r_{\text{mLJ}}=2^{1/6}$. 
%
%

Notamos en esta figura (\ref{FIGLJn6}) que a medida que $n$ disminuye, la
intensidad de las fuerzas tanto atractivas como repulsivas tambi\'{e}n
disminuye; esto es, los respectivos perfiles se hacen m\'{a}s \textit{suaves}%
, o lo que es lo mismo, se reduce el valor de la derivada local. En las
siguientes secciones precisaremos el concepto de suavidad.

Uno de nuestros objetivos es usar la teor\'{\i}a ANC para cuantificar la
suavidad de diversos modelos moleculares con interacci\'{o}n dipolar y
estudiar su relaci\'{o}n con las propiedades termodin\'{a}micas.

\subsubsection{El fluido de Lennard-Jones}

De manera general podemos describir cualquier interacci\'{o}n entre mol\'{e}%
culas neutras por complicada que sea, usando la forma b\'{a}sica del
potencial binario, $w(r)=A/r^{n}-B/r^{6}$. Siendo $A$ y $B$ par\'{a}metros
proporcionales, respectivamente, a la intensidad de las interacciones
repulsiva y atractiva; mientras que el exponente $n$ es una cantidad muy
relacionada al concepto de suavidad que desarrollaremos posteriormente en
este cap\'{\i}tulo. Por ahora basta decir que $n$ es una cantidad
proporcional a la fuerza de interacci\'{o}n. El potencial binario de
Lennard-Jones (LJ), $\varphi _{\text{LJ}}$, lo obtenemos haciendo $n=12$: 
\begin{equation}
\varphi _{\text{LJ}}(r)=\epsilon _{\text{LJ}}\left\{ \left( \frac{r_{\text{%
mLJ}}}{r}\right) ^{12}-2\left( \frac{r_{\text{mLJ}}}{r}\right) ^{6}\right\} 
\text{,}  \label{potencial LJ}
\end{equation}
donde $r_{\text{mLJ}}$ y $\epsilon _{\text{LJ}}$ son respectivamente la
posici\'{o}n del m\'{\i}nimo y su profundidad. La distancia $r_{\text{mLJ}}$
marca la frontera entre las regiones repulsiva y atractiva. En la Fig. \ref%
{FIGLJn6}, vemos como el exponente $n$ modifica el perfil de la funci\'{o}n
dada por la Ec. (\ref{LJ n/6 reducido}) haci\'{e}ndola m\'{a}s o menos
suave. La l\'{\i}nea continua representa la funci\'{o}n (\ref{potencial LJ}%
), en donde la unidad de distancia est\'{a} reducida como $z=r$/$\sigma _{%
\text{LJ}}$ con $\sigma _{\text{LJ}}=r_{\text{mLJ}}/2^{1/6}$.

Existe una gran variedad de modelos de potenciales binarios: el de esferas
duras, el de esferas repulsivas suaves, el de pozo cuadrado, el de
Born-Mayer y algunos otros que son generalizaciones al de LJ como el de
Kihara. El objetivo de este trabajo es obtener potenciales modelo binarios
que sean capaces de reproducir propiedades termodin\'{a}micas de fluidos
dipolares.

\subsubsection{El fluido dipolar de Stockmayer}

Podemos construir una representaci\'{o}n muy simple de la interacci\'{o}n
entre dos mol\'{e}culas esf\'{e}ricas separadas por una distancia $r$ y cada
una con un momento dipolar central reducido $\mathbf{\mu }^{\ast }=\mathbf{%
\mu }/\sqrt{\epsilon _{\text{LJ}}r_{\text{mLJ}}^{3}}$, si a un potencial
binario y esf\'{e}rico $\varphi _{\text{s}}(r)$ le sumamos el t\'{e}rmino de
interacci\'{o}n entre dos momentos dipolares, $\varphi _{\text{dip}}$.%
\footnote{%
Este t\'{e}rmino lo definiremos posteriormente en la secci\'{o}n `Fluidos
polares no conformales', Ec. (\ref{dipolo-dipolo})} 
\begin{equation}
\varphi _{\text{dipolar}}(\mathbf{r,\mu }_{0}^{\ast })\equiv \varphi _{\text{%
s}}(r)-\varphi _{\text{dip}}(\mathbf{r,\mu }^{\ast })\text{.}
\label{Udipolar general}
\end{equation}%
A la parte no polar en un potencial de interacci\'{o}n --la que representa
las fuerzas repulsivas y las de van der Waals-- nos referimos como \textit{%
kernel}\footnote{%
En ausencia de un t\'{e}rmino en castellano para referirnos a la parte no
polar en un potencial de interacci\'{o}n, usamos \textit{kernel}, palabra de
origen alem\'{a}n que significa n\'{u}cleo.}. Seg\'{u}n el kernel que usemos
como funci\'{o}n $\varphi _{\text{s}}(r)$ para definir la Ec.(\ref{Udipolar
general}), podemos citar diversos modelos de interacci\'{o}n dipolar:
esferas duras dipolares \cite{Henderson 1999}, esferas suaves dipolares \cite%
{Kusalik 1990, Stevens 1995}, Yukawas dipolares \cite{Henderson 1999}, pozos
cuadrados dipolares \cite{Ana Laura 1994, Ana Laura 1995}, entre otros. Si
usamos $\varphi _{\text{s}}(r)=\varphi _{\text{LJ}}(r)$, obtenemos 
\begin{equation}
\varphi _{\text{SM}}(\mathbf{r,\mu }_{0}^{\ast })\equiv \varphi _{\text{LJ}%
}(r)-\varphi _{\text{dip}}(\mathbf{r,\mu }^{\ast })\text{.}
\end{equation}%
A la funci\'{o}n $\varphi _{\text{SM}}$ se le conoce como \textit{potencial
de Stockmayer,} (SM) \cite{Stockmayer 1959, Maitland, Gubbins 1997}, y es
quiz\'{a} la representaci\'{o}n m\'{a}s difundida en la literatura de la
interacci\'{o}n molecular en los fluidos polares (ver Fig. (\ref%
{FigureDiagSM})). Sin embargo, el potencial SM no es el mejor debido a que
su kernel no permite variaci\'{o}n en la forma del perfil del potencial.

%
 \begin{figure}[h]
        \begin{center}
        \includegraphics[width=.8\hsize]{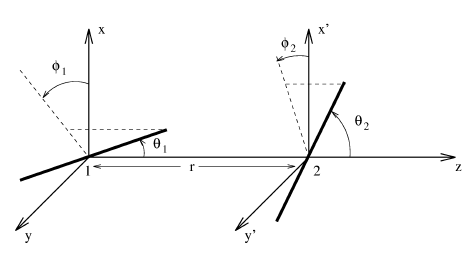}
        \end{center}
        \caption{%
Representaci\'{o}n esquem\'{a}%
tica de la interacci\'{o}n de Stockmayer entre las mol\'{e}culas 1 y 2.
Siendo $\protect\theta _{i}$ y $\protect\phi _{i}$, respectivamente, los 
\'{a}ngulos zenital y azimutal de la mol\'{e}cula $i$.
                     }%
       \label{FigureDiagSM}
\end{figure}

\subsection{Segundo coeficiente virial}

La \textit{ecuaci\'{o}n de estado} virial la escribimos representando la
presi\'{o}n como una serie infinita de potencias del inverso del volumen 
\begin{equation}
\left. PV^{\ast }\right/ kT=1+\sum_{n=2}^{\infty }B_{n}(T)\frac{1}{V^{\ast n}%
}\text{,}  \label{Ec Virial}
\end{equation}
siendo $B_{n}(T)$ el $n$-\'{e}simo coeficiente virial, el cual depende s\'{o}%
lo de las interacciones entre $n$ mol\'{e}culas \footnote{%
El comportamiento de la fase vapor puede aproximarse bastante bien usando
unos cuantos coeficientes viriales, sin embargo, la serie definida por la
Ec. (\ref{Ec Virial}) no converge para la fase l\'{\i}quida.}.

\section{Conformalidad y Principio de estados correspondientes}

El principio de estados correspondientes (PEC) es una ley de universalidad 
\cite{Ben-Amotz 2002, Ben-Amotz 2004, Hirsch}. Una sustancia que se
encuentra en un estado definido por una temperatura y un volumen, tiene una
presi'on determinada por su ecuaci'on de estado (EOS). Para otra sustancia
la presi'on es diferente aunque se encuentre en el mismo estado. El PEC
establece que al escalar las variables termodin'amicas de las EOS de ambas
sustancias con sus respectivos valores cr\'{\i}ticos, ($P_{\text{c}}$, $T_{%
\text{c}}$, $V_{\text{c}}$), las EOS de ambas sustancias son equivalentes.
Esto es, en unidades reducidas una amplia familia de sustancias comparten
una misma EOS, digamos dada por la funci\'{o}n $f$, 
\begin{equation}
P/P_{\text{c}}=f\{\left( T/T_{\text{c}}\right) \text{, }\left( V/V_{\text{c}%
}\right) \}  \label{eos pec}
\end{equation}
Esta condici\'{o}n de equivalencia funcional se cumple con una alta precisi%
\'{o}n\ para sustancias molecularmente sim\'{e}tricas como son los gases
nobles y el CO y el CH$_{4}$, entre otras. Sin embargo falla cuando se
aplica en mol\'{e}culas demasiado ligeras (como el H$_{2}$ \'{o} el He), o
cuando hay interacciones polares de magnitud considerable (como en el caso
del CO$_{2}$ y SF$_{6}$) y tampoco se cumple cuando la geometr\'{\i}a
molecular se vuelve considerablemente no esf\'{e}rica, como en el caso de
los alkanos y los alkenos \cite{Guggenheim}.

La interpretaci\'{o}n molecular del PEC est\'{a} restringida a sistemas en
los que la interacci\'{o}n molecular total entre $N$ part\'{\i}culas puede
escribirse como una suma de interacciones binarias, $u$, \cite{Frenkel 2000,
McQuarry} 
\begin{equation}
U(1,2,...,N)=\frac{1}{2}\sum_{ij}\epsilon u\left( r_{ij}/r_{\text{m}}\right)
\label{potencial por pares}
\end{equation}
donde $\epsilon $ y $r_{\text{m}}$ son respectivamente, par\'{a}metros de
energ\'{\i}a y longitud, caracter\'{\i}sticos de cada sistema en particular.
Entre dos sustancias A y B se cumple el PEC si sus respectivas funciones de
potencial, $U_{\text{A}}$ y $U_{\text{B}}$, son \textit{conformales}, i.e.,
cuando sus perfiles se pueden superponer al ajustar los valores de $\epsilon 
$ y $r_{\text{m}}$. Dos fluidos que cumplen con el PEC se dice que son
conformales \cite{Rowlinson}. La funci\'{o}n de partici\'{o}n, Ec. (\ref{Z
factorizada}), asociada a un potencial dado por la Ec. (\ref{potencial por
pares}) es, 
\begin{equation}
Z(\rho ^{\ast },T^{\ast })=\frac{Z_{\text{id}}}{V^{N}}\int \left\{ d\mathbf{r%
}_{N}\right\} \exp \left[ -\beta \frac{1}{2}\sum_{ij}\epsilon u\left( \frac{%
\left\vert \mathbf{r}_{i}-\mathbf{r}_{j}\right\vert }{r_{\text{m}}}\right) %
\right] \text{,}  \label{Z pec}
\end{equation}
donde $\rho ^{\ast }=N/Vr_{\text{m}}^{3}$ y $T^{\ast }=kT/\epsilon $. Un
grupo de sustancias conformales comparten una misma funci\'{o}n de partici%
\'{o}n (\ref{Z pec}), as\'{\i} como tambi\'{e}n la misma forma funcional de
sus ecuaciones de estado, Ec.(\ref{eos pec}). De esta forma, podemos usar el
PEC como gu\'{\i}a cuando se necesita una estimaci\'{o}n r\'{a}pida de las
propiedades termodin\'{a}micas de un sistema dado.

\subsection{Extensiones emp\'{\i}ricas para fluidos no conformales}

Desafortunadamente, las interacciones entre las mol\'{e}culas reales no son
completamente aditivas por pares, ni tampoco, las interacciones binarias
entre diferentes mol\'{e}culas son conformales. Aun en los gases nobles, la
conformalidad entre ellos es s\'{o}lo aproximada. Solo una peque\~{n}a
familia de sustancias puede ser descrita usando la forma original del PEC.

Debido a estas circunstancias, se han desarrollado diversas l\'{\i}neas de
investigaci\'{o}n para superar esta limitaci\'{o}n \cite{Betts 1971, Pitzer
1990, Dunikov 2001}. Entre las m\'{a}s populares podemos citar aquellas que
introducen un tercer par\'{a}metro, $x$, que caracteriza la no
conformalidad. El factor de compresibilidad $z$ --en t\'{e}rminos de este
tercer par\'{a}metro-- lo podemos expresar como 
\begin{equation*}
z=\left. \beta P\right/ \rho =f\left( T^{\ast },\rho ^{\ast },x\right) \text{%
.}
\end{equation*}

En la teor\'{\i}a de Pitzer, $x$ es un par\'{a}metro emp\'{\i}rico,\ $\omega 
$, llamado \textit{factor ac\'{e}ntrico }\cite{Prausnitz 2001, Rowlinson}.
Para definir el factor ac\'{e}ntrico de un fluido se aprovecha la caracter%
\'{\i}stica de que el logaritmo decimal de la presi\'{o}n de saturaci\'{o}n
reducida, $P_{\sigma }^{\ast }=P_{\sigma }/P_{\sigma \text{c}} $, y el
inverso de la temperatura reducida mantienen una relaci\'{o}n
aproximadamente lineal, 
\begin{equation*}
\log P_{\sigma }^{\ast }(T^{\ast })=\alpha _{k}\left( 1-\left. 1\right/
T^{\ast }\right) \text{.}
\end{equation*}
La no conformalidad entre dos sustancias se relaciona con la diferencia de
sus constantes $\alpha _{k}$ \cite{Orlando tesis}. En el caso del arg\'{o}n $%
\alpha _{k}\simeq 7/3$, de manera que $\log P_{\sigma \text{Ar}}^{\ast }($0.7%
$)=-1$. El factor ac\'{e}ntrico de un fluido etiquetado con el sub\'{\i}%
ndice \textquotedblleft k\textquotedblright\ se define como el logaritmo
decimal del cociente $\left. P_{\sigma \text{Ar}}^{\ast }(\text{0.7})\right/
P_{\sigma \text{A}}^{\ast }($0.7$)$, es decir, 
\begin{equation*}
\omega =-1-P_{\sigma \text{k}}^{\ast }(\text{0.7})\text{.}
\end{equation*}
Podemos usar el factor ac\'{e}ntrico de Pitzer para calcular $z$ como, 
\begin{equation*}
z\left( T^{\ast },\rho ^{\ast },\omega \right) =z_{0}\left( T^{\ast },\rho
^{\ast }\right) +\omega z_{1}\left( T^{\ast },\rho ^{\ast }\right) \text{.}
\end{equation*}
El primer t\'{e}rmino en esta expresi\'{o}n para $z$ es la ecuaci\'{o}n de
estado del arg\'{o}n.

En otros tratamientos de utilidad en la descripci\'{o}n de sistemas
coloidales \cite{Frenkel 2000}, se ha identificado el segundo coeficiente
virial reducido como el par\'{a}metro de no conformalidad $x$. A pesar de su
utilidad pr\'{a}ctica \cite{Weber:1994}, las extensiones emp\'{\i}ricas al
PEC son insatisfactorias desde el punto de vista de la mec\'{a}nica estad%
\'{\i}stica porque sus par\'{a}metros de forma no se definen directamente en
t\'{e}rminos moleculares sino macrosc\'{o}picos.

En este trabajo estamos interesados en la extensi\'{o}n misma del PEC a
nivel molecular, de manera que el par\'{a}metro $x$ tenga un significado f%
\'{\i}sico en el potencial de interacci\'{o}n. Nuestro prop\'{o}sito
consiste en constuir un \textit{potencial modelo o efectivo} que sea capaz
de reproducir --a trav\'{e}s de una ecuaci\'{o}n de estado-- las propiedades
termodin\'{a}micas de sustancias cuya no conformalidad se debe a la
presencia de interacciones polares entre las mol\'{e}culas.

\subsection{Teor\'{\i}a Aproximada No Conformal}

La teor\'{\i}a ANC se propuso como una extensi\'{o}n rigurosa al principio
de estados correspondientes para los gases. Aqu\'{\i} haremos una breve
revisi\'{o}n de los principales resultados de la teor\'{\i}a ANC \cite{del
Rio 1998} que son relevantes para este trabajo de investigaci\'{o}n. La teor%
\'{\i}a ANC introduce una familia de potenciales efectivos esf\'{e}ricos $%
\varphi _{\text{ANC}}(r;\epsilon ,r_{\text{m}},s)$ con m\'{\i}nimo en $r_{%
\text{m}}$ y de profundidad $\epsilon $. Adicionalmente a estos dos par\'{a}%
metros, tambi\'{e}n se introduce un nuevo par\'{a}metro molecular, $s $,
--la suavidad del potencial-- este nuevo par\'{a}metro toma en cuenta la
forma del perfil del potencial.\cite{del Rio 1998 II}.

\subsubsection{Potenciales efectivos}

En esta secci\'{o}n discutiremos el concepto y los requisitos para
desarrollar un potencial efectivo que refleje lo mejor posible las caracter%
\'{\i}sticas de la interacci\'{o}n exacta. Existen muchas razones por las
cuales se prefiere el uso de un potencial efectivo en lugar del potencial
\textquotedblleft real\textquotedblright\ o exacto. En primer lugar, para
obtener el potencial exacto para una mol\'{e}cula en particular se requiere
resolver la ecuaci\'{o}n de Schr$\ddot{o}$dinger para todos los electrones,
al menos para el estado base. En la pr\'{a}ctica, esta es una tarea
sumamente complicada \cite{Stillinger}. En segundo lugar, un problema puede
ser sencillamente inabordable si se usa el potencial exacto \cite{Gao}.

La idea b\'{a}sica de potencial modelo o efectivo consiste en reemplazar el
potencial exacto, $u$, de muchos cuerpos que caracteriza a un sistema con
uno que sea matem\'{a}ticamente m\'{a}s simple y accesible, $u_{\text{ef}}$;
i.e., 
\begin{equation}
u_{\text{ef}}\cong u\text{ .}  \label{Def pot efectivo}
\end{equation}
El problema de caracterizar la interacci\'{o}n se transforma en hacer una
selecci\'{o}n adecuada de la funci\'{o}n $u_{\text{ef}}$. Los resultados
generalmente depender\'{a}n de la temperatura $T$, del volumen del sistema $%
V $, y del n\'{u}mero de part\'{\i}culas $N$ si \ tratamos con un sistema
cerrado, o bien del potencial qu\'{\i}mico $\mu $ en el caso de que el
sistema sea abierto \cite{Stillinger}. En todos los casos que aqu\'{\i}
consideraremos, $u_{\text{ef}}$, representa la interacci\'{o}n efectiva por
pares.

Desde el punto de vista termodin\'{a}mico requerimos que $u_{\text{ef}}$
reproduzca lo mejor posible alguna propiedad termodin\'{a}mica. \textit{e.g.}%
, la presi\'{o}n $P(V,T)$. En un sistema cuyo potencial exacto sabemos que
es $u$, \textit{i.e.}, requerimos que 
\begin{equation}
P(V,T;[u_{\text{ef}}])\cong P(V,T;[u])  \label{P con u efe}
\end{equation}
sea una buena aproximaci\'{o}n. En general, $u_{\text{ef}}$ adem\'{a}s de
depender del estado, tambi\'{e}n depender\'{a} de la propiedad escogida \cite%
{Orlando tesis}. Nosotros, en este\ trabajo, estamos interesados en obtener
potenciales efectivos ANC que puedan usarse para tratar diferentes
propiedades termodin\'{a}micas. El hecho de que $u_{\text{ef}}$ puede
depender del estado, es delicado \cite{Louis 2002, Stillinger} y aqu\'{\i}
nos ocupamos de potenciales independientes de la densidad que se obtienen a
partir de la Ec. (\ref{P con u efe}) para el gas diluido.

En este trabajo, desarrollamos un novedoso m\'{e}todo para construir
potenciales efectivos esf\'{e}ricos para fluidos polares. M\"{u}ller and
Gelb \cite{Muller} han sugerido que esta clase de potenciales isotr\'{o}%
picos pueden ser suficientes para tratar con fluidos en regiones de
temperatura suficientemente altas.

Finalmente, hay que mencionar que los potenciales efectivos pueden definirse
de manera diferente, dependiendo del contexto en el que se van a usar. Por
ejemplo, se puede definir la interacci\'{o}n efectiva entre dos part\'{\i}%
culas coloidales (grandes) en un medio compuesto por part\'{\i}culas peque%
\~{n}as, en t\'{e}rminos de la fuerza requerida para desplazar las part\'{\i}%
culas peque\~{n}as de la regi\'{o}n entre las part\'{\i}culas grandes \cite%
{ramon}. Otro ejemplo m\'{a}s ocurre cuando se tratan interacciones con m%
\'{a}s de un m\'{\i}nimo, por ejemplo en el doblamiento de prote\'{\i}nas.
En esos casos, la barrera central del potencial puede eliminarse mediante
una transformaci\'{o}n de variables que conduce a sustituir el potencial
exacto por uno efectivo, que resulta ser mucho m\'{a}s sencillo \cite%
{TuckermanPRL2002}.

\subsubsection{Elementos de la teor\'{\i}a ANC}

El potencial ANC que m\'{a}s ampliamente se ha usado es \cite{del Rio 1998}: 
\begin{equation}
\varphi _{\text{ANC}}(z;\epsilon ,s)=\epsilon \ \left[ \left( \frac{1-a}{%
\zeta (z,s)-a}\right) ^{12}-2\left( \frac{1-a}{\zeta (z,s)-a}\right) ^{6}%
\right] \text{,}  \label{potencialANC}
\end{equation}
donde $\zeta (z,s)=(1+(z^{3}-1)/s)^{1/3}$, $z=r/r_{\text{m}}$, $a$ es el di%
\'{a}metro de n\'{u}cleo duro y $r$ es la distancia centro a centro. El
potencial $\varphi _{1}(z)=\varphi _{\text{ANC}}(z,s=1)$ tiene $\zeta =z$ \
y tiene la forma de una funci\'{o}n de Kihara esf\'{e}rica con di\'{a}metro
de n\'{u}cleo duro $a$. El potencial $\varphi _{1}(z)$, es el potencial ANC
de referencia y escogemos $a=$0.0957389 de tal manera que $\varphi _{1}(z)$
sea aproximadamente conformal al potencial del\ arg\'{o}n. \cite{del Rio
1998 III} En general para un potencial esf\'{e}rico, $\varphi (z)$, si $%
z_{1}(\varphi )$ y $z(\varphi )$ son respectivamente el inverso de $\varphi
_{1}(z)$ y $\varphi (z)$, la suavidad local $S(\varphi )$ mide la raz\'{o}n
de la presiones locales; \textit{i.e.}, 
\begin{equation}
S(\varphi _{0})=\left[ \frac{1}{z_{1}^{2}}\frac{\partial \varphi _{1}(z_{1})%
}{\partial z_{1}}\right] _{z_{1}(\varphi _{0})}\left/ \left[ \frac{1}{z^{2}}%
\frac{\partial \varphi (z)}{\partial z}\right] _{z(\varphi _{0})}\right. 
\text{.}  \label{suavidad}
\end{equation}

De (\ref{suavidad}) podemos ver que $S(\varphi _{0})$ es inversamente
proporcional a la \textit{fuerza} intermolecular a una distancia $z(\varphi
_{0})$. Cualquier potencial con $s=1$ para todo valor de $\varphi _{0}$ ser%
\'{a} conformal a $\varphi _{1}(z)$.

\begin{figure}[h]
        \begin{center}
        \includegraphics[width=.8\hsize]{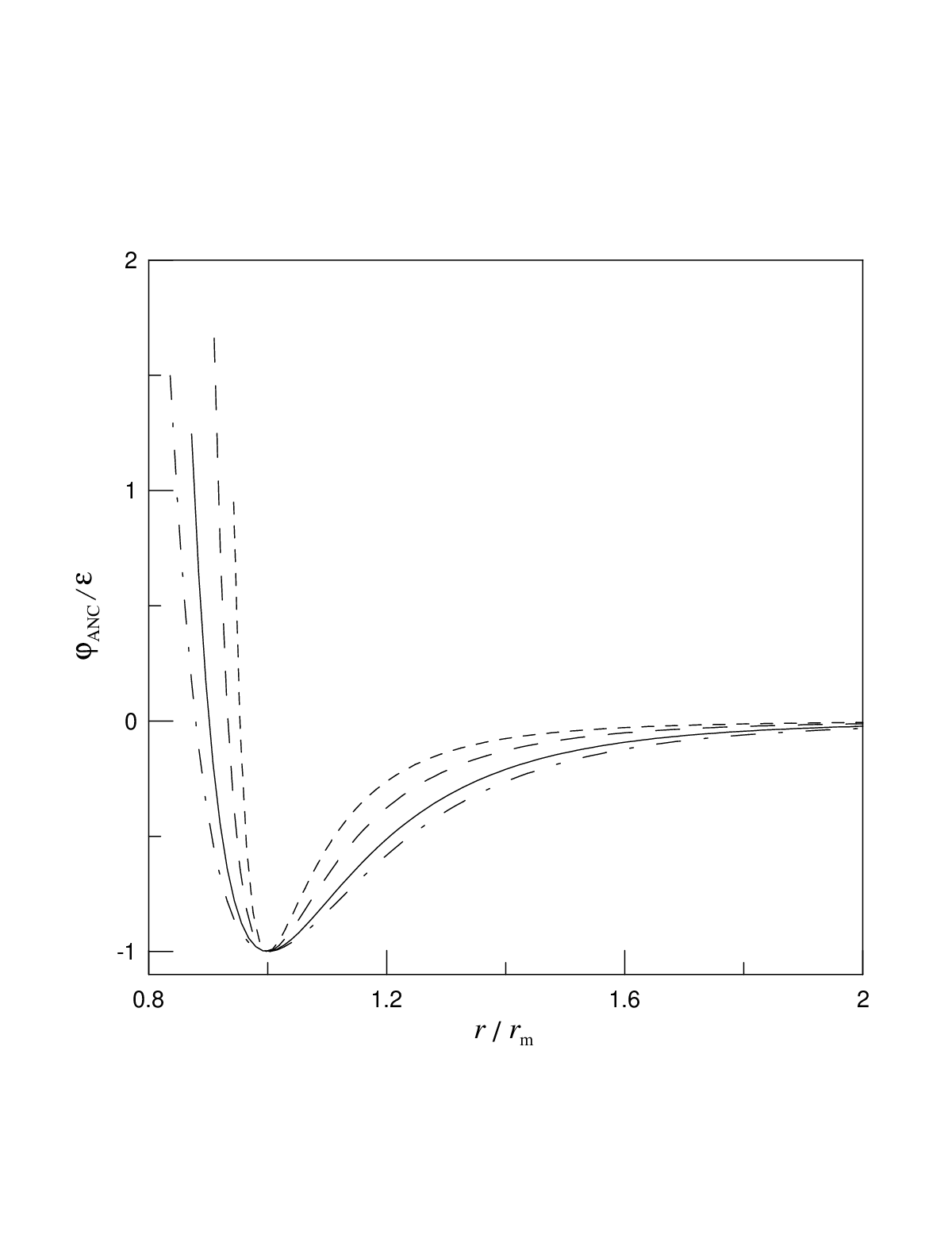}
        \end{center}
        \caption{%
El potencial $\protect\varphi _{\text{ANC}}(z)$
graficado contra $z$ para diferentes valores de suavidad $s$. La l\'{\i}nea
continua representa el potencial de referencia $s=$1.0. Tambi\'{e}n se
muestran los casos con: $s=$0.5 (segmentos cortos), $s=$0.7 (segmentos
largos) y $s=$1.2 (segmentos y puntos).
                     }%
       \label{FIGpotANC}
\end{figure}

Como el lector puede verificar, el potencial $\varphi _{\text{ANC}}(z,s)$
definido en la Ec. (\ref{potencialANC}) tiene un valor constante de suavidad
local, $S(\varphi _{\text{0}})=s$, para todos los valores de la energ\'{\i}a 
$\varphi _{0}$. Variando $s$ en $\varphi _{\text{ANC}}(z,s)$ cambiamos la
forma del perfil del potencial de una manera muy simple: disminuyendo $s$ se
hace el potencial m\'{a}s duro, y veceversa, incrementando $s$ el perfil se
hace m\'{a}s suave (ver Fig. \ref{FIGpotANC}). 
%
%

\subsubsection{Aproximaci\'{o}n $1$-s}

Cuando la sustancia en consideraci\'{o}n tiene un potencial esf\'{e}rico sim%
\'{e}trico $\varphi (z)$ con $S(\varphi )=$cte. en sus ramas atractiva y
repulsiva, la teor\'{\i}a ANC provee una relaci\'{o}n lineal exacta entre
los segundos coeficientes viriales de la sustancia $B^{\ast }(T^{\ast })$ y
la referencia $B_{\text{1}}^{\ast }(T^{\ast })$ \cite{del Rio 1998}:

\begin{equation}
B^{\ast }(T^{\ast })=1-s+sB_{1}^{\ast }(T^{\ast })\text{,}  \label{Bstar}
\end{equation}
donde $B^{\ast }(T^{\ast })=B(T)/(2\pi r_{\text{m}}^{3}/3)$ y $T^{\ast
}=kT/\epsilon $, con $r_{\text{m}}$ y $\epsilon $ que son respectivamente,
la posici\'{o}n del m\'{\i}nimo y \ la profundidad del potencial de inter%
\'{e}s. M\'{a}s en general, $S$ es una funci\'{o}n de la energ\'{\i}a $%
\varphi $; de manera que, en este caso, el resultado (\ref{Bstar}) ser\'{a}
una buena aproximaci\'{o}n cuando consideramos que $s$ es el promedio de $%
S(\varphi )$ sobre el intervalo apropiado de energ\'{\i}a \cite{del Rio 1998}

Adicionalmente, cuando el potencial por pares $\varphi (z,\Omega )$ depende
de la orientaci\'{o}n intermolecular $\Omega $, la suavidad local en (\ref%
{suavidad}) se vuelve dependiente de los \'{a}ngulos y (\ref{Bstar}) se
cumple nuevamente s\'{o}lo de manera aproximada cuando consideramos que $s$
es un promedio angular de $S(\Omega )$. En todos estos casos cuando $%
S(\varphi ,\Omega )$ lo remplazamos por una sola $s$ constante, la Ec. (\ref%
{Bstar}) opera de manera aproximada y uno habla de la aproximaci\'{o}n de 1-$%
s$.

\subsubsection{Aproximaci\'{o}n $2$-s}

Una aproximaci\'{o}n de segundo orden considera que el promedio de $%
S(\varphi ,\Omega )$ es diferente en ambos lados de $\varphi (z,\Omega )$,
en el lado atractivo ($z<1)$ y en el repulsivo ($z>1)$; en este caso el
potencial efectivo ANC aun est\'{a} bien definido por (\ref{potencialANC})
pero por partes, es decir, con $s=s_{\text{A}}$ para $z<1$ y $s=s_{\text{R}}$
para $z>1$. En este caso, en la aproximaci\'{o}n de 2-$s$ el segundo
coeficiente virial est\'{a} dado por: \cite{del Rio 1998 II} 
\begin{equation}
B^{\ast }(T^{\ast })=b^{\ast }(T^{\ast })e^{\beta \epsilon }-\Lambda ^{\ast
}(T^{\ast })\left( e^{\beta \epsilon }-1\right) \text{,}  \label{Bstar2}
\end{equation}
donde los vol\'{u}menes de colisi\'{o}n repulsivo $b^{\ast }(T^{\ast })$ y
atractivo $\Lambda ^{\ast }(T^{\ast })$ son 
\begin{eqnarray}
b^{\ast } &=&1-s_{\text{R}}+s_{\text{R}}b_{\text{1}}^{\ast }(T^{\ast })\text{%
,}  \label{be} \\
\Lambda ^{\ast } &=&1-s_{\text{A}}+s_{\text{A}}\Lambda _{\text{1}}^{\ast
}(T^{\ast })\text{.}  \label{lambda}
\end{eqnarray}
donde $b_{\text{1}}^{\ast }(T^{\ast })$ y $\Lambda _{\text{1}}^{\ast
}(T^{\ast })$\ son los vol\'{u}menes de colisi\'{o}n del sistema de
referencia. Estos vol\'{u}menes se necesitan en la teor\'{\i}a para expresar 
$B^{\ast }(T^{\ast })$ ya sea usando (\ref{Bstar}) o (\ref{Bstar2}) y est%
\'{a}n dados expl\'{\i}citamente en la Ref. \cite{del Rio 1998 III}. En su
aplicaci\'{o}n a las sustancias reales, la aproximaci\'{o}n de 1-$s$ es lo
suficientemente buena como para reproducir $B(T)$ dentro del error
experimental para muchas sustancias, sin embargo, cuando estamos trabajando
con modelos te\'{o}ricos para los que $B(T)$ puede ser calculado con la
precisi\'{o}n deseada, la aproximaci\'{o}n de 2-$s$ puede ser necesaria.

\subsubsection{Par\'{a}metros ANC para un fluido de Lennard-Jones}

Aqu\'{\i} estimamos la suavidad de un fluido de LJ que le corresponde a un
potencial ANC en la aproximaci\'{o}n $1$-s. El potencial cl\'{a}sico LJ se
obtiene haciendo $n=12$ (C$_{n}=4$) en la Ec. (\ref{LJ n/6}). Al ser esf\'{e}%
rico, la no conformalidad del potencial LJ con respecto a la referencia ANC
se debe por completo a la forma del perfil del potencial. Podemos hacer
diferentes aproximaciones para estimar los par\'{a}metros ANC de un fluido
de LJ 12/6, en todos los casos suponemos que son independientes del estado 
\cite{Orlando tesis}. Aqu\'{\i} mostramos resultados de tres estimaciones
que hicimos a estos par\'{a}metros. En el primer caso a partir de datos de $%
B(T)$ de un fluido de LJ 12/6, $\varphi _{\text{LJ}}$, ajustamos \'{u}%
nicamente el par\'{a}metro $s$ usando la Ec. (\ref{Bstar}), suponiendo que $%
r_{\text{m}}$ y $\epsilon $ son los mismos que en el modelo de LJ. En el
segundo caso ajustamos los tres par\'{a}metros y finalmente en el \'{u}ltimo
caso invertimos datos de $B(T)$ de un potencial ANC\ con $s=s_{\text{LJ}}=$%
1.1315.\footnote{%
El valor $s=s_{\text{LJ}}=$1.1315 es m\'{a}s ampliamente usado en la
literatura \cite{del Rio 1998 III}. Sin embargo, este valor puede variar con
un error razonable, por distintas razones. Por un lado, los resultados
obtenidos son sensibles al extremo inferior de temperatura en que
desarrollamos el ajuste. Por otro lado, podemos estimar el valor de $s$
usando un m\'{e}todo geom\'{e}trico; no por inversi\'{o}n de datos de $B(T)$
sino midiendo la pendiente asociada a la relaci\'{o}n lineal dada por la Ec.
(\ref{Bstar}).} En el Cuadro \ref{aprox s de LJ} mostramos los resultados
que obtuvimos. 
\begin{table}[tbp] \centering%
\begin{tabular}{|lllll|}
\hline
Aproximaci\'{o}n & $s$ & $\epsilon /\epsilon _{\text{LJ}}$ & $r_{\text{m}%
}/r_{\text{mLJ}}$ & $q$ \\ \cline{1-4}\cline{2-5}
$\varphi _{\text{LJ}}$ con $r_{\text{m}}$ y $\epsilon $ fijos & 1.1299 & 1 & 
1 & 0.0192 \\ 
$\varphi _{\text{LJ}}$ con $r_{\text{m}}$ y $\epsilon $ libres & 1.1306 & 
0.9895 & 1.0101 & 0.0054 \\ 
$\varphi _{\text{ANC}}(s_{\text{LJ}})$ con $r_{\text{m}}$ y $\epsilon $
libres & 1.1040 & 0.9951 & 1.0137 & 0.0154 \\ \hline
\end{tabular}
\caption{Estimaci\'{\o}n de s para el potencial de Lennard-Jones.} \label{aprox s
de LJ} 
\end{table}%

\section{Fluidos polares no conformales}

Para fluidos no polares y ligeramente polares, usando la teor\'{\i}a ANC en
su aproximaci\'{o}n 1-$s$ con par\'{a}metros constantes $\left( \epsilon
_{0},r_{\text{m}_{0}},s_{0}\right) $, se han construido potenciales
efectivos muy precisos. Esto se ha hecho para mol\'{e}culas lineales tan
enlongadas como C$_{8}$H$_{18}$ y C$_{7}$F$_{16}$. \cite{Ramos 2000} Para mol%
\'{e}culas ligeramente polares y cuasi-esf\'{e}ricas como el NO and CO la
teor\'{\i}a es una buena aproximaci\'{o}n pero falla cuando tratamos con
sustancias con car\'{a}cter fuertemente polar como los alcoholes. Dada la
importancia de las mol\'{e}culas considerablemente polares, en este trabajo
de investigaci\'{o}n est\'{a} concentrado en obtener potenciales efectivos
para mol\'{e}culas polares con kernel esf\'{e}rico y enlongado.

\subsection{Potencial esf\'{e}rico efectivo entre dos momentos dipolares
puntuales}

En el cap\'{\i}tulo anterior se\~{n}alamos que la interacci\'{o}n entre
momentos dipolares permanentes cuando se describe en un intervalo adecuado
de temperaturas, decae como $1/r^{6}$. En esta secci\'{o}n demostraremos expl%
\'{\i}citamente esta importante caracter\'{\i}stica que permite incluir la
interacci\'{o}n entre momentos dipolares en el mismo grupo que el resto de
las interacciones de dispersi\'{o}n; esto a su vez permite dise\~{n}ar un
potencial efectivo de forma simple.

Recordemos que la interacci\'{o}n entre dos momentos dipolares puntuales, $%
\mathbf{\mu }_{1}$ y $\mathbf{\mu }_{2}$, colocados en $\mathbf{r}_{1}$%
\textbf{\ y }$\mathbf{\mathbf{r}}_{2}$ respectivamente, cuyo vector de posici%
\'{o}n relativo es $\mathbf{r=r}_{1}\mathbf{-\mathbf{r}}_{2}=r\mathbf{n}$,
est\'{a} dada por, \cite{ENFE} 
\begin{equation}
\varphi _{\text{dip}}(\mathbf{r,\mu }_{1},\mathbf{\mu }_{2})=-3(\mathbf{\mu }%
_{1}\cdot \mathbf{r})(\mathbf{\mu }_{2}\cdot \mathbf{r})/r^{5}+(\mathbf{\mu }%
_{1}\cdot \mathbf{\mu }_{2})/r^{3}\text{.}  \label{dipolo-dipolo}
\end{equation}%
En un sistema formado por $N$ momentos dipolares puntuales que interaccionan
por pares con una energ\'{\i}a, $\varphi _{\text{dip}}(ij)$, la funci\'{o}n
de partici\'{o}n, $Z$, en el ensamble can\'{o}nico la obtenemos usando la
Ec. (\ref{Z factorizada}), 
\begin{equation}
Z=\frac{1}{N!\lambda ^{3N}}\int \left\{ d\mathbf{r}_{N}\right\} \left\{ d%
\mathbf{\omega }_{N}\right\} \exp \left[ -\beta \frac{1}{2}\sum_{ij}\varphi
_{\text{dip}}(ij)\right] \text{,}
\end{equation}%
donde la integraci\'{o}n incluye las coordenadas angulares asociadas a los
vectores del momento angular, $\left\{ \mathbf{\omega }_{N}\right\} $. La
energ\'{\i}a interna, $E$, se obtiene de la relaci\'{o}n 
\begin{equation}
E=-\frac{1}{Z}\frac{\partial Z}{\partial \beta }\text{.}
\end{equation}%
Tenemos que $E=3NkT/2+\left\langle U\right\rangle _{\text{conf}}$,\ en donde 
$\left\langle U\right\rangle _{\text{conf}}$ es el promedio configuracional
de la energ\'{\i}a potencial del sistema, que est\'{a} dada por 
\begin{equation}
\left\langle U\right\rangle _{\text{conf}}=\frac{\int \left\{ d\mathbf{r}%
_{N}\right\} \left\{ d\mathbf{\omega }_{N}\right\} \Gamma e^{-\beta \Gamma }%
}{\int \left\{ d\mathbf{r}_{N}\right\} \left\{ d\mathbf{\omega }_{N}\right\}
e^{-\beta \Gamma }}  \label{U conf 1}
\end{equation}%
donde 
\begin{equation}
\Gamma =\frac{1}{2}\sum_{ij}\varphi _{\text{dip}}(ij)
\end{equation}%
Como se\~{n}alamos en el Cap\'{\i}tulo 2, podemos asumir que en las
condiciones en las que se encuentra un fluido real, la temperatura es lo
suficientemente alta como para que la energ\'{\i}a t\'{e}rmica, $kT$, sea
mucho mayor que la energ\'{\i}a de interacci\'{o}n entre momentos dipolares%
\footnote{%
En otras palabras, al hacer esta aproximaci\'{o}n, estamos suponiendo que
los momentos dipolares pueden rotar libremente. Esta restricci\'{o}n que
hacemos a la orientaci\'{o}n que los momentos dipolares pueden tener, puede
ser limitante al describir fases fluidos asociados como el agua, donde si
aparecen restricciones debidas a la configuraci\'{o}n angular de su momento
dipolar \cite{ENFE}.}, podemos aproximar la funci\'{o}n exponencial en (\ref%
{U conf 1}) como, $\exp -\beta \Gamma \simeq 1-\beta \Gamma +\cdots $. Adem%
\'{a}s en este caso, 
\begin{equation}
\Gamma =\frac{1}{2}N(N-1)\varphi _{\text{dip}}(\mathbf{r}_{1}\mathbf{,%
\mathbf{r}}_{2}\mathbf{\mathbf{,}\mu }_{1},\mathbf{\mu }_{2})  \label{gama 2}
\end{equation}%
y 
\begin{equation}
\int d\mathbf{\omega }_{1}d\mathbf{\omega }_{2}\varphi _{\text{dip}}(\mathbf{%
r}_{1}\mathbf{,\mathbf{r}}_{2}\mathbf{,\mu }_{1},\mathbf{\mu }_{2})=0\text{.}
\end{equation}%
Entonces la Ec. (\ref{U conf 1}) queda como 
\begin{equation}
\left\langle U\right\rangle _{\text{conf}}=-\beta \frac{\int \left\{ d%
\mathbf{r}_{N}\right\} \left\{ d\mathbf{\omega }_{N}\right\} \Gamma ^{2}}{%
\int \left\{ d\mathbf{r}_{N}\right\} \left\{ d\mathbf{\omega }_{N}\right\} }
\label{U conf 2}
\end{equation}%
Podemos definir una energ\'{\i}a potencial efectiva del sistema mediante el
promedio angular 
\begin{equation}
\left\langle \Gamma ^{2}\right\rangle _{\Omega }=\frac{\int \left\{ d\mathbf{%
\omega }_{N}\right\} \Gamma ^{2}}{\int \left\{ d\mathbf{\omega }_{N}\right\} 
}  \label{gama cuadrada}
\end{equation}%
Sustituyendo (\ref{gama cuadrada}) en (\ref{U conf 2}) tenemos que 
\begin{eqnarray}
\left\langle U\right\rangle _{\text{conf}} &=&-\beta \frac{\int \left\{ d%
\mathbf{r}_{N}\right\} \left\langle \Gamma ^{2}\right\rangle _{\Omega }}{%
\int \left\{ d\mathbf{r}_{N}\right\} }  \notag \\
&=&-\frac{\beta }{V^{N}}\int \left\{ d\mathbf{r}_{N}\right\} \left\langle
\Gamma ^{2}\right\rangle _{\Omega }  \label{U conf 3}
\end{eqnarray}%
donde $V$ es el volumen accesible a los momentos dipolares. Para $N>>1$,
usamos (\ref{gama 2}) en la Ec. (\ref{U conf 3}) obteniendo 
\begin{eqnarray}
\left\langle U\right\rangle _{\text{conf}} &=&-\frac{\beta N^{2}}{4V^{N}}%
\int \left\{ d\mathbf{r}_{N}\right\} \left\langle \varphi _{\text{dip}%
}^{2}\right\rangle _{\Omega } \\
&=&-\frac{1}{4}\beta \rho ^{2}\int d\mathbf{r}_{1}d\mathbf{r}%
_{2}\left\langle \varphi _{\text{dip}}^{2}\right\rangle _{\Omega }\text{.}
\end{eqnarray}%
En donde $\varphi _{\text{dip}}$ viene dado por la Ec. (\ref{dipolo-dipolo}%
). De manera que, las part\'{\i}culas interaccionan con un potencial
efectivo de la forma $1/r^{6}$, independientemente de que la naturaleza de
la interacci\'{o}n dipolar sea por fluctuaci\'{o}n de carga o por interacci%
\'{o}n entre momentos dipolares permanentes o inducidos.

\subsection{Segundo coeficiente virial de un fluido polar}

El segundo coeficiente virial, $B\equiv B_{2}$ en la Ec. (\ref{Ec Virial}),
para un fluido cuyas mol\'{e}culas interact\'{u}an a trav\'{e}s de un
potencial por pares $u(r,\Omega )$; lo podemos calcular como, 
\begin{equation}
B(T)=-2\pi \int d\Omega \int_{0}^{\infty }dr\ f\ r^{2}\text{,}
\label{segundo Coef virial}
\end{equation}
donde $r$ es la distancia entre los centros de masa moleculares, $\Omega $
representa la orientaci\'{o}n relativa entre dos mol\'{e}culas y finalmente
la expresi\'{o}n $f=\exp (-u(r,\Omega )/kT-1)$, es la \textit{funci\'{o}n de
Mayer}. Las interacciones entre dos mol\'{e}culas dipolares que trataremos
en esta tesis son un ejemplo t\'{\i}pico de potenciales que involucran la
dependencia angular $\Omega $.

Brevemente el procedimiento que usamos ampliamente en esta tesis para
integrar $B(T)$ de un potencial $u$ con contribuci\'{o}n polar es como
sigue. Integramos por partes la Ec. (\ref{segundo Coef virial}), para
obtener, 
\begin{equation}
B(T)=\frac{2\pi }{3}\int_{0}^{\infty }r^{3}dr\int d\Omega \ \frac{\partial }{%
\partial r}e^{^{-u(r,\Omega )/kT}}  \label{B para u polar}
\end{equation}%
Factorizamos la funci\'{o}n exponencial en un factor esf\'{e}rico y otro no
esf\'{e}rico, 
\begin{eqnarray}
e^{^{-u(r,\Omega )/kT}} &=&e^{^{-u_{\text{esf}}(r)/kT}}e^{^{-u_{\text{ang}%
}(r,\Omega )/kT}}  \label{factor exp} \\
&=&e^{^{-u_{\text{esf}}(r)/kT}}\sum_{l=0}^{\infty }\frac{1}{l!}\left[
-\left. u_{\text{ang}}(r,\Omega )\right/ kT\right] ^{l}\text{.}  \notag
\end{eqnarray}%
Al sustituir la Ec. (\ref{factor exp}) en la Ec. (\ref{B para u polar})
notamos que podemos escribir la integral angular independiente de la radial.
Podemos definir un potencial auxiliar, $w(r,T)$, como el logaritmo natural
de la integral angular \footnote{%
Si se emplea una funci\'{o}n $12/6$ --como Lennard Jones-- para $u_{\text{esf%
}}(r)$, el t\'{e}rmino resultante $w(r,T)$ es una generalizaci\'{o}n del
potencial de Keesom que considera todas las potencias de orden superior para 
$1/r^{6}$.}, de manera que 
\begin{eqnarray}
B(T) &=&\frac{2\pi }{3}\int_{0}^{\infty }dr\ r^{3}\frac{\partial }{\partial r%
}\left[ e^{^{-u_{\text{esf}}(r)/kT-w(r,T))/kT}}\right]  \notag \\
&=&\frac{2\pi }{3}\int_{0}^{\infty }dr\ r^{3}\frac{\partial }{\partial r}%
e^{^{-u_{\text{ef}}(r,T)/kT}}\text{,}
\end{eqnarray}%
donde $u_{\text{ef}}$ es un potencial efectivo esf\'{e}rico \textit{%
dependiente de la temperatura} que es capaz de reproducir $B(T)$ de un
potencial $u$ con contribuci\'{o}n polar. Esta idea de promediar la parte
angular para aproximar la interacci\'{o}n total con una efectiva esf\'{e}%
rica fue empleada por Singh y Singh \cite{Singh 1971, Singh 1972} para el
caso del potencial de Stockmayer\footnote{%
Existen otros tratamientos en donde se calcula el potencial esf\'{e}rico
efectivo correspondiente a una interacci\'{o}n anisotr\'{o}pica. La
aproximaci\'{o}n RAM (Reference Average Mayer-function) desarrollada por
Smith y Nezbeda \cite{Smith Nez 1978, Smith Nez 1981}, permite calcular
alguna funcional de un potencial no esf\'{e}rico como una perturbaci\'{o}n
alrededor de un sistema con un potencial esf\'{e}rico de referencia. Si esa
funcional es la funci\'{o}n de Mayer, $\exp \left[ ^{-u(r,\Omega )/kT}\right]
-1$, el potencial esf\'{e}rico de referencia resulta ser un promedio angular
dependiente de la temperatura equivalente al que acabamos de describir
arriba como, $w(r,T)$.}. A grandes rasgos, \'{e}ste es el m\'{e}todo que
usaremos a lo largo de este trabajo para construir un potencial efectivo
para las interacciones dipolares.

En la Fig. (\ref{FIGBSMantes}) --del cap\'{\i}tulo anterior-- graficamos $%
B(T)$ para un potencial de SM con distintos valores del momento dipolar.
Notamos que para un mismo valor de temperatura el segundo coeficiente virial
es m\'{a}s negativo a medida que consideramos momentos dipolares m\'{a}s
intensos; es decir, siguendo la interpretaci\'{o}n de la Ec. (\ref{Ec Virial}%
), la presi\'{o}n en los gases polares es menor que en los no polares.

\subsection{Efectos de la forma del potencial de interacci\'{o}n en el
equilibrio L-V}

En esta secci\'{o}n vamos a estudiar como influye la forma del potencial de
interacci\'{o}n con una contribuci\'{o}n dipolar fija en las propiedades de
coexistencia de fases l\'{\i}quido-gas. El potencial de interacci\'{o}n que
consideramos lo obtenemos al sumar un t\'{e}rmino de interacci\'{o}n dipolar
de intensidad $\widetilde{\mu }$ a un kernel de Lennard-Jones generalizado, $%
\varphi _{\text{LJ}n\text{/6}}$, (LJ$n$6) en donde $n$ es un exponente que
indica la intensidad de las fuerzas repulsivas en la interacci\'{o}n.

Realizamos un estudio de simulaci\'{o}n computacional en colaboraci\'{o}n
con S. Calero y S. Lago para estudiar el comportamiento de las cantidades
que caracterizan el equilibrio de fases como funci\'{o}n del exponente $n$
en el potencial $\varphi _{\text{LJ}n\text{/6}}$. En particular hemos
considerado los casos con $n=$18, 12 y 10.5 con un valor fijo de la interacci%
\'{o}n dipolar, $\widetilde{\mu }^{2}=\mu ^{2}/\epsilon _{\text{LJ}}\sigma _{%
\text{LJ}}^{3}=$1.07,\footnote{%
El valor, $\widetilde{\mu }^{2}=$1.07, que estamos usando corresponde a la
intensidad del momento dipolar reducido de la metilamina (CH$_{3}$NH$_{2}$).}
con $\sigma _{\text{LJ}}=r_{\text{mLJ}}/2^{1/6}$. Donde $\epsilon _{\text{LJ}%
}$ y $r_{\text{m}}$ son respectivamente la profundidad y posici\'{o}n del m%
\'{\i}nimo del potencial de LJ definido por la Ec. (\ref{potencial LJ}) y $%
\mu $ es el momento dipolar en \textit{Debyes}. En este estudio de simulaci%
\'{o}n las propiedades termodin\'{a}micas coronadas con el s\'{\i}mbolo
\textquotedblleft $\symbol{126}$\textquotedblright\ est\'{a}n reducidas con $%
\sigma _{\text{LJ}}$ y $\epsilon _{\text{LJ}}$.

Usamos un potencial dipolar de traslape gaussiano modificado \cite{Sofia
2003} en el que escogimos adecuadamente sus par\'{a}metros para reproducir
un potencial LJ$n$6 dipolar. La simulaci\'{o}n la realizamos en el ensamble
de Gibbs usando dos cajas de simulaci\'{o}n con 256 part\'{\i}culas en cada
una\footnote{%
La caja de simulaci\'{o}n es un cubo con condiciones peri\'{o}dicas a la
frontera cuya arista tiene una longitud $L_{x}=L_{y}=L_{z}=4r_{\text{m}%
}/2^{1/6}$. Las simulaciones se realizan truncando y desplazando el
potencial con un valor de cuttof de $3r_{\text{m}}/2^{1/6}$.}.

\begin{figure}[h]
        \begin{center}
        \includegraphics[width=.8\hsize]{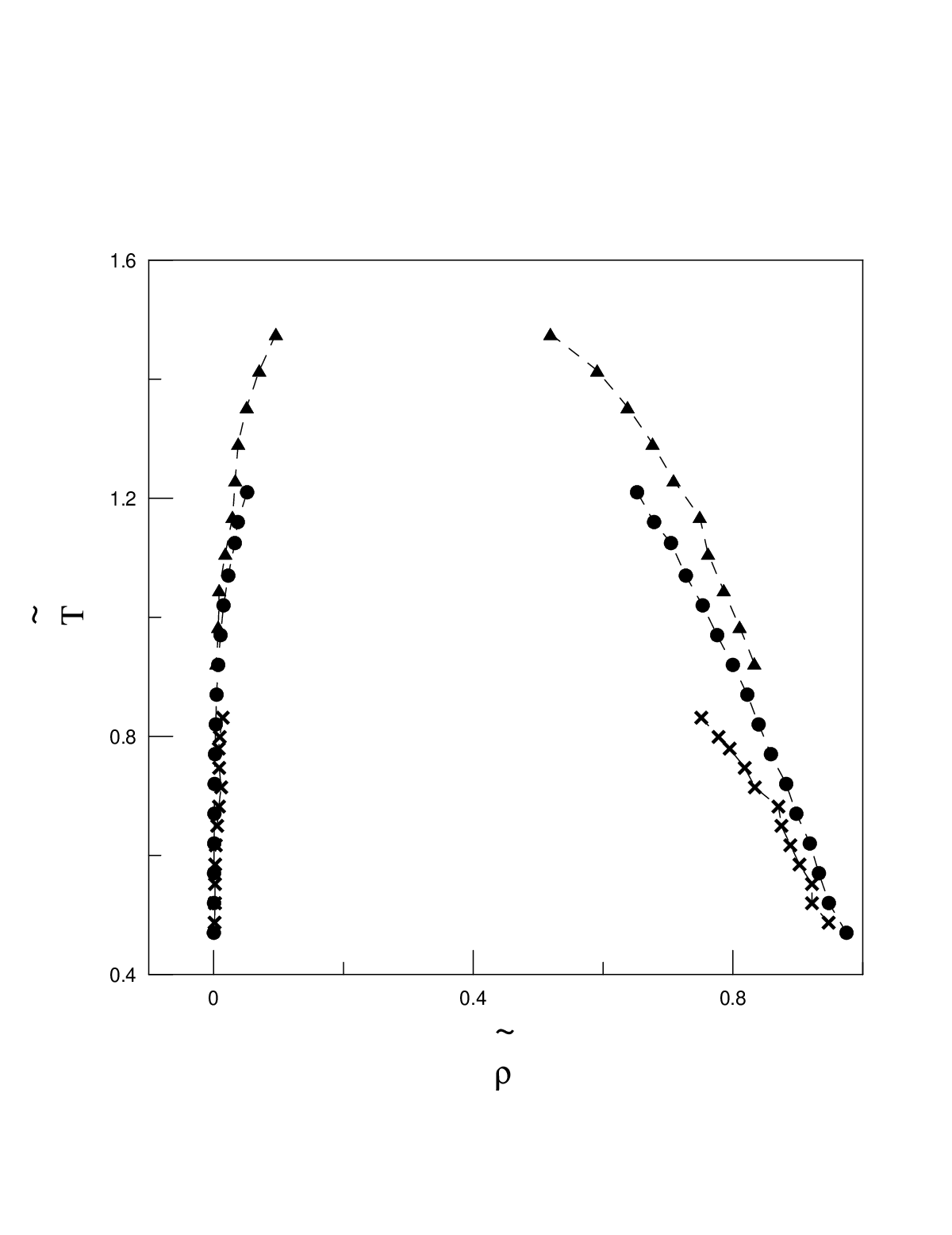}
        \end{center}
        \caption{%
Puntos de coexistencia de un
fluido cuyas mol\'{e}culas interact\'{u}an con un potencial LJ$n$6 m\'{a}s
un dipolo de magnitud $\widetilde{\protect\mu }^{2}=$1.07. Los distintos s%
\'{\i}mbolos corresponden al valor del exponente $n$ como sigue: $n=$10.5
(tri\'{a}ngulos), $n=$12 (c\'{\i}rculos) y $n=$18 (cruces).
                     }%
       \label{FIGcoexisLJn6}
\end{figure}

Para los distintos sistemas calculamos un total de 36 estados para cada
fase. En la Fig. (\ref{FIGcoexisLJn6}) mostramos los resultados de
simulaci'on para los sistemas considerados. La posici\'{o}n del punto
critico ($\widetilde{T}_{\text{c}}$, $\widetilde{\rho }_{\text{c}}$) lo
obtenemos a partir de los resultados de simulaci\'{o}n, asumiendo una expansi%
\'{o}n de Wegner \cite{Wegner 1972}, 
\begin{equation}
\widetilde{\rho }_{\text{L,V}}(t)=\widetilde{\rho }_{\text{c}}\pm \frac{1}{2}%
(B_{\text{0}}t^{\beta }+B_{\text{1}}t^{\beta +\Delta _{\text{1}}}+C_{\text{2}%
}t)\text{ ,}  \label{Wegner}
\end{equation}
donde $t=1-\widetilde{T}/\widetilde{T}_{\text{c}}$, con amplitudes $B_{\text{%
0}}$, $B_{\text{1}}$ y $C_{\text{2}}$ e \'{\i}ndices cr\'{\i}ticos
universales $\beta =$0.325 y $\Delta _{\text{1}}=$0.5. En cada sistema
definido por un valor de $n$, ajustamos las amplitudes $B_{\text{0}}$, $B_{%
\text{1}}$ y $C_{\text{2}}$ en la Ec. (\ref{Wegner}) a los cuatro estados m%
\'{a}s cercanos al punto cr\'{\i}tico. En el Cuadro (\ref{PC LJ n6})
mostramos los puntos cr\'{\i}ticos que obtuvimos. 
\begin{table}[tbp] \centering%
\begin{tabular}{|l|l|l|}
\hline
$n$ & $\widetilde{\rho }_{\text{c}}$ & $\widetilde{T}_{\text{c}}$ \\ \hline
18 & 0.3269 & 0.9822 \\ 
12 & 0.3188 & 1.3975 \\ 
10.5 & 0.2934 & 1.5403 \\ \hline
\end{tabular}
\caption{Punto cr\'{i}tico del potencial LJ$n6$ con dipolo.} \label{PC LJ n6} 
\end{table}
%

\subsection{Efectos de la suavidad en la termodin\'{a}mica}

Diversas son las propiedades termodin\'{a}micas que la teor\'{\i}a ANC\ nos
permite analizar, \textit{e.g.} en su aproximaci\'{o}n de $1$-s, la expresi%
\'{o}n (\ref{Bstar}) la podemos escribir como $B^{\ast }(T^{\ast })-1=s\left[
B_{1}^{\ast }(T^{\ast })-1\right] $, es decir, $B^{\ast }(T^{\ast })$ es una
cantidad directamente proporcional a $s$, de manera que la presi\'{o}n de un
fluido ANC --que podemos obtener a partir de la Ec. (\ref{Ec Virial})--
disminuye a medida que la suavidad aumenta. En el Cap\'{\i}tulo 4, cuando
estudiemos la forma del potencial efectivo GSM, abordaremos el caso general
cuando a un kernel esf\'{e}rico le agregamos un t\'{e}rmino de interacci\'{o}%
n dipolar. Por el momento, notemos en la Fig. \ref{FIGBmu08} que en efecto, $%
B^{\ast }(T^{\ast })$ se hace m\'{a}s negativo a medida que el kernel se
hace m\'{a}s suave, i.e., la presi\'{o}n se reduce a medida que el potencial
se suaviza. En ese caso hay un t\'{e}rmino dipolar constante en el potencial.

\subsubsection{Efectos de la suavidad en el equilibrio l\'{\i}quido-vapor:
Un estudio de simulaci\'{o}n}

En esta secci\'{o}n exhibimos los efectos de la forma del potencial interacci%
\'{o}n en las propiedades termodin\'{a}micas asociadas al equil\'{\i}brio l%
\'{\i}quido-vapor (LV) de un fluido ANC. Estudiamos el equilibrio LV de un
sistema de una sola especie suponiendo que interact\'{u}a con un potencial
ANC de suavidad variable (Ec.(\ref{potencialANC})). Realizamos en colaboraci%
\'{o}n con F. del R\'{\i}o, E. D\'{\i}az\ y J. Alejandre \cite{DiazMC 2004}
simulaciones de din\'{a}mica molecular (DM) isot\'{e}rmicas para calcular,
entre otras cosas, los valores de la densidad en coexistencia de fases a una
determinada temperatura para distintos valores de suavidad. Encontramos
curvas de coexistencia bien comportadas y adem\'{a}s puntos cr\'{\i}ticos,
tensiones superficiales \footnote{%
En esta tesis no incluimos el an\'{a}lisis de la tensi\'{o}n superficial. Si
el lector lo desea puede consultar la Ref. \cite{DiazMC 2004}, en donde est%
\'{a}n condensados los resultados que incluyen esta propiedad.} y presiones
de vapor, para distintos valores de la suavidad.

Esperamos que un fluido ANC --cuyo potencial es capaz de reproducir las
propiedades de una sustancia real-- sea un modelo razonable para describir
la regi\'{o}n densa. Una de las razones de nuestra confianza es una sencilla
correlaci\'{o}n entre la temperatura cr\'{\i}tica, $T_{\text{c}}$, de un
fluido real, el par\'{a}metro de energ\'{\i}a, $\epsilon $, y la suavidad
que resulta ser adecuada en m\'{a}s de 40 sustancias \cite{Del Rio 2001} 
\footnote{%
Entre las pocas sustancias que \ muestran claras desviaciones a esta relaci%
\'{o}n son H$_{2}$, D$_{2}$ y He. Estas sustancias a\'{u}n en estado fluido
requieren un tratamiento cu\'{a}ntico.}. \'{E}ste es un resultado notable
debido a que es un indicio de que la teor\'{\i}a ANC puede extender su
cobertura hacia estados densos.

Se sabe que una gran cantidad de sustancias estudiadas en su fase vapor, est%
\'{a}n construidas con mol\'{e}culas cuyas suavidades caen en el intervalo,
0.5$\leq s\leq $1.0 \cite{del Rio 1998, del Rio 1998 II, del Rio 1998 III,
Ramos 2000, Del Rio 2001}. Este intervalo cubre desde los gases nobles con $%
s\cong $1.0, hasta cadenas moleculares como el hexano con $s\cong $0.5.

\paragraph{Detalles de la simulaci\'{o}n}

Los resultados que aqu\'{\i} se muestran son para los sistemas con $s=$1.0,
0,7 y 0.5, los cuales corresponden aproximadamente al arg\'{o}n ($s=$%
0.9993), propano ($s=$0.7008) y hexano ($s=$0.5119), respectivamente.

Para estos sistemas realizamos simulaciones de din\'{a}mica molecular NVT
con $N=2000$ mol\'{e}culas usando condiciones peri\'{o}dicas a la frontera
en las tres direcciones de un paralelep\'{\i}pedo. El \'{a}rea de la
superficie es $A=L_{x}L_{y}$ con $L_{x}=L_{y}=9r_{\text{m}}$ y $L_{z}=50r_{%
\text{m}}$. Todas las unidades est\'{a}n reducidas con la distancia $r_{%
\text{m}}$ y la energ\'{\i}a $\epsilon $. Las simulaciones se hicieron
truncando la energ\'{\i}a de interacci\'{o}n en 3.5$r_{\text{m}}$. La
distancia a la cual se corta el potencial de interacci\'{o}n se conoce como
radio de corte. El efecto de este corte es mayor para el potencial ANC m\'{a}%
s suave, $s=1$, debido a que es el que tiene el pozo m\'{a}s amplio. Este
potencial es muy similar a una funci\'{o}n LJ $12/6$. Truncando esta funci%
\'{o}n LJ$12/6$ con un radio de corte en $r=$3.5$r_{\text{m}}\cong 4\sigma _{%
\text{LJ}}$ se pueden calcular propiedades termodin\'{a}micas bastante
cercanas a las que se obtienen usando el potencial completo \cite{andrij}.
De manera que, los resultados ser\'{a}n mejores para los potenciales m\'{a}s
duros. El paso reducido de tiempo que se uso para integrar las ecuaciones de
movimiento es de 0.005, el equilibrio lo alcanzamos despu\'{e}s de 50000
pasos\ y el promedio de las propiedades lo calculamos a lo largo de los
siguientes 400000 pasos adicionales.

\paragraph{Resultados de la simulaci\'{o}n}

El Cuadro \ref{tabla ANC MC} muestra las propiedades reducidas: densidad de
coexistencia del vapor y el l\'{\i}quido, $\rho _{V}^{\ast }$ y $\rho
_{L}^{\ast }$, y presi\'{o}n ortob\'{a}rica, $P_{\sigma }^{\ast }$, a
diferentes temperaturas para los sistemas con $s=$1.0, 0.7 y 0.5, que fueron
calculadas directamente a partir de de las simulaciones. En la Fig. \ref%
{FIGperfilmcancs07} mostramos\ un grupo t\'{\i}pico de perfiles de densidad
--que en este caso corresponde a $s=$0.7-- calculado a distintas
temperaturas. Como se muestra en la figura, los perfiles tienen
\textquotedblleft hombros\textquotedblright\ bastante afilados, lo cual
indica que la frontera entre las fases vapor y l\'{\i}quido est\'{a} bien
definida. Adicionalmente, las fluctuaciones en la densidades --en cualquiera
de las dos fases-- son lo bastante peque\~{n}as; raz\'{o}n por la que el c%
\'{a}lculo de las densidades se hace con plena confianza. 
\begin{table}[tbp] \centering%
\begin{tabular}{lllll}
\hline
\multicolumn{1}{|l}{$T^{\ast }$} & $\rho _{V}^{\ast }$ & $\rho _{L}^{\ast }$
& $\frac{\text{Pasos}}{2000}$ & \multicolumn{1}{l|}{$P_{\sigma }^{\ast }$}
\\ \hline
&  & s=0.5 &  &  \\ \hline
\multicolumn{1}{|l}{0.45} & 0.0120 & 1.0754 & 14 & \multicolumn{1}{l|}{0.0042
} \\ 
\multicolumn{1}{|l}{0.50} & 0.0355 & 0.9956 & 12 & \multicolumn{1}{l|}{0.0151
} \\ 
\multicolumn{1}{|l}{0.55} & 0.0859 & 0.8889 & 28 & \multicolumn{1}{l|}{0.0346
} \\ 
\multicolumn{1}{|l}{0.57} & 0.1139 & 0.8341 & 17 & \multicolumn{1}{l|}{0.0458
} \\ \hline
&  & s=0.7 &  &  \\ \hline
\multicolumn{1}{|l}{0.50} & 0.0037 & 1.1114 & 15 & \multicolumn{1}{l|}{0.0016
} \\ 
\multicolumn{1}{|l}{0.55} & 0.0108 & 1.0612 & 15 & \multicolumn{1}{l|}{0.0057
} \\ 
\multicolumn{1}{|l}{0.60} & 0.0257 & 1.0035 & 16 & \multicolumn{1}{l|}{0.0122
} \\ 
\multicolumn{1}{|l}{0.65} & 0.0483 & 0.9362 & 15 & \multicolumn{1}{l|}{0.0261
} \\ \hline
&  & s=1.0 &  &  \\ \hline
\multicolumn{1}{|l}{0.7} & 0.0081 & 1.0939 & 20 & \multicolumn{1}{l|}{0.0061}
\\ 
\multicolumn{1}{|l}{0.8} & 0.0285 & 1.0141 & 20 & \multicolumn{1}{l|}{0.0188}
\\ 
\multicolumn{1}{|l}{0.9} & 0.0625 & 0.9227 & 20 & \multicolumn{1}{l|}{0.0461}
\\ 
\multicolumn{1}{|l}{1.0} & 0.1375 & 0.7924 & 20 & \multicolumn{1}{l|}{0.0862}
\\ \hline
\end{tabular}
\caption{Resultados de simulaci\'{o}n de DM para el potencial ANC} \label{tabla
ANC MC} 
\end{table}
%
%
 \begin{figure}[h]
        \begin{center}
        \includegraphics[width=.8\hsize]{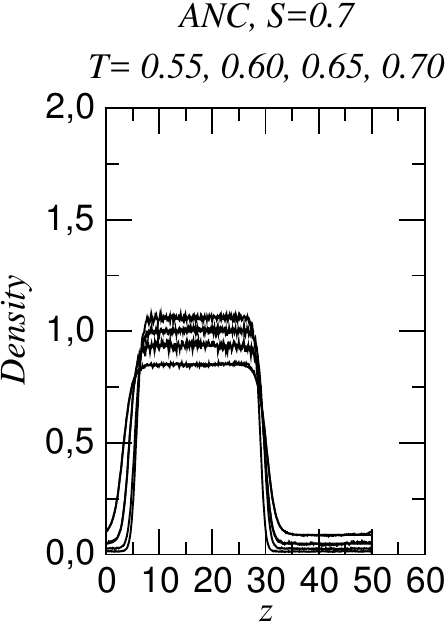}
        \end{center}
        \caption{%
Densidad de coexistencia para
el modelo ANC con $s=$0.7. Se muestran los perfiles para las temperaturas $%
T^{\ast }=$0.55, 0.60, 0.65 y 0.70. El eje de las abscisas representa la
coordenada $z$ de la caja de simulaci\'{o}n. N\'{o}tese que la frontera
entre las fases vapor y l\'{\i}quido esta bien definida. Las densidad de
cada estado se calcula promediando a lo largo del perfil de cada fase.
Debido a que las fluctuaciones en la densidad son m\'{\i}nimas, el valor de
la densidad promedio es bastante confiable.
                     }%
       \label{FIGperfilmcancs07}
\end{figure}

\begin{figure}[h]
        \begin{center}
        \includegraphics[width=.8\hsize]{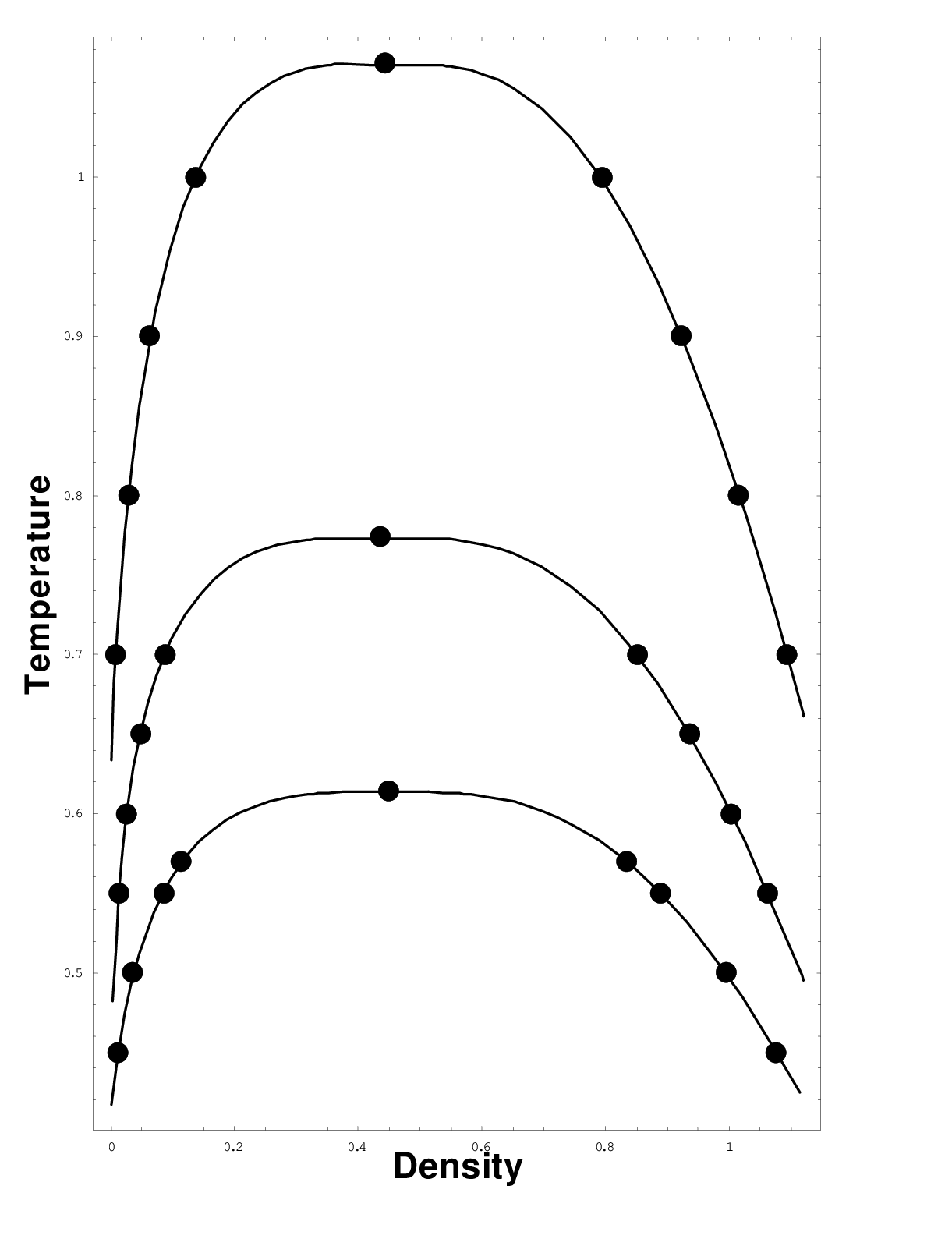}
        \end{center}
        \caption{%
Curvas ortob\'{a}ricas con el
punto cr\'{\i}tico estimado para el modelo ANC. Se muestran de arriba a
abajo los casos con $s=$1.0, 0.7 y 0.5. N\'{o}tese que la curva que
corresponde al potencial m\'{a}s duro ($s=$0.5) es la que tiene la
temperatura cr\'{\i}tica menor.
                     }%
       \label{FIGcoexisancmca}
\end{figure}

Las curvas ortob\'{a}ricas con el punto cr\'{\i}tico estimado los mostramos
en la Fig. \ref{FIGcoexisancmca}. Como el lector puede notar en esta figura,
la temperatura cr\'{\i}tica crece proporcionalmente con el valor de la
suavidad. Los valores cr\'{\i}ticos que encontramos est\'{a}n condensados en
el Cuadro \ref{tabla PC ANC MC}. Podemos ajustar una funci\'{o}n sencilla de 
$s$ a los valores cr\'{\i}ticos; por ejemplo la temperatura cr\'{\i}tica
sigue una tendencia bastante regular que queda bien representada por la
siguiente expresi\'{o}n, \cite{DiazMC 2004} 
\begin{equation}
T_{\text{c}}^{\ast }(s)=\text{0.314432}+\text{0.438392}\ s+\text{0.320100}\
s^{2}\text{.}  \label{TcAnc}
\end{equation}
%
%
%
\begin{table}[tbp] \centering%
\begin{tabular}{|llll|}
\hline
$s$ & $\rho _{\text{c}}^{\ast }$ & $T_{\text{c}}^{\ast }$ & $P_{\text{c}%
}^{\ast }$ \\ \hline
0.5 & 0.4495 & 0.6141 & 0.0912 \\ 
0.7 & 0.4372 & 0.7767 & 0.1226 \\ 
1.0 & 0.4438 & 1.0719 & 0.1349 \\ \hline
\end{tabular}
\caption{Punto cr\'{i}tico del potencial ANC para distintas suavidades.} \label%
{tabla PC ANC MC} 
\end{table}%

\chapter{\textbf{Fluidos polares con kernel esf\'{e}rico}}

\section{Introducci\'{o}n}

Este cap\'{\i}tulo est\'{a} dedicado a los modelos moleculares dipolares con
kernel esf\'{e}rico. Los modelos moleculares dipolares con kernel esf\'{e}%
rico se caracterizan por tener un kernel independiente de la orientaci\'{o}n
angular. El car\'{a}cter dipolar se obtiene sumando un t\'{e}rmino de
interacci\'{o}n momento dipolar-momento dipolar al n\'{u}cleo esf\'{e}rico.
Definimos el potencial de Stockmayer generalizado (GSM) que --como su nombre
lo indica-- es una generalizaci\'{o}n al bien conocido potencial dipolar de
Stokmayer (SM). El potencial GSM es una generalizaci\'{o}n en el sentido de
que su kernel es m\'{a}s vers\'{a}til que el del SM ya que permite realizar
cambios de forma en el perfil del potencial. Usaremos el potencial GSM para
construir los par\'{a}metros de una funci\'{o}n ANC efectiva cuya termodin%
\'{a}mica reproduzca la del potencial original GSM. El objetivo de este cap%
\'{\i}tulo es mostrar que para la familia GSM, la teor\'{\i}a ANC nos
permite construir potenciales efectivos ANC con par\'{a}metros $\left(
\epsilon ,r_{\text{m}},s\right) $ que son funciones de la temperatura $T$.
Mostramos que los potenciales efectivos ANC que desarrollaremos en este cap%
\'{\i}tulo reproducen muy bien diversas propiedades termodin\'{a}micas.
Estudiaremos el caso especial de GSM con suavidad unitaria y calcularemos
anal\'{\i}ticamente $B(T)$. Analizaremos los efectos de la temperatura en el
potencial GSM. Calcularemos dos aproximaciones --una anal\'{\i}tica y una num%
\'{e}rica-- de los par\'{a}metros de una funci\'{o}n ANC que reproduzca
datos de integraci\'{o}n num\'{e}rica de $B(T)$ de un fluido de SM.
Desarrollaremos el c\'{a}lculo formal de los par\'{a}metros efectivos ANC
correspondientes al potencial GSM. Los par\'{a}metros que describen la
suavidad los encontraremos usando dos distintas representaciones --con una y
con dos suavidades. En cada representaci\'{o}n mostraremos que podemos
reproducir con excelente aproximaci\'{o}n los datos de integraci\'{o}n num%
\'{e}rica de $B(T)$. Analizaremos la termodin\'{a}mica de un fluido GSM
concentr\'{a}ndonos en la temperatura de Boyle, las propiedades cr\'{\i}%
ticas y de equilibrio, y describiremos como la densidad afecta a los par\'{a}%
metros ANC y en consecuencia a la ecuaci\'{o}n de estado de un fluido
dipolar.

\section{El potencial GSM}

Definimos el potencial GSM al sumar un kernel esf\'{e}rico ANC de suavidad
variable, $\varphi _{\text{ANC}}(z;\epsilon _{0},s_{0})$,\ a un t\'{e}rmino
de interacci\'{o}n dipolar. La interacci\'{o}n dipolar contiene una
dependencia orientacional que al integrar sobre los \'{a}ngulos, la
transformamos en una dependencia t\'{e}rmica obteniendo un potencial
efectivo ANC con par\'{a}metros $\left( \epsilon ,r_{\text{m}},s\right) $
que ser\'{a}n funciones de la temperatura $T$.

En esta secci\'{o}n vamos a considerar el modelo molecular de Stockmayer
generalizado que interact\'{u}a con otro igual v\'{\i}a un potencial GSM.
Este potencial lo definimos como la suma de un potencial ANC cuyo m\'{\i}%
nimo tiene una profundidad $\epsilon _{0}$ localizado en $r_{\text{m0}}$, y
con suavidad $s_{0}$, m\'{a}s la interacci\'{o}n entre dos momentos
dipolares permanentes de momento $\mathbf{\mu }$, denotada por $\varphi _{%
\text{DD}}$ \cite{Edgar 2003}. De esta forma, el potencial GSM\ lo denotamos
como $\varphi _{\text{GSM}}$. En unidades reducidas lo escribimos como 
\begin{equation}
\varphi _{\text{GSM}}(z,\Omega ;\mu _{\text{0}}^{\ast },\epsilon
_{0},s_{0})=\varphi _{\text{ANC}}(z;\epsilon _{0},s_{0})+\varphi _{\text{DD}%
}(z,\Omega ;\mu _{0}^{\ast })\text{,}  \label{GSM}
\end{equation}
donde $\varphi _{\text{ANC}}(z;\epsilon _{0},s_{0})$ est\'{a} dado por (\ref%
{potencialANC}) con $z=r/r_{\text{m0}}$ y 
\begin{equation*}
\varphi _{\text{DD}}(z,\Omega ;\mu _{0}^{\ast })/\epsilon _{0}=-\mu
_{0}^{\ast 2}g(\Omega )/z^{3}\text{,}
\end{equation*}
es el t\'{e}rmino que porta la contribuci\'{o}n dipolar, donde $g(\Omega )$
es una abreviaci\'{o}n de 
\begin{equation}
g(\Omega )=g(\theta _{1},\theta _{2},\phi _{2}-\phi _{1})=2\cos \theta
_{1}\cos \theta _{2}-\sin \theta _{1}\sin \theta _{2}\cos (\phi _{2}-\phi
_{1})\text{,}  \label{g}
\end{equation}
y $\mu _{0}^{\ast 2}=\mu ^{2}/\epsilon _{0}r_{\text{m0}}^{3}$, mientras que $%
\theta _{i}$ y $\phi _{i}$ son, respectivamente, los \'{a}ngulos zenital y
azimutal de la mol\'{e}cula $i$. El potencial GSM es en efecto una
generalizaci\'{o}n del de Stockmayer debido a que se reduce a este \'{u}%
ltimo al hacer $s_{0}=s_{\text{LJ}}$.

Para eliminar la dependencia orientacional, integraremos angularmente el
potencial GSM siguiendo el procedimiento descrito por Hirschfelder \textit{%
et al}. \cite{Hirsch} para calcular el segundo coeficiente $B(T)$ para $%
\varphi _{\text{GSM}}$. Definimos $B^{\ast }=B/(2\pi r_{\text{m0}}^{3}/3)$ y 
$T_{0}^{\ast }=kT/\epsilon _{0}$ para escribir el segundo coeficiente
virial, en unidades reducidas como 
\begin{equation}
B^{\ast }(T_{0}^{\ast })=\frac{1}{8\pi }\int_{\Omega \ }\,d\Omega
\int_{z=a_{c}}^{\infty }dz\frac{\partial }{\partial z}\left\{ \exp \left[ -%
\frac{\varphi _{\text{GSM}}(z,\Omega ;\mu _{0}^{\ast },s_{0})}{\epsilon
_{0}T_{0}^{\ast }}\right] \right\} z^{3}
\end{equation}
donde $d\Omega \equiv \sin \theta _{1}d\theta _{1}\sin \theta _{2}d\theta
_{2}d\phi $. Factorizamos 
\begin{equation*}
\exp \left[ -\varphi _{\text{ANCDD}}/\epsilon _{0}T_{0}^{\ast }\right] =\exp
(-\varphi _{\text{ANC}}/\epsilon _{0}T_{0}^{\ast })\times \exp (\varphi _{%
\text{DD}}/\epsilon _{0}T_{0}^{\ast })\text{,}
\end{equation*}
desarrollamos la funci\'{o}n exponencial $\exp (\varphi _{\text{DD}%
}/\epsilon _{0}T_{0}^{\ast })$ como una suma infinita e integramos sobre $%
\Omega $ para obtener 
\begin{equation}
\frac{1}{8\pi }\int d\Omega \exp (-\varphi _{\text{DD}}(z,\Omega ;\mu
_{0}^{\ast })/\epsilon _{0}T_{0}^{\ast })=\sum_{m=0}^{\infty }\left\{ \frac{1%
}{\left( 2m\right) !}\left[ \mu _{0}^{\ast 2}/T_{0}^{\ast }\ z^{3}\right]
^{2m}G_{m}\right\} \text{,}  \label{Udd integrado en los angulos}
\end{equation}
donde 
\begin{equation*}
G_{m}=(1/8\pi )\int d\Omega \ g^{2m}(\Omega )\text{,}
\end{equation*}
puede calcularse anal\'{\i}ticamente para cualquier $m$ finita usando $%
g(\Omega )$ definida en (\ref{g}). Definimos el potencial esf\'{e}rico
auxiliar $\varphi _{\text{DD}}^{\text{sphe}}$ como 
\begin{equation*}
\exp [\varphi _{\text{DD}}^{\text{sphe}}(z,T_{0}^{\ast },\mu _{0}^{\ast
})/\epsilon _{0}T_{0}^{\ast }]=\frac{1}{8\pi }\int d\Omega \exp (\varphi _{%
\text{DD}}(z,\Omega ;\mu _{0}^{\ast })/\epsilon _{0}T_{0}^{\ast })
\end{equation*}
el cual, usando (\ref{Udd integrado en los angulos}) lo escribimos como 
\begin{equation}
\varphi _{\text{DD}}^{\text{sphe}}(z,T_{0}^{\ast },\epsilon _{0},\mu
_{0}^{\ast })=-\epsilon _{0}T_{0}^{\ast }\ln \sum_{m=0}^{\infty }\left\{ 
\frac{1}{\left( 2m\right) !}\left[ \mu _{0}^{\ast 4}/T_{0}^{\ast 2}z^{6}%
\right] ^{m}G_{m}\right\} \text{.}  \label{potDDsphe}
\end{equation}
Una vez integrada la contribuci\'{o}n angular en la interacci\'{o}n momento
dipolar-momento dipolar, mostramos que el potencial original $\varphi _{%
\text{GSM}}$ puede reemplazarse por un potencial sin dependencia angular que
depende de los par\'{a}metros $T_{0}^{\ast }$ y $\mu _{0}^{\ast }$: 
\begin{eqnarray}
\varphi _{\text{GSM}}(z,\Omega ;\mu _{\text{0}}^{\ast },\epsilon _{0},s_{0})
&\rightarrow &\varphi _{\text{GSM}}^{\text{sphe}}(z,T_{0}^{\ast };\mu _{%
\text{0}}^{\ast },\epsilon _{0},s_{0})  \notag \\
&=&\varphi _{\text{ANC}}(z;\epsilon _{0},s_{0})+\varphi _{\text{DD}}^{\text{%
sphe}}(z,T_{0}^{\ast },\mu _{0}^{\ast })\text{.}  \label{ANCdipSphe}
\end{eqnarray}
En t\'{e}rminos de este potencial sin dependencia angular podemos escribir,

\begin{equation}
B^{\ast }(T_{0}^{\ast })=\int_{z=a_{c}}^{\infty }dz\frac{\partial }{\partial
z}\left\{ \exp \left[ -\frac{\varphi _{\text{GSM}}^{\text{sphe}%
}(z;T_{0}^{\ast },\mu _{0}^{\ast },s_{0})}{\epsilon _{0}T_{0}^{\ast }}\right]
\right\} z^{3}\text{.}  \label{B usado el esfericalizado}
\end{equation}

Este potencial sin dependencia angular (\ref{ANCdipSphe}) que hemos
construido consiste de una parte ANC, $\varphi _{\text{ANC}}$, m\'{a}s un t%
\'{e}rmino de interacci\'{o}n entre momentos dipolares permanentes $\varphi
_{\text{DD}}^{\text{sphe}}$ que opera como una serie de contribuciones
alternativamente atractivas y repulsivas de magnitud decreciente \footnote{%
A una temperatura moderada ($T_{0}^{\ast }\approx 1.0$) y para un valor
aproximado de $z\lesssim 1.0$, el segundo t\'{e}rmino en la expresi\'{o}n (%
\ref{potDDsphe}) --el de orden $1/z^{12}$-- es aproximadamente el 20\% del
primero --el de orden $1/z^{6}$ \cite{Singh 1972}.}. Nosotros consideramos
todos las contribuciones para asegurar un c\'{a}lculo m\'{a}s riguroso de
las propiedades macrosc\'{o}picas de los fluidos dipolares. M\'{a}s
adelante, en esta secci\'{o}n, remplazaremos $\varphi _{\text{GSM}}^{\text{%
sphe}}$ por un ANC cuyos par\'{a}metros efectivos ser\'{a}n $\epsilon $, $r_{%
\text{m}}$, y $s$ ahora ser\'{a}n, funciones de $\epsilon _{0}$, $r_{\text{m0%
}}$, and $s_{0}$.

La \'{u}ltima expresi\'{o}n (\ref{B usado el esfericalizado}) muestra que $%
\varphi _{\text{GSM}}^{\text{sphe}}$ es en efecto un potencial efectivo, en
el sentido de que puede ser usado en lugar del potencial original $\varphi _{%
\text{GSM}}$ para calcular la misma propiedad termodin\'{a}mica, $B(T)$. Sin
embargo, el nuevo potencial tiene la ventaja de que es independiente de la
orientaci\'{o}n.

Este nuevo potencial (\ref{ANCdipSphe}) que hemos construido, depende de la
temperatura y del momento dipolar. La presencia del momento dipolar hace que
cambie el tama\~{n}o y la posici\'{o}n del m\'{\i}nimo de $\varphi _{\text{%
GSM}}^{\text{sphe}}$ conforme baja la temperatura. En la Figura \ref%
{FIGspheGSM} est\'{a} representado el potencial $\varphi _{\text{GSM}}^{%
\text{sphe}}$. El lector puede notar que, para un momento dipolar dado y
usando una suavidad fija en el kernel, a medida que la temperatura disminuye,%
$\ \varphi _{\text{GSM}}^{\text{sphe}}$ alcanza un m\'{\i}nimo m\'{a}s
profundo que $\varphi _{\text{ANC}}$ y su posici\'{o}n se desplaza . Por su
parte, la suavidad efectiva crece conforme la temperatura disminuye, para un
momento dipolar dado.
%
 \begin{figure}[h]
        \begin{center}
        \includegraphics[width=.8\hsize]{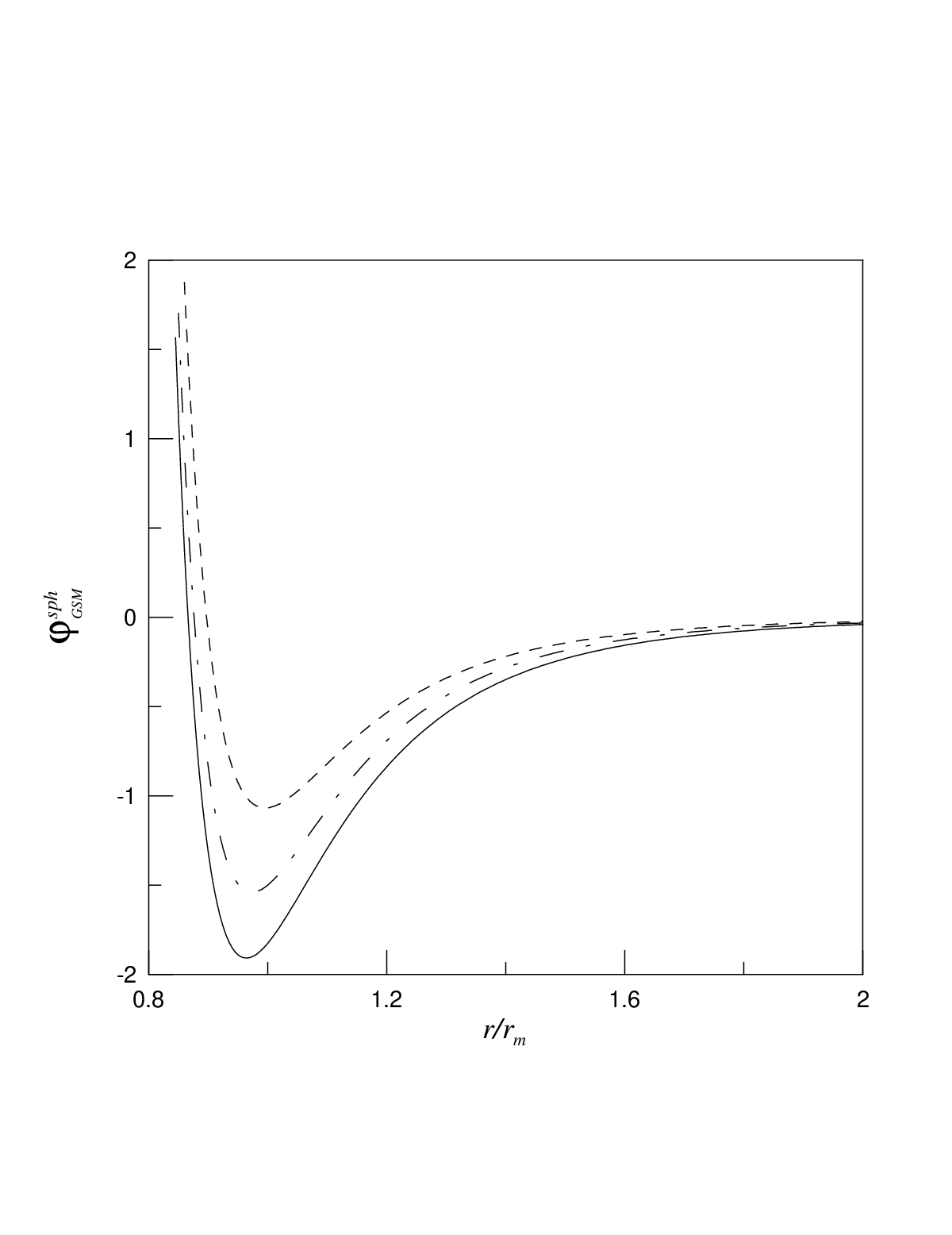}
        \end{center}
        \caption{%
Dependencia con la temperatura del potencial Stockmayer esfericalizado, $%
\protect\varphi _{\text{GSM}}^{\text{sphe}}(z)$. Las l\'{\i}neas
corresponden a un GSM con momento dipolar $\protect\mu _{0}^{\ast }=1.0$ y
un kernel esf\'{e}rico con $s_{0}=1.0$; con las siguientes temperaturas: $%
T_{0}^{\ast }=0.3$ (l\'{\i}nea continua ), $T_{0}^{\ast }=0.6$ (l\'{\i}nea
de segmentos y puntos) y $T_{0}^{\ast }=5.0$ (l\'{\i}nea de segmentos).
                     }%
       \label{FIGspheGSM}
\end{figure}

\subsection{GSM con suavidad unitaria}

Antes de continuar, mostramos el c\'{a}lculo de $B(T)$ para el caso especial
de las mol\'{e}culas GSM con $s_{0}=1$. Este caso es notable puesto que $%
\varphi _{\text{GSM}_{1}}=\varphi _{\text{GSM}}(z,T_{0}^{\ast };\mu _{\text{0%
}}^{\ast },\epsilon _{0},s_{0}=1)$ es un potencial dipolar cuyo kernel es la
funci\'{o}n esf\'{e}rica de Kihara $\varphi _{\text{K}}(z)$ con di\'{a}metro
de n\'{u}cleo duro $a$, 
\begin{equation}
\varphi _{\text{K}}(z)=\ \epsilon _{\text{0}}\left[ \left( \frac{1-a}{z-a}%
\right) ^{12}-2\left( \frac{1-a}{z-a}\right) ^{6}\right] \text{ ,}
\label{Kihara}
\end{equation}
de manera que 
\begin{equation}
\varphi _{\text{GSM}_{1}}=\varphi _{\text{K}}(z)+\varphi _{\text{DD}}\text{ .%
}
\end{equation}
\bigskip

No tenemos conocimiento de c\'{a}lculos reportados para $B(T)$ de este
potencial, de tal manera que hicimos el c\'{a}lculo y aqu\'{\i} mostraremos
el resultado de la integraci\'{o}n anal\'{\i}tica. Integrando para calcular
el segundo coeficiente virial obtenemos, \footnote{%
En c\'{a}lculo explicito se encuentra en un ap\'{e}ndice al final de esta
tesis.} 
\begin{eqnarray}
B^{\ast }\left( T_{0}^{\ast },\mu _{0}^{\ast }\right) &=&-\overset{\infty }{%
\underset{n=0}{\sum }}\underset{k=0}{\overset{k\leq \frac{n}{2}}{\sum }}%
\overset{\infty }{\underset{l=0}{\sum }}\frac{2^{n-2k}}{n!}\frac{a^{l}}{%
(1-a)^{6k+l}}\left( \frac{1}{T_{0}^{\ast }}\right) ^{\frac{6n+6k-l}{12}} 
\notag \\
&&\times \binom{n}{2k}\QATOPD\{ \} {6k}{l}\mu _{0}^{\ast ^{4k}}G_{k}\left\{
(1-a)^{3}\frac{1}{4}\left( \frac{1}{T_{0}^{\ast }}\right) ^{\frac{1}{4}%
}\right.  \notag \\
&&\times \Gamma \QOVERD( ) {6n-6k+l-3}{12}+(1-a)^{2}a\frac{1}{2}\left( \frac{%
1}{T_{0}^{\ast }}\right) ^{\frac{1}{6}}  \notag \\
&&\times \Gamma \QOVERD( ) {6n-6k+l-2}{12}+(1-a)a^{2}\frac{1}{4}\left( \frac{%
1}{T_{0}^{\ast }}\right) ^{\frac{1}{12}}  \notag \\
&&\left. \times \Gamma \QOVERD( ) {6n-6k+l-1}{12}\right\} \text{,}
\label{BMSM1_data}
\end{eqnarray}
siendo$\binom{n}{2k}$ el coeficiente del desarrollo binomial, $\Gamma $, la
funci\'{o}n gama, $G_{k}$ es la misma que la que usamos arriba y 
\begin{equation}
\QATOPD\{ \} {6k}{l}=(-1)^{l}\frac{(6k+l-1)!}{l!(6k-1)!}\text{.}
\end{equation}
Truncando las series que aparecen en (\ref{BMSM1_data}) podemos calcular $%
B(T)$ con la precisi\'{o}n que se desee para un modelo GSM con un kernel de
tipo Kihara (\ref{Kihara}). A\'{u}n m\'{a}s, notemos que al hacer $a=0$, el
kernel de Kihara se vuelve uno de LJ y por tanto recuperamos el bien
conocido resultado de Stokmayer.

Como una primera ilustraci\'{o}n del uso de los potenciales efectivos ANC,
podemos construir un potencial ANC que sea efectivo al potencial $\varphi _{%
\text{GSM}_{1}}$ que reproduzca correctamente los datos de $B(T)$ (\ref%
{BMSM1_data}). B\'{a}sicamente la idea consiste en calcular los par\'{a}%
metros que hacen que un potencial ANC sea conformal al $\varphi _{\text{GSM}%
_{1}}$. Calcularemos los par\'{a}metros efectivos que correspondan al
potencial $\varphi _{\text{GSM}_{1}}^{\text{sphe}}$, siendo este,
simplemente un kernel tipo Kihara esf\'{e}rico sumado a un t\'{e}rmino
dipolar sin dependencia angular $\varphi _{\text{DD}}^{\text{sphe}}$. El
efecto total de la contribuci\'{o}n $\varphi _{\text{K}}(z)$ y de la momento
dipolar-momento dipolar, $\varphi _{\text{DD}}(z,\Omega )$, queda bien
descrito por $\varphi _{\text{GSM}_{1}}^{\text{sphe}}$.\ Una vez logrado un
potencial esf\'{e}rico, lo que sigue es representarlo con un ANC efectivo.
Para hacer esto debemos encontrar los par\'{a}metros efectivos ANC $r_{\text{%
m}}$, $\epsilon $ y $s$. En primer lugar, notamos que debido a la influencia
de la interacci\'{o}n dipolar el m\'{\i}nimo del potencial $\varphi _{\text{%
GSM}_{1}}$ es m\'{a}s profundo que el de\ $\varphi _{\text{K}}(z)$,
alcanzando un valor de $\epsilon $ y su posici\'{o}n es ahora un poco m\'{a}%
s cercana al origen de coordenadas, localiz\'{a}ndose en $r_{\text{m}}$.

Para calcular num\'{e}ricamente este desplazamiento, fijamos valores para $%
\mu _{0}^{\ast }$ y para $T^{\ast }$, reescalamos $\varphi _{\text{GSM}%
_{1}}^{\text{sphe}}$ con los valores de la posici\'{o}n del m\'{\i}nimo $r_{%
\text{m}}$ y de su profundidad $\epsilon $ de tal manera que $\varphi _{%
\text{GSM}_{1}}^{\text{sphe}}(z_{\text{ef}})/\epsilon $ tenga un valor de $%
-1 $ en $z_{\text{ef}}=1$, donde $z_{\text{ef}}=r/r_{\text{m}}$. Haciendo
este rescalamiento encontramos las funciones $r_{\text{m}}$ y $\epsilon $.

Adicionalmente, comparando los perfiles de los potenciales reescalados $%
\varphi _{\text{GSM}_{1}}^{\text{sphe}}(z_{\text{ef}})/\epsilon $ dentro de
un intervalo de temperaturas de 0.3$\leq T_{0}^{\ast }\leq $10 y para
intensidades del momento dipolar $\mu _{0}^{\ast }\leq $1.0, encontramos que
todos empalman uno sobre otro a muy buena aproximaci\'{o}n, \textit{i.e.}
todos tienen el mismo valor de $s_{0}=1$; lo cual nos indica que la
precencia de la interacci\'{o}n dipolar no afecta significativamente la
suavidad.

Podemos ajustar los par\'{a}metros $\epsilon $ y $r_{\text{m}}$ como
funciones apropiadas de $T_{0}^{\ast }$ y $\mu _{0}^{\ast }$. \footnote{%
Para calcular estas funciones seguimos el procedimiento anterior. En primer
lugar buscamos el m\'{\i}nimo de $\varphi _{\text{GSM}_{1}}^{\text{sphe}}(z_{%
\text{ef}},T_{0}^{\ast },\mu _{0}^{\ast })$ fijando el valor de $\mu
_{0}^{\ast }$ para valores de temperatura en un intervalo $0.3\leq
T_{0}^{\ast }\leq 10$ en pasos de 0.05. Consideramos cinco valores de $\mu
_{0}^{\ast }$, a saber,1.0, 0.9, 0.7, 0.5 y 0.25. Para cada valor de $\mu
^{\ast }$ ajustamos los datos de $r_{\text{mef}}(T_{0}^{\ast })$ y $\epsilon
_{\text{ef}}(T_{0}^{\ast })$, a funciones de la forma $C_{1}+C_{2}/T_{0}^{%
\ast }+C_{3}/T_{0}^{\ast 2}$. Obtenemos cinco funciones de esta forma de las
cuales ajustaremos los puntos $C_{2}(\mu _{0}^{\ast })$ y $C_{3}(\mu
_{0}^{\ast })$ respectivamente a funciones de la forma$\ C_{2}=D_{1}+$ $D$ $%
_{2}\mu _{0}^{\ast 4}$ y $C_{3}=D_{3}+$ $D_{4}\mu _{0}^{\ast 8}$. Siguiendo
este procedimiento descrito para $s_{0}=$1.0, obtenemos las siguientes
funciones, las cuales mostramos aqu\'{\i} con tres cifras significativas, 
\begin{eqnarray*}
\epsilon _{\text{ef}}(T_{0}^{\ast },\mu _{0}^{\ast })/\epsilon _{0} &=&1+%
\frac{-\text{0.00187}+\text{0.363}\mu _{0}^{\ast 4}}{T^{\ast }}-\frac{\text{%
0.000139}+\text{0.0247}\mu _{0}^{\ast 8}}{T^{\ast 2}}\text{,} \\
r_{\text{mef}}(T_{0}^{\ast },\mu _{0}^{\ast })/r_{0} &=&1-\frac{-\text{%
0.0000444}+\text{0.0231}\mu _{0}^{\ast 4}}{T^{\ast }}+\frac{\text{0.000116}+%
\text{0.00417}\mu _{0}^{\ast 8}}{T^{\ast 2}}\text{.}
\end{eqnarray*}%
} En este caso, el potencial ANC apropiado ser\'{a} 
\begin{equation}
\varphi _{\text{ANC}}(z,\epsilon ,s_{0}=1)\cong \varphi _{\text{GSM}_{1}}^{%
\text{sphe}}(z_{\text{ef}},T_{0}^{\ast },\mu _{0}^{\ast })\text{.}
\end{equation}
De la teor\'{\i}a ANC sabemos que el segundo coeficiente virial lo
escribimos como, 
\begin{equation}
B^{\ast }(T^{\ast })\equiv \frac{1}{\frac{2}{3}\pi r_{\text{m}_{\text{0}%
}}^{3}}B(T^{\ast })=1-s_{0}+s_{0}B_{0}^{\ast }(T^{\ast })\text{.}
\end{equation}
Esta expresi\'{o}n --escrita en t\'{e}rminos de los par\'{a}metros que
dependen de $T_{0}^{\ast }$ y $\mu _{0}^{\ast }$-- queda como, 
\begin{eqnarray}
B^{\ast }(T_{0}^{\ast }) &=&\frac{r_{\text{m}}^{3}(T_{0}^{\ast },\mu ^{\ast
})}{r_{\text{m}_{\text{0}}}^{3}}\frac{B(T_{0}^{\ast })}{\frac{2}{3}\pi r_{%
\text{m}}^{3}(T_{0}^{\ast },\mu ^{\ast })}  \notag \\
&=&\frac{r_{\text{m}}^{3}(T_{0}^{\ast },\mu ^{\ast })}{r_{\text{m}_{\text{0}%
}}^{3}}\{1-s_{0}+s_{0}\ B_{0}^{\ast }\left( T_{0}^{\ast }/(\epsilon
(T_{0}^{\ast },\mu ^{\ast })/\epsilon _{0}\right) )\}.  \label{BMSM1anc}
\end{eqnarray}

\begin{figure}[h]
        \begin{center}
        \includegraphics[width=.8\hsize]{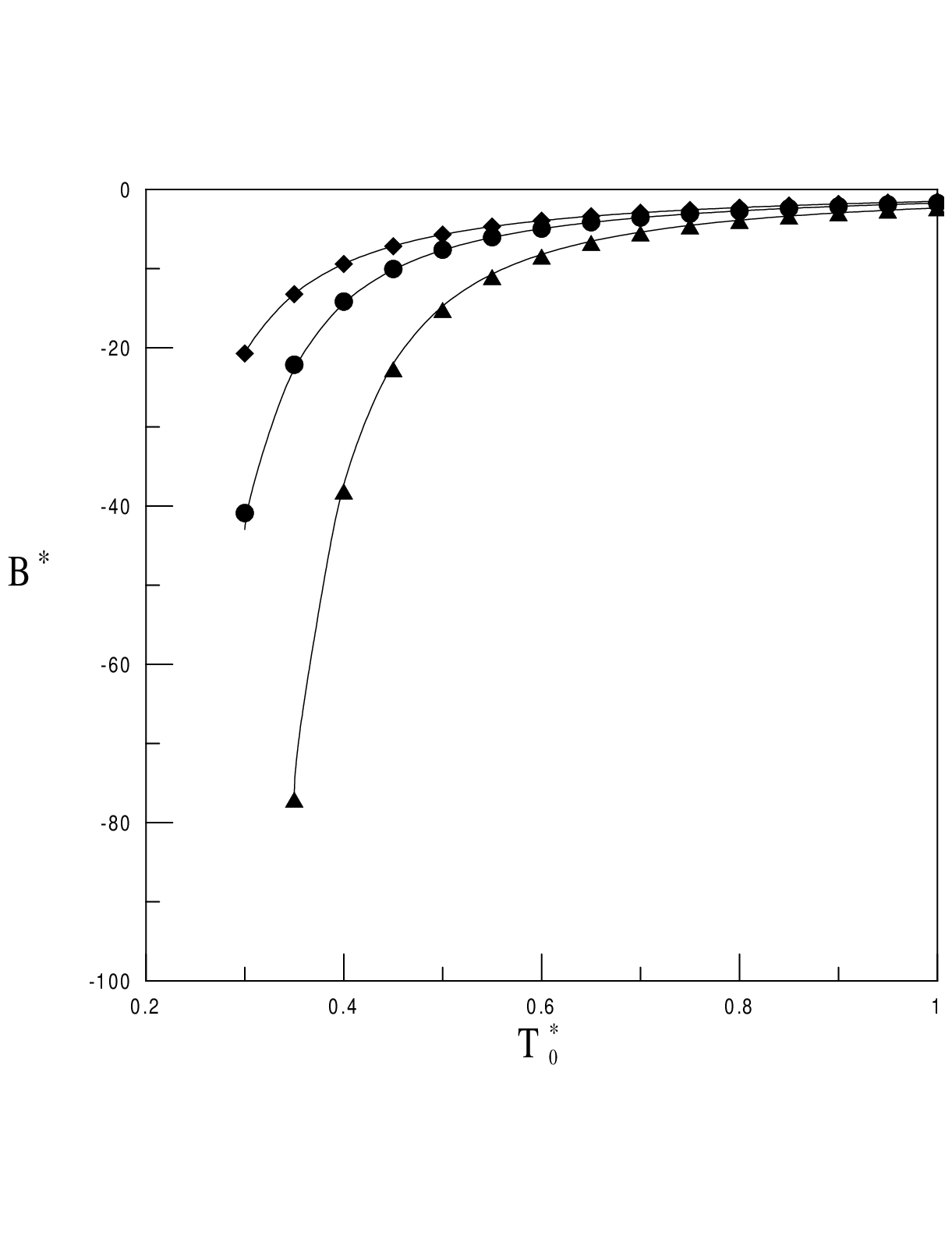}
        \end{center}
        \caption{%
Segundo coeficiente
virial del modelo ANC dipolar con suavidad unitaria para distintos momentos
dipolares, de arriba abajo, $\protect\mu _{0}^{\ast }=$0.5, 0.75 y 1.0. N%
\'{o}tese que al fluido que tiene el momento dipolar mayor (tri\'{a}ngulos)
le corresponden valores del segundo coeficiente virial m\'{a}s negativos y
en consecuencia, la presi\'{o}n de este sistema es menor que la de aquellos
con dipolos m\'{a}s peque\~{n}os. Las l\'{\i}neas continuas representan la
predicci\'{o}n de la teor\'{\i}a ANC.
                     }%
       \label{ancdipm107505s1}
\end{figure}

En la Fig. (\ref{ancdipm107505s1}) mostramos que al evaluar directamente la
expresi\'{o}n anal\'{\i}tica definida por la Ec. (\ref{BMSM1_data}), con $%
n_{\max }=$50 y $l_{\max }=$30, para valores de $\mu _{0}^{\ast }=$0.5, 0.75
y 1.0, en un intervalo de temperatura de $T_{0}^{\ast }\in \lbrack $0.3$,$10$%
)$, encontramos que concuerda muy bien con la predicci\'{o}n de la teor\'{\i}%
a ANC (\ref{BMSM1anc}). El potencial ANC es bastante preciso para el caso $%
s_{0}=$1.0.
%

\section{Preludio al potencial efectivo}

En esta secci\'{o}n hacemos un an\'{a}lisis cualitativo de los efectos de la
interacci\'{o}n momento dipolar-momento dipolar (DD) en la forma de $\varphi
_{\text{GSM}}^{\text{sphe}}(z)$ y despu\'{e}s --para el caso particular del
fluido de Stokmayer-- calculamos los par\'{a}metros de un potencial efectivo
que sea conformal al de Stockmayer. Este c\'{a}lculo es importante debido a
que puede abordarse de manera anal\'{\i}tica.

\subsection{El modelo GSM: Efectos de la temperatura}

Podemos caracterizar la intensidad de la interacci\'{o}n DD respecto a la
energ\'{\i}a cin\'{e}tica del fluido como la raz\'{o}n 
\begin{equation}
\delta _{0}=\frac{\mu _{0}^{\ast 2}}{T_{0}^{\ast }}=\frac{\mu ^{2}}{kTr_{%
\text{m0}}^{3}}\text{.}  \label{delta0}
\end{equation}

Para poder ver anal\'{\i}ticamente el efecto de la interacci\'{o}n DD en $%
\varphi _{\text{GSM}}^{\text{sphe}}$, asumimos que la temperatura es lo
suficientemente alta, o bien, que la intensidad de la interacci\'{o}n DD es
lo suficientemente baja para que 
\begin{equation*}
\delta _{0}<1\text{,}
\end{equation*}
de tal manera que podemos truncar la suma infinita en (\ref{potDDsphe}) y
--en este caso-- los t\'{e}rminos dominantes en $\varphi _{\text{DD}}^{\text{%
sphe}}$ son 
\begin{equation}
\varphi _{\text{DD}}^{\text{sphe}}(z)=-\epsilon _{0}T_{0}^{\ast }\ln \left[
1+\frac{\mu _{0}^{\ast 4}}{3T_{0}^{\ast 2}}\frac{1}{z^{6}}+\frac{\mu
_{0}^{\ast 8}}{25T_{0}^{\ast 4}}\frac{1}{z^{12}}+...\right] \text{,}
\label{U dd expandido}
\end{equation}
donde hemos usado que $G_{0}=1$, $G_{1}=2/3$ y $G_{2}=24/25$. Series del
tipo (\ref{U dd expandido}) han sido analizadas por by Massih y Mansoori 
\cite{Massih: 1983}.

Veamos como son los cambios en la magnitud de DD debidos a la temperatura.
En la expresi\'{o}n (\ref{U dd expandido}) vemos que $\varphi _{\text{DD}}^{%
\text{sphe}}(z)\rightarrow 0$ cuando $T_{0}^{\ast }\rightarrow \infty $. En
este caso $\varphi _{\text{GSM}}^{\text{sphe}}(z)\rightarrow \varphi _{\text{%
ANC}}(z)$ por lo que la interacci\'{o}n dipolar no tiene\ efecto en este l%
\'{\i}mite. El efecto relativo de la interacci\'{o}n dipolar se incrementa a
medida que la temperatura disminuye.

Para analizar los detalles de los efectos dipolares en $\varphi _{\text{GSM}%
}^{\text{sphe}}(z)$, recordemos que este potencial est\'{a} formado al sumar
el t\'{e}rmino $\varphi _{\text{DD}}^{\text{sphe}}(z)$ al kernel $\varphi _{%
\text{ANC}}(z=r/r_{\text{m0}})$ que tiene un m\'{\i}nimo de profundidad\ $%
\epsilon _{0}$ en $r=r_{\text{m0}}$. Notemos en (\ref{U dd expandido}) que
el t\'{e}rmino dipolar promediado es negativo, $\varphi _{\text{DD}}^{\text{%
sphe}}(z)<0$, y adem\'{a}s su magnitud decrece monot\'{o}nicamente a medida
que\ $z$ se incrementa. Como resultado, $\varphi _{\text{GSM}}^{\text{sphe}%
}(z)$, mantiene un m\'{\i}nimo bien definido que ocurre a una distancia $r_{%
\text{m}}$ y tiene una profundidad $\epsilon $. A estas cantidades $r_{\text{%
m}}$ y $\epsilon $ nos vamos a referir como par\'{a}metros efectivos y en
general $r_{\text{m}}\neq r_{\text{m0}}$ y $\epsilon $ $\neq \epsilon _{0}$.
Como notamos ya en la secci\'{o}n anterior al estudiar el caso particular de 
$s_{0}=1$, debido a que $\varphi _{\text{DD}}^{\text{sphe}}(z)<0$, el sumar $%
\varphi _{\text{DD}}^{\text{sphe}}(z)$ a $\varphi _{\text{ANC}}(z)$ hace que
el pozo del potencial se vuelva m\'{a}s profundo. --i.e. $\epsilon >\epsilon
_{0}$. Adicionalmente la posici\'{o}n del m\'{\i}nimo se mueve a $r_{\text{m}%
}<r_{\text{m0}}$. En otras palabras, la interacci\'{o}n DD profundiza el
pozo de potencial y la posici\'{o}n del m\'{\i}nimo la hace m\'{a}s cercana
al eje $r=0$. Ambos efectos se hacen m\'{a}s evidentes a medida que $\delta
_{0}=\mu _{0}^{\ast 2}/T_{0}^{\ast }$ se incrementa.

\subsection{El modelo SM: Aproximaci\'{o}n anal\'{\i}tica}

Antes de desarrollar el c\'{a}lculo general de los par\'{a}metros efectivos
ANC para el potencial $\varphi _{\text{GSM}}^{\text{sphe}}$, en esta secci%
\'{o}n usamos un modelo muy simple para calcular anal\'{\i}ticamente los par%
\'{a}metros efectivos que le corresponden a un potencial de Stockmayer, el
cual puede ser considerado como un caso especial de GSM con $s_{0}=1$ y $a_{%
\text{c}}=0$. Mostramos tambi\'{e}n que el potencial $\varphi _{\text{SM}}^{%
\text{sphe}}(z)$ puede aproximarse muy bien con un potencial efectivo y
mostramos que los podemos aproximar muy bien con un potencial efectivo, $u_{%
\text{ef}}$ con par\'{a}metros efectivos $r_{\text{mef}}$ y $\epsilon _{%
\text{ef}}$ calculados anal\'{\i}ticamente. El an\'{a}lisis que hacemos aqu%
\'{\i} es muy simple, sin embargo, es importante pues servir\'{a} como
introducci\'{o}n cuando abordemos el\ caso general GSM.

En potencial de Stockmayer es 
\begin{equation}
\varphi _{\text{SM}}(z,\mu _{0}^{\ast },\Omega )\equiv \varphi _{\text{LJ}%
}(z;\epsilon _{0})-\epsilon _{0}\frac{\mu _{0}^{\ast 2}}{z^{3}}g(\Omega )%
\text{.}  \label{SMpotential}
\end{equation}
donde 
\begin{equation}
\varphi _{\text{LJ}}(z;\epsilon _{0})=\epsilon _{0}\left\{ \frac{1}{z^{12}}-2%
\frac{1}{z^{6}}\right\}  \label{uLJ}
\end{equation}
Como ya hemos dicho arriba, el potencial de SM, es un caso especial del GSM
pues lo podemos escribir como

\begin{equation*}
\varphi _{\text{SM}}(z,\mu _{0}^{\ast },\Omega )=\varphi _{\text{GSM}%
}(z,\Omega ;\mu _{0}^{\ast },s_{0}=1,a_{c}=0)\text{,}
\end{equation*}
de manera que el potencial sin dependencia angular de SM, $\varphi _{\text{SM%
}}^{\text{sphe}}$, lo obtenemos simplemente haciendo $s_{0}=1$ y $a_{\text{c}%
}=0$ en (\ref{ANCdipSphe}), como sigue, 
\begin{eqnarray}
\varphi _{\text{SM}}^{\text{sphe}}(z;\epsilon _{0},T_{0}^{\ast },\mu
_{0}^{\ast }) &=&\varphi _{\text{GSM}}^{\text{sphe}}(z;\epsilon
_{0},T_{0}^{\ast },\mu _{0}^{\ast },s_{0}=1)  \notag \\
&=&u_{\text{LJ}}(z;\epsilon _{0})+\varphi _{\text{DD}}^{\text{sphe}%
}(z;T_{0}^{\ast },\mu _{0}^{\ast })\text{.}  \label{SM esferical}
\end{eqnarray}
Desarrollando el logaritmo en (\ref{U dd expandido}) hasta orden $\delta
_{0}^{4}$ obtenemos 
\begin{equation}
\varphi _{\text{DD}}^{\text{sphe}}(z,T_{0}^{\ast },\mu _{0}^{\ast })\cong
\epsilon _{\text{0}}T_{0}^{\ast }\left[ -\frac{1}{3}\frac{\mu _{0}^{\ast 4}}{%
z^{6}T_{0}^{\ast 2}}+\frac{7}{450}\frac{\mu _{0}^{\ast 8}}{z^{12}T_{0}^{\ast
4}}\right]  \label{dd sphe desarollado}
\end{equation}
El primer t\'{e}rmino --el de orden $1/z^{6}$-- dentro del par\'{e}ntesis
cuadrado, en el lado derecho, es el potencial de Keesom \footnote{%
El potencial de Keesom es la energ\'{\i}a con la que inteact\'{u}an dos
dipolos con momentos $\mu _{1}$ y $\mu _{2}$ \ que rotan en libertad, $V=-%
\frac{2}{3k_{B}T}\left( \frac{\mu _{1}\mu _{2}}{4\pi \varepsilon _{0}}%
\right) ^{2}\frac{1}{d^{6}}$ . A una temperatura de 25$^{\circ }$C la energ%
\'{\i}a de Keesom promedio de interacci\'{o}n entre pares de mol\'{e}culas
con $\mu =1$ D es alrededor de $-$1.4 \ kJ/mol cuando est\'{a}n separadas
por 0.3 nm. La energ\'{\i}a cin\'{e}tica promedio molar a 25$^{\circ }$C es $%
\frac{3}{2}\ RT=$3.7 \ kJ/mol. N\'{o}tese que la energ\'{\i}a de Keesom es
diferente al t\'{e}rmino de interacci\'{o}n dipolo-dipolo que aparece en la
Ec. (\ref{SMpotential}). La energ\'{\i}a de Keesom corresponde a dos dipolos
que rotan libremente, mientras que $\varphi _{\text{dd}}(x,\Omega ;%
\widetilde{\mu })$ es la energ\'{\i}a entre dos dipolos fijos.} \cite%
{French, Gao 1997}. Entonces al sustituir (\ref{dd sphe desarollado}) en (%
\ref{SM esferical}) --despu\'{e}s de reunir los t\'{e}rminos comunes--
encontramos que $\varphi _{\text{SM}}^{\text{sphe}}$ 
\begin{equation}
\varphi _{\text{SM}}^{\text{sphe}}(z)\cong \epsilon _{\text{0}}T_{0}^{\ast }%
\left[ \left( \frac{1}{T_{0}^{\ast }}+\frac{7}{450}\frac{\mu _{0}^{\ast 8}}{%
T_{0}^{\ast 4}}\right) \frac{1}{z^{12}}-2\left( \frac{1}{T_{0}^{\ast }}+%
\frac{1}{6}\frac{\mu _{0}^{\ast 4}}{T_{0}^{\ast 2}}\right) \frac{1}{z^{6}}%
\right]  \label{U SM esfericalizado}
\end{equation}
Hasta este momento, la \'{u}nica aproximaci\'{o}n que hemos hecho ha sido
cortar la serie en (\ref{dd sphe desarollado}). Si observamos la forma de la
Ec. (\ref{U SM esfericalizado}), podemos ver que $\varphi _{\text{SM}}^{%
\text{sphe}}(z)$ puede remplazarse con un potencial efectivo $u_{\text{ef}%
}(z_{\text{ef}})$ de la forma de uno de Lennard-Jones ordinario, 
\begin{equation}
u_{\text{ef}}(z_{\text{ef}})=\epsilon _{\text{ef}}\left( \frac{1}{z_{\text{ef%
}}^{12}}-2\frac{1}{z_{\text{ef}}^{6}}\right) \text{,}  \label{Uefe LJ}
\end{equation}
donde $z_{\text{ef}}=z/r_{\text{m}}$ y con m\'{\i}nimo en $z_{\text{ef}}=1$.
Para hacer que $u_{\text{eff}}$ sea conformal a $\varphi _{\text{SM}}^{\text{%
sphe}}$, es decir que 
\begin{equation*}
\varphi _{\text{SM}}^{\text{sphe}}(z)\cong u_{\text{ef}}(z_{\text{ef}})\text{%
,}
\end{equation*}
los par\'{a}metros efectivos del valor de la posici\'{o}n del m\'{\i}nimo, $%
r_{\text{m}}$, y de su profundidad, $\epsilon $, deben ser 
\begin{eqnarray}
\epsilon &=&\epsilon _{\text{0}}\left\{ 1+\frac{\mu _{0}^{\ast 4}}{%
3T_{0}^{\ast }}+\left( \frac{1}{2}-\frac{7}{25T_{0}^{\ast }}\right) \frac{%
\mu _{0}^{\ast 8}}{18T_{0}^{\ast 2}}\right\} \text{,}  \label{epsSMteo} \\
r_{\text{m}} &=&r_{\text{m0}}\left\{ 1-\frac{\mu _{0}^{\ast 4}}{%
36T_{0}^{\ast }}+\left( \frac{1}{24}+\frac{1}{25T_{0}^{\ast }}\right) \frac{%
7\mu _{0}^{\ast 8}}{108T_{0}^{\ast 2}}\right\} \text{.}  \label{rmSMteo}
\end{eqnarray}

\begin{figure}[h]
        \begin{center}
        \includegraphics[width=.8\hsize]{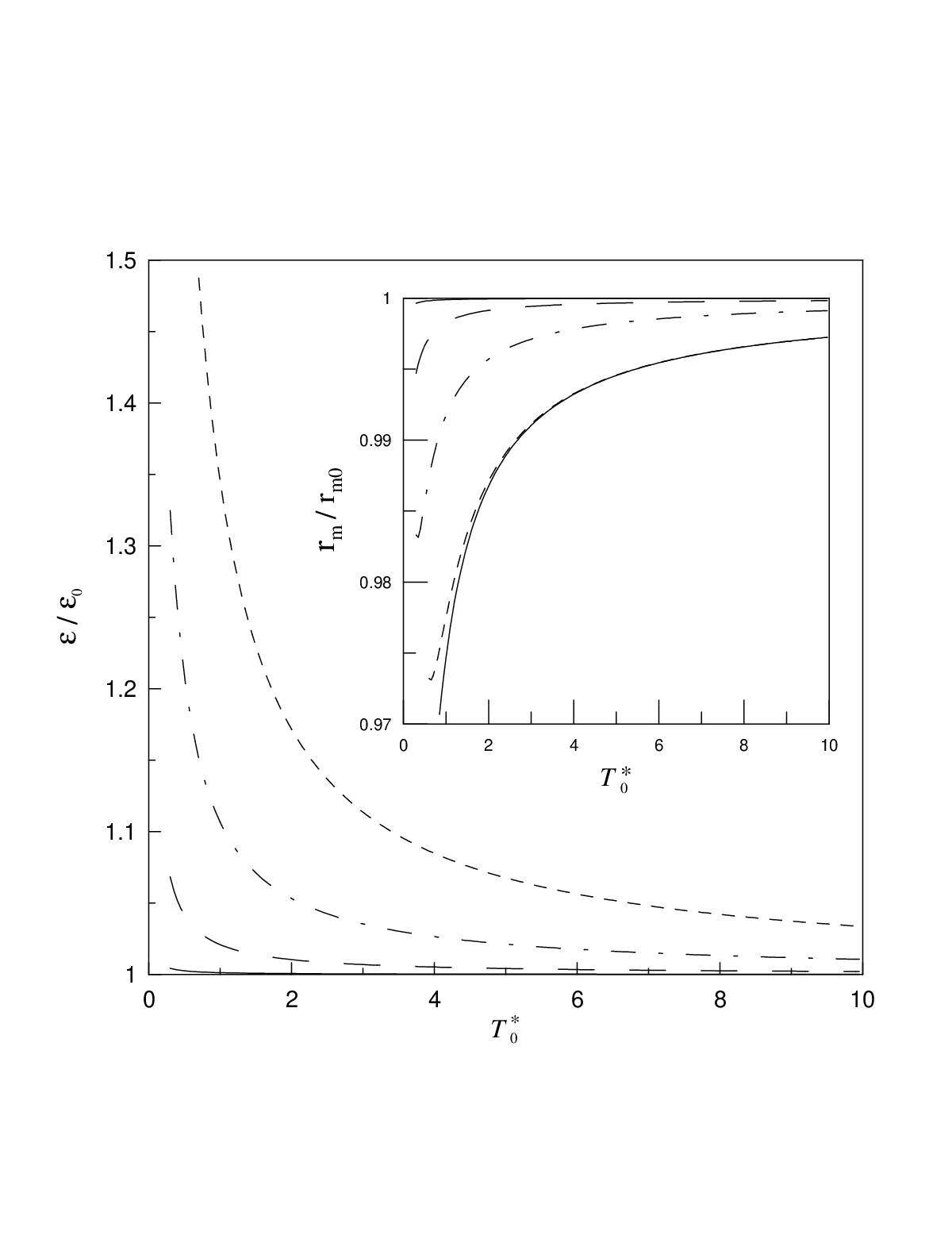}
        \end{center}
        \caption{%
La figura externa muestra la dependencia con
la temperatura de la profundidad efectiva relativa del potencial, $\protect%
\varepsilon /\protect\varepsilon _{0}$, para la interacci\'{o}n de
Stockmayer ordinaria de acuerdo a la aproximaci\'{o}n anal\'{\i}tica dada
por la Ec. (\protect\ref{epsSMteo}). \ Las l\'{\i}neas corresponden a los
siguientes momentos dipolares: $\protect\mu _{0}^{\ast }=0.25$ (l\'{\i}nea
continua), $\protect\mu _{0}^{\ast }=0.5$ (segmentos largos), $\protect\mu %
_{0}^{\ast }=0.75$ (segmentos y puntos) y $\protect\mu _{0}^{\ast }=1.0$
(segmentos cortos). La l\'{\i}nea correspondiente a $\protect\mu _{0}^{\ast
}=0.25$ est\'{a} muy abajo y apenas se distingue del eje horizontal. La
figura interna muestra la dependencia con la temperatura del di\'{a}metro
efectivo relativo, $r_{\text{m}}/r_{\text{m0}}$, para la interacci\'{o}n de
Stockmayer ordinaria de acuerdo con la aproximaci\'{o}n anal\'{\i}tica, Ec.(%
\protect\ref{rmSMteo}). \ Las l\'{\i}neas tienen el mismo significado que en
la figura externa, salvo que aqu\'{\i}, la l\'{\i}nea continua que aparece
debajo de la de segmentos cortos ($\protect\mu _{0}^{\ast }=1.0$),
representa un aproximante de Pade para este momento dipolar.
                     }%
       \label{FIGepsilonrmSM}
\end{figure}

Estos resultados confirman los argumentos cualitativos que explicamos al
principio de la secci\'{o}n: A temperaturas finitas, la interacci\'{o}n
dipolar incrementa la profundidad del pozo de potencial y mueve la posici%
\'{o}n del m\'{\i}nimo. En la Fig. (\ref{FIGepsilonrmSM}) se muestran estos
efectos en $\varphi _{\text{SM}}^{\text{sphe}}(z)$, en donde hemos graficado
las razones $\epsilon /\epsilon _{\text{0}}$ y $r_{\text{m}}/r_{\text{m0}}$
para diferentes valores de la intensidad del momento dipolar $\mu _{0}^{\ast
}$ y dentro del intervalo de temperaturas 0.3$<T_{0}^{\ast }<$10.0. Para un
valor\ constante de $\mu _{0}^{\ast }$ la profundidad efectiva $\epsilon _{%
\text{ef}}/\epsilon _{\text{0}}$ se incrementa monot\'{o}nicamente a medida
que la temperatura disminuye. En estas figuras podemos notar que el efecto
neto de la interacci\'{o}n dipolar se hace bastante fuerte cuando $\mu
_{0}^{\ast }\geq $0.75 y es casi imperceptible cuando $\mu _{0}^{\ast }\leq $%
0.25. Para valores $\mu _{0}^{\ast }\leq $1.0 y dentro del rango de
temperatura que hemos considerado, la aproximaci\'{o}n (\ref{epsSMteo}) da
una respuesta confiable. Por su parte, el di\'{a}metro efectivo $r_{\text{m}%
}/r_{\text{m0}}$, disminuye sistem\'{a}ticamente con $T_{0}^{\ast }$. Este
efecto, sin embargo, es bastante peque\~{n}o; de hecho, $r_{\text{m}}/r_{%
\text{m0}}$ disminuye menos del 3\% a un valor de temperatura de $%
T_{0}^{\ast }=$0.65 y $\mu _{0}^{\ast }=$1.0. Para este valor de la
intensidad del momento dipolar, la relaci\'{o}n (\ref{rmSMteo}) deja de
funcionar adecuadamente cuando la temperatura disminuye por debajo de $%
T_{0}^{\ast }=$0.65 debido a que fueron truncados los t\'{e}rminos de orden
superior en la contribuci\'{o}n dipolar. Es importante notar que al truncar
la suma en (\ref{U dd expandido}) obtenemos la misma forma funcional que
tiene el potencial de LJ haciendo que $\varphi _{\text{SM}}^{\text{sphe}}(z)$
sea conformal a $\varphi _{\text{LJ}}(z)$. El considerar m\'{a}s t\'{e}%
rminos de la serie (\ref{U dd expandido}) encontramos que $\varphi _{\text{SM%
}}^{\text{sphe}}(z)$ difiere en suavidad respecto a la de $\varphi _{\text{LJ%
}}(z)$ por una cantidad despreciable dentro de los intervalos considerados
de $\mu _{0}^{\ast }$ y $T_{0}^{\ast }$.

\paragraph{Aproximante de Pad\'{e}}

Para que el potencial $\varphi _{\text{SM}}^{\text{sphe}}(z)$ pueda
remplazarse con un potencial efectivo $u_{\text{ef}}(z_{\text{ef}})$, es
necesario truncar los t\'{e}rminos de orden superior (mayores que $\mathcal{O%
}\mu _{0}^{\ast 8}$) en el c\'{a}lculo de los par\'{a}metros efectivos $%
\epsilon $ y $r_{\text{m}}$. Este corte ocasiona que $r_{\text{m}}/r_{\text{%
m0}}$ --tal y como est\'{a} expresado en la Ec. (\ref{rmSMteo})-- se desv%
\'{\i}e del comportamiento mon\'{o}tono a bajas temperaturas cuando se
aplica a momentos dipolares de magnitud considerable (v\'{e}ase el sector
interno de la Fig. (\ref{FIGepsilonrmSM}). Este inconveniente los podemos
remediar reemplazando la Ec. (\ref{rmSMteo}) por un aproximante de Pad\'{e}
adecuado. En particular para el caso de $\mu _{0}^{\ast }=1$, encontramos
que podemos usar un aproximante $1,1$ de la forma 
\begin{equation}
r_{\text{m}}/r_{\text{m0}}=\frac{1+a/T}{1+b/T}\text{,}  \label{PadeRmSMteo}
\end{equation}
donde los coeficientes $a$ y $b$ los obtenemos al requerir que, al
desarrollar esta expresi\'{o}n (Ec. (\ref{PadeRmSMteo})) con una serie de
Taylor alrededor del origen, los primeros t\'{e}rminos coincidan con los del
desarrollo de (\ref{rmSMteo}). Para este caso $a=5/72$ y $b=7/72$. En la
Fig. (\ref{FIGepsilonrmSM}) se muestra el aproximante de Pad\'{e} comparado
con el par\'{a}metro efectivo $r_{\text{m}}/r_{\text{m0}}$. 

El lector puede revisar en la Fig. (\ref{FIGBSM}) la predicci\'{o}n de la
teor\'{\i}a ANC para $B(T)$ --que est\'{a} representada en esa figura con
cruces ($\times $) para el caso $\mu _{0}^{\ast }=1/2^{1/4}$. En esa figura
se muestra la predicci\'{o}n realizada con los par\'{a}metros calculados anal%
\'{\i}ticamente, Ecs. (\ref{epsSMteo}) y (\ref{rmSMteo}), comparada con los
resultados obtenidos de la integraci\'{o}n num\'{e}rica de Vega \textit{et al%
}, ($\cdot $) \cite{carloses}. Como el lector puede apreciar, la aproximaci%
\'{o}n anal\'{\i}tica, ($\times $), pr\'{a}cticamente coincide con los
resultados de la integraci\'{o}n num\'{e}rica ($\cdot $).

\subsection{El modelo SM: Aproximaci\'{o}n num\'{e}rica}

En la secci\'{o}n anterior calculamos una aproximaci\'{o}n anal\'{\i}tica a
los par\'{a}metros efectivos $\epsilon $ y $r_{\text{m}}$. Mostramos que
usando estos par\'{a}metros podemos predecir con una excelente aproximaci%
\'{o}n los datos de Hirschfelder \textit{et al}.\cite{Hirsch} para $B(T)$.
En esta secci\'{o}n calculamos una aproximaci\'{o}n num\'{e}rica a $\epsilon 
$ y $r_{\text{m}}$ lo cual corrobora que la aproximaci\'{o}n anal\'{\i}tica
es en efecto correcta. Este c\'{a}lculo es similar al que hicimos para el
caso especial de $s_{0}=$1.0; es decir, calculamos por diferenciaci\'{o}n el
m\'{\i}nimo del potencial SM dentro del intervalo de temperatura\ 0.3$\leq
T_{0}^{\ast }\leq $10.

Comparando los perfiles de los potenciales rescalados $\varphi _{\text{SM}}^{%
\text{sphe}}(r/r_{\text{m}})/\epsilon $ dentro del intervalo de temperaturas
seleccionado y para $\mu _{0}^{\ast }\leqslant 1/2^{1/4}$, encontramos que
todos son conformales, \textit{i.e.} todos tienen el mismo valor de $s$ que
el potencial de LJ. De esta manera, el potencial ANC apropiado (\ref%
{potencialANC}) tendr\'{a} suavidad $s=s_{\text{LJ}}$, profundidad $\epsilon
(T_{0}^{\ast },\mu _{0}^{\ast })$ y m\'{\i}nimo en $r_{\text{m}}(T_{0}^{\ast
},\mu _{0}^{\ast })$, i.e. 
\begin{equation}
\varphi _{\text{ANC}}(z,s_{\text{LJ}},\epsilon )\cong \varphi _{\text{SM}}^{%
\text{sphe}}(z,T_{0}^{\ast },\mu _{0}^{\ast })
\end{equation}
donde ahora $z=r/r_{\text{m}}$.

Aqu\'{\i} presentamos resultados para $\mu _{0}^{\ast }=1/2^{1/4}$. Para
calcular $\epsilon _{\text{eff}}(T_{0}^{\ast },1/2^{1/4})$ y $r_{\text{mef}%
}(T_{0}^{\ast },1/2^{1/4})$ buscamos el m\'{\i}nimo de $\varphi _{\text{SM}%
}^{\text{sphe}}(z,T_{0}^{\ast },1/2^{1/4})$ recorriendo el intervalo de
temperaturas seleccionado en pasos de $0.1$. Los resultados los ajustamos\ a
funciones de la forma $\epsilon (T_{0}^{\ast })/\epsilon _{0}=$ $%
1+C_{2}/T_{0}^{\ast }+C_{3}/T_{0}^{\ast 2}$ $r_{\text{m}}(T_{0}^{\ast })/r_{%
\text{0}}=1+C_{2}^{^{\prime }}/T_{0}^{\ast }+C_{3}^{^{\prime }}/T_{0}^{\ast
2}$ para obtener los coeficientes $C_{i}$ y $C_{i}^{\prime }$ \footnote{%
En este caso los coeficientes que obtuvimos fueron $C_{2}=$1.183047, $%
C_{3}=- $0.009916, $C_{2}^{^{\prime }}=-$0.054776$/2^{1/6}$, $%
C_{3}^{^{\prime }}=$0.001641$/2^{1/6}$.}.

Usando estos ajustes podemos predecir $B^{\ast }(T_{\text{ef}}^{\ast },\mu
_{0}^{\ast }=1/2^{1/4})$. En la teor\'{\i}a ANC $B(T)$ se reduce como

\begin{equation}
B^{\ast }(T_{\text{ef}}^{\ast })=\left( r_{\text{m}}/r_{\text{0}}\right)
^{3}\left\{ 1-s_{\text{LJ}}+s_{\text{LJ}}\ B_{1}^{\ast }\left( T_{\text{ef}%
}^{\ast }\right) \right\} \text{,}  \label{BstarANC}
\end{equation}
donde $T_{\text{ef}}^{\ast }=kT/\epsilon $. En Fig. \ref{FIGBSM} los
resultados de la predicci\'{o}n $B(T)$ para $\mu _{0}^{\ast }=1/2^{1/4}$
realizada con los par\'{a}metros que aqu\'{\i} ajustamos num\'{e}ricamente,
no se distinguen de los resultados obtenidos de la integraci\'{o}n num\'{e}%
rica de Vega \textit{et al}. \cite{carloses}.

\section{Par\'{a}metros efectivos para el potencial GSM}

En esta secci\'{o}n nos centramos en el potencial GSM. Usaremos $a_{\text{c}%
}=$0.0957389, que es el valor del n\'{u}cleo duro del kernel ANC y un valor
variable de su suavidad $s_{0}$ dentro del intervalo que corresponde a las
sustancias reales y no polares \cite{del Rio 1998 III}. El potencial total
esf\'{e}rico $\varphi _{\text{GSM}}^{\text{sphe}}(z)$ est\'{a} dado aqu\'{\i}
por (\ref{ANCdipSphe}), el cual tiene las caracter\'{\i}sticas que ya
discutimos en la secci\'{o}n anterior. Lo que haremos aqu\'{\i} para $%
\varphi _{\text{GSM}}^{\text{sphe}}(z)$ es justamente lo mismo que hicimos
para $\varphi _{\text{SM}}^{\text{sphe}}(z)$, \textit{i.e.}, vamos a
calcular los par\'{a}metros efectivos de un potencial $u_{\text{ef}%
}(z;\epsilon ,r_{\text{m}},s)$ y mostratremos que este potencial se puede
representar correctamente con una funci\'{o}n ANC (\ref{potencialANC}). Al
igual que en el caso de SM, los par\'{a}metros efectivos son funciones de $%
T_{0}^{\ast }$ y $\mu _{0}^{\ast }$, pero adicionalmente ahora dependen tambi%
\'{e}n $s_{0}$. En consecuencia, tendremos que calcular la suavidad del
potencial efectivo $u_{\text{ef}}(z)$. Al examinar algunos cuantos ejemplos
num\'{e}ricos nos damos cuenta de que el efecto de sumar $\varphi _{\text{DD}%
}^{\text{sphe}}(z)$ a $\varphi _{\text{ANC}}(z,s_{0})$ para obtener $\varphi
_{\text{GSM}}^{\text{sphe}}(z)$ es diferente en la rama repulsiva que en la
atractiva. Por lo que, en rigor, un potencial efectivo deber\'{\i}a
incorporar dos diferentes suavidades $s_{\text{R}}$ y $s_{\text{A}}$. Para
denotar esta dependencia introducimos las funciones $f_{\epsilon }$, $f_{r}$%
, $f_{s_{\text{A}}}$ y $f_{s_{\text{R}}}$, las cuales administran los
cambios debidos a la interacci\'{o}n dipolar\ y en general dependen de $%
T_{0}^{\ast }$, $\mu _{0}^{\ast }$ and $s_{0}$. Respectivamente estas
funciones operan como sigue, 
\begin{eqnarray}
\epsilon &=&\epsilon _{0}f_{\epsilon }(T_{0}^{\ast },\mu _{0}^{\ast },s_{0})%
\text{,}  \label{EpsEff1} \\
r_{\text{m}} &=&r_{\text{m0}}f_{r}(T_{0}^{\ast },\mu _{0}^{\ast },s_{0})%
\text{,}  \label{RmEff1} \\
s_{\text{A}} &=&s_{0}f_{s_{\text{A}}}(T_{0}^{\ast },\mu _{0}^{\ast },s_{0})%
\text{,}  \label{SAEff1} \\
s_{\text{R}} &=&s_{0}f_{s_{\text{R}}}(T_{0}^{\ast },\mu _{0}^{\ast },s_{0})%
\text{.}  \label{SREff1}
\end{eqnarray}
En donde debemos recordar que $\epsilon _{0}$, $r_{\text{m0}}$ y $s_{0}$ son
la profundidad, di\'{a}metro y suavidad del kernel en el potencial GSM.

\subsection{La posici\'{o}n del m\'{\i}nimo y su profundidad}

\begin{figure}[h]
        \begin{center}
        \includegraphics[width=.8\hsize]{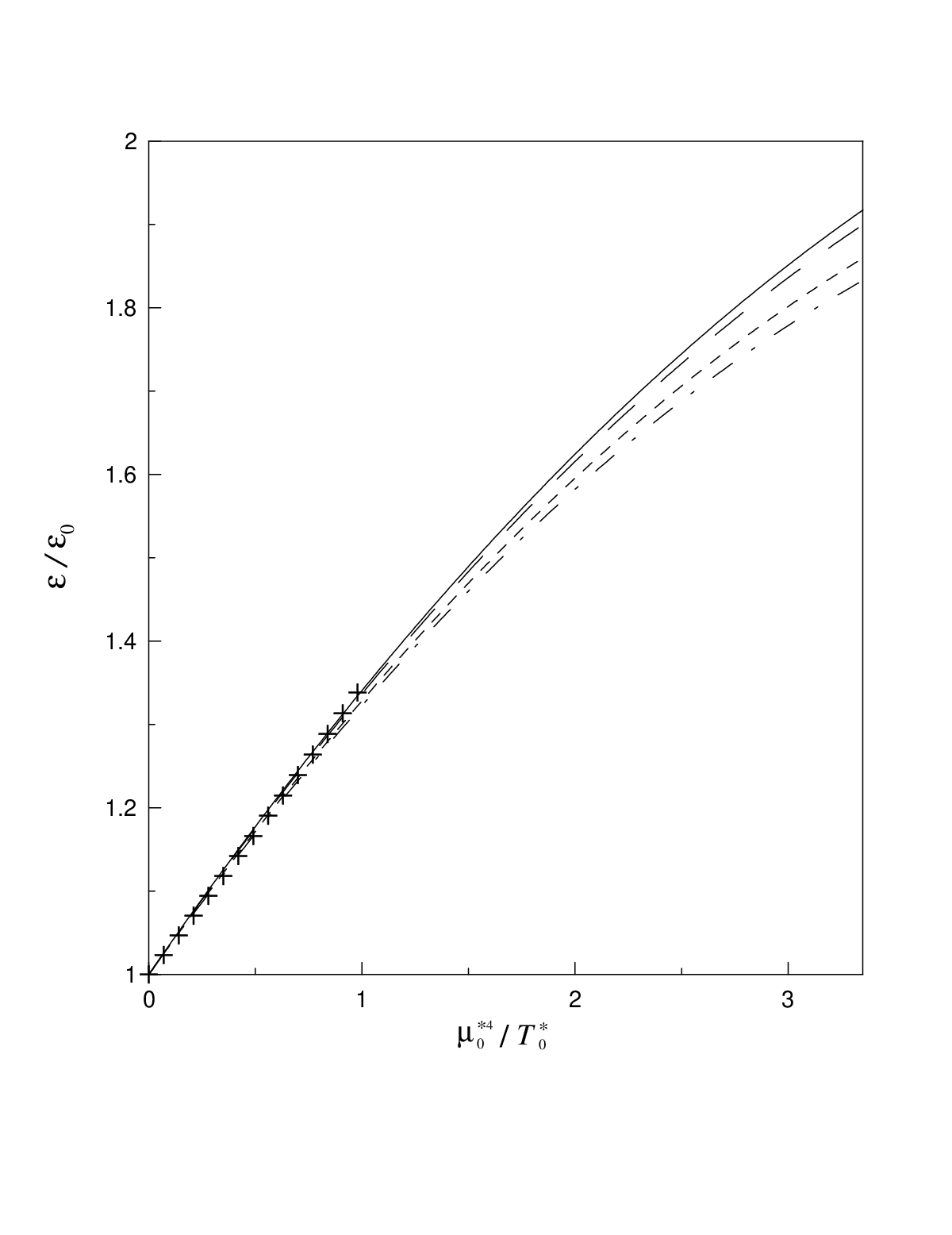}
        \end{center}
        \caption{%
Profundidad efectiva relativa del potencial, $f_{\protect%
\epsilon }=\protect\epsilon /\protect\epsilon _{0}$ para interacci\'{o}n de
Stockmayer generalizada como funci\'{o}n de $\protect\mu _{0}^{\ast
4}/T_{0}^{\ast }$. Las l\'{\i}neas corresponden a diferentes suavidades $%
s_{0}$ del kernel del GSM: $s_{0}=$0.5 (segmentos y puntos), $s_{0}=$0.7
(segmentos cortos), $s_{0}=$1.0 (segmentos largos) y $s_{0}=$1.1315 (l\'{\i}%
nea continua). Los s\'{\i}mbolos ($+$) muestran valores de la aproximaci\'{o}%
n anal\'{\i}tica para el caso SM ordinario, i.e. Eq. (\protect\ref{epsSMteo}%
) \ para $T_{0}^{\ast }=$1.0; la cual es muy cercana a la l\'{\i}nea que
corresponde a $s_{0}=$1.1315.
                     }%
       \label{FIGfepsilon}
\end{figure}

\begin{figure}[h]
        \begin{center}
        \includegraphics[width=.8\hsize]{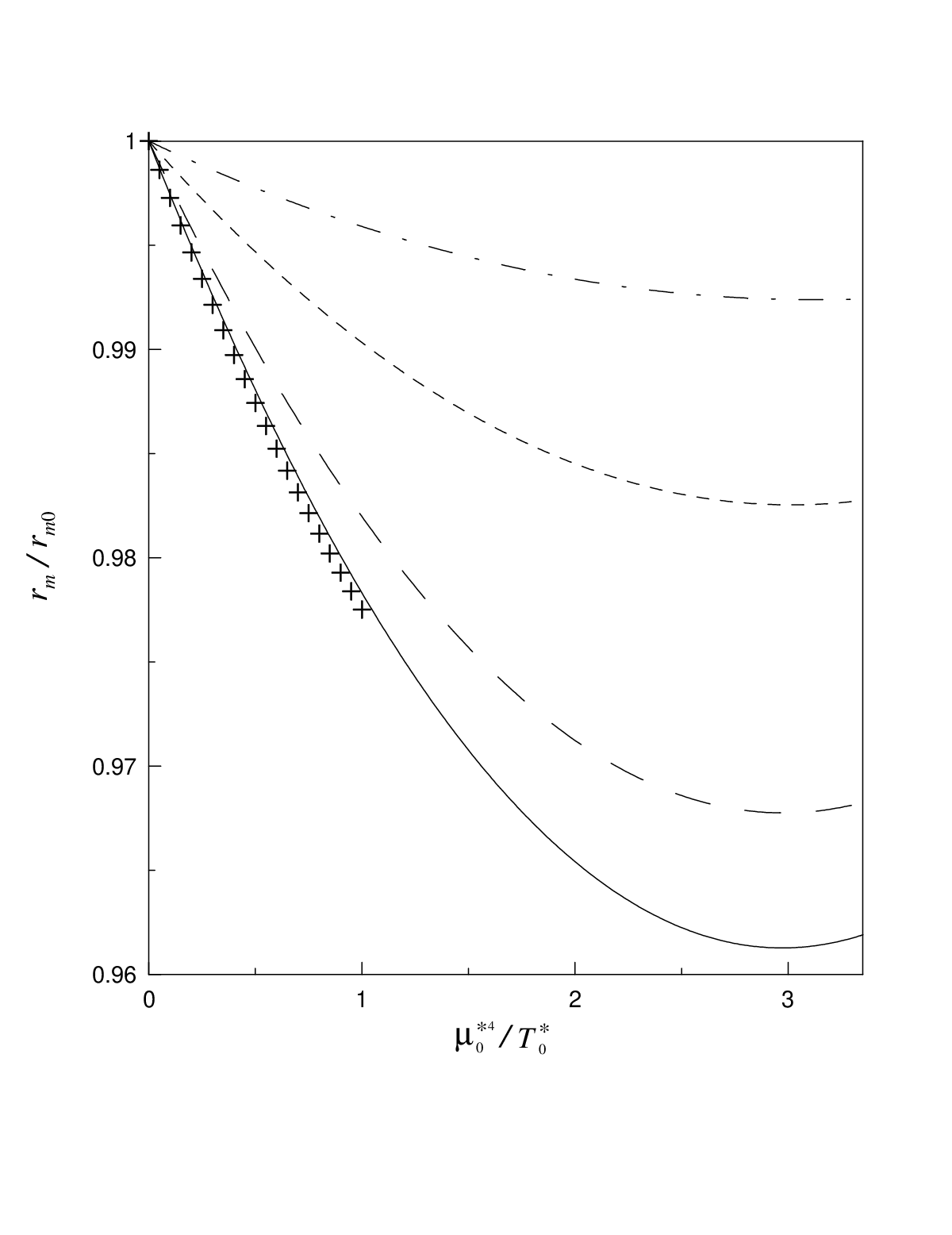}
        \end{center}
        \caption{%
Di\'{a}metro efectivo relativo, $f_{r}=r_{%
\text{m}}/r_{\text{m0}}$, para interacci\'{o}n de Stockmayer generalizada
como funci\'{o}n de $\protect\mu _{0}^{\ast 4}/T_{0}^{\ast }$. Las l\'{\i}%
neas tienen el mismo significado que en la Fig. \protect\ref{FIGfrm}.
                     }%
       \label{FIGfrm}
\end{figure}

Para calcular las funciones $f_{\epsilon }$ y $f_{r}$ fijamos valores para $%
\mu _{0}^{\ast }$, $s_{0}$ y $T_{0}^{\ast }$ en el potencial $\varphi _{%
\text{GSM}}^{\text{sphe}}(z)$ y localizamos su m\'{\i}nimo por diferenciaci%
\'{o}n. Los valores que usamos para la suavidad del kernel y el momento
dipolar, los tomamos dentro de los intervalos\ 0.5$\leq s_{0}<$ 1.0 y 0.0$%
\leq \mu _{0}^{\ast }\leq $ 1.0, respectivamente. El intervalo seleccionado
para $s_{0}$ cubre la mayor\'{\i}a de las sustancias reales y no polares que
han sido ya estudiadas \cite{del Rio 1998 III} y deber\'{\i}a ser apropiado
para los kernels de las mol\'{e}culas polares reales. Para cada modelo
definido por una pareja $(\mu _{0}^{\ast },s_{0})$, escogemos valores de
temperatura dentro del intervalo 0.3$\leq T_{0}^{\ast }<$10.0, y calculamos
y registramos los valores de $\epsilon $ y $r_{\text{m}}$. Para representar
la variaci\'{o}n con la temperatura de estos resultados, asumimos que la
dependencia con $T$ es similar a la del caso m\'{a}s simple de SM, que est%
\'{a} ilustrada por las ecuaciones (\ref{epsSMteo}) y (\ref{rmSMteo}), i.e., 
\begin{equation*}
f_{\epsilon }=1+C_{1}/T_{0}^{\ast }+C_{2}/T_{0}^{\ast 2}\text{,}
\end{equation*}
\begin{equation*}
f_{r}=1+C_{3}/T_{0}^{\ast }+C_{4}/T_{0}^{\ast 2}\text{.}
\end{equation*}
Estos par\'{a}metros $C_{i}$ los ajustamos a los resultados num\'{e}ricos,
dentro del intervalo considerado para $T_{0}^{\ast }$, para cada valor de $%
(\mu _{0}^{\ast },s_{0})$. La dependencia de los coeficientes $C_{i}$ con $%
\mu _{0}^{\ast }$ la encontramos que sigue un comportamiento similar al
encontrado en (\ref{epsSMteo}) y (\ref{rmSMteo}); mientras que el efecto de
la variac\'{o}n de $s_{0}$ encontramos que sigue un comportamiento lineal.
Entonces las expresiones que obtenemos finalmente son 
\begin{equation}
f_{\epsilon }=1+(c_{10}+c_{11}s_{0})\mu _{0}^{\ast 4}/T_{0}^{\ast
}+(c_{20}+c_{21}s_{0})\mu _{0}^{\ast 8}/T_{0}^{\ast 2}\text{,}
\label{EfeEps}
\end{equation}
\begin{equation}
f_{r}=1+(c_{30}+c_{31}s_{0})\mu _{0}^{\ast 4}/T_{0}^{\ast
}+(c_{40}+c_{41}s_{0})\mu _{0}^{\ast 8}/T_{0}^{\ast 2}\text{,}  \label{EfeRm}
\end{equation}
donde las constantes $c_{i,j}$ est\'{a}n dadas en el Cuadro \ref{fx 3}\ en
un ap\'{e}ndice al final de este trabajo. La conducta que siguen (\ref%
{EfeEps}) y (\ref{EfeRm}) por efecto de $\mu _{0}^{\ast }$ y $T_{0}^{\ast }$
es muy similar a la que se observa en el fluido SM ordinario en su aproximaci%
\'{o}n anal\'{\i}tica, (\ref{epsSMteo}) y (\ref{rmSMteo}). La caracter\'{\i}%
stica novedosa que encontramos en (\ref{EfeEps}) y (\ref{EfeRm}) es su
dependencia con $s_{0}$. Las Figs. (\ref{FIGfepsilon}) y (\ref{FIGfrm})
muestran $f_{\epsilon }$ and $f_{r}$ como funciones de $\mu _{0}^{\ast
4}/T_{0}^{\ast }$ para $s_{0}=$0.5, 0.7, 1.0 y 1.1315. Las mismas figuras
muestran una comparaci\'{o}n las aproximaciones anal\'{\i}ticas del caso SM
ordinario, (\ref{epsSMteo}) y (\ref{rmSMteo}), para $\mu _{0}^{\ast } $
variable pero con $T_{0}^{\ast }=$1.0. Este \'{u}ltimo caso, como esperamos,
es muy cercano al resultado del fluido GSM con $s_{0}=$1.1315 debido a que
es el valor apropiado de la suavidad efectiva el fluido de LJ (que es el
kermel del SM) en la aproximaci\'{o}n 1-$s$. 
%

\subsection{La forma del potencial efectivo}

Los cambios en la forma del perfil del potencial debidos al momento dipolar
los podemos describir ya sea con la aproximaci\'{o}n de dos suavidades o con
la de una.

\subsubsection{La forma del potencial efectivo: aproximaci\'{o}n 2-$s$}

\begin{figure}[h]
        \begin{center}
        \includegraphics[width=.8\hsize]{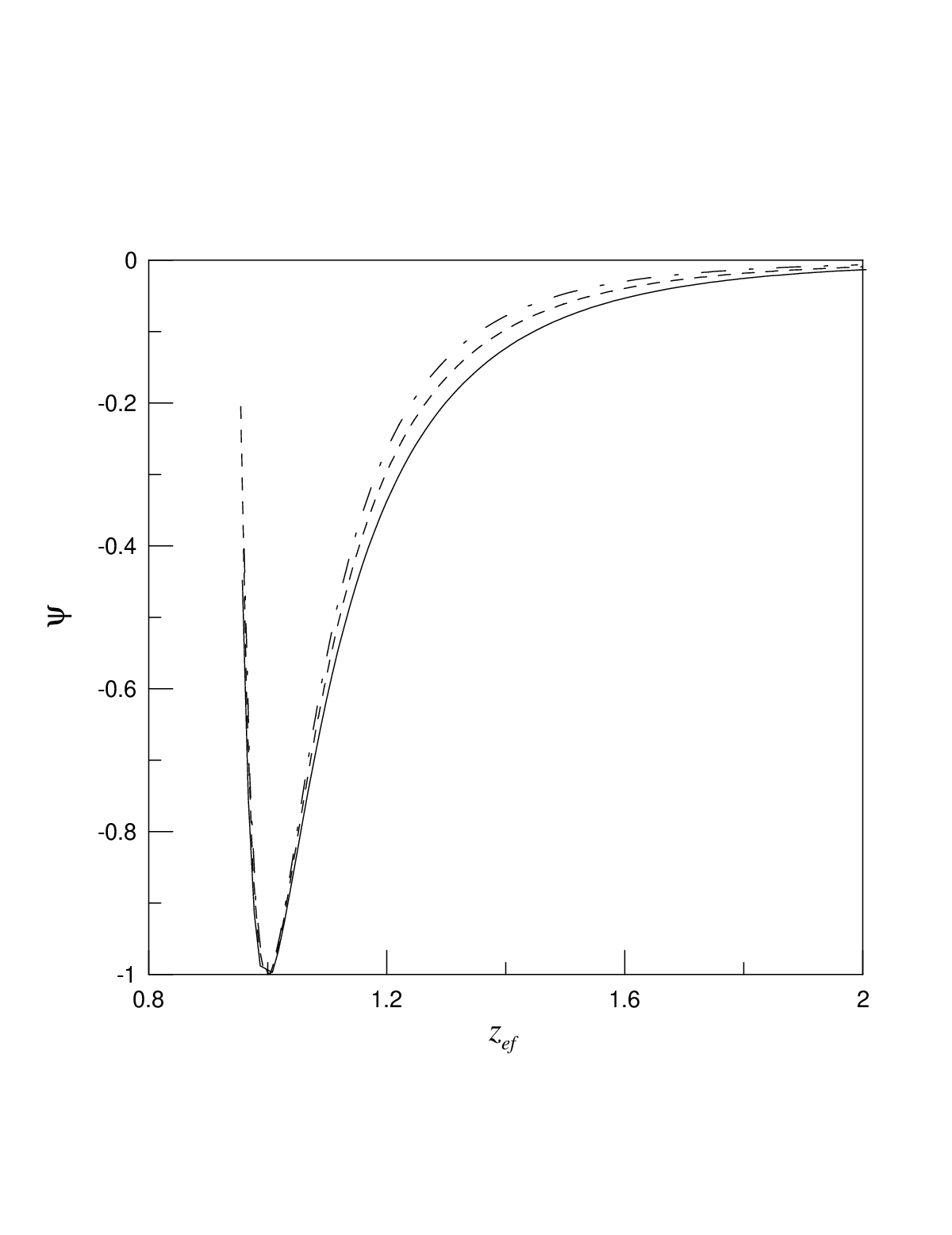}
        \end{center}
        \caption{%
Efecto de la interacci\'{o}n dipolar ($\protect\mu %
_{0}^{\ast }=1$) en la forma del potencial esfericalizado $\protect\psi (z_{%
\text{ef}})$ para el potencial de Stockmayer generalizado con un kernel de
suavidad $s_{0}=0.5$. El potencial $\protect\psi (z)$ est\'{a} escalado de
tal maneta que si m\'{\i}nimo se encuentre en $\protect\psi (z_{\text{ef}%
}=1)=-1.0$. Las l\'{\i}neas corresponden a diferentes valores de la
temperatura: $T_{0}^{\ast }=0.3$ (l\'{\i}nea continua), $T_{0}^{\ast }=0.8$
(segmentos) y $T_{0}^{\ast }=10.0$ (segmentos y puntos). El perfil del
potencial se hace m\'{a}s suave a medida que la temperatura baja.
                     }%
       \label{FIGpsidz}
\end{figure}

Con las funciones $f_{r}$ y $f_{\epsilon }$ es posible normalizar $\varphi _{%
\text{GSM}}^{\text{sphe}}$ para definir la interacci\'{o}n efectiva dada por 
\begin{equation}
\psi (z_{\text{ef}})=\left. \varphi _{\text{GSM}}^{\text{sphe}}\right/
\epsilon =\left. \varphi _{\text{GSM}}^{\text{sphe}}(z=z_{\text{ef}})\right/
\epsilon _{0}f_{\epsilon }
\end{equation}
donde $z_{\text{ef}}=r/r_{\text{m}}=r/(r_{\text{m0}}f_{r})$. Para todas las
temperaturas y para valores arbitrarios de $\mu _{0}^{\ast }$ y $s_{0}$, el
potencial $\psi (z_{\text{ef}})$ tiene su m\'{\i}nimo en $z_{\text{ef}}=1$ y 
$\psi (z_{\text{ef}}=1)=-1$. Sin embargo, la forma del perfil de $\psi (z_{%
\text{ef}})$ depende del peso relativo de sus dos componentes: $\varphi _{%
\text{DD}}^{\text{sphe}}(z)$ y $\varphi _{\text{ANC}}(z)$, y debido a que un
cambio en $T_{0}^{\ast }$ afecta \textit{\'{u}nicamente} la parte dipolar,
la forma final de $\psi (z_{\text{ef}})$ tambi\'{e}n va a depender de $%
T_{0}^{\ast }$. Esto est\'{a} ilustrado en la Fig. \ref{FIGpsidz}: para
valores fijos de $\mu _{0}^{\ast }$ y $s_{0}$, el perfil de $\psi (z_{\text{%
ef}})$ es distinto cuando lo graficamos para diferentes valores de
temperatura, $T_{0}^{\ast }=$0.3, 0.8 y 10.0. Estos perfiles s\'{o}lo
coinciden --por construcci\'{o}n-- en sus m\'{\i}nimos. Esto significa que $%
\psi (z_{\text{ef}};T_{01}^{\ast })$\ es no conformal a $\psi (z_{\text{ef}%
};T_{02}^{\ast }\neq T_{01}^{\ast })$. En esta figura tambi\'{e}n podemos
ver que el cambio en la forma del perfil de $\psi (z_{\text{ef}})$ es m\'{a}%
s apreciable en la rama atractiva que en la repulsiva. Las funciones, $f_{s_{%
\text{A}}}$, y $f_{s_{\text{R}}}$, introducidas en (\ref{SAEff1}) y (\ref%
{SREff1}), respectivamente, describen como cambian en forma las ramas
atractiva y repulsiva de $\psi (z_{\text{ef}})$. Formalmente, estas
funciones administran los cambios en suavidad debidos a $T_{0}^{\ast }$, $%
\mu _{0}^{\ast }$ y $s_{0}$. Aqu\'{\i} describimos el procedimiento para
calcular $f_{s_{\text{A}}}$ y $f_{s_{\text{R}}}$.
%
%

\paragraph{Parte atractiva}

Empezamos considerando la funci\'{o}n que corresponde a la suavidad
atractiva $f_{s_{\text{A}}}=s_{\text{A}}/s_{0}$. Para calcular esta funci%
\'{o}n asumimos que una vez que ha sido definido un modelo fijando valores
para $\mu _{0}^{\ast }$ y $s_{0}$, $\psi $ es \textit{conformal} a un
potencial $\varphi _{\text{ANC}}$, para cada valor de temperatura dentro de
un intervalo $\{T_{0}^{\ast }\}$.

Para calcular $f_{s_{\text{A}}}(T_{0}^{\ast })$ o equivalentemente $s_{\text{%
A}}(T_{0}^{\ast })$, definimos un modelo fijando valores para $\mu
_{0}^{\ast }$ y $s_{0}$, y recurrimos a una caracter\'{\i}stica b\'{a}sica
de los potenciales ANC: para $\varphi _{\text{ANC}}(z,s_{\text{A}})$ con $s_{%
\text{A}}=$ constante, se cumple exactamente la relaci\'{o}n lineal que
aparece en la Ec. (\ref{lambda}) entre los vol\'{u}menes de colisi\'{o}n $%
\Lambda ^{\ast }$ y $\Lambda _{1}^{\ast }$; entonces asumimos que, para un
valor fijo de $T_{0}^{\ast }$ de inter\'{e}s, la rama atractiva de $\psi (z_{%
\text{ef}};T_{0}^{\ast })$ es \textit{conformal} a la misma rama de $\varphi
_{\text{ANC}}(z_{\text{ef}};s_{\text{A0}})$ donde $s_{\text{A0}}=s_{\text{A}%
}(T_{0}^{\ast })$ es la suavidad del potencial a la temperatura escogida $%
T_{0}^{\ast }$. Entonces la expresi\'{o}n (\ref{lambda}) se interpreta como 
\begin{equation}
\Lambda ^{\ast }(T^{\ast })=1-s_{\text{A0}}+s_{\text{A0}}\Lambda _{1}^{\ast
}(T^{\ast })  \label{lambdaSA0}
\end{equation}
donde $\Lambda _{0}^{\ast }$ es el volumen atractivo de referencia y $%
\Lambda ^{\ast }(T^{\ast })$ est\'{a} dada por

\begin{equation}
\Lambda ^{\ast }(T^{\ast })=1+\frac{3}{\exp \left( 1/T^{\ast }\right) -1}%
\int_{1}^{\infty }\,dz\ z^{2}\left\{ \exp \left[ -\left. \psi (z,T_{0}^{\ast
})\right/ T^{\ast }\right] -1\right\} \text{.}  \label{lambdacalc}
\end{equation}
Debido a que $\Lambda _{1}^{\ast }(T^{\ast })$ es una funci\'{o}n bien
conocida, para obtener $s_{\text{A0}}$ a partir de (\ref{lambdaSA0}) s\'{o}%
lo necesitamos calcular $\Lambda ^{\ast }(T^{\ast })$ usando (\ref%
{lambdacalc}) a un valor de $T^{\ast }$. De hecho, calculamos num\'{e}%
ricamente $\Lambda ^{\ast }(T^{\ast })$ en el intervalo $1\leq z\leq 100$
usando diferentes valores de $T^{\ast }$ para asegurarnos que la relaci\'{o}%
n lineal (\ref{lambdaSA0}) sea valida como una buena aproximaci\'{o}n. Una
vez que confirmamos que se cumple el comportamiento lineal, calculamos el
valor final de $s_{\text{A0}}$ como el promedio de dos valores usando $%
\Lambda ^{\ast }(T^{\ast }=$0.5$)$ y $\Lambda ^{\ast }(T^{\ast }=$5.0$)$ en (%
\ref{lambdaSA0}).

Siguiendo este procedimiento para cada modelo con $\mu _{0}^{\ast }$ y $%
s_{0} $ fijos y repiti\'{e}ndolo para 0.3$\leq T_{0}^{\ast }<$ 10.0 en pasos
de 0.05, calculamos $s_{\text{A0}}=s_{\text{A}}(T_{0}^{\ast })$. Los valores
que usamos para la intensidad del momento dipolar son los siguientes: $\mu
_{0}^{\ast }=$0.25, 0.5, 0.75, 0.9 y 1.0 y para la suavidad del kernel: $%
s_{0}=$0.5, 0.7, 0.8, 0.9 y 1.0.

Para representar el comportamiento de $s_{\text{A}}$, usamos formas
funcionales similares a las que usamos para $f_{r}$ y $f_{\epsilon }$, y
encontramos la expresi\'{o}n 
\begin{equation}
s_{\text{A}}=s_{0}+a_{1}/T_{0}^{\ast }+a_{2}/T_{0}^{\ast 2}\text{ .}
\end{equation}

Tenemos 5 valores de $s_{0}$ y 5 de $\mu _{0}^{\ast }$, de manera que
tenemos 50 funciones de estas para calcular $f_{s_{\text{A}}}$ y $f_{s_{%
\text{R}}}$. Evaluamos un total de 9700 integrales num\'{e}ricas. Estos par%
\'{a}metros $a_{i}$ los ajustamos a los resultados num\'{e}ricos, dentro del
intervalo considerado para $T_{0}^{\ast }$, para cada valor de $(\mu
_{0}^{\ast },s_{0})$. La dependencia de los coeficientes $a_{i}$ con $\mu
_{0}^{\ast }$ encontramos que sigue un comportamiento similar al encontrado
en (\ref{EfeEps}) y (\ref{EfeRm}); entonces la expresi\'{o}n que obtenemos
finalmente para $f_{s_{\text{A}}}=s_{\text{A}}/s_{0}$ es 
\begin{equation}
f_{s_{\text{A}}}=1+p_{1}(s_{0})\mu _{0}^{\ast 4}/T_{0}^{\ast
}+p_{2}(s_{0})\mu _{0}^{\ast 6}/T_{0}^{\ast 2}  \label{SA}
\end{equation}%
donde $p_{i}(s_{0})$ son polinomios c\'{u}bicos en $s_{0}$, que cuando los
ajustamos a los resultados num\'{e}ricos, (\ref{SA}) reproduce adecuadamente
la dependencia de $s_{\text{A}}$ con $T_{0}^{\ast }$ y $\mu _{0}^{\ast }$.
Los polinomios resultantes $p_{1}(s_{0})$ y $p_{2}(s_{0})$ los reportamos en
el ap\'{e}ndice C.

\paragraph{Parte repulsiva}

El procedimiento para calcular $f_{s_{\text{R}}}=s_{\text{R}}/s_{0}$,
correspondiente a la regi\'{o}n repulsiva de $\psi (z_{\text{ef}%
};T_{0}^{\ast })$, es el mismo que el desarrollado para $f_{s_{\text{A}}}$.
Para calcular esta funci\'{o}n asumimos que una vez que ha sido definido un
modelo fijando valores para $\mu _{0}^{\ast }$ y $s_{0}$, la rama repulsiva
de $\psi (z_{\text{ef}};T_{0}^{\ast })$ es \textit{conformal} a la misma
rama de $\varphi _{\text{ANC}}(z_{\text{ef}};s_{\text{R0}})$. La relaci\'{o}%
n lineal an\'{a}loga a (\ref{lambdaSA0}) involucra ahora a $s_{\text{R0}}=s_{%
\text{R}}(T_{0}^{\ast })$ y $b^{\ast }(T^{\ast })$ en lugar de $s_{\text{A0}%
} $ y $\Lambda ^{\ast }(T^{\ast })$, donde el volumen de colisi\'{o}n
repulsivo $b^{\ast }(T^{\ast })$ est\'{a} dado por 
\begin{equation}
b^{\ast }(T^{\ast })=1-3\exp \left( -1/T^{\ast }\right) \int_{0}^{1}\,dz\
z^{2}\exp \left[ -\left. \psi (z,T_{0}^{\ast })\right/ T^{\ast }\right] 
\text{.}  \label{becalc}
\end{equation}
La expresi\'{o}n final para $f_{s_{\text{R}}}$ es 
\begin{equation}
f_{s_{\text{R}}}=1+q_{1}(s_{0})\mu _{0}^{\ast 4}/T_{0}^{\ast
}+q_{2}(s_{0})\mu _{0}^{\ast 6}/T_{0}^{\ast 2}+q_{3}(s_{0})\mu _{0}^{\ast
8}/T_{0}^{\ast 3}+q_{4}(s_{0})\mu _{0}^{\ast 10}/T_{0}^{\ast 4}\text{.}
\label{SR}
\end{equation}
donde los polinomios c\'{u}bicos $q_{i}$ los reportamos en el ap\'{e}ndice C.

\paragraph{C\'{a}lculo de $B(T)$ con la aproximaci\'{o}n 2-$s$}

Tenemos entonces el potencial efectivo del GSM con un kernel de profundidad $%
\epsilon _{0}$, di\'{a}metro $r_{\text{m0}}$ y suavidad $s_{0}$ y con un
momento dipolar $\mu _{0}^{\ast }$. El correspondiente potencial efectivo
tiene la forma de un ANC con par\'{a}metros efectivos apropiados: 
\begin{equation}
u_{\text{ef}}(z_{\text{ef}})=\varphi _{\text{ANC}}(z_{\text{ef}};\epsilon
,s_{\text{R}},s_{\text{A}})  \label{uANCeff}
\end{equation}
donde $\varphi _{\text{ANC}}$ est\'{a} dado por (\ref{potencialANC}), $%
\epsilon =\epsilon _{0}f_{\epsilon }$, $s_{\text{R}}=s_{0}f_{s_{\text{R}}}$, 
$s_{\text{A}}=s_{0}f_{s_{\text{A}}}$ y $z_{\text{ef}}=r/r_{\text{m}}$ con $%
r_{\text{m}}=r_{\text{m0}}f_{r}$ y las funciones $f$ est\'{a}n dadas por (%
\ref{EfeEps}), (\ref{EfeRm}), (\ref{SR}) y (\ref{SA}). La diferencia m\'{a}s
importante de este $u_{\text{ef}}$ comparado con uno para sistemas no
polares es que sus par\'{a}metros efectivos dependen de la temperatura. Este
potencial reproduce las propiedades termodin\'{a}micas del gas diluido GSM y
lo podemos usar para estimar valores de las propiedades termodin\'{a}micas
en otras regiones. En particular, el segundo coeficiente virial del gas GSM, 
$B_{\text{GSM}}^{\ast }(T_{0}^{\ast })$, debe ser igual al coeficiente
virial ANC evaluado con par\'{a}metros efectivos, i.e., 
\begin{equation}
B_{\text{GSM}}^{\ast }(T_{0}^{\ast };s_{0},\mu _{0}^{\ast })=B_{\text{ANC}%
}^{\ast }(T_{\text{ef}}^{\ast };f_{r},s_{\text{R}},s_{\text{A}})
\label{BGSM}
\end{equation}
donde de (\ref{Bstar2}), (\ref{be}) y (\ref{lambda}),

\begin{eqnarray}
B_{\text{ANC}}^{\ast }(T_{\text{ef}}^{\ast }) &=&f_{r}^{3}\left\{ b^{\ast
}(T_{\text{ef}}^{\ast },s_{\text{R}})\exp \left( 1/T_{\text{ef}}^{\ast
}\right) \right.  \notag \\
&&\left. -\Lambda ^{\ast }(T_{\text{ef}}^{\ast },s_{\text{A}})\left[ \exp
\left( 1/T_{\text{ef}}^{\ast }\right) -1\right] \right\} \text{,}
\label{Befff}
\end{eqnarray}
aqu\'{\i} $T_{\text{ef}}^{\ast }=kT/\epsilon =T_{0}^{\ast }/f_{\epsilon }$
con $b^{\ast }$ y $\Lambda ^{\ast }$ definidas en (\ref{be}) y (\ref{lambda}%
).

\begin{figure}[h]
        \begin{center}
        \includegraphics[width=.8\hsize]{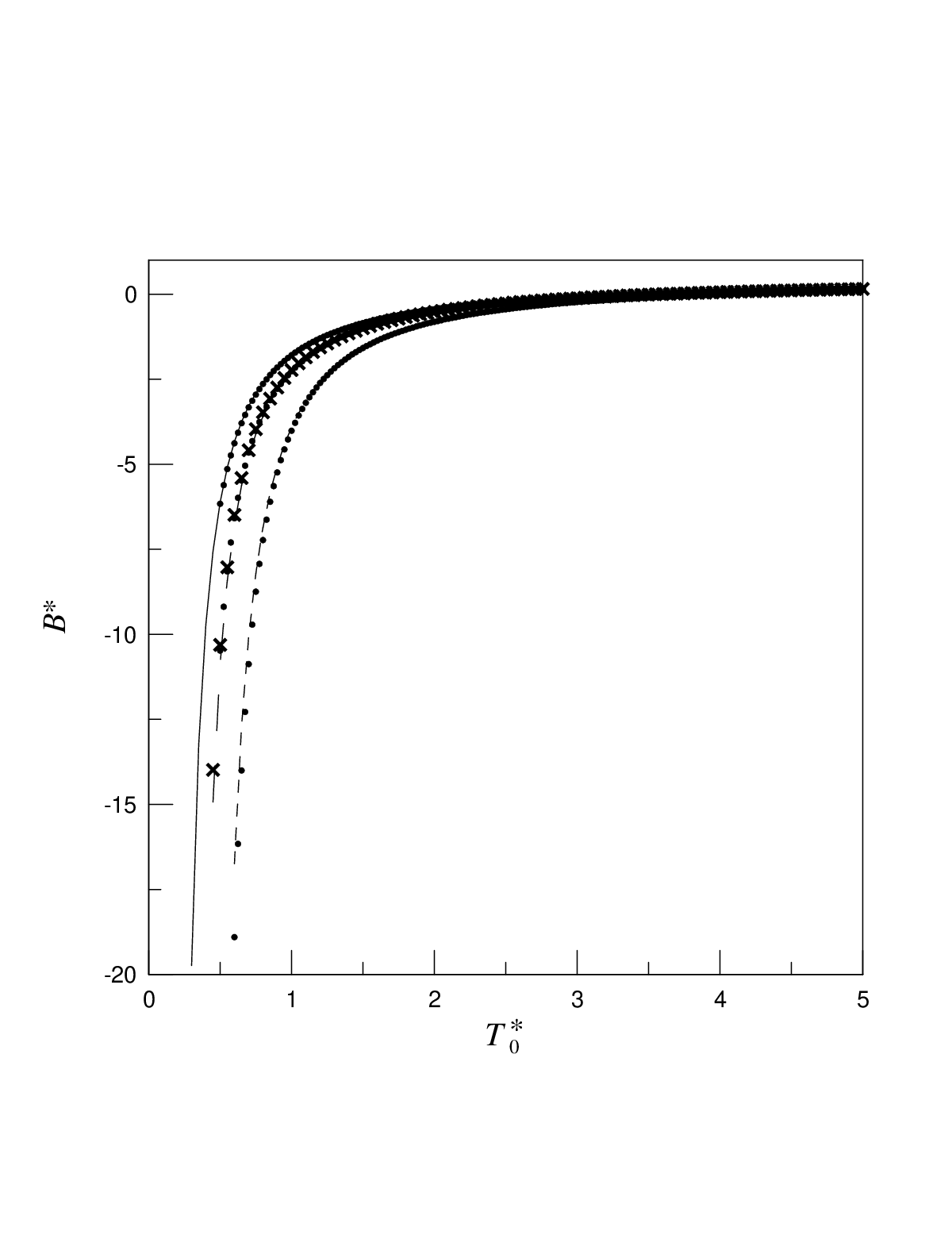}
        \end{center}
        \caption{%
Prueba
de la predicci\'{o}n de segundo coeficiente virial $B^{\ast }(T_{0}^{\ast })$
para el potencial de Stockmayer ordinario ($s_{0}=s_{\text{LJ}}$, $L^{\ast
}=0$) para varios momentos dipolares usando la aproximaci\'{o}n 2-$s$. Los
puntos corresponden a los valores num\'{e}ricos de C. Vega et al. 
\protect\cite{carloses} Los casos que se muestran son: $\protect\mu %
_{0}^{\ast }=$0.0 (l\'{\i}nea continua), $\protect\mu _{0}^{\ast }=$0.841
(segmentos largos) y $\protect\mu _{0}^{\ast }=$1.189 (segmentos cortos).
Las cruces indican la predicci\'{o}n realizada usando los par\'{a}metros
calculados anal\'{\i}ticamente para el caso, $\protect\mu _{0}^{\ast }=$%
0.841, Ecs. (\protect\ref{epsSMteo}) y (\protect\ref{rmSMteo}). En la escala
de esta gr\'{a}fica no se notan las diferencias cuando comparamos con los
resultados de la integraci\'{o}n directa siguiendo el procedimiento de
Hirschfelder \textit{et al}.\protect\cite{Hirsch}.
                     }%
       \label{FIGBSM}
\end{figure}

La ecuaci\'{o}n (\ref{Befff}) --con la aproximaci\'{o}n de 2-$s$ incorporada
en (\ref{SR}) y (\ref{SA})-- reproduce de manera bastante precisa el segundo
coeficiente virial del fluidos de SM ordinario en el intervalo 0.0$\leq \mu
_{0}^{\ast }\leq $ 1.0. En la Fig. \ref{FIGBSM} se muestra la gr\'{a}fica de 
$B_{\text{GSM}}^{\ast }$ contra $T_{0}^{\ast }$ --donde $B_{\text{GSM}%
}^{\ast }$ expl\'{\i}citamente est\'{a} dado en la ecuaci\'{o}n (\ref{BGSM}%
)-- para el fluido de SM ordinario (i.e. con $s_{0}=s_{\text{LJ}}=$1.1315)
comparada con los resultados num\'{e}ricos que C. Vega \textit{et al}
obtuvieron recientemente para tres momentos dipolares. \cite{carloses} Esta
figura muestra un muy buen acuerdo entre el modelo ANC de 2-$s$ (\ref{BGSM})
y los resultados num\'{e}ricos. La Fig. \ref{FIGBSM} tambi\'{e}n muestra que
podemos usar este modelo para extrapolar valores de $B(T)$ en regiones bajas
de temperatura en las que no hay datos disponibles as\'{\i} como tambi\'{e}n
para interpolar valores arbitrarios de $\mu _{0}^{\ast }$ dentro del
intervalo considerado.
%
%

\subsubsection{La forma del potencial efectivo: aproximaci\'{o}n 1-$s$}

\paragraph{Aproximaci\'{o}n 1-$s$ anal\'{\i}tica}

Los potenciales modelo esf\'{e}ricos tales como (\ref{uANCeff}) deber\'{\i}%
an ser \'{u}tiles en la caracterizaci\'{o}n de fluidos GSM usando
simulaciones computacionales o teor\'{\i}as de ecuaciones integrales. Otras
aplicaciones tienen que ver con las sustancias reales, donde la teor\'{\i}a
ANC ha mostrado ser de gran utilidad para generar par\'{a}metros moleculares
que son confiables para esas sustancias. Sin embargo, en el \'{u}ltimo caso
la precisi\'{o}n que se tiene en los datos termodin\'{a}micos m\'{a}s
recientes, no permite distinguir entre las aproximaciones de 2-$s$ y la de 1-%
$s$. Teniendo esto en cuenta, desarrollamos tambi\'{e}n una aproximaci\'{o}n
al potencial GSM\ usando s\'{o}lo una suavidad. En este caso, debido a que
la energ\'{\i}a y el di\'{a}metro efectivos, $\epsilon $ y $r_{\text{m}}$,
son independientes de la forma del potencial, estas cantidades son las
mismas que en el modelo de 2-$s$; pero ahora tenemos la misma suavidad
efectiva, $s=s_{0}f_{s}$ para los lados atractivo y repulsivo de $u_{\text{ef%
}}$ . Aqu\'{\i} mostramos el procedimiento b\'{a}sico para calcular la funci%
\'{o}n $f_{s}$ m\'{a}s simple que cumpla con su prop\'{o}sito. Solo
necesitamos calcular una suavidad efectiva $s_{\text{ef}}$ para cada sistema
--definido por un par de valores $s_{0}$ y $\mu _{0}^{\ast }$. Para hacer
esto recurrimos al hecho de que la temperatura de Boyle, $T_{0\text{B}%
}^{\ast }=kT_{\text{B}}/\epsilon _{0}$, de un sistema ANC es una funci\'{o}n
conocida de su suavidad; i.e., $T_{0\text{B}}^{\ast }=T_{0\text{B}}^{\ast
}(s_{\text{ef}})$. \cite{del Rio 1998 III} De esta manera, el conocer $T_{0%
\text{B}}^{\ast }$ para un sistema dado GSM --que est\'{a} disponible a
partir del modelo de 2-$s$-- equivale a conocer su suavidad $s$ a $%
T_{0}^{\ast }=T_{0\text{B}}^{\ast }$. Una vez que aplicamos este
procedimiento a un amplio conjunto de sistemas GSM con diferentes valores de 
$s_{0}$ y $\mu _{0}^{\ast }$, encontramos la funci\'{o}n $s(s_{0},\mu
_{0}^{\ast },T_{0}^{\ast })$ que sigue un comportamiento muy simple. Luego
si $s=s_{0}f_{s}$, la funci\'{o}n $f_{s}$ la podemos escribir como 
\begin{equation}
f_{s}=1+A(s_{0})\left. \mu _{0}^{\ast 4}\right/ T_{0}^{\ast }  \label{oneS}
\end{equation}
donde $A(s_{0})=$0.44784$-$0.88078 $s_{0}+$0.44504 $s_{0}^{2}\simeq $0.44$($1%
$-s_{0})^{2}$. La funci\'{o}n $f_{s}$ es creciente con el momento dipolar
cuando $s_{0}\lesssim $1.026. Esta es una condici\'{o}n usual en las mol\'{e}%
culas reales. Las consecuencias termodin\'{a}micas de los cambios de forma
est\'{a}n descritas en la siguiente secci\'{o}n.

\paragraph{C\'{a}lculo de $B(T)$ con la aproximaci\'{o}n 1-$s$ anal\'{\i}tica%
}

En la aproximaci\'{o}n de 1-$s$, el segundo coeficiente virial del fluido
GSM, al igual que en la Ec. (\ref{BGSM}), tambi\'{e}n est\'{a} dado en t\'{e}%
rminos de $B_{\text{ANC}}^{\ast }$. A partir de la Ec. (\ref{Bstar}), $B_{%
\text{ANC}}^{\ast }$ ahora sigue la siguiente relaci\'{o}n\ de 1-$s$, 
\begin{equation}
B_{\text{ANC}}^{\ast }(T_{\text{ef}}^{\ast })=f_{r_{\text{m}}}^{3}\left[
1-s+sB_{\text{1}}^{\ast }(T_{\text{ef}}^{\ast })\right]  \label{BeffOneS}
\end{equation}
En la Fig. \ref{FIGBSM} mostramos tambi\'{e}n resultados para $B_{\text{GSM}%
}^{\ast }(T^{\ast })$ basados en el modelo de 1-$s$. En los casos que se
consideran en esta figura no observamos una diferencia significativa entre
los modelos de $1$-s y el de $2$-s; por lo que podemos decir que el de $1$-s
constituye una muy buena predicci\'{o}n de $B(T)$.

\paragraph{Aproximaci\'{o}n $1$-s num\'{e}rica}

Usando la aproximaci\'{o}n $1$-s que desarrollamos arriba podemos para
representar muy bien los resultados de $B_{\text{GSM}}^{\ast }(T^{\ast })$
para distintos modelos; lo cual por s\'{\i} solo contituye una confirmaci%
\'{o}n de que la aproximaci\'{o}n es exelente. Sin embargo, tambi\'{e}n
desarrollamos una aproximaci\'{o}n num\'{e}rica para verificar que en efecto
la aproximaci\'{o}n $1$-s anal\'{\i}tica es correcta. Para esto asumimos que
la dependencia de $s$ es similar a la de $s_{\text{A}}$ \'{o} $s_{\text{R}}$%
, i.e., 
\begin{equation}
s_{\text{ef}}(s_{0},\mu _{0}^{\ast },T_{0}^{\ast })=s_{0}(1+j(s_{0})\mu
_{0}^{\ast 4}/T_{0}^{\ast }+(k(s_{0})\mu _{0}^{\ast 6}+l(s_{0})\mu
_{0}^{\ast 8})/T_{0}^{\ast 2})\text{;}  \label{s numerica}
\end{equation}
donde $j$, $k$ y $l$ son polinomios cuadr\'{a}ticos en $s_{0}$, que
reportamos en un Ap\'{e}ndice al final de este trabajo; as\'{\i} como tambi%
\'{e}n los detalles del c\'{a}lculo. La aproximaci\'{o}n $1$-s num\'{e}rica
a la funci\'{o}n $f_{s}$ la obtenemos sabiendo que $s_{\text{ef}}=s_{0}f_{s}$%
.

\section{Termodin\'{a}mica de un fluido GSM}

\subsection{Efectos de la forma en $B(T)$ de un modelo GSM}

\begin{figure}[h]
        \begin{center}
        \includegraphics[width=.8\hsize]{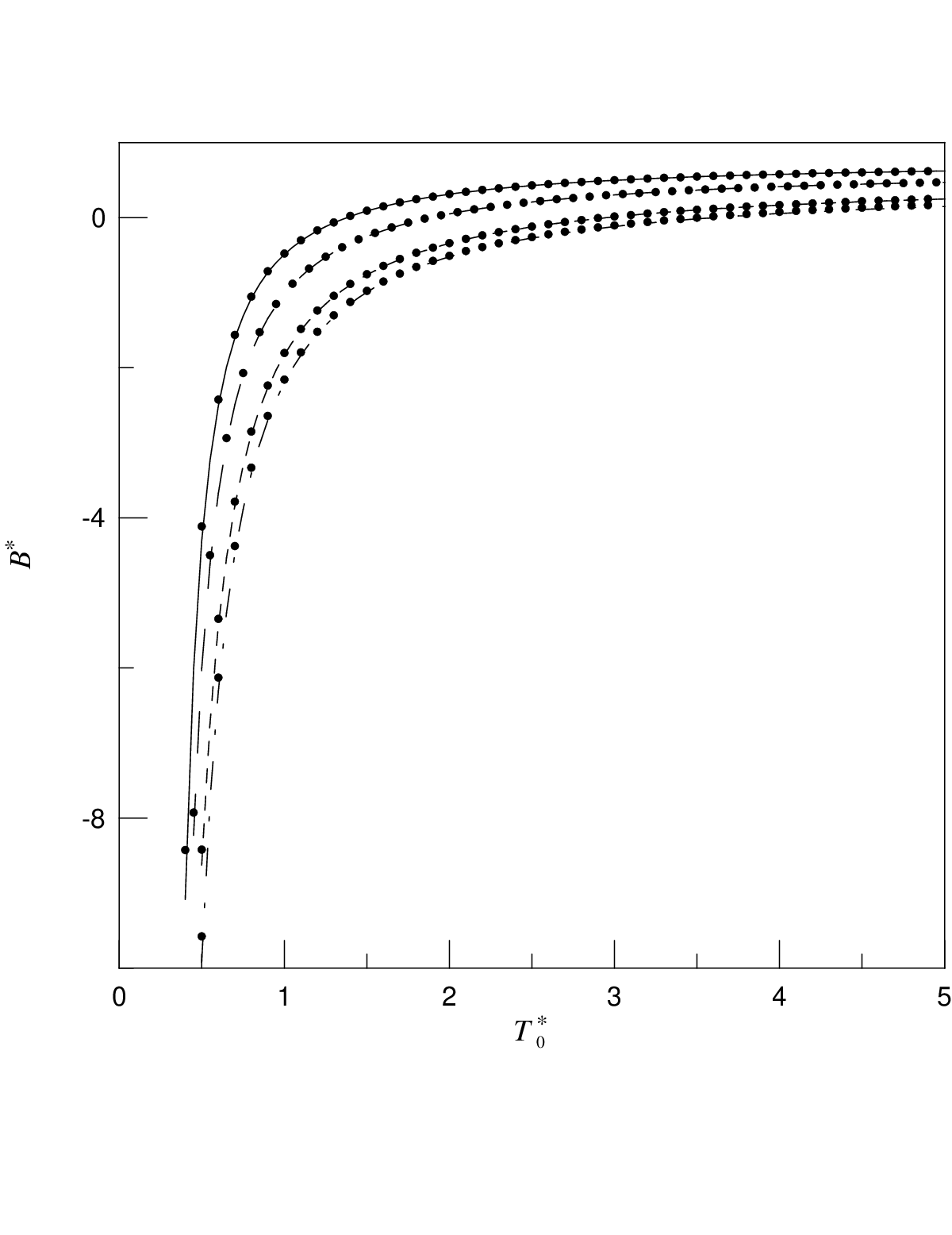}
        \end{center}
        \caption{%
Efecto
de un kernel de suavidad variable, $s_{0}$, en el segundo coeficiente virial 
$B^{\ast }(T_{0}^{\ast })$ para el potencial de Stockmayer generalizado ($%
L^{\ast }=0$) con momento dipolar $\protect\mu _{0}^{\ast }=0.8$. Las l\'{\i}%
neas corresponden a diferentes valores de la suavidad del kernel del GSM: $%
s_{0}=0.5$ (arriba, l\'{\i}inea continua), $s_{0}=0.7$ (segmentos largos), $%
s_{0}=1.0$ (segmentos cortos) y $s_{0}=1.1315$ (abajo, segmentos y puntos).
Las l\'{\i}neas corresponden a la aproximaci\'{o}n 1-$s$. La aproximaci\'{o}%
n 2-$s$ es indistinguible de la 1-$s$ en la escala de esta gr\'{a}fica. Los
puntos representan resultados de integraci\'{o}n num\'{e}rica para fines de
comparaci\'{o}n.
                     }%
       \label{FIGBmu08}
\end{figure}

Nuestro modelo tambi\'{e}n nos permite predecir $u_{\text{ef}}$ y el
correspondiente segundo coeficiente virial $B(T)$ para las mol\'{e}culas GSM
que tengan un kernel diferente al de LJ. En la Fig. \ref{FIGBmu08}
graficamos $B_{\text{GSM}}^{\ast }$ contra $T_{0}^{\ast }$ usando $\mu
_{0}^{\ast }=$0.8 y cuatro diferentes valores para la suavidad, a saber, $%
s_{0}=$1.1315, 1.0, 0.7 y 0.5. El primero de estos valores --el m\'{a}s
grande-- corresponde al kernel de LJ y da lugar a la mayor temperatura de
Boyle de los cuatro en el grupo. En esta figura se hace una comparaci\'{o}n
con resultados de integraci\'{o}n num\'{e}rica y podemos observar que hay un
acuerdo excelente. Notemos que a medida que la suavidad se incrementa, $B(T)$
se hace m\'{a}s negativo.
%
%

\subsection{Punto cr\'{\i}tico de un fluido GSM}

En esta secci\'{o}n usaremos el potencial modelo que hemos desarrollado para
calcular el efecto de la interacci\'{o}n dipolar sobre las cantidades que
definen el punto critico, a saber, la temperatura y densidad cr\'{\i}ticas.
Nos concentraremos en estudiar el modelo SM. Sin embargo, el m\'{e}todo que
aqu\'{\i} desarrollamos es v\'{a}lido para cualquier fluido dipolar de la
familia GSM. Como aproximaci\'{o}n inicial, asumimos que el potencial modelo
es independiente de la densidad; sin embargo, nos introducimos en el an\'{a}%
lisis de los efectos de densidad en el potencial.

\subsubsection{Temperatura cr\'{\i}tica}

Aqu\'{\i} aprovechamos los resultados de simulaci\'{o}n de din\'{a}mica
molecular del modelo ANC que obtuvimos previamente. La relaci\'{o}n que hay
entre la temperatura cr\'{\i}tica $T_{\text{c}}^{\ast }$ (Ec. (\ref{TcAnc}%
))y los par\'{a}metros ANC\ para el fluido de SM, es como sigue, 
\begin{eqnarray}
T_{\text{c}}^{\ast } &=&kT_{\text{c}}/\epsilon  \notag \\
&=&kT_{\text{c}}/\epsilon _{\text{0}}f_{\epsilon }  \notag \\
&=&T_{0\text{c}}^{\ast }/f_{\epsilon }(T_{0\text{c}}^{\ast },\mu _{0}^{\ast
},s_{0})\text{.}  \label{Ec for Tc}
\end{eqnarray}
La Ec. (\ref{TcAnc}) nos permite calcular la temperatura cr\'{\i}tica de un
fluido ANC de suavidad $s$ \footnote{%
La Ec. (\ref{Ec for Tc}) tiene la misma forma que la Ec. (\ref{Tef}),
discutida en la secci\'{o}n anterior. El lector incluso puede comprobar f%
\'{a}cilmente que (\ref{TcAnc}) representa una temperatura cr\'{\i}tica en
el espacio de coordenadas efectivas $(\rho _{\text{ef}}^{\ast },T_{\text{ef}%
}^{\ast })$.}. Al sustituir (\ref{TcAnc}) en (\ref{Ec for Tc}) obtenemos una
ecuaci\'{o}n polinomial de tercer orden en $T_{0\text{c}}^{\ast }$ cuyas
soluciones reales conducen a la relaci\'{o}n deseada 
\begin{equation}
T_{0\text{c}}^{\ast }=T_{0\text{c}}^{\ast }(\mu _{0}^{\ast },s_{0})\text{,}
\label{TcCero}
\end{equation}
donde $T_{0\text{c}}^{\ast }(\mu _{0}^{\ast },s_{0})$ es una funci\'{o}n
explicita aunque complicada de $\mu _{0}^{\ast 2}$ cuyos coeficientes
dependen de $s_{0}$. La \ Ec. (\ref{TcCero}) predice que, para un kernel con
suavidad constante $s_{0}$, la temperatura cr\'{\i}tica $T_{0\text{c}}^{\ast
}$ se incrementa con $\mu _{0}^{\ast }$; lo cual es de esperarse ya que el
potencial se hace m\'{a}s intenso con $\mu _{0}^{\ast }$.

\begin{figure}[h]
        \begin{center}
        \includegraphics[width=.8\hsize]{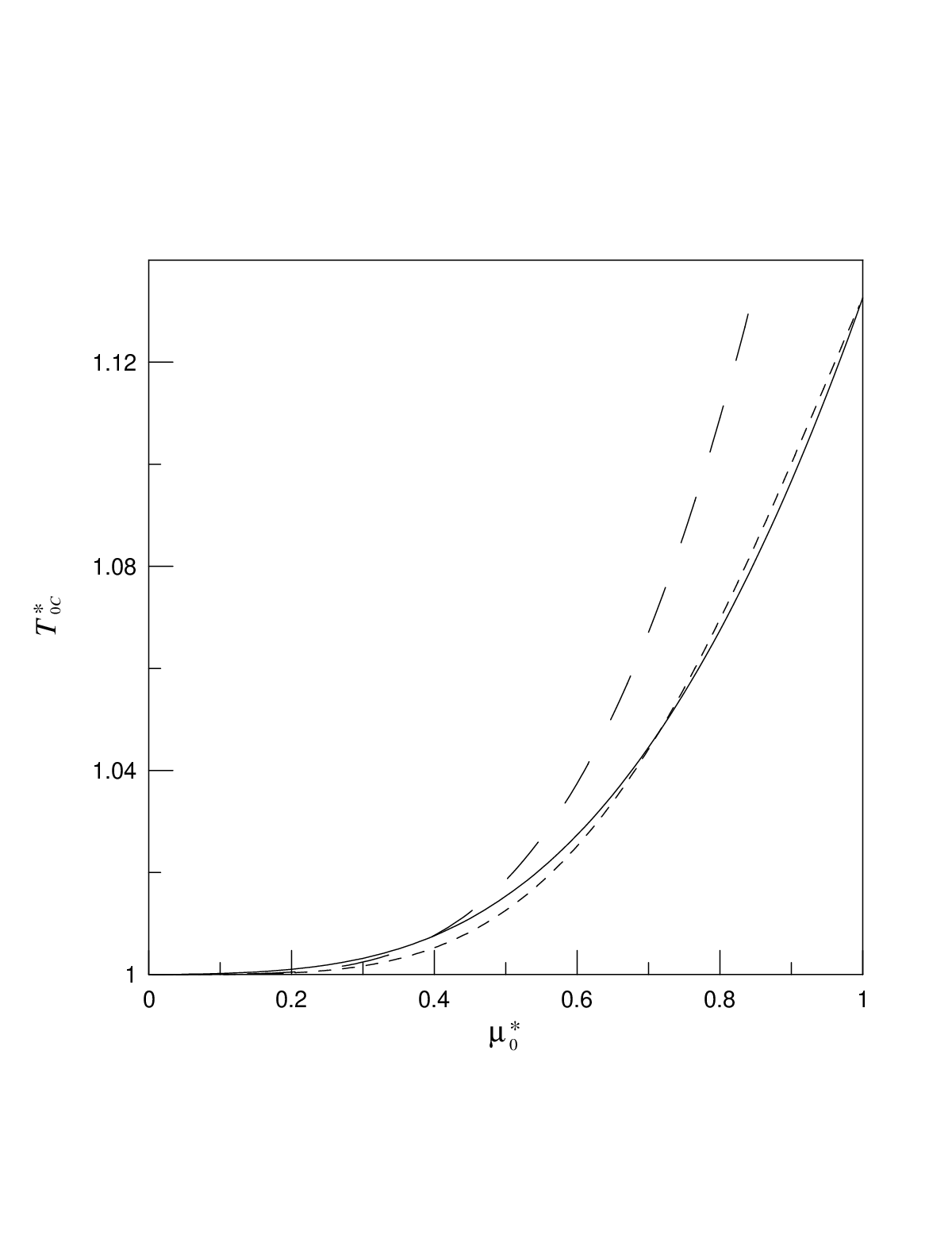}
        \end{center}
        \caption{%
Temperatura cr\'{\i}tica de un fluido de
Stockmayer ($s_{\text{0}}=s_{\text{LJ}}$), como funci\'{o}n del momento
dipolar $\protect\mu _{0}^{\ast }$. La predicci\'{o}n realizada con el
potencial efectivo desarrollado en este trabajo (l\'{\i}nea de segmentos
largos) la comparamos con un ajuste a los datos de simulaci\'{o}n (l\'{\i}%
nea continua) de van Leeuwen \protect\cite{Leeuwen 1994 mp}. Tambi\'{e}n se
muestra una comparaci\'{o}n con la predicci\'{o}n realizada con un potencial
efectivo que depende de la densidad (l\'{\i}nea de segmentos cortos).
                     }%
       \label{FIGTcCero}
\end{figure}

En la Fig. \ref{FIGTcCero} se ilustra el comportamiento que hemos descrito
para $s_{0}=s_{\text{LJ}}$ donde este modelo lo comparamos con los
resultados de simulaci\'{o}n en el ensamble de Gibbs (GEMC) que obtuvo van
Leeuwen y sus colaboradores para un fluido SM con $0\leq \mu _{0}^{\ast
}\leq 1$ \cite{Leeuwen 1993, Leeuwen 1994 mp, Leeuwen 1994 fpe}. Debido a
que los resultados te\'{o}rico y de simulaci\'{o}n se obtuvieron usando
diferentes modelos, no necesariamente tienen que coincidir en $\mu
_{0}^{\ast }=0$. Para evidenciar que ambos resultados siguen la misma
tendencia, normalizamos los resultados de tal manera que --para fines de
comparaci\'{o}n-- ambos empiezan en $T_{0\text{c}}^{\ast }=$1.0. Como se
muestra en la figura, la tendencia se\~{n}alada por el potencial efectivo est%
\'{a} confirmada por los resultados de simulaci\'{o}n --a pesar de que el
modelo desarrollado en este trabajo estima excesivamente el efecto de un
momentos dipolares de intensidad progresivamente mayor. 
%
%

\paragraph{Par\'{a}metros que dependen de la densidad}

En el caso que aqu\'{\i} estamos analizando encontramos un error m\'{a}ximo
en la estimaci\'{o}n de la temperatura cr\'{\i}tica de alrededor de un 4\%
presente en el extremo superior del intervalo considerado. Esta sobreestimaci%
\'{o}n la podemos corregir si consideramos que la densidad tiene efecto en
los par\'{a}metros efectivos. En otras palabras, debemos usar un potencial
cuyos par\'{a}metros --adem\'{a}s de depender de la temperatura-- tambi\'{e}%
n incluyan un factor que dependa de la densidad. El uso de potenciales que
dependen de la densidad y en general del estado termodin\'{a}mico, es
delicado \cite{Louis 2002, Stillinger}; aqu\'{\i} nuestro inter\'{e}s
consiste en exprorar la posibilidad de incluir los efectos de la densidad en
la descripci\'{o}n del estado cr\'{\i}tico de un fluido GSM.

Para calcular correctamente la temperatura cr\'{\i}tica de un fluido GSM,
incluimos los efectos de la densidad escribiendo el par\'{a}metro efectivo
ANC que define a la profundidad como el producto $f_{\epsilon }$\ $%
j_{\epsilon }$, siendo $j_{\epsilon }=1+c_{\epsilon }\rho _{0}^{\ast }\mu
_{0}^{\ast 4}$ el factor de correcci\'{o}n \footnote{%
La dependencia en $\mu _{0}^{\ast 4}$ que asumimos para $j_{\epsilon }$ la
conjeturamos al analizar la tendencia que sigue $T_{0\text{c}}^{\ast }(\mu
_{0}^{\ast })/T_{0\text{cL}}^{\ast }(\mu _{0}^{\ast })$, donde $T_{0\text{c}%
}^{\ast }(\mu _{0}^{\ast })$ y $T_{0\text{cL}}^{\ast }(\mu _{0}^{\ast })$,
son respectivamente la temperatura cr\'{\i}tica seg\'{u}n nuestro modelo,
Ec. (\ref{TcCero}) y seg\'{u}n los resultados de simulaci\'{o}n de van
Leeuwen y sus colaboradores \cite{Leeuwen 1993,Leeuwen 1994 mp,Leeuwen 1994
fpe}.} que depende de la densidad.

Usando el producto $f_{\epsilon }$\ $j_{\epsilon }$ como par\'{a}metro
efectivo, la expresi\'{o}n equivalente a (\ref{Ec for Tc}) ahora es, 
\begin{equation}
T_{\text{cc}}^{\ast }=T_{0\text{cc}}^{\ast }/(f_{\epsilon }\ j_{\epsilon })%
\text{,}  \label{Ec for Tc Corr}
\end{equation}
donde usamos una ``c''\ adicional en el sub\'{\i}ndice como nemot\'{e}cnica
de ``correcci\'{o}n''. Podemos calcular el valor de la constante $%
c_{\epsilon }$ para un fluido SM al requerir que la temperatura cr\'{\i}tica 
$T_{0\text{cc}}^{\ast }$ (Ec. (\ref{Ec for Tc Corr})) sea una buena
aproximaci\'{o}n a los resultados de simulaci\'{o}n de Leeuwen \textit{et al.%
} \cite{Leeuwen 1993,Leeuwen 1994 mp,Leeuwen 1994 fpe}. La temperatura cr%
\'{\i}tica de un fluido SM es $T_{\text{cc}}^{\ast }(s_{\text{LJ}})$ (Ec. (%
\ref{TcAnc})). La constante $c_{\epsilon }$ resulta ser $-$0.22. \footnote{%
Al realizar un estudio de simulaci\'{o}n computacional en el ensamble de
Gibbs (GEMC) \cite{Guzman 2001, Eloy Tesis}, se encontraron tambi\'{e}n
correlaciones entre los par\'{a}metros efectivos de un modelo ANC ($\epsilon 
$, $r_{\text{m}}$ y $s$) y sus valores cr\'{\i}ticos. Usando esos resultados
de simulaci\'{o}n GEMC (Ec. \ref{CorrANC_T}), la constante $c_{\epsilon }$
resulta ser $-$0.20. En general las correlaciones encontradas con el m\'{e}%
todo GEMC difieren de las que encontramos nosotros (Ej. Ec. (\ref{TcAnc})), b%
\'{a}sicamente debido a que la simulaci\'{o}n de din\'{a}mica molecular
requiere truncar la parte atractiva del potencial de interacci\'{o}n (cutoff
en $r=$3.5$r_{\text{m}}$). Este corte da lugar a una subestimaci\'{o}n de la
temperatura cr\'{\i}tica \cite{Frenkel libro}.} Siguiendo el mismo
procedimiento que usamos antes de incluir la correcci\'{o}n debida a la
densidad, sustituimos (\ref{TcAnc}) en (\ref{Ec for Tc Corr}) obteniendo
nuevamente una ecuaci\'{o}n polinomial de tercer orden en $T_{0\text{c}%
}^{\ast }$ cuya soluci\'{o}n requiere conocer el valor de la densidad cr%
\'{\i}tica de un fluido de SM como funci\'{o}n del momento dipolar, $\rho _{0%
\text{c}}^{\ast }(\mu _{0}^{\ast })$ \cite{Leeuwen 1994 mp} \footnote{%
La funci\'{o}n que cita van Leeuwen y sus colaboradores en la Ref. \cite%
{Leeuwen 1994 mp} para $\rho _{0\text{c}}^{\ast }(\mu _{0}^{\ast })$ no
corresponde a los datos que estos autores encontraron usando simulaci\'{o}n
computacional. Usamos los resultados de simulaci\'{o}n citados para ajustar
la siguiente expresi\'{o}n 
\begin{equation}
\rho _{0\text{c}}^{\ast }=0.42733+0.011288\mu _{0}^{\ast 2}-0.0047144\mu
_{0}^{\ast 2}\text{.}
\end{equation}
Esta es la expresi\'{o}n que usamos para calcular la temperatura cr\'{\i}%
tica definida por la Ec. (\ref{TcCero corr}).}. La soluci\'{o}n que
obtuvimos la escribimos como es 
\begin{equation}
T_{0\text{cc}}^{\ast }=T_{0\text{cc}}^{\ast }(\mu _{0}^{\ast },s_{0})\text{,}
\label{TcCero corr}
\end{equation}
En la Fig. \ref{FIGTcCero} --para el caso SM ($s_{0}=s_{\text{LJ}}$)--
incluimos el resultado que obtuvimos al usar un potencial que depende de la
densidad (l\'{\i}nea continua). Usando la constante $c_{\epsilon }$ adecuada
podemos lograr la aproximaci\'{o}n deseada.

\subsubsection{Densidad cr\'{\i}tica}

Para calcular la evoluci\'{o}n de la densidad cr\'{\i}tica con $\mu
_{0}^{\ast }$, procedemos de manera similar al caso de la temperatura cr%
\'{\i}tica. Sabemos que el volumen cr\'{\i}tico, $V_{\text{c}}$, de un
fluido ANC, queda bien representado por la siguiente expresi\'{o}n, \cite%
{Del Rio 2001} 
\begin{equation}
V_{\text{c}}=\text{1.748\ }r_{\text{m0}}^{3}\text{.}  \label{Vc anc}
\end{equation}
An\'{a}logamente a la expresi\'{o}n que obtuvimos para la relaci\'{o}n entre 
$T_{\text{c}}^{\ast }$ y los par\'{a}metros ANC, Ec.(\ref{Ec for Tc}); en el
caso de la densidad cr\'{\i}tica tenemos que, 
\begin{eqnarray}
\rho _{\text{c}}^{\ast } &=&\rho _{\text{c}}r_{\text{m}}^{3}  \notag \\
&=&\rho _{\text{c}}r_{\text{m0}}^{3}f_{r}^{3}  \notag \\
&=&\rho _{0\text{c}}^{\ast }f_{r}^{3}(T_{0\text{c}}^{\ast },\mu _{0}^{\ast
},s_{0})\text{,}  \label{Ec for RhoC}
\end{eqnarray}
donde $\rho _{\text{c}}=N/V_{\text{c}}$ ($N$ es el n\'{u}mero de part\'{\i}%
culas, que es una cantidad sin dimensiones), de manera que $\rho _{\text{c}%
}^{\ast }=N/$1.748. Resolvemos la Ec. (\ref{Ec for RhoC}) para $\rho _{0%
\text{c}}^{\ast }$, 
\begin{equation}
\rho _{0\text{c}}^{\ast }=\rho _{0\text{c}}^{\ast }(\mu _{0}^{\ast },s_{0})%
\text{,}  \label{RhocCero}
\end{equation}
para lo cual, aprovechamos la soluci\'{o}n que encontramos en la secci\'{o}n
anterior para $T_{0\text{c}}^{\ast }$.

La soluci\'{o}n de $\rho _{0\text{c}}^{\ast }(\mu _{0}^{\ast },s_{0})$ para
el caso particular de $s_{0}=s_{\text{LJ}}$, la podemos ver en la Fig. \ref%
{FIGRhocCero}, en donde tambi\'{e}n estamos comparando con la parametrizaci%
\'{o}n encontrada por van Leeuwen y sus colaboradores para un fluido SM con 0%
$\leq \mu _{0}^{\ast }\leq $1.0. \cite{Leeuwen 1994 mp} Al igual que en el
caso de $T_{0\text{c}}^{\ast }$, hemos renormalizado los resultados para que
ambos empiecen en $\rho _{\text{c}}^{\ast }=$1.0; pues debido a que hemos
usado diferentes modelos, los resultados no necesariamente coinciden al
principio, pero s\'{\i} siguien la misma tendencia. 
%
 \begin{figure}[h]
        \begin{center}
        \includegraphics[width=.8\hsize]{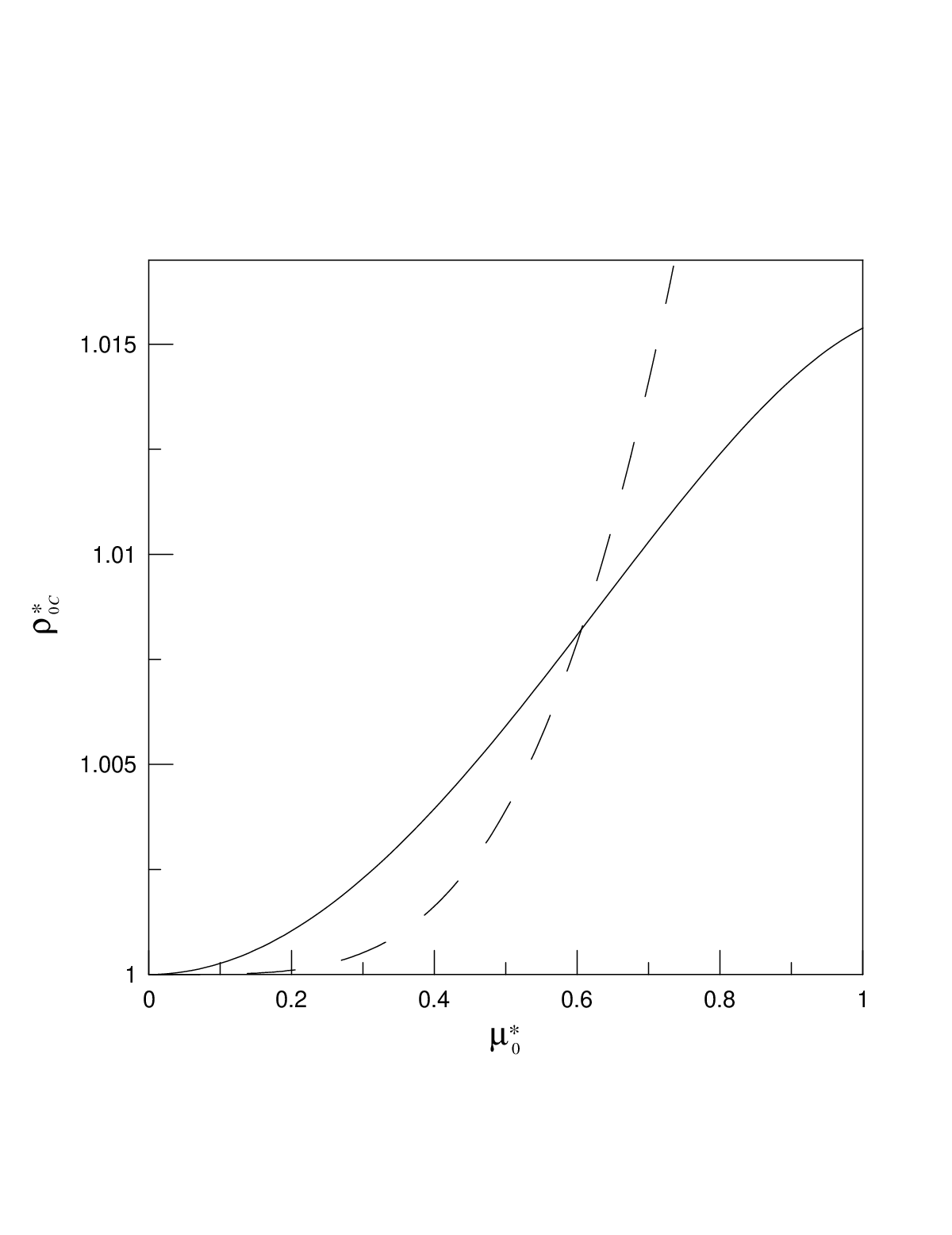}
        \end{center}
        \caption{%
Densidad cr\'{\i}tica de un fluido de Stockmayer como
funci\'{o}n del momento dipolar $\protect\mu _{0}^{\ast }$. Las l\'{\i}neas
tienen el mismo significado que en la Fig. \protect\ref{FIGTcCero}.
                     }%
       \label{FIGRhocCero}
\end{figure}

\subsection{Equilibrio l\'{\i}quido-vapor de un fluido GSM}

En esta secci\'{o}n describimos c\'{o}mo obtener datos del equilibrio l\'{\i}%
quido-vapor (LV) de un fluido GSM. Sabemos que este modelo consiste de una
interacci\'{o}n dipolar sumada a un kernel ANC; de manera que nuestro
objetivo aqu\'{\i} es calcular el efecto de momento dipolar sobre este
kernel y sobre la coexistencia de un fluido de GSM. Mostraremos c\'{o}mo
calcular los datos del equilibrio LV para un fluido GSM a partir de los
datos de coexistencia de un fluido ANC con par\'{a}metros efectivos, i.e., a
partir de un fluido ANC efectivo.

Al igual que en la secci\'{o}n anterior, como una primera aproximaci\'{o}n,
asumimos que el potencial modelo es independiente de la densidad. Sin
embargo, introducimos el efecto de densidad en el potencial en un apartado
al final de esta secci\'{o}n.

Sea $(\rho _{0}^{\ast }=\rho r_{m0}^{3},T_{0}^{\ast }=kT/\epsilon _{0})$ un
punto de coexistencia de un fluido de GSM (con momento dipolar $\mu
_{0}^{\ast }$ y suavidad $s_{0}$) representado en las coordenadas de un
fluido ANC, Ec. (\ref{potencialANC}); donde $r_{m0}$ y $\epsilon _{0}$\ son
respectivamente la posici\'{o}n y profundidad\ del m\'{\i}nimo del potencial
de ANC y $s_{0}$ su suavidad. A cada punto $(\rho _{0}^{\ast },T_{0}^{\ast
}) $ le podemos asociar uno en el espacio de coordenadas efectivas $(\rho _{%
\text{ef}}^{\ast }=\rho r_{\text{m}}^{3},T_{\text{ef}}^{\ast }=kT/\epsilon )$%
, donde $r_{\text{m}}=r_{\text{m}0}f_{r}$ y $\epsilon =\epsilon
_{0}f_{\epsilon }$. Definimos la transformaci\'{o}n de coordenadas, 
\begin{equation}
\rho _{\text{ef}}^{\ast }=\rho _{0}^{\ast }f_{r}^{3}(T_{0}^{\ast },\mu
_{0}^{\ast },s_{0})  \label{rhoEf}
\end{equation}
\begin{equation}
T_{\text{ef}}^{\ast }=T_{0}^{\ast }/f_{\epsilon }(T_{0}^{\ast },\mu
_{0}^{\ast },s_{0})\text{,}  \label{Tef}
\end{equation}
que representa cualquier conjunto de datos de coexistencia de un fluido GSM
con momento dipolar $\mu _{0}^{\ast }$ de tal manera que todas las curvas de
coexistencia coincidir\'{a}n con la de un fluido ANC efectivo. Entonces, a
cada punto de coexistencia del fluido ANC efectivo, $(\rho _{\text{ef}%
}^{\ast }$, $T_{\text{ef}}^{\ast })$, le corresponde uno de coexistencia del
fluido GSM $(\rho _{0}^{\ast }$, $T_{0}^{\ast })$, que lo encontramos
resolviendo simult\'{a}neamente (\ref{rhoEf}) y (\ref{Tef}). Este m\'{e}todo
los podemos usar para todos los valores posibles de $\mu _{0}^{\ast }$
dentro del intervalo de validez de las funciones $f_{\epsilon }(T_{0}^{\ast
},\mu _{0}^{\ast },s_{0})$ y $f_{r}(T_{0}^{\ast },\mu _{0}^{\ast },s_{0})$.

\paragraph{Par\'{a}metros que dependen de la densidad}

Para el caso particular de $s_{\text{0}}=s_{\text{LJ}}$ --que corresponde al
modelo de SM-- la predicci\'{o}n de las densidades de coexistencia del vapor
es bastante buena. Al usar los par\'{a}metros efectivos ANC y un momento
dipolar podemos construir con muy buena aproximaci\'{o}n los datos de
coexistencia de este fluido a partir de los datos de coexistencia de uno de
LJ. Sin embargo, del lado del l\'{\i}quido, la predicci\'{o}n de los datos
de coexistencia excede a los datos de simulaci\'{o}n en promedio por un 5\%.
Podemos mejorar la predicci\'{o}n al incluir el efecto de la densidad
escribiendo el par\'{a}metro efectivo ANC que define a la profundidad como
el producto $f_{\epsilon }$\ $j_{\epsilon }$; donde $j_{\epsilon }$, es la
misma funci\'{o}n que definimos en la secci\'{o}n anterior.

Los valores de coexistencia del fluido de Stockmayer, $(\rho _{0}^{\ast }$, $%
T_{0}^{\ast })$, los obtenemos resolviendo simult\'{a}neamente las Ecs. (\ref%
{rhoEf}) y (\ref{Tef con densidad}). Esta \'{u}ltima ecuac\'{\i}\'{o}n la
usamos en lugar de la Ec (\ref{Tef}). 
\begin{equation}
T_{\text{ef}}^{\ast }=T_{0}^{\ast }/f_{\epsilon }j_{\epsilon }
\label{Tef con densidad}
\end{equation}

\begin{figure}[h]
        \begin{center}
        \includegraphics[width=.8\hsize]{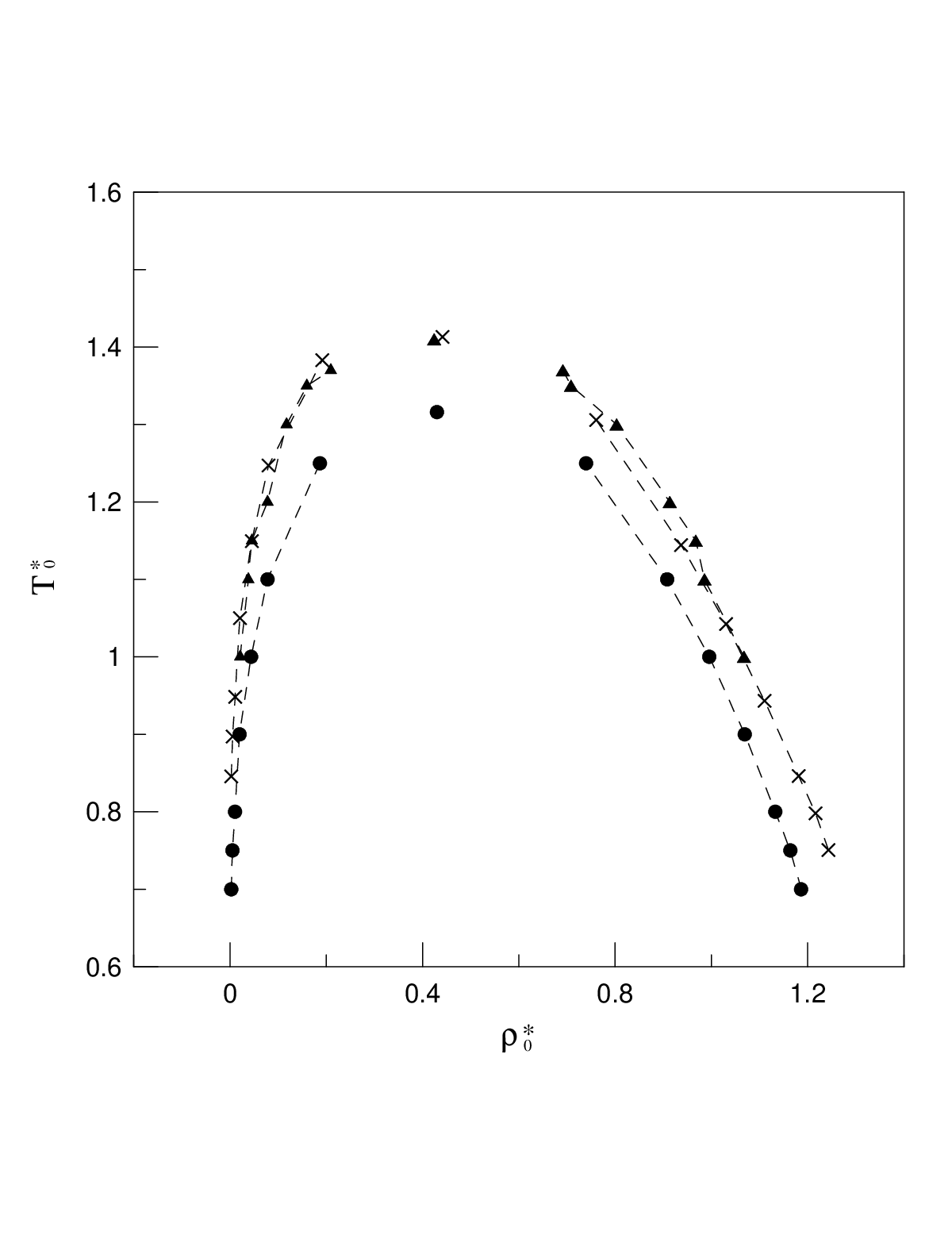}
        \end{center}
        \caption{%
Coexistencia de un fluido de Stockmayer (SM) y de un
Lennard-Jones (LJ). Se muestran los datos de coexistencia obtenidos por
simulaci\'{o}n computacional\ de un fluido de SM (tri\'{a}ngulos) para $%
\protect\mu _{0}^{\ast }=0.841$ \protect\cite{Smit 1989} y tambi\'{e}n de un
fluido de LJ (c\'{\i}rculos) (\protect\cite{Mezei 1992}, \protect\cite%
{Harismiadis 1996}). A partir de los datos LJ y del valor de $\protect\mu %
_{0}^{\ast }$, hacemos la predicci\'{o}n ANC (cruces) para el fluido de SM.
                     }%
       \label{FIGcoexisLJSM}
\end{figure}

La Fig. \ref{FIGcoexisLJSM} representa la coexistencia l\'{\i}quido-vapor
del modelo de SM. Mostramos datos de coexistencia representados en las
coordenadas, $(\rho _{0}^{\ast },T_{0}^{\ast })$, obtenidos de simulaci\'{o}%
n computacional\ de un fluido de SM para $\mu _{0}^{\ast }=$0.841 \cite{Smit
1989}. Mostramos tambi\'{e}n los datos de coexistencia, de un fluido de LJ 
\cite{Mezei 1992, Harismiadis 1996} a partir de los cuales, conociendo el
valor de $\mu _{0}^{\ast }$, hacemos la predicci\'{o}n ANC del fluido SM.
Incluimos el resultado que obtuvimos al usar un potencial que depende tambi%
\'{e}n de la densidad. El valor de la constante $c_{\epsilon }$ es el mismo
que obtuvimos en la secci\'{o}n anterior que resulto ser $-$0.22. Notamos en
la figura que la predicci\'{o}n de los datos de coexistencia es excelente
incluso tambi\'{e}n en el punto cr\'{\i}tico.
%

\subsection{Ecuaci\'{o}n de estado de un fluido GSM}

En la secci\'{o}n anterior mostramos que podemos predecir el equilibrio de
un fluido GSM a partir del equilibrio de un fluido de ANC con par\'{a}metros
efectivos adecuados. En particular obtuvimos el equilibrio de un fluido SM a
partir de los datos de equilibrio de un fluido de LJ y el valor del momento
dipolar correspondiente. De manera que es posible obtener una EOS\ para un
fluido GSM. En esta secci\'{o}n --para el modelo de SM como caso particular
del GSM-- presentamos una ecuaci\'{o}n de estado basada en la ecuaci\'{o}n
de Johnson \textit{et al}. \cite{Jonson 1993, Mulero 1999} con la que
obtenemos propiedades del equilibrio l\'{\i}quido-vapor de un fluido SM.

Johnson \textit{et al. }hicieron una revisi\'{o}n de los par\'{a}metros de
la EOS modificada de Benedict, Webb y Rubin \cite{Nicolas 1979}, que
representa a la presi\'{o}n de un fluido de LJ, $P^{\ast }(\rho _{0}^{\ast
},T_{0}^{\ast })$, como, 
\begin{equation}
P^{\ast }=\rho _{0}^{\ast }T_{0}^{\ast }+\sum_{i=1}^{8}a_{i}\rho _{0}^{\ast
(i+1)}+F\sum_{i=1}^{6}b_{i}\rho _{0}^{\ast (2i+1)}\text{,}  \label{EOS LJ}
\end{equation}
donde $\rho _{0}^{\ast }=\rho \sigma _{0}^{3}$ y $T_{0}^{\ast }=kT/\epsilon
_{0}$ son respectivamente la densidad y temperatura reducidas con los par%
\'{a}metros de longitud y energ\'{\i}a del modelo de LJ Eq. (\ref{potencial
LJ}) y $F=\exp \left( -3\rho _{0}^{\ast 2}\right) $. Los coeficientes $a_{i}$%
\ y $b_{i}$ son funciones sencillas de $T_{0}^{\ast }$ que se pueden
consultar en la fuente original, Ref. \cite{Jonson 1993}.

Sabemos c\'{o}mo evolucionan los par\'{a}metros $\sigma $ y $\epsilon $ al
introducir un momento dipolar; de manera que obtenemos una EOS para un
fluido de SM simplemente al considerar efectivos los par\'{a}metros de LJ en
la Ec. (\ref{EOS LJ}); \textit{i.e.}, la expresi\'{o}n, 
\begin{equation}
P^{\ast }(\rho _{\text{ef}}^{\ast },T_{\text{ef}}^{\ast })\text{,}
\label{EOS SM}
\end{equation}
con $\rho _{\text{ef}}^{\ast }=\rho \sigma ^{3}$ y $T_{\text{ef}}^{\ast
}=kT/\epsilon $, donde $\sigma =\sigma _{0}f_{r}$ y $\epsilon =\epsilon
_{0}f_{\epsilon }$; representa la presi\'{o}n de un fluido de SM. $f_{r}$ es
el mismo factor que hace efectivo a $r_{\text{m}}$. La expresi\'{o}n. (\ref%
{EOS SM}), nos permite encontrar el equilibrio LV de un fluido de SM.

\subsubsection{C\'{a}lculo de la curva de coexistencia}

En esta secci\'{o}n obtenemos la curva de coexistencia LV de un fluido SM
calculada a partir de la EOS dada por la expresi\'{o}n (\ref{EOS SM}). Para
obtener la curva de coexistencia implementamos el m\'{e}todo de las \'{a}%
reas iguales de Maxwell \cite{Quintales 1988} que a continuaci\'{o}n
recordamos brevemente. Para cada modelo SM definido por un valor $\mu
_{0}^{\ast }$, las isotermas subcr\'{\i}ticas de la Ec. (\ref{EOS SM})
forman un \textquotedblleft bucle\textquotedblright\ con un m\'{\i}nimo
relativo (con presi\'{o}n reducida $P_{\text{m}}^{\ast }$) y un m\'{a}ximo
relativo (con presi\'{o}n reducida $P_{\text{M}}^{\ast }$), de tal manera
que para alguna presi\'{o}n en el intervalo $P_{\text{m}}^{\ast }<P_{\text{p}%
}^{\ast }<P_{\text{M}}^{\ast }$, la Ec. (\ref{EOS SM}) tiene tres ra\'{\i}%
ces en $v^{\ast }=1/\rho _{0}^{\ast }$, a saber $v_{\text{1}}^{\ast } $, $v_{%
\text{2}}^{\ast }$ y $v_{\text{3}}^{\ast }$. La intersecci\'{o}n de una
isobara con una isoterma genera dos \'{a}reas I y II dadas por 
\begin{equation}
\text{\'{A}rea I}=(v_{\text{2}}^{\ast }-v_{\text{1}}^{\ast })P_{\text{p}%
}^{\ast }-\int_{v_{\text{1}}^{\ast }}^{v_{\text{2}}^{\ast }}P^{\ast }dv
\end{equation}
\begin{equation}
\text{\'{A}rea II}=\int_{v_{\text{2}}^{\ast }}^{v_{\text{3}}^{\ast }}P^{\ast
}dv-(v_{\text{3}}^{\ast }-v_{\text{2}}^{\ast })P_{\text{p}}^{\ast }
\end{equation}
Como es bien sabido, seg\'{u}n el m\'{e}todo de Maxwell, la presi\'{o}n
reducida, $P_{\text{eq}}^{\ast }$, para la cual las dos fases coexisten en
equil\'{\i}brio es aquella para la cual ambas \'{a}reas son iguales y los
correspondientes valores $v_{\text{1}}^{\ast }$ y $v_{\text{3}}^{\ast }$ son
respectivamente los vol\'{u}menes reducidos de las fases l\'{\i}quido y
vapor en coexistencia.

Implementamos el siguiente algoritmo iterativo para desarrollar el c\'{a}%
lculo del equil\'{\i}brio a la temperatura $T_{0\text{eq}}^{\ast }$;
comenzamos con un valor inicial, 
\begin{equation}
P_{\text{p}(0)}^{\ast }=\frac{1}{2}(P_{\text{m}}^{\ast }+P_{\text{M}}^{\ast
})  \label{p maxwell prom}
\end{equation}%
Para este valor $P_{\text{p}(0)}^{\ast }$, obtenemos las ra\'{\i}ces $v_{%
\text{1}(0)}^{\ast }$ y $v_{\text{3}(0)}^{\ast }$ al resolver (\ref{EOS SM}%
); con estas raices calculamos las \'{a}reas I y II. Para simplificar el c%
\'{a}lculo y evaluar \'{u}nicamente una integral por cada temperatura,
requerimos que el \'{a}rea total, $A_{T}$ 
\begin{equation}
A_{T}=\int_{v_{\text{1}}^{\ast }}^{v_{\text{3}}^{\ast }}\left( P^{\ast }-P_{%
\text{p}(0)}^{\ast }\right) dv,
\end{equation}%
sea cero. Si $A_{T}\neq 0$ hacemos una siguiente iteraci\'{o}n con el
criterio: Si $A_{T}>0$, definimos un nuevo promedio $P_{\text{p}(1)}^{\ast }$
--siguiendo la Ec. (\ref{p maxwell prom})-- en el que $P_{\text{m}}^{\ast
}=P_{\text{p}(0)}^{\ast }$; pero si $A_{T}<0$, entonces $P_{\text{M}}^{\ast
}=P_{\text{p}(0)}^{\ast }$. Continuamos este proceso $i$-veces hasta lograr
que el valor de $A_{T}$ sea tan cercano a cero como se desee y obtendremos
los valores de equilibrio a la temperatura $T_{0\text{eq}}^{\ast }$; $P_{%
\text{eq}}^{\ast }=P_{\text{p}(i)}^{\ast }$,\ $\rho _{0\text{V}}^{\ast
}=1/v_{\text{3}(i)}^{\ast }$\ y $\rho _{0\text{L}}^{\ast }=1/v_{\text{1}%
(i)}^{\ast }$. Nosotros en particular requerimos que ambas \'{a}reas (I y
II) difieran a lo m\'{a}s en una parte en 100 000; es decir que $\left\vert
A_{T}\right\vert <$0.00001. Para lograr esta precisi\'{o}n se necesitaron
entre 10 y 15 iteraciones ($10\leq i\leq 15$) por cada valor de $T_{0\text{eq%
}}^{\ast }$.

\begin{figure}[h]
        \begin{center}
        \includegraphics[width=.8\hsize]{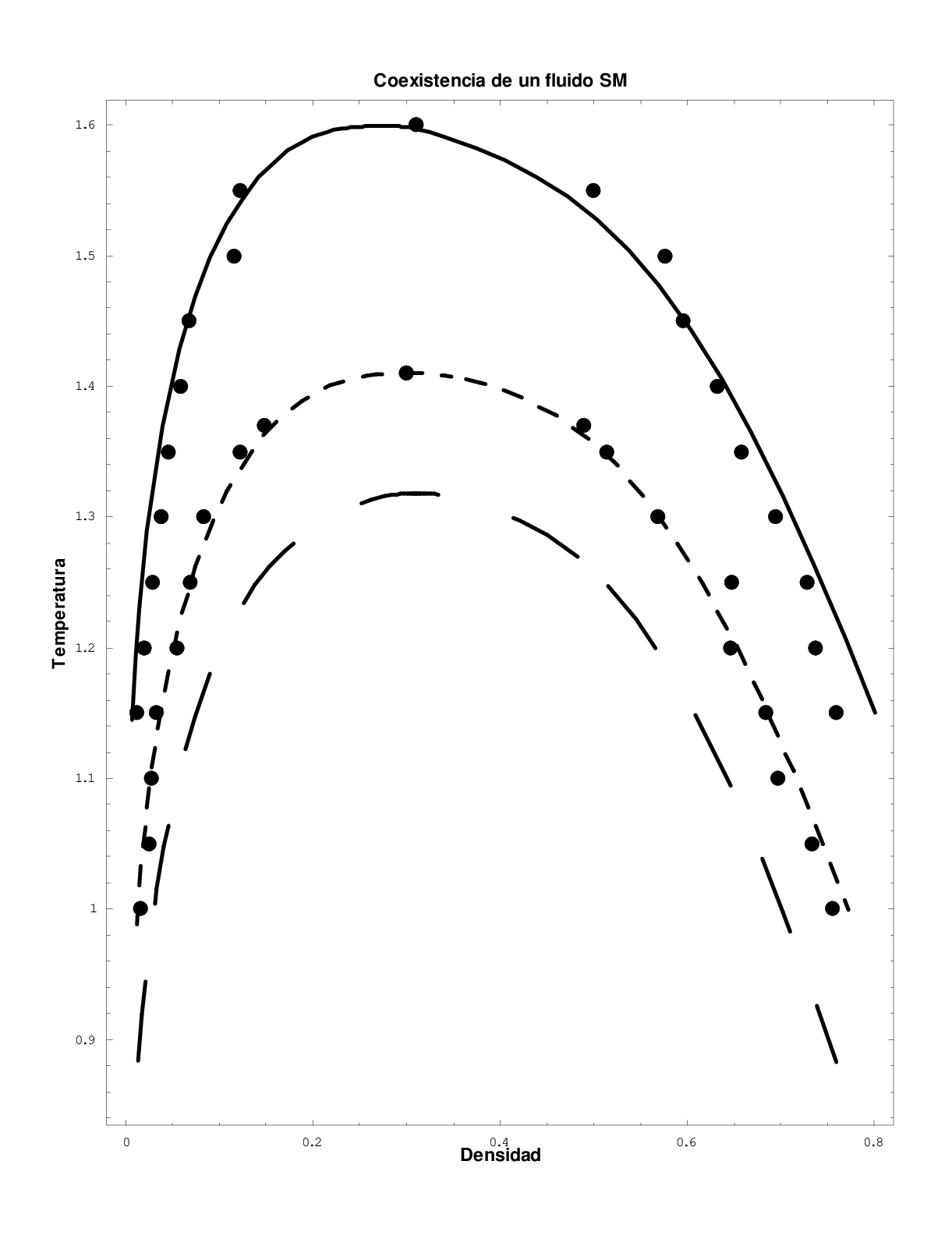}
        \end{center}
        \caption{%
Curvas de coexistencia de un fluido de Stockmayer
obtenidas a partir de la Ec. (\protect\ref{EOS SM}), con $\protect\mu %
_{0}^{\ast }=$0.841 (segmentos cortos), 1.1892 (l\'{\i}nea continua) y
Lennard-Jones (segmentos largos). Para fines de comparaci\'{o}n, se muestran
tambi\'{e}n datos de simulaci\'{o}n.
                     }%
       \label{FIGcoexistenciaedosm}
\end{figure}

En la Figura \ref{FIGcoexistenciaedosm} se muestran dos curvas de
coexistencia\footnote{%
Cada curva de coexistencia se obtuvo al interpolar hasta 15 parejas de datos
de equilibrio, $(\rho _{0\text{V}}^{\ast },T_{0}^{\ast })$ y $(\rho _{0\text{%
L}}^{\ast },T_{0}^{\ast })$ para cada valor de $\mu _{0}^{\ast }$.}
obtenidas a partir de la Ec. (\ref{EOS SM}), con $\mu _{0}^{\ast }=$0.841 y
1.1892. En la construcci\'{o}n de las curvas de coexistencia incluimos el
efecto de la densidad escribiendo el par\'{a}metro efectivo ANC que define a
la profundidad efectiva como el producto $f_{\epsilon }$\ $j_{\epsilon }$, 
\textit{i.e.}, $\epsilon _{\text{ef}}=\epsilon _{0}f_{\epsilon }j_{\epsilon
} $, con $j_{\epsilon }=1+c_{\epsilon }\rho _{0}^{\ast }\mu _{0}^{\ast 4}$.
La constante $c_{\epsilon }$ la ajustamos de tal manera que la temperatura
critica obtenida con la Ec. (\ref{EOS SM}), coincida con los resultados de
simulaci\'{o}n. Con $\mu _{0}^{\ast }=$0.841 y 1.1892, respectivamente
usamos $c_{\epsilon }=-$0.179\ y $-$0.137069.
%

\section{Resumen}

Este cap\'{\i}tulo lo dedicamos a los modelos moleculares dipolares con
kernel esf\'{e}rico. Definimos el potencial de Stockmayer generalizado (GSM)
como una generalizaci\'{o}n del potencial de Stokmayer (SM). Mostramos que
el GSM, al permitir realizar cambios de forma en el perfil del potencial,
resulta ser m\'{a}s vers\'{a}til que el del SM. Con el fin de eliminar la
dependencia angular en la definici\'{o}n del potencial GSM, definimos una
versi\'{o}n promediada angularmente de la interacci\'{o}n original, $\varphi
_{\text{GSM}}^{\text{sphe}}$. Remplazamos el potencial $\varphi _{\text{GSM}%
}^{\text{sphe}}$ por una funci\'{o}n ANC con par\'{a}metros efectivos $%
\left( \epsilon ,r_{\text{m}},s\right) $ --dependientes de la temperatura--
capaz de reproducir la termodin\'{a}mica del potencial original GSM.
Analizamos los efectos de la temperatura en el potencial GSM. Para el caso
especial del GSM con suavidad unitaria, calculamos anal\'{\i}ticamente $B(T)$%
. Usando una aproximaci\'{o}n anal\'{\i}tica y otra num\'{e}rica para los par%
\'{a}metros de una funci\'{o}n ANC, logramos reproducir $B(T)$ de un fluido
de SM. Los par\'{a}metros ANC que describen la suavidad del GSM los
encontramos usando dos distintas representaciones --con una y con dos
suavidades. Finalmente, analizamos la termodin\'{a}mica de un fluido GSM,
describiendo su temperatura de Boyle, sus propiedades cr\'{\i}ticas y de
equilibrio. Describiremos c\'{o}mo la densidad afecta a los par\'{a}metros
ANC y en consecuencia a la ecuaci\'{o}n de estado de un fluido dipolar.
Mostramos que podemos encontrar una ecuaci\'{o}n de estado para fluidos
dipolares que, con la inclusi\'{o}n de un factor semiemp\'{\i}rico de
correcci\'{o}n por efectos de densidad finita, es capaz de reproducir datos
de simulaci\'{o}n computacional.

\chapter{\textbf{Fluidos polares con kernel alargado}}

\section{Introducci\'{o}n}

En el cap\'{\i}tulo anterior desarrollamos potenciales tipo ANC que son
capaces de reproducir diversas propiedades termodin\'{a}micas de fluidos
cuya interacci\'{o}n consiste de un kernel esf\'{e}rico sumado a un t\'{e}%
rmino dipolar. Sin embargo, debido a que la presencia del momento dipolar
permanente en una mol\'{e}cula es una consecuencia de la falta de simetr%
\'{\i}a esf\'{e}rica, los modelos GSM son s\'{o}lo una primera aproximaci%
\'{o}n a la interacci\'{o}n real entre dos mol\'{e}culas dipolares. Los
modelos dipolares con kernel no esf\'{e}rico son modelos m\'{a}s realistas 
\cite{Lago 1997}. Por este motivo, en este cap\'{\i}tulo estudiaremos los
potenciales efectivos de mol\'{e}culas dipolares \textit{lineales} c\'{o}mo
medio para investigar sus propiedades termodin\'{a}micas. Previamente ya se
han estudiado los potenciales efectivos ANC de las mol\'{e}culas no polares
lineales con la ayuda de dos modelos: El multicentros de LJ ($n$-CLJ) y el
Kihara con n\'{u}cleo esferocil\'{\i}ndrico, (KSC). \cite{del Rio 1998 II}

Los potenciales efectivos de las mol\'{e}culas lineales dependen de su
elongaci\'{o}n. Para el modelo KSC, el potencial efectivo se puede obtener
directamente a partir del correspondiente potencial de Kihara, mientras que
en el caso del modelo $n$-CLJ el potencial efectivo se obtiene por inversi%
\'{o}n del segundo coeficiente virial.

En este cap\'{\i}tulo desarrollaremos una extensi\'{o}n de la teor\'{\i}a
ANC para incluir el tratamiento de diversos modelos dipolares de LJ con dos
centros ($2$-CLJ) a los que tambi\'{e}n nos referiremos simplemente con el
nombre de\textit{\ mol\'{e}culas diat\'{o}micas dipolares}. Desarrollaremos
un potencial modelo ANC con par\'{a}metros que dependen de la temperatura
que es capaz de reproducir la termodin\'{a}mica de diversos modelos $2$-CLJ 
\cite{carloses}. El esquema de este cap\'{\i}tulo es como sigue: primero,
haremos una breve descripci\'{o}n de los modelos $2$-CLJ. Despu\'{e}s
obtendremos los par\'{a}metros que dependen de la temperatura y de las
cantidades que definen un modelo $2$-CLJ como el momento dipolar y la
elongaci\'{o}n. Este problema lo abordaremos de manera progresiva en
complejidad: primero el caso del momento dipolar alineado con el eje
molecular y despu\'{e}s el caso m\'{a}s general cuando el momento dipolar
hace un \'{a}ngulo $\alpha $ con el eje molecular. Finalmente ponemos a
prueba de nuestro m\'{e}todo y su versatilidad para predecir datos del
segundo coeficiente virial.

\section{Mol\'{e}culas diat\'{o}micas dipolares}

\subsection{El modelo}

Estos modelos consisten de dos mol\'{e}culas homo-diat\'{o}micas cada una
con distancia inter-nuclear $\ell $ y un momento dipolar centrado en el eje
molecular y con una inclinaci'on $\alpha $ respecto a este eje. Vega \textit{%
et al.} \cite{carloses}, reportan alrededor de $10^{5}$ datos del segundo
coeficiente virial como funci\'{o}n de la temperatura correspondientes a
diversos modelos moleculares con distintos valores de elongaci\'{o}n, $0\leq 
\widetilde{L}\leq 1$, intensidad de momento dipolar, $0\leq \widetilde{\mu }%
^{2}\leq 10$, e inclinaci'on, $0\leq \alpha \leq 90^{\text{o}}$. La
elongaci'on, $\ell $, y el momento dipolar, $\mu $, en unidades de $\sigma $%
, se reducen como

\begin{equation}
\widetilde{L}=\frac{\ell }{\sigma }
\end{equation}

\begin{equation}
\widetilde{\mu }^{2}=\frac{\mu ^{2}}{\epsilon \sigma ^{3}}
\end{equation}%
Es importante hacer la distinci'on entre los c'alculos que hagamos con los
modelos con y sin elongaci'on ya que en el primer caso corresponden a un
modelo de LJ con un centro y en el segundo, a uno con dos centros. Esto es,
el modelo $\widetilde{L}=0$ no es igual al l'imite de las diat'omicas en el
l'imite cuando $\widetilde{L}$ se aproxima a cero. En el modelo $\widetilde{L%
}=0$, tenemos un centro --s\'{o}lo un centro LJ interaccionando con otro
centro. Sin embargo, en una diat'omica con $\widetilde{L}$ tendiendo hacia
cero tenemos $2$-CLJ (uno encima del otro) interaccionando con otra
mol'ecula (formada por $2$-CLJ uno encima del otro). Por lo tanto en el
modelo $\widetilde{L}=0$, tenemos la energ'ia de $1$LJ. Sin embargo, en el
modelo $\widetilde{L}\rightarrow 0$, tenemos la energ'ia de $4$LJ (los
cuatro al mismo valor de $r$) \cite{Vega:2003}. Esto hace que para comparar
valores de $B^{\ast }(T^{\ast })$ del modelo $2$-CLJ con un $1$-CLJ, debemos
dividir entre $4$ las temperaturas reducidas del primero para compararlas
con el segundo, o en nuestro caso, con un ANC --que tambi'en es un modelo de 
$1$C. An'alogamente, para comparar el momento dipolar del modelo $2$-CLJ con
un $1$-CLJ, lo que hacemos es dividir el primero entre $2$. Esto es,

\begin{equation}
T_{2C}^{\ast }=\frac{T}{\epsilon _{2C}/k}=\frac{T}{4\epsilon _{1C}/k}=\frac{1%
}{4}T_{1C}^{\ast }
\end{equation}
\begin{equation}
\widetilde{\mu }_{2C}^{2}=\frac{\mu ^{2}}{\epsilon _{2C}\sigma ^{3}}=\frac{%
\mu ^{2}}{4\epsilon _{1C}\sigma ^{3}}=\frac{1}{4}\widetilde{\mu }_{1C}^{2}
\end{equation}
De manera que, para el caso $\widetilde{L}=0$ ($1$-CLJ), la relaci\'{o}n
entre las unidades en $\sigma $\ con $r_{\text{m}}$, es, 
\begin{equation}
\mu ^{\ast }=\frac{\widetilde{\mu }}{2^{\frac{1}{4}}}
\end{equation}
\begin{equation}
B^{\ast }(T^{\ast })=\frac{\widetilde{B}_{Car}(T^{\ast })}{\frac{2\pi }{3}2^{%
\frac{1}{2}}}
\end{equation}
Y para los casos en que $\widetilde{L}\neq 0$ ($2$-CLJ), 
\begin{equation}
\mu ^{\ast }=\frac{\left. \widetilde{\mu }\right/ 2}{2^{\frac{1}{4}}}
\end{equation}
\begin{equation}
B^{\ast }(T^{\ast })=\frac{\widetilde{B}_{Car}(\frac{1}{4}T^{\ast })}{\frac{%
2\pi }{3}2^{\frac{1}{2}}}
\end{equation}
En ambos casos, $\widetilde{B}_{Car}$, es segundo coeficiente virial tal
como est\'{a} reducido por Vega \textit{et al.} \cite{carloses}

\subsection{Diat\'{o}micas con momento dipolar alineado}

Consideremos un modelo sitio-sitio para la interacci\'{o}n entre dos mol\'{e}%
culas homodiat\'{o}micas. Sea $\varphi _{\text{at}}(r)$ el potencial \'{a}%
tomo-\'{a}tomo, donde $r$ es la distancia entre dos diferentes mol\'{e}%
culas; asumimos que $\varphi _{\text{at}}(r)$ es esf\'{e}rico y
caracterizado por un pozo de profundidad $\epsilon _{\text{at}}$, un di\'{a}%
metro $r_{\text{at}}$ (definido en la posici\'{o}n del m\'{\i}nimo) y una
suavidad $s_{\text{at}}$. La elongaci\'{o}n de una mol\'{e}cula diat\'{o}%
mica con distancia \'{\i}nter-nuclear $\ell $ es reducida en t\'{e}rminos
del di\'{a}metro at\'{o}mico como $L^{\ast }=\ell /r_{\text{at}}$, pero
nosotros usamos $\widetilde{L}=2^{1/6}L^{\ast }$, para comparar con la
literatura \cite{carloses}. El segundo coeficiente virial de esta diat\'{o}%
mica $B_{\text{dia}}^{\ast }=B_{\text{dia}}/r_{\text{at}}^{3}$ ser\'{a} una
funci\'{o}n de todos estos par\'{a}metros, 
\begin{equation*}
B_{\text{dia}}^{\ast }=B_{\text{dia}}^{\ast }(T_{\text{at}}^{\ast },s_{\text{%
at}},\widetilde{L})
\end{equation*}%
con $T_{\text{at}}^{\ast }=kT/\epsilon _{\text{at}}$. Para obtener el
potencial efectivo \textit{intermolecular }ANC invertimos el segundo
coeficiente virial de las diat\'{o}micas, $B_{\text{dia}}(T)$, para ajustar
los par\'{a}metros del potencial ANC esf\'{e}rico (\ref{potencialANC})
usando la aproximaci\'{o}n de $1$-$s$. Los par\'{a}metros intermoleculares
resultantes $\epsilon _{\text{mol}}$, $r_{\text{mol}}$ y $s_{\text{mol}}$
depender\'{a}n de $s_{\text{at}}$ y $\widetilde{L}$. Sin embargo, la
dependencia con $s_{\text{at}}$ es muy d\'{e}bil, por lo que podemos asumir
con seguridad que, 
\begin{equation}
\epsilon _{\text{mol}}=\epsilon _{\text{at}}g_{\epsilon }(\widetilde{L})%
\text{,}  \label{epsmol}
\end{equation}%
\begin{equation}
r_{\text{mol}}=r_{\text{at}}g_{r}(\widetilde{L})\text{,}  \label{rmmol}
\end{equation}%
\begin{equation}
s_{\text{mol}}=s_{\text{at}}g_{s}(\widetilde{L})\text{.}  \label{smol}
\end{equation}%
Donde $g_{s}(\widetilde{L})$ describe la evoluci\'{o}n de $s_{0}$ conforme $%
\widetilde{L}$ cambia y es una cantidad tal que $\lim g_{s}(\widetilde{L})=1$
cuando $\widetilde{L}\longrightarrow 0$. El sistema que m\'{a}s
extensivamente se ha estudiado es el de las mol\'{e}culas diat\'{o}micas $2$%
-CLJ para las que se ha calculado su segundo coeficiente virial en un cierto
intervalo de elongaciones. Adem\'{a}s de los resultados originales de
Boublik, C. Vega y sus colegas reportaron recientemente valores bastante
precisos de $B_{\text{2CLJ}}^{\ast }(T^{\ast })$ \cite{carloses}. Para este
modelo, $s_{\text{at}}=s_{\text{LJ}}$ y los par\'{a}metros efectivos
dependen s\'{o}lo de la elongaci\'{o}n. Esta dependencia es muy suave, de
tal manera que $g_{\epsilon }(\widetilde{L})$, $g_{r}(\widetilde{L})$ y $%
g_{s}(\widetilde{L})$ quedan bien representados por polinomios en $%
\widetilde{L}$ cuyas expresiones expl\'{\i}citas y los detalles de su c\'{a}%
lculo se encuentran al final de este reporte en un ap\'{e}ndice. Estos
resultados, en lo general, est\'{a}n de acuerdo con los que obtuvieron
previamente Ramos \textit{et al}., \cite{del Rio 1998 II} pero los que
encontramos en esta investigaci\'{o}n son m\'{a}s confiables debido a que
los resultados num\'{e}ricos de los datos de $B_{\text{2CLJ}}^{\ast
}(T^{\ast })$ calculados por Vega y sus colegas, \cite{carloses} son m\'{a}s
precisos que los anteriores. Usando $\epsilon _{\text{mol}}$, $r_{\text{mol}%
} $ y $s_{\text{mol}}$ como par\'{a}metros de un sistema ANC podemos
reproducir los valores num\'{e}ricos de $B_{\text{2CLJ}}^{\ast }(T^{\ast })$
con bastante precisi\'{o}n.

Para poder calcular el potencial efectivo y el segundo coeficiente virial de
mol\'{e}culas dipolares diat\'{o}micas, en este trabajo estamos considerando
el caso restrictivo de que el momento dipolar est\'{a} localizado en el
centro de la mol\'{e}cula. El caso m\'{a}s simple es cuando el momento
dipolar est\'{a} orientado paralelo al eje molecular. El caso del momento
dipolar formando un \'{a}ngulo $\alpha $ con el eje lo consideramos m\'{a}s
adelante en una secci\'{o}n posterior.

En una mol\'{e}cula diat\'{o}mica dipolar el potencial efectivo ser\'{a}
afectado tanto por la presencia del momento dipolar como tambi\'{e}n por la
elongaci\'{o}n del kernel molecular. Como una restricci\'{o}n adicional,
supondremos que la no esfericidad del kernel no afecta la influencia del
momento dipolar. En otras palabras, suponemos, que los efectos del momento
dipolar y de la elongaci\'{o}n del kernel no est\'{a}n correlacionados y por
tanto los podemos abordar separadamente.

Dentro de esta suposici\'{o}n, el potencial efectivo de la mol\'{e}cula
dipolar diat\'{o}mica es la versi\'{o}n 1-$s$ de (\ref{uANCeff}) 
\begin{equation}
u_{\text{ef}}(z_{\text{ef}})=\varphi _{\text{ANC}}(z_{\text{ef}},\epsilon ,s)
\label{uANCeffOneS}
\end{equation}%
donde los par\'{a}metros efectivos los obtenemos superponiendo los efectos
de la polaridad y los de la elongaci\'{o}n. Esto lo hacemos en dos pasos:
Primero usamos (\ref{epsmol}), (\ref{rmmol}) y (\ref{smol}) para obtener un
potencial efectivo esf\'{e}rico de la mol\'{e}cula diat\'{o}mica y, segundo,
usamos este potencial como kernel para la interacci\'{o}n dipolar. De esta
manera, los par\'{a}metros que deben usarse en (\ref{uANCeffOneS}) son 
\begin{equation}
\epsilon =\epsilon _{\text{mol}}\ f_{\epsilon }(T_{\text{mol}}^{\ast },\mu _{%
\text{mol}}^{\ast },s_{\text{mol}})  \label{EpsEff2CLJ}
\end{equation}%
\begin{equation}
s=s_{\text{mol}}\ f_{s}(T_{\text{mol}}^{\ast },\mu _{\text{mol}}^{\ast },s_{%
\text{mol}}),  \label{sEff2CLJ}
\end{equation}%
y $z_{\text{ef}}=r/r$ donde

\begin{equation}
r=r_{\text{mol}}\ f_{r}(T_{\text{mol}}^{\ast },\mu _{\text{mol}}^{\ast },s_{%
\text{mol}}).  \label{rmEff2CLJ}
\end{equation}
donde $\epsilon _{\text{mol}}$, $s_{\text{mol}}$ y $r_{\text{mol}}$ est\'{a}%
n dadas en t\'{e}rminos de $\epsilon _{\text{at}}$, $s_{\text{at}}$, $r_{%
\text{at}}$\ y $\widetilde{L}$ por (\ref{epsmol}), (\ref{rmmol}) y (\ref%
{smol}); adem\'{a}s $T_{\text{mol}}^{\ast }=T_{\text{at}}^{\ast
}/g_{\epsilon }(\widetilde{L})$, $\mu _{\text{mol}}^{\ast 2}=\left. \mu _{%
\text{at}}^{\ast 2}\right/ \left[ g_{\epsilon }(\widetilde{L})g_{r}^{3}(%
\widetilde{L})\right] $ y $\mu _{\text{at}}^{\ast 2}=\left. \mu ^{2}\right/
\left( \epsilon _{\text{at}}r_{\text{at}}^{3}\right) $.

Los par\'{a}metros en (\ref{EpsEff2CLJ}), (\ref{sEff2CLJ}) y (\ref{rmEff2CLJ}%
) junto con (\ref{uANCeffOneS}) determinan el potencial efectivo para las mol%
\'{e}culas dipolares diat\'{o}micas. Estos mismos par\'{a}metros usados en
conjunci\'{o}n con (\ref{BeffOneS}) determinan $B(T)$ para estas mol\'{e}%
culas. No olvidemos que estos resultados los hemos obtenido asumiendo que
los efectos de polaridad y los de elongaci\'{o}n no est\'{a}n
correlacionados. Aqu\'{\i} mostraremos predicciones de $B^{\ast }(T)$ para
mol\'{e}culas dipolares del tipo $2$-CLJ, la precisi\'{o}n de estas
predicciones es una indicaci\'{o}n de la veracidad del potencial efectivo
correspondiente. Escogimos el modelo de $2$-CLJ debido a que disponemos de
los resultados de los c\'{a}lculos de $B_{\text{D2CLJ}}^{\ast }(T)$ que
recientemente calcularon C. Vega y sus colegas.

\begin{figure}[h]
        \begin{center}
        \includegraphics[width=.8\hsize]{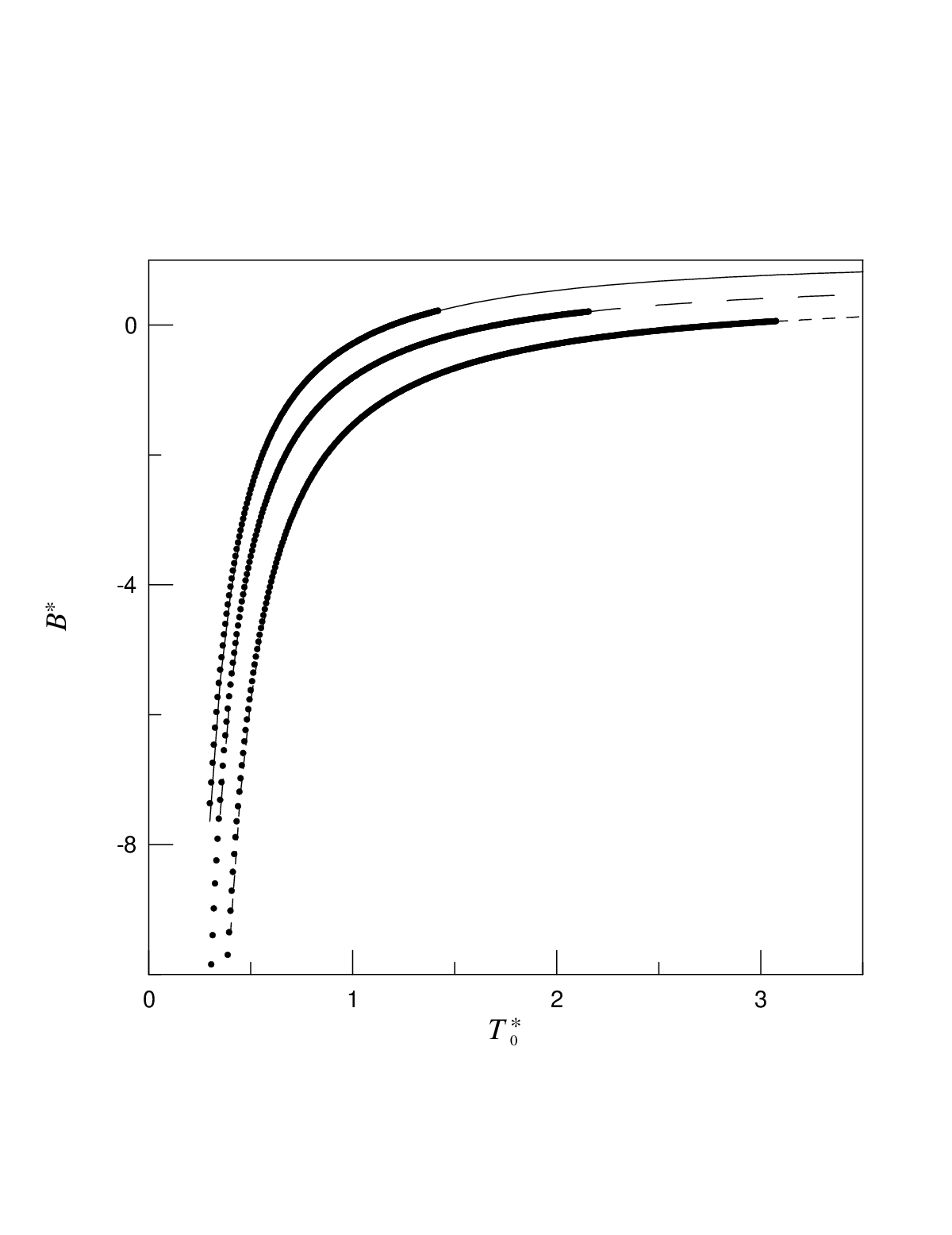}
        \end{center}
        \caption{%
Prueba de la predicci\'{o}n del segundo coeficiente
virial $B^{\ast }(T_{0}^{\ast })$ para las mol\'{e}culas diat\'{o}micas LJ
polares ($s_{\text{at}}=s_{\text{LJ}}$) con dipolo $\protect\mu _{\text{at}%
}^{\ast }=0.595$, alineado con el eje molecular, para varias enlongaci\'{o}%
nes $\widetilde{L}$. Los puntos corresponden a los valores num\'{e}ricos de
C. Vega \textit{et al}. \protect\cite{carloses} Los casos que se muestran
son para $\widetilde{L}=0.2$ (segmentos cortos, abajo), $\widetilde{L}=0.5$
(segmentos largos, en medio) y $\widetilde{L}=0.8$ (l\'{\i}nea continua,
arriba). Las l\'{\i}neas corresponden a la aproximaci\'{o}n 1-$s$; la
aproximaci\'{o}n 2-$s$ es indistinguible de la 1-$s$ en la escala de esta gr%
\'{a}fica.
                     }%
       \label{FIGBmu0595}
\end{figure}

En la Fig. \ref{FIGBmu0595} mostramos la dependencia con la temperatura de $%
B_{\text{D2CLJ}}^{\ast }$ para $\mu _{\text{at}}^{\ast }=$0.595 con
elongaciones $\widetilde{L}=$0.2, 0.5 y 0.8. El coeficiente virial $B_{\text{%
D2CLJ}}^{\ast }(T_{\text{at}}^{\ast })$ lo calculamos con las dos
aproximaciones que desarrollamos; la de $1$-$s$ y la de $2$-$s$. Comparando
con los resultados num\'{e}ricos de Vega \textit{et al.}, observamos un
acuerdo excelente. Respecto a\ los datos num\'{e}ricos con los que
comparamos, encontramos un muy buen acuerdo incluso para valores grandes
tanto del momento dipolar como de la elongaci\'{o}n, $\mu _{\text{at}}^{\ast
}\leq $0.841\emph{\ }y\emph{\ }$\widetilde{L}=$1.0. A\'{u}n cuando no hay
datos disponibles de $B^{\ast }$ para mol\'{e}culas diat\'{o}micas con $s_{%
\text{at}}\neq s_{\text{LJ}}$, nosotros esperamos que la predicci\'{o}n ANC
sea precisa para valores de suavidad en el intervalo, 0.5$\leq s\leq $1.0,
que es el que estamos usando en este trabajo. 
%

Para terminar debemos mencionar que este modelo puede ser usado para obtener
potenciales efectivos y coeficientes viriales de mol\'{e}culas polares con
kernel diferente al del tipo Kihara lineal, por ejemplo para kernels
esferocil\'{\i}ndricos suaves.

\subsection{Diat\'{o}micas con momento dipolar inclinado}

\begin{figure}[h]
        \begin{center}
        \includegraphics[width=.8\hsize]{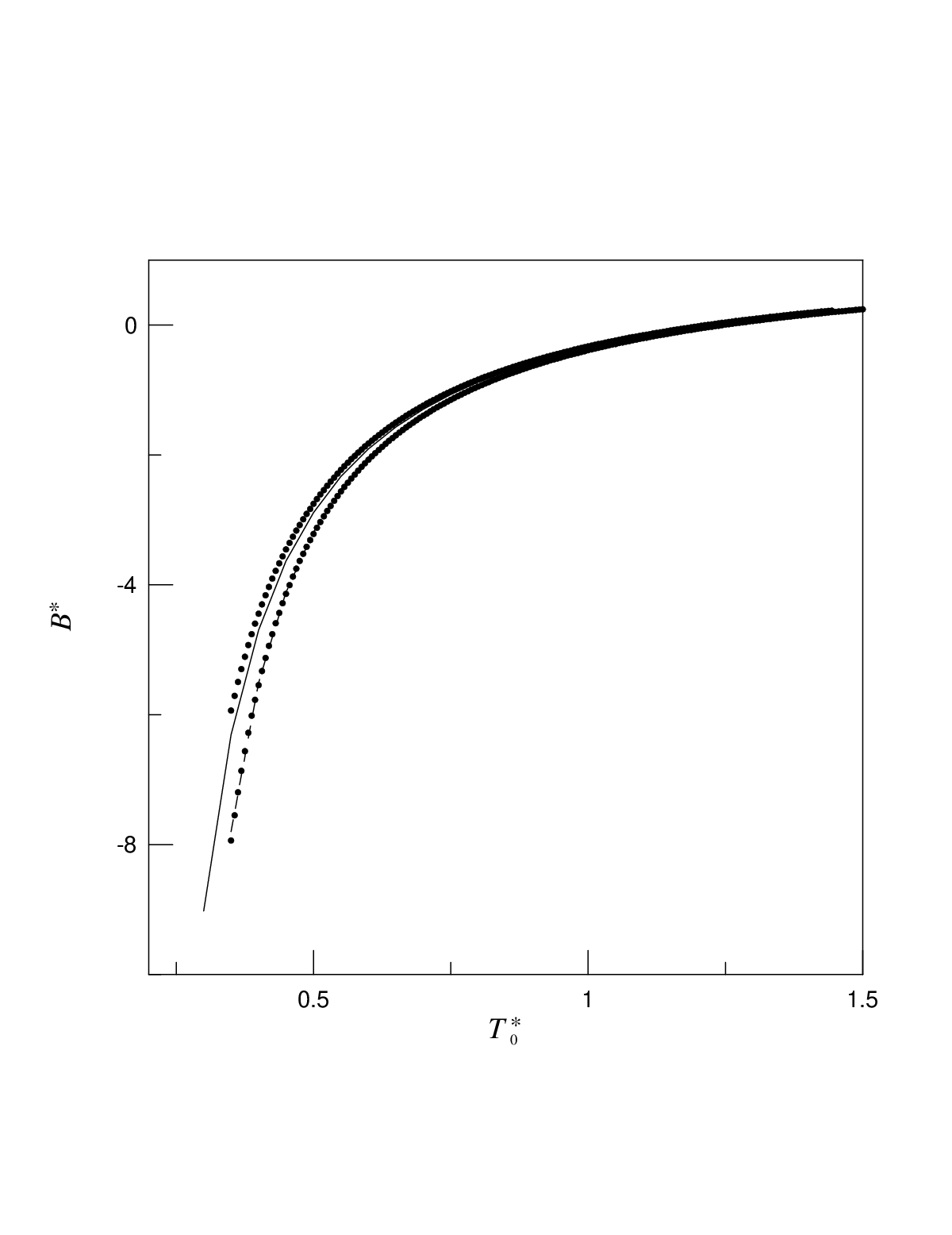}
        \end{center}
        \caption{%
Prueba del potencial modelo para calcular el segundo coeficiente virial, $%
B^{\ast }(T_{0}^{\ast })$, de mol\'{e}culas diat\'{o}micas LJ ($s_{\text{at}%
}=s_{\text{LJ}}$) con dipolo $\protect\mu _{\text{at}}^{\ast }=$0.728,
inclinado un \'{a}ngulo $\protect\alpha $ respecto al eje molecular y con
una enlongaci\'{o}n $\widetilde{L}=$0.8. Los puntos corresponden a los
valores num\'{e}ricos de C. Vega \textit{et al}. \protect\cite{carloses} Los
casos que se muestran son: $\protect\alpha =0$ (l\'{\i}nea solida,) and $%
\protect\alpha =90^{\circ }$ (segmentos cortos). Las l\'{\i}neas
corresponden a la aproximaci\'{o}n$1$-$s$. La aproximaci\'{o}n $2$-$s$ \ es
indistinguible de la $1$-$s$ en la escala de esta gr\'{a}fica.
                     }%
       \label{FIGBmu0728chueco}
\end{figure}

Ahora consideraremos el caso m\'{a}s general de las mol\'{e}culas polares
lineales en las que el momento dipolar no est\'{a} alineado con el eje
molecular, sino que hace un \'{a}ngulo $\alpha $ con este. El efecto de
variar $\alpha $ es bastante significativo como se muestra en la Fig. \ref%
{FIGBmu0728chueco} en la que vemos $B(T)$ para dos fluidos tipo $2$-CLJ,
ambos con momento dipolar $\mu _{\text{at}}^{\ast }=$0.841 y elongaci\'{o}n $%
\widetilde{L}=$0.8. Uno con el momento dipolar alineado con el eje ($\alpha
=0$) y el otro ortogonal al eje ($\alpha =\pi /2$).\ En ese caso notamos que
--en un cierto intervalo de temperatura-- $B(T)$ es ligeramente menor en el
modelo definido con el momento dipolar ortogonal al eje que cuando est\'{a}
alineado. Este efecto es perceptible sobre todo a bajas temperaturas. Como
podemos ver hay una peque\~{n}a pero evidente diferencia entre los dos
casos, lo cual significa que cuando $\alpha \neq 0$ los efectos de la
polaridad y de la elongaci\'{o}n no est\'{a}n simplemente superpuestos. En
otras palabras, estos efectos est\'{a}n correlacionados. Desafortunadamente
en este momento no disponemos de una manera de modelar te\'{o}ricamente la
correlaci\'{o}n subyacente entre $\mathbf{\mu }$ y $L$. Por ahora no hemos
desarrollado un potencial modelo basado en hip\'{o}tesis simples y que a su
vez sea capaz de reproducir los datos que se conocen de $B_{\text{D2CLJ}%
}^{\ast }(T)$. Para superar este problema asumimos que el efecto de hacer $%
\alpha \neq 0$ en cada uno de los par\'{a}metros efectivos lo podemos
incorporar mediante el uso de factores de correlaci\'{o}n apropiados, $%
h_{\epsilon }$, $h_{r}$, $h_{sA}$ y $h_{sR}$. Como est\'{a} ilustrado en la
Fig. \ref{FIGBmu0728chueco}, este efecto de la desalineaci\'{o}n es lo
suficientemente peque\~{n}o como para asumir dos simplificaciones
adicionales: la primera, que los factores de correlac\'{o}n son
independientes de la temperatura y la segunda, que estos factores son
iguales para ambas suavidades, \textit{i.e.}, hacemos $h_{s\text{A}}=h_{s%
\text{R}}=h_{s}$. De esta manera, los factores s\'{o}lo dependen de $\mu _{%
\text{at}}^{\ast }$, $\widetilde{L}$ y $\alpha $. Estos factores los
evaluamos por inversi\'{o}n de los datos num\'{e}ricos de $B_{\text{D2CLJ}%
}^{\ast }(T,\mu _{\text{at}}^{\ast },\widetilde{L},\alpha )$ en el intervalo
reportado de temperaturas. Formalmente, considerando las hip\'{o}tesis y
simplificaciones que hemos considerado, estos factores alteran los par\'{a}%
metros efectivos como

\begin{eqnarray}
\epsilon &=&\epsilon _{\text{mol}}\ f_{\epsilon }(T_{\text{mol}}^{\ast },\mu
_{\text{mol}}^{\ast },s_{\text{mol}})h_{\epsilon }(\mu _{\text{at}}^{\ast },%
\widetilde{L},\alpha )\text{,}  \label{EfeEpsH} \\
r &=&r_{\text{mol}}\ f_{r}(T_{\text{mol}}^{\ast },\mu _{\text{mol}}^{\ast
},s_{\text{mol}})h_{r}(\mu _{\text{at}}^{\ast },\widetilde{L},\alpha )\text{,%
}  \label{EfeRmH} \\
s &=&s_{\text{mol}}\ f_{s}(T_{\text{mol}}^{\ast },\mu _{\text{mol}}^{\ast
},s_{\text{mol}})h_{s}(\mu _{\text{at}}^{\ast },\widetilde{L},\alpha )\text{,%
}  \label{EfeSH}
\end{eqnarray}
donde por simplicidad hemos escrito la versi\'{o}n de $1$-s. La aproximaci%
\'{o}n de $2$-s involucra $s_{\text{R}}$ y $s_{\text{A}}$ que obtenemos
usando $f_{s_{\text{R}}}(T_{\text{mol}}^{\ast },\mu _{\text{mol}}^{\ast },s_{%
\text{mol}})$ y $f_{s_{\text{A}}}(T_{\text{mol}}^{\ast },\mu _{\text{mol}%
}^{\ast },s_{\text{mol}})$, respectivamente, de las Ecs. (\ref{SA}) y(\ref%
{SR}). Como es de esperarse, la alineaci\'{o}n no es relevante cuando el
kernel es esf\'{e}rico, por lo que $\lim h_{x}=1$ cuando $\widetilde{L}%
\longrightarrow 0$.\ En la inversi\'{o}n de los datos de $B_{\text{D2CLJ}%
}^{\ast }(T)$ \cite{carloses} usamos la aproximaci\'{o}n de 2-$s$ (\ref%
{Befff}) con valores fijos de $\mu _{\text{at}}^{\ast }$, $\widetilde{L} $ y$%
\alpha $. Aqu\'{\i} $\epsilon _{\text{mol}}$, $r_{\text{mol}}$ y $s_{\text{%
mol}}$ nuevamente est\'{a}n dadas por las Ecs. (\ref{epsmol}), (\ref{rmmol})
y (\ref{smol}) y a\'{u}n $L^{\ast }=\ell /r_{\text{at}}$. Siguiendo estas
instrucciones evaluamos los factores $h_{x}$ para $\mu _{\text{at}}^{\ast }=$
0.0, 0.420, 0.595, 0.728, 0.841, $\widetilde{L}=$ 0.2, 0.4, ..., 1.0 y $%
\alpha =$ $0$, $\pi /6$, $\pi /3$, $\pi /2$. Los valores resultantes los
usamos para ajustar expresiones de la forma

\begin{eqnarray}
h_{\epsilon }(\mu _{\text{at}}^{\ast },\widetilde{L},\alpha )
&=&1+d_{11}(\mu _{\text{at}}^{\ast },\alpha )\widetilde{L}+d_{12}(\mu _{%
\text{at}}^{\ast },\alpha )\widetilde{L}^{2}  \label{HEps} \\
h_{r}(\mu _{\text{at}}^{\ast },\widetilde{L},\alpha ) &=&1+d_{21}(\mu _{%
\text{at}}^{\ast },\alpha )\widetilde{L}+d_{22}(\mu _{\text{at}}^{\ast
},\alpha )\widetilde{L}^{2}+d_{23}(\mu _{\text{at}}^{\ast },\alpha )%
\widetilde{L}^{3}  \label{HRm} \\
h_{s}(\mu _{\text{at}}^{\ast },\widetilde{L},\alpha ) &=&1+d_{31}(\mu _{%
\text{at}}^{\ast },\alpha )\widetilde{L}+d_{32}(\mu _{\text{at}}^{\ast
},\alpha )\widetilde{L}^{2}\text{,}  \label{Hs}
\end{eqnarray}
donde las funciones $d_{ij}$ son polinomios sencillos en $\mu _{\text{at}%
}^{\ast }$ y $\alpha $ dados en el ap\'{e}ndice. Los factores de correlaci%
\'{o}n que definen estas ecuaciones satisfacen el requerimiento de que $\lim
h_{x}=1$ cuando $\widetilde{L}\longrightarrow 0$, $\mu _{\text{at}}^{\ast
}\longrightarrow 0$ y por supuesto, cuando $\alpha \longrightarrow 0$.El
modelo te\'{o}rico para $u_{\text{ef}}(z_{\text{ef}})$ y $B(T)$ para las mol%
\'{e}culas dipolares diat\'{o}micas --con un \'{a}ngulo $\alpha $ entre el
momento dipolar y el eje molecular-- est\'{a} dado como antes por las Ecs. (%
\ref{uANCeffOneS}) y (\ref{BeffOneS}) en la aproximaci\'{o}n de $1$-$s$ pero
ahora los par\'{a}metros efectivos est\'{a}n dados por las Ecs. (\ref%
{EfeEpsH}), (\ref{EfeRmH}) y (\ref{EfeSH}).

Usamos la teor\'{\i}a ANC para calcular $B(T)$ de mol\'{e}culas diat\'{o}%
micas dipolares con alineaci\'{o}n arbitraria de su momento dipolar para
interacciones cuyo kernel sea del tipo LJ sitio-sitio (haciendo $s_{\text{at}%
}=s_{\text{LJ}}$). Aqu\'{\i} comparamos con los resultados de $B_{\text{D2CLJ%
}}^{\ast }(T)$ \cite{carloses} para momentos dipolares\ no demasiado grandes
y observamos que el acuerdo es excelente. Esto est\'{a} ilustrado en la Fig. %
\ref{FIGBmu0728chueco}, la cual muestra el caso con $\mu _{\text{at}}^{\ast
}=$0.728, $\widetilde{L}=$0.8 y dos orientaciones del momento dipolar ($%
\alpha =0$ y $\alpha =\pi /2$). 
%

\section{Resumen}

Este cap\'{\i}tulo fue dedicado al an\'{a}lisis de modelos dipolares con
kernel no esf\'{e}rico, en particular a los modelos dipolares de LJ con dos
centros ($2$-CLJ). Desarrollamos los par\'{a}metros ANC dependientes de la
temperatura que fueron capaces de reproducir la termodin\'{a}mica de
diversos modelos $2$-CLJ. En el caso del momento dipolar alineado con el eje
molecular, encontramos que la elongaci\'{o}n, $\widetilde{L}$, afecta los par%
\'{a}metros ANC en una cantidad $g_{x}$ que depende de $\widetilde{L}$. En
el caso m\'{a}s general cuando el momento dipolar hace un \'{a}ngulo $\alpha 
$ con el eje molecular, el efecto sobre los par\'{a}metros fue introducido
--a manera de correlaci\'{o}n-- usando las funciones $h_{\epsilon }$. Para
diversos modelos, observamos un acuerdo excelente entre las predicciones
realizadas y los datos de integraci\'{o}n num\'{e}rica.

\chapter{\textbf{Sustancias polares reales}}

\section{Introducci\'{o}n}

En este cap\'{\i}tulo abordamos las sustancias reales. Obtenemos los par\'{a}%
metros de la funci\'{o}n ANC que sea capaz de predecir el segundo
coeficiente virial de una sustancia dipolar t\'{\i}pica. A partir de
informaci\'{o}n experimental del momento dipolar y de propiedades cr\'{\i}%
ticas, y con la ayuda de correlaciones entre resultados de simulaci\'{o}n
del modelo ANC, calcularemos esos par\'{a}metros. Compararemos con datos
experimentales y tambi\'{e}n con el m\'{e}todo emp\'{\i}rico basado en el
factor ac\'{e}ntrico de Pitzer.

\section{Aplicaci\'{o}n a sustancias reales}

Requerimos de algunos resultados previos acerca de las propiedades del
fluido ANC. La ecuaci\'{o}n de estado y las propiedades cr\'{\i}ticas se han
estudiado v\'{\i}a simulaci\'{o}n Monte Carlo NVT y tambi\'{e}n en el Gibbs
ensamble \cite{Guzman 2001, Eloy Tesis}. Aqu\'{\i} s\'{o}lo mostramos los
resultados relevantes. La temperatura cr\'{\i}tica reducida $T_{\text{c}%
}^{\ast }=kT_{\text{c}}/\epsilon _{\text{c}}$, la presi\'{o}n $P_{\text{c}%
}^{\ast }=P_{\text{c}}r_{\text{c}}^{3}/\epsilon _{\text{c}}$ \ y el volumen $%
V_{\text{c}}^{\ast }=V_{\text{c}}/r_{\text{c}}^{3}$\ \ del fluido ANC son
una funci\'{o}n de $s_{\text{c}}$. Escribimos $\epsilon _{\text{c}},$ $r_{%
\text{c}}$ y $s_{\text{c}}$\ para indicar que estas cantidades est\'{a}n
evaluadas en el punto cr\'{\i}tico. Para 0.5$\leq s\leq $1.2 se encontraron
las siguientes correlaciones \cite{Del Rio}, 
\begin{eqnarray}
\frac{T_{\text{c}}}{\epsilon _{\text{c}}/k} &=&\text{0.348915}+\text{0.326038%
}\ s+\text{0.443771}\ s^{2}\text{,}  \label{CorrANC_T} \\
r_{\text{c}}^{3}/V_{\text{c}} &=&\text{0.74226,}  \label{CorrANC_V} \\
\frac{P_{\text{c}}r_{\text{c}}^{3}}{\epsilon _{\text{c}}} &=&-\text{4.2179}+%
\text{25.1353}\ s_{\text{c}}\text{,}  \label{CorrANC_P}
\end{eqnarray}%
donde $\epsilon _{\text{c}}/k$ y $r_{\text{c}}$ est\'{a}n dados
respectivamente en kelvin (K) y amstong (\AA ). La suavidad efectiva, $s$ es
un par\'{a}metro sin dimensiones. Las cantidades cr\'{\i}ticas $T_{\text{c}}$%
, $V_{\text{c}}$ y $P_{\text{c}}$ est\'{a}n dadas respectivamente en K, cm$^{%
\text{3}}/$mol y atm.

\subsection{HCl}

Una t\'{\i}pica sustancia polar es el HCl (\textit{\'{a}cido clorh\'{\i}drico%
}). Las mol\'{e}culas de esta sustancia tienen un momento dipolar cuya
magnitud es de $\mu =$1.08 debyes \cite{Prausnitz 2001}. Las propiedades cr%
\'{\i}ticas del HCl medidas en el equilibrio l\'{\i}quido-vapor son, $T_{%
\text{c}}=$324.7$\ K$, $V_{\text{c}}=$81 $cm^{3}/mol$ y $P_{\text{c}}=$82$\
atm$ \cite{Ambrose}. Asumiendo que es valido un potencial ANC, primero
determinaremos $\epsilon _{\text{c}}$, $r_{\text{c}}$ y $s_{\text{c}}$ a
partir de las propiedades cr\'{\i}ticas y despu\'{e}s obtendremos los par%
\'{a}metros del kernel en el gas diluido. Finalmente mostraremos que usando
esos par\'{a}metros podemos calcular una predicci\'{o}n del segundo
coeficiente virial que concuerda muy bien con los datos experimentales; as%
\'{\i} como tambi\'{e}n con la correlaci\'{o}n emp\'{\i}rica que Tsonopoulos
hace \cite{Weber:1994} usando el factor ac\'{e}ntrico de Pitzer.

Con los datos cr\'{\i}ticos experimentales del HCl resolvemos simult\'{a}%
neamente las Ecs. (\ref{CorrANC_T}), (\ref{CorrANC_V}) y (\ref{CorrANC_P})
obteniendo $\epsilon _{\text{c}}/k=$468.9 K, $r_{\text{c}}=$3.918 \AA\ y $s_{%
\text{c}}=$\ 0.5861. Este valor de $\epsilon _{\text{c}}$ incluye el efecto
de la interacci\'{o}n dipolar. En los fluidos reales subyacen efectos
adicionales. Entre estos, los m\'{a}s importantes podemos mencionar --por un
lado-- los asociados a las correlaciones estad\'{\i}sticas que introducen
una dependencia en la densidad como la que discutimos en una secci\'{o}n
previa. Por otro lado, est\'{a} la presencia de fuerzas de muchos cuerpos
que no son relevantes en el estado de gas diluido. Debido a que el
procedimiento anterior involucra cantidades que fueron obtenidas
experimentalmente, los par\'{a}metros $\epsilon _{\text{c}}$, $r_{\text{c}}$
y $s_{\text{c}}$ contienen todos estos efectos. Entre las fuerzas de muchos
cuerpos, la m\'{a}s importante es la de tres cuerpos de tipo Axilrod-Teller
(AT) \cite{Guzman 2001}. El efecto neto de la fuerza AT es adicionar una
repulsi\'{o}n entre las mol\'{e}culas, de tal manera que disminuye la
profundidad del pozo del potencial. Esto significa que el valor de $\epsilon
_{0}/k$ para el HCl a bajas densidades --es decir, en una regi\'{o}n que no
es \textit{apreciablemente} afectada por los efectos de muchos cuerpos--
debe ser mayor que el que justo hemos calculado, $\epsilon _{\text{K}}/k=$%
468.9 K. El efecto de las fuerzas de AT ha sido exhibido, entre otros
investigadores, por Anta y sus colaboradores para el caso del arg\'{o}n \cite%
{Anta 1997}. La inclusi\'{o}n del t\'{e}rmino AT se hace necesaria si se
desea hacer una predicci\'{o}n cuantitativamente adecuada de la curva ortob%
\'{a}rica del arg\'{o}n. De los resultados de Anta y de sus colaboradores
uno encuentra que $\epsilon _{0}/\epsilon _{c}\approx $1.07. M\'{a}s
adelante veremos la teor\'{\i}a ANC permite recuperar este mismo valor con
buena aproximaci\'{o}n. Aun cuando el arg\'{o}n no es una mol\'{e}cula
polar, el poder predecir esta raz\'{o}n como un caso especial, nos da
confianza en que nuestro m\'{e}todo es correcto. El efecto de las fuerzas de
AT est\'{a} presente cuando las propiedades cr\'{\i}ticas $T_{\text{c}}$ y $%
V_{\text{c}}$ de sustancias reales son reducidas con los par\'{a}metros ANC
de baja densidad $\epsilon _{0}$, $r_{\text{m}0}$ y $s_{0}$. Esas cantidades
cr\'{\i}ticas satisfacen con buena aproximaci\'{o}n las siguientes
correlaciones: \cite{Del Rio 2001}

\begin{eqnarray}
\frac{T_{\text{c}}}{\epsilon _{0}/k} &=&\text{0.1206337}+\text{0.759946}%
s_{0}+\text{0.151390}s_{0}^{2}\text{,}  \label{CorrReal_T} \\
r_{\text{m}0}^{3}/V_{\text{c}} &=&\text{0.426}  \label{CorrReal_rho}
\end{eqnarray}%
donde $\rho _{\text{c}}$ est\'{a} dado en \textit{part\'{\i}culas }$/$\AA $%
^{3}$. El factor de forma no es apreciablemente afectado ni por la presencia
de las correlaciones (al igual que en el modelo GSM) ni tampoco por las
fuerzas de tres cuerpos \cite{Guzman 2001}. La diferencia entre las Ecs. (%
\ref{CorrANC_T}), y \ref{CorrANC_V} y las Ecs. (\ref{CorrReal_T}) y (\ref%
{CorrReal_rho}) es una medida de los efectos de muchos cuerpos y de correlaci%
\'{o}n sobre los potenciales efectivos. Por ejemplo, calculando $\epsilon
_{0}/\epsilon _{c}$ a partir de (\ref{CorrANC_T}) y (\ref{CorrReal_T}) para $%
s_{0}=$1.0, uno encuentra que $\epsilon _{0}/\epsilon _{c}=$1.08 para el arg%
\'{o}n --valor muy cercano al de Anta y sus colaboradores.

Usando las Ecs. (\ref{CorrReal_T}) y (\ref{CorrReal_rho}) y asumiendo que $%
s_{\text{c}}=s_{0}$, encontramos que los par\'{a}metros a baja densidad para
el HCl kernel son: $\epsilon _{0}/k=$525.4 K, $r_{\text{m}0}=$3.855 \AA\ y $%
s_{0}=$0.5861. Estos valores dan $\epsilon _{0}/\epsilon _{\text{c}}=$1.12
para el HCl siendo mayor que el caso del arg\'{o}n, pero del mismo orden de
magnitud.

El siguiente paso consiste en obtener los par\'{a}metros del kernel a baja
densidad que ser\'{a}n denotados por $\epsilon _{\text{K}}$, $r_{\text{K}}$
y $s_{\text{K}}$. Debido a que el HCl no es una mol\'{e}cula
(apreciablemente) elongada, asumimos que su kernel es esf\'{e}rico y usamos
el modelo ANC para el fluido GSM. Rescribiendo las expresiones (\ref{EpsEff1}%
), (\ref{RmEff1}) y (\ref{oneS}) como, 
\begin{eqnarray}
\epsilon _{\text{0}}/k &=&(\epsilon _{\text{K}}/k)f_{\epsilon }\text{,} \\
r_{\text{m0}} &=&\epsilon _{\text{K}}f_{r}\text{,} \\
s_{\text{0}} &=&s_{\text{K}}f_{\text{s}}\text{,}
\end{eqnarray}%
y usando $\mu =$1.08 debyes, finalmente obtenemos los par\'{a}metros del
kernel, $\epsilon _{\text{K}}/k=$501.9, $r_{\text{K}}=$3.859 \AA\ y $s_{%
\text{K}}=$0.579.

\begin{figure}[h]
        \begin{center}
        \includegraphics[width=.8\hsize]{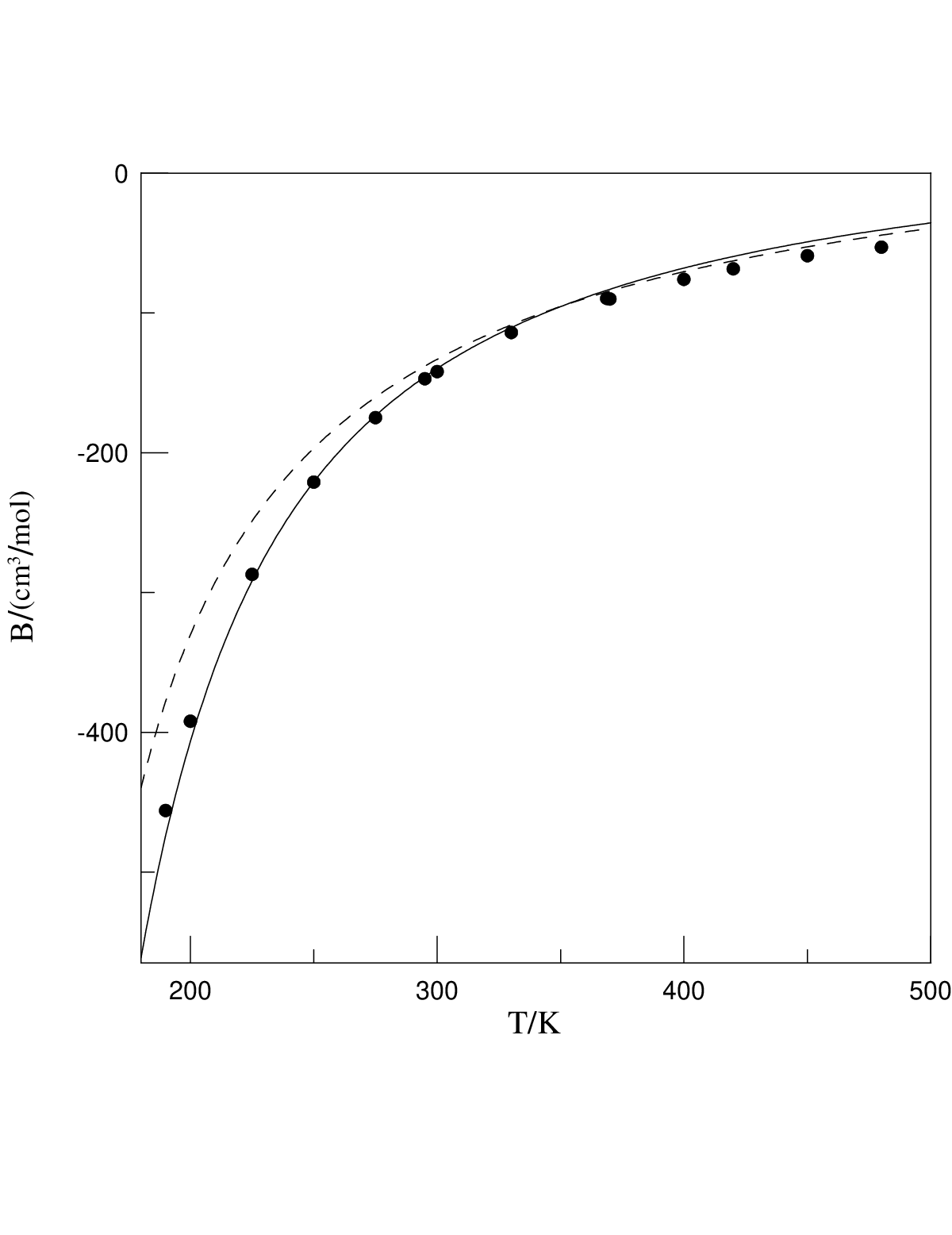}
        \end{center}
        \caption{%
Prueba del potencial modelo ANC para
predecir el segundo coeficiente virial del HCl. Los puntos representan datos
experimentales, la l\'{\i}nea continua es la predicci\'{o}n ANC y la l\'{\i}%
nea de segmentos es la correlaci\'{o}n de Tsonopoulos.
                     }%
       \label{FIGpredicBancHCL}
\end{figure}

Hemos usado estos valores en el modelo GSM para hacer una predicci\'{o}n del
segundo coeficiente virial del HCl. En la Fig. (\ref{FIGpredicBancHCL}) se
muestra un acuerdo excelente de los resultados de esta predicci\'{o}n con
datos experimentales \cite{Dymond 1980} dentro del intervalo de temperatura
disponible, $190\ $K$<T<480\ $K \footnote{%
El punto de fusi\'{o}n del HCl es 114.8 K y la temperatura cr\'{\i}tica es
324.7 K.}. Tambi\'{e}n comparamos con una predicci\'{o}n realizada usando
correlaciones de Tsonopoulos para calcular $B(T)$ --basadas en el factor ac%
\'{e}ntrico dePitzer. La predicci\'{o}n realizada con el potencial efectivo
ANC supera en precisi\'{o}n a las correlaciones de Tsonopoulos.
%

\section{Resumen}

En este cap\'{\i}tulo aplicamos la extensi\'{o}n al tratamiento de
sustancias polares. Calculamos los par\'{a}metros de la funci\'{o}n ANC
capaz de predecir el segundo coeficiente virial de una sustancia dipolar t%
\'{\i}pica como el HCl. A partir de informaci\'{o}n experimental del momento
dipolar y propiedades cr\'{\i}ticas, y con la ayuda de resultados de simulac%
\'{o}n\ de Monte Carlo NVT, encontramos una primera aproximaci\'{o}n a los p%
\'{a}ramentos ANC ($\epsilon _{\text{c}}$, $r_{\text{c}}$ y $s_{\text{c}}$).
Excluimos los efectos debidos a la influencia de las fuerzas de muchos
cuerpos sobre los p\'{a}ramentos ANC de baja densidad, obteniendo $\epsilon
_{0}$, $r_{\text{m}0}$ y $s_{0}$. Finalmente, usamos los par\'{a}metros
efectivos ANC para descontar del kernel, los efectos debidos al momento
dipolar. Usando los par\'{a}metros resultantes del kernel ($\epsilon _{\text{%
K}}$, $r_{\text{K}}$ y $s_{\text{K}}$) y el momento dipolar correspondiente,
realizamos una prediccion del segundo coeficiente virial del HCl. La predicci%
\'{o}n result\'{o} ser muy buena e incluso mejor que la realizada por un m%
\'{e}todo emp\'{\i}rico basado en el factor ac\'{e}ntrico de Pitzer.

\chapter{\textbf{Conclusiones y perspectivas}}

\section{Conclusiones}

En este trabajo hemos construido un modelo de interacci\'{o}n molecular
efectivo para mol\'{e}culas con momento dipolar permanente. Este modelo
tiene la forma de un potencial ANC cuyos par\'{a}metros de energ\'{\i}a, di%
\'{a}metro y suavidad dependen del momento dipolar de las mol\'{e}culas, as%
\'{\i} como del di\'{a}metro y de la suavidad del kernel de las mol\'{e}%
culas --\textit{i.e.}, de su parte no polar. Para un kernel diat\'{o}mico,
encontramos tambi\'{e}n una dependencia de la elongaci\'{o}n.

El modelo que desarrollamos es capaz de predecir el segundo coeficiente
virial $B(T)$ de gran variedad modelos de interacci\'{o}n dipolar con
distinto kernel --esf\'{e}rico y alargado-- en los que incluso el cuyo
momento dipolar puede no estar alineado con el eje molecular. Como se
esperaba, al comparar con datos de simulaci\'{o}n de Vega \textit{et al.},
estas predicciones resultan ser muy buenas para momentos dipolares no muy
grandes ($\mu _{\text{at}}^{\ast }\lesssim 1$). Vega \textit{et al.} simulan
modelos dipolares incluyendo la dependencia angular subyacente y en nuestro
modelo hemos promediado esa dependencia. De manera que cuando se tratan
momentos dipolares muy intensos, la interacci\'{o}n de largo alcance se hace
evidente. Sin embargo, dentro del dominio de validez, este modelo sintetiza
con mucha precisi\'{o}n y de manera compacta una gran cantidad de datos
disponibles de B(T) de modelos $2$-CLJ dipolares. Hasta el momento, ha sido
posible estudiar la fase vapor as\'{\i} como tambi\'{e}n el equilibrio l%
\'{\i}quido-vapor.

El modelo puede usarse en mol\'{e}culas dipolares cuyos kernels sean no
conformales a una funci\'{o}n LJ-12/6. Debido a que una gran cantidad de
sustancias reales son notablemente no conformales a los fluidos de LJ, el
modelo que aqu\'{\i} desarrollamos, debe representar efectivamente de manera
precisa y simple los modelos dipolares descritos en el p\'{a}rrafo anterior.
Los par\'{a}metros ANC que describen la suavidad del modelo propuesto, los
encontramos usando dos distintas representaciones --con una y con dos
suavidades. Encontramos que en la gran mayor\'{\i}a de las aplicaciones, la
representaci\'{o}n con una sola suavidad es m\'{a}s que suficiente para los
casos que consideramos.

Analizamos la termodin\'{a}mica de los modelos propuestos, describiendo su
temperatura de Boyle, sus propiedades cr\'{\i}ticas y de equilibrio, y
describimos --de manera aproximada-- como la densidad afecta a los par\'{a}%
metros ANC y en consecuencia a la ecuaci\'{o}n de estado de un fluido
dipolar. Estudiamos por simulaci\'{o}n, los efectos de cambio de forma del
potencial en la termodin\'{a}mica del equilibrio de los fluidos polares.
Para el potencial ANC estudiamos --tambi\'{e}n por simulaci\'{o}n-- la
influencia del par\'{a}metro de suavidad en el equilibrio l\'{\i}quido-vapor.

Realizamos con \'{e}xito una prueba en una sustancia polar t\'{\i}pica, el
HCl. La predicci\'{o}n result\'{o} ser muy buena e incluso mejor que la
realizada por un m\'{e}todo emp\'{\i}rico basado en el factor ac\'{e}ntrico
de Pitzer.

\section{Perspectivas}

Existen varias l\'{\i}neas de investigaci\'{o}n con las que se pueden
ampliar los resultados de este trabajo. Antes de mencionar los temas en los
que estamos interesados a mediano plazo, describir\'{e} un problema en el
que hemos hecho ya una breve incursi\'{o}n para aplicar el modelo de
interacci\'{o}n que desarrollamos en esta tesis.

La constante diel\'{e}ctrica, $\overline{\epsilon }$, es una propiedad
macrosc\'{o}pica que puede obtenerse a partrir de variables microsc\'{o}%
picas. La ecuaci\'{o}n de Kirkwood (Ec. (\ref{Kirkwood})), relaci\'{o}na $%
\overline{\epsilon }$ con una proyecci\'{o}n de la funci\'{o}n de distribuci%
\'{o}n par, $h(1,2)$, como sigue \cite{Ballenegger 2004, valisko 2001}, 
\begin{equation}
\frac{\left( \overline{\epsilon }-1\right) \left( 2\overline{\epsilon }%
+1\right) }{9\overline{\epsilon }}=y\left( 1+\frac{4\pi \rho }{3}%
\int_{0}^{\infty }h^{110}(r)r^{2}dr\right) \ \text{.}  \label{Kirkwood}
\end{equation}
\ En esta ecuaci\'{o}n, $y$ es un par\'{a}metro molecular dependiente de la
temperatura, $T$, densidad, $\rho $, y momento dipolar, $\mu $, de las part%
\'{\i}culas, $y=\left( 4\pi /9\right) \left( \rho \mu ^{2}/kT\right) $.
Existen diversos estudios de simulaci\'{o}n en donde se ha calculado $%
\overline{\epsilon }$ usando distintos modelos \cite{Szalai 1999, Skaf 1995}%
. Desde el punto de vista te\'{o}rico quiz\'{a} la clave para calcular $%
\overline{\epsilon }$ es hacer una aproximaci\'{o}n adecuada de $h^{110}$.
Ivanov y Kuznetsova \cite{Ivanov 2001} han avanzado ya en este problema y
uno de nuestros objetivos a corto plazo es generalizar sus resultados para
incluir los efectos de la suavidad.

Una de las l\'{\i}neas de investigaci\'{o}n que ser\'{\i}a deseable abordar
en el mediano plazo, consiste en averiguar como modelar el efecto de la
densidad en los par\'{a}metros ANC. Hasta ahora s\'{o}lo intentamos un
acercamiento semi-emp\'{\i}rico a este problema. El resolver esta cuesti\'{o}%
n nos permitir\'{\i}a completar la descripci\'{o}n de una ecuaci\'{o}n de
estado te\'{o}rica para un fluido dipolar. En particular, ser\'{\i}a
deseable revisar la teor\'{\i}a de perturbaciones para dise\~{n}ar alg\'{u}n
m\'{e}todo formal que nos permita obtener potenciales esf\'{e}ricos que sean
de utilidad en regiones de alta densidad --comparables con la de un l\'{\i}%
quido. Una de las alternativas para avanzar en esta direcci\'{o}n consiste
en desarrollar la funci\'{o}n de distribuci\'{o}n --angularmente
dependiente-- como una serie en t\'{e}rminos de invariantes rotacionales 
\cite{Canessa 1991}.

Otro problema importante --y quiz\'{a} m\'{a}s ambicioso-- consiste en
incorporar al modelo los efectos de la polarizabilidad molecular, as\'{\i}
como los debidos a los puentes de hidr\'{o}geno (HB). En este trabajo
abordamos con \'{e}xito el HCl. Esta sustancia en realidad s\'{\i} forma HB.
Sin embargo, el valor de la energ\'{\i}a asociada al HB en HCl ($-$1.64 kcal$%
/$mol) es peque\~{n}a con respecto a una sustancia con HB intensos como el
agua ($-$6.02 kcal$/$mol) \cite{Poater 2003}. La inclusi\'{o}n de HB
intensos en nuestro modelo representa un reto considerable debido a que
estos --adem\'{a}s de la t\'{\i}pica descripci\'{o}n electrost\'{a}tica--
tienen un car\'{a}cter parcialmente covalente, lo cual abre un camino enorme
hacia el entendimiento de las sustancias polares.

\appendix

\chapter{\textbf{C\'{a}lculo de }$\mathbf{B(T)}$\textbf{\ para SM}}

Para calcular el segundo coeficiente virial, $B(T)$, de un fluido de
Stockmayer, recordemos que el potencial de interacci\'{o}n molecular
correspondiente es,

\begin{equation}
\varphi (r,\theta _{1},\theta _{2},\phi ;\widetilde{\mu })=4\epsilon \left[
\left( \frac{\sigma }{r}\right) ^{12}-\left( \frac{\sigma }{r}\right) ^{6}%
\right] -\epsilon \frac{\mu ^{2}}{r^{3}}g(\theta _{1},\theta _{2},\phi )
\label{SMoriginal}
\end{equation}
con 
\begin{equation*}
g(\Omega )=g(\theta _{1},\theta _{2},\phi )=2\cos \theta _{1}\cos \theta
_{2}-\sin \theta _{1}\sin \theta _{2}\cos \phi
\end{equation*}
La forma general del c\'{a}lculo de $B(T)$ para un potencial que tiene
dependencia angular es \cite{Hirsch},

\begin{equation}
B(T)=-\frac{\widetilde{N}}{4}\underset{0}{\overset{+\infty }{\int }}\underset%
{0}{\overset{2\pi }{\int }}\underset{0}{\overset{2\pi }{\int }}\underset{0}{%
\overset{2\pi }{\int }}f\sin \theta _{1}d\theta _{1}\sin \theta _{2}d\theta
_{2}d\phi r^{2}dr\text{ con }f=e^{-\frac{\varphi (r)}{kT}}-1
\end{equation}
Despu\'{e}s de integrar por partes, uno de los t\'{e}rminos se anula debido
a que $\varphi \rightarrow 0$ cuando $r\rightarrow \infty $, y nos queda,

\begin{equation}
B(T)=-\frac{\widetilde{N}}{4}\frac{1}{3kT}\underset{T}{\int }d\Omega \frac{%
\partial \phi }{\partial r}e^{-\frac{\varphi (r)}{kT}}r^{3}dr
\label{Bgeneral1}
\end{equation}
En donde hemos usado $\underset{T}{\int }\equiv \underset{0}{\overset{%
+\infty }{\int }}\underset{0}{\overset{2\pi }{\int }}\underset{0}{\overset{%
2\pi }{\int }}\underset{0}{\overset{2\pi }{\int }}$ y $d\Omega \equiv \sin
\theta _{1}d\theta _{1}\sin \theta _{2}d\theta _{2}d\phi $ para simplificar
la notaci\'{o}n. Notemos que

\begin{equation*}
\frac{\partial }{\partial r}e^{-\frac{\varphi (r)}{kT}}=e^{-\frac{\varphi (r)%
}{kT}}\frac{\partial }{\partial r}\left( -\frac{\varphi (r)}{kT}\right) =-%
\frac{1}{kT}\frac{\partial \varphi (r)}{\partial r}e^{-\frac{\varphi (r)}{kT}%
}
\end{equation*}
Entonces 
\begin{equation}
B(T)=\frac{\widetilde{N}}{12}\underset{T}{\int }d\Omega \frac{\partial }{%
\partial r}\left( e^{-\frac{\varphi (r)}{kT}}\right) r^{3}dr\text{.}
\label{VirialConderivada}
\end{equation}
Si reducimos la unidad de longitud con $\sigma $ definimos $z\equiv r/\sigma 
$ y la temperatura como $\widetilde{T}=kT/\epsilon $ obtenemos, 
\begin{equation}
\widetilde{B}(\widetilde{T})=\frac{B(\widetilde{T})}{\frac{2}{3}\pi 
\widetilde{N}\sigma ^{3}}=\frac{1}{8\pi }\underset{T}{\int }d\Omega \frac{%
\partial }{\partial z}\left( e^{-\frac{\varphi (z)}{\epsilon \widetilde{T}}%
}\right) z^{3}dz  \label{Breducido}
\end{equation}
Para integrar esta expresi\'{o}n debemos escribir el potencial original SM (%
\ref{SMoriginal}) reduciendo el momento dipolar como $\widetilde{\mu }=\mu /%
\sqrt{\epsilon \sigma ^{3}}$. Simplificamos la notaci\'{o}n definiendo $%
\Omega \equiv (\theta _{1},\theta _{2},\phi ).$%
\begin{equation}
\varphi (z)=4\epsilon \left[ \frac{1}{z^{12}}-\frac{1}{z^{6}}\right]
-\epsilon \frac{\widetilde{\mu }^{2}}{z^{3}}g(\Omega )  \label{SMreducido}
\end{equation}
Substituyendo el potencial SM reducido (\ref{SMreducido}) en la funci\'{o}n
exponencial que aparece en la integral (\ref{Breducido}) obtenemos, 
\begin{equation}
\exp \left( -\frac{\varphi (z)}{\epsilon \widetilde{T}}\right) =-\exp \left(
-\frac{4}{\widetilde{T}}\frac{1}{z^{12}}\right) \exp \left( \frac{4}{%
\widetilde{T}}\frac{1}{z^{6}}+\frac{\widetilde{\mu }^{2}}{\widetilde{T}}%
\frac{1}{z^{3}}g(\Omega )\right)
\end{equation}
donde 
\begin{equation}
\exp \left( \frac{4}{\widetilde{T}}\frac{1}{z^{6}}+\frac{\widetilde{\mu }^{2}%
}{\widetilde{T}}\frac{1}{z^{3}}g(\Omega )\right) =\overset{\infty }{\underset%
{n=0}{\sum }}\frac{1}{n!}\left( \frac{4}{\widetilde{T}}\frac{1}{z^{6}}+\frac{%
\widetilde{\mu }^{2}}{\widetilde{T}}\frac{1}{z^{3}}g(\Omega )\right) ^{n}
\end{equation}
Desarrollando el binomio dentro de la suma lo podemos desarrollar como $%
(a+b)^{n}=\overset{\infty }{\underset{n=0}{\sum }}\binom{n}{k}a^{n-k}b^{k}$,
obtenemos 
\begin{equation}
\exp \left( +\frac{4}{\widetilde{T}}\frac{1}{z^{6}}+\frac{\widetilde{\mu }%
^{2}}{z^{3}}g(\Omega )\right) =\overset{\infty }{\underset{n=0}{\sum }}\frac{%
1}{n!}\overset{n}{\underset{m=0}{\sum }}\binom{n}{m}\left( \frac{4}{%
\widetilde{T}}\frac{1}{z^{6}}\right) ^{n-m}\left( \frac{\widetilde{\mu }^{2}%
}{\widetilde{T}}\frac{1}{z^{3}}g(\Omega )\right) ^{m}
\end{equation}
Entonces, 
\begin{eqnarray}
\exp \left( -\frac{\varphi (z)}{\epsilon \widetilde{T}}\right) &=&-\exp
\left( -\frac{4}{\widetilde{T}}\frac{1}{z^{12}}\right)  \notag \\
&&\times \ \overset{\infty }{\underset{n=0}{\sum }}\frac{1}{n!}\overset{n}{%
\underset{m=0}{\sum }}\binom{n}{m}\left( \frac{4}{\widetilde{T}}\frac{1}{%
z^{6}}\right) ^{n-m}\left( \frac{\widetilde{\mu }^{2}}{\widetilde{T}}\frac{1%
}{z^{3}}g(\Omega )\right) ^{m}
\end{eqnarray}
Sustituyendo esto en (\ref{Breducido}) y realizando el cambio de variable, $%
t=4/\widetilde{T}z^{12}$, despu\'{e}s de desarrollar y simplificar, la
integral de $\widetilde{B}(\widetilde{T})$ queda como 
\begin{eqnarray}
\widetilde{B}(\widetilde{T}) &=&\frac{1}{8\pi }\left( \frac{4}{\widetilde{T}}%
\right) ^{\frac{1}{4}}\overset{\infty }{\underset{n=0}{\sum }}\overset{n}{%
\underset{m=0}{\sum }}\widetilde{\mu }^{2m}\frac{1}{n!}\binom{n}{m}%
4^{-m}\left( \frac{4}{\widetilde{T}}\right) ^{\frac{2n+m}{4}}  \notag \\
&&\times \underset{T}{\int }d\Omega \frac{\partial }{\partial t}\left[
e^{-t}t^{\frac{2n-m}{4}}g^{m}\left( \Omega \right) \right] t^{-\frac{1}{4}}dt
\label{BSMantes de integrar}
\end{eqnarray}
La integral en esta expresi\'{o}n, que denominaremos $I$,\ la integraremos
primero en su parte radial, para lo cual, la desarrollamos y escribimos como 
\begin{equation}
I=-\underset{\Omega }{\int }d\Omega g^{m}\left( \Omega \right) \underset{0}{%
\overset{\infty }{\int }}\left( \frac{2n-m}{4}t^{\frac{2n-m}{4}-1}-t^{\frac{%
2n-m+3}{4}-1}\right) e^{-t}dt
\end{equation}
La integral cambia de signo al invertir los l\'{\i}mites de integraci\'{o}n
por haber sustituido la variable $z$ por la variable $t$. $\Omega $ debajo
de la primera integral indica que la integraci\'{o}n se realiza sobre todo
el dominio de integraci\'{o}n para $\Omega $, respectivamente. Los factores
de la forma $t^{\alpha -1}e^{-t}$, se integran como\footnote{$%
\int_{0}^{\infty }$ $t^{z-1}e^{-t}dt$ es la definici\'{o}n de la \textit{%
funci\'{o}n gamma}, $\Gamma \left( z\right) $, cuando la variable compleja \ 
$z$ tiene una parte real positiva, $\mathbf{Re}$ $z>0$. Una propiedad de a
funci\'{o}n gamma\emph{\ }importante es $\Gamma \left( z+1\right) =z\Gamma
\left( z\right) $. Adem\'{a}s la funci\'{o}n gamma esta relacionada con la
funci\'{o}n binomial como 
\begin{equation*}
\left( 
\begin{array}{c}
n \\ 
m%
\end{array}
\right) =\frac{\Gamma (n+1)}{\Gamma (m+1)\Gamma (n-m+1)}
\end{equation*}%
\par
\bigskip} $\int_{0}^{\infty }t^{\alpha -1}e^{-t}dt=\Gamma (\alpha )$. Usando
las propiedades de la funci\'{o}n gamma y definiendo $G_{m}=(1/8\pi )\int
d\Omega \ g^{2m}(\Omega )$ para toda $m$ finita, la expresi\'{o}n (\ref%
{BSMantes de integrar}) finalmente la escribimos como 
\begin{eqnarray}
\widetilde{B}(\widetilde{T}) &=&\left( \frac{4}{\widetilde{T}}\right) ^{%
\frac{1}{4}}\left\{ \Gamma \left( \frac{3}{4}\right) -\frac{1}{4}\overset{%
\infty }{\underset{n=1}{\sum }}\overset{k\leq \frac{n}{2}}{\underset{k=0}{%
\sum }}\frac{2^{n-3k}\ G_{m}}{n!}\right.  \notag \\
&&\left. \times \ \binom{n}{2k}\Gamma \left( \frac{2n-2k-1}{4}\right) \frac{%
\widetilde{\mu }^{4k}}{\widetilde{T}^{\frac{n+k}{2}}}\right\}
\label{B Stockmayer}
\end{eqnarray}

\section{Las constantes $G_{m}$}

Las constantes $G_{m}$ se calculan como sigue, 
\begin{eqnarray}
G_{k} &=&\frac{1}{8\pi }\int d\Omega \ g^{2k}(\Omega )  \notag \\
&&\frac{1}{8\pi }\int_{0}^{2\pi }\int_{0}^{\pi }\int_{0}^{\pi }\left( 2\cos
\theta _{1}\cos \theta _{2}-\sin \theta _{1}\sin \theta _{2}\cos \phi
\right) ^{2k}  \notag \\
&&\times \sin \theta _{1}\sin \theta _{2}\ d\theta _{1}d\theta _{1}d\phi
\label{Ge}
\end{eqnarray}
La cantidad entre par\'{e}ntesis la desarrollamos como un binomio; 
\begin{equation}
\left( 2\cos \theta _{1}\cos \theta _{2}-\sin \theta _{1}\sin \theta
_{2}\cos \phi \right) ^{2k}=\sum_{l=0}^{2k}\binom{2k}{l}a^{2k-l}b^{l}\text{,}
\label{binomio}
\end{equation}
donde $a$ y $b$ son respectivamente el primero y segundo t\'{e}rminos de
este binomio. Desarrollando la Ec. (\ref{binomio}) podemos rescribir la Ec. (%
\ref{Ge}) como, 
\begin{eqnarray}
G_{k} &=&\frac{1}{8\pi }\sum_{l=0}^{2k}\frac{\left( 2k\right) !}{\left(
2k-l\right) !l!}\left( -1\right) ^{l}2^{2k-l}  \notag \\
&&\times \int_{0}^{\pi }\left( \cos \theta _{1}\right) ^{2k-l}\left( \sin
\theta _{1}\right) ^{l+1}d\theta _{1}\int_{0}^{2\pi }\left( \cos \phi
\right) ^{l}d\phi  \label{Ge computo}
\end{eqnarray}
Podemos evaluar la Ec. (\ref{Ge computo}) para toda $m$ finita. La evaluaci%
\'{o}n de $G_{k}$ correspondiente a las primeras $k=0$, $1$, $2$ y $3 $ es $%
1 $, $2/3$, $24/25$ y $464/245$, respectivamente. Nosotros usamos $100 $ t%
\'{e}rminos para evaluar la Ec. (\ref{B Stockmayer}) ($k=50$); sin embargo,
usando los $31$ primeros t\'{e}rminos podemos recuperar la precisi\'{o}n que
usa Maitland \textit{et. al.} \cite{Maitland}.

\chapter{\textbf{C\'{a}lculo de }$\mathbf{B(T)}$\textbf{\ para GSM con} $%
s_{0}=1$}

Este c\'{a}lculo es m\'{a}s elaborado que el anterior y no tenemos
referencia de que se haya hecho previamente. Para calcular el segundo
coeficiente virial, $B(T)$, de un fluido GSM con $s_{0}=1$, recordemos que,
en unidades reducidas, $\varphi _{\text{MSM}_{1}}(z,\Omega ,\mu ^{\ast })$,
lo escribimos sumando el potencial ANC de referencia, $\varphi _{\text{ANC}%
}(z,s=1)$ con profundidad $\epsilon _{1}$ y m\'{\i}nimo en $r_{\text{m}1}$,
i.e. en $z=r/r_{\text{m}1}$, y un t\'{e}rmino dipolar, $\varphi _{\text{dd}%
}(z,\Omega ,\mu ^{\ast })=\epsilon _{1}\mu ^{\ast 2}g(\Omega )/z^{3}$,
siendo $\mu ^{\ast }=\mu /\sqrt{\epsilon _{1}r_{\text{m1}}^{3}}$. 
\begin{equation}
\varphi _{\text{GSM}_{1}}(z,\Omega ,\mu ^{\ast })=\ \epsilon _{1}\left\{
\left( \frac{1-a}{z-a}\right) ^{12}-2\left( \frac{1-a}{z-a}\right)
^{6}\right\} -\epsilon _{1}\frac{\mu ^{\ast 2}}{z^{3}}g(\Omega )
\label{MSMreducido}
\end{equation}
En donde abreviamos con $\Omega $ la dependencia angular ($\theta
_{1},\theta _{2},\phi $). $g(\Omega )=g(\theta _{1},\theta _{2},\phi )=2\cos
\theta _{1}\cos \theta _{2}-\sin \theta _{1}\sin \theta _{2}\cos \phi $.
Este c\'{a}lculo lo hacemos de manera muy similar a como lo hicimos para el
potencial de Stockmayer. Calculamos $B(T)$ usando la expresi\'{o}n (\ref%
{VirialConderivada}) que usamos en el caso anterior.\ 
\begin{equation}
B(T)=\frac{\widetilde{N}}{12}\underset{T}{\int }d\Omega \frac{\partial }{%
\partial r}\left( e^{-\frac{\varphi }{kT}}\right) r^{3}dr
\end{equation}
En donde hemos usado $\underset{T}{\int }\equiv \underset{0}{\overset{%
+\infty }{\int }}\underset{0}{\overset{2\pi }{\int }}\underset{0}{\overset{%
2\pi }{\int }}\underset{0}{\overset{2\pi }{\int }}$ y $d\Omega \equiv \sin
\theta _{1}d\theta _{1}\sin \theta _{2}d\theta _{2}d\phi $ para simplificar
la notaci\'{o}n. Notemos que, debido a que en este modelo hay un n\'{o}dulo
duro en $r=a$, el l\'{\i}mite inferior de la integral en $r$ es $r=a$. Si
reducimos la temperatura como $T^{\ast }=kT/\epsilon _{1}$ obtenemos, 
\begin{equation}
B^{\ast }\left( T^{\ast }\right) =\frac{B(\widetilde{T})}{\frac{2}{3}\pi 
\widetilde{N}r_{\text{m}1}^{3}}=\frac{1}{8\pi }\underset{T}{\int }d\Omega 
\frac{\partial }{\partial z}\left( e^{-\frac{\varphi _{\text{GSM}%
_{1}}(z,\Omega ,\mu ^{\ast })}{\epsilon _{1}T^{\ast }}}\right) z^{3}dz
\label{BreducidoMSM}
\end{equation}
El l\'{\i}mite inferior de la integral en $z$ es $z=a$. Para integrar esta
expresi\'{o}n definimos $c=1-a$ y $y=(z-a)/c$, de manera que $z=cy+a$.
Usando estas variables podemos simplificar el c\'{a}lculo definiendo un
nuevo potencial, $w(y,\Omega ,\mu ^{\ast })=\varphi _{\text{GSM}%
_{1}}(z,\Omega ,\mu ^{\ast })/\epsilon _{1}=1/y^{12}-2/y^{6}-\mu ^{\ast
2}g(\Omega )/z^{3}$. N\'{o}tese que $\mu ^{\ast }$ es un par\'{a}metro y no
una variable de integraci\'{o}n. En estas nuevas variables (\ref%
{BreducidoMSM}) queda como 
\begin{equation}
B^{\ast }\left( T^{\ast }\right) =\frac{1}{8\pi }\underset{T}{\int }d\Omega 
\frac{\partial }{\partial y}\left( e^{-\frac{w(y,\Omega ,\mu ^{\ast })}{%
\epsilon _{1}T^{\ast }}}\right) \left( cy+a\right) ^{3}dy\text{.}
\label{BreducidoMSM2}
\end{equation}
Substituyendo el potencial MSM reducido (\ref{MSMreducido}) en la funci\'{o}%
n exponencial que aparece en la integral (\ref{BreducidoMSM2}) obtenemos, 
\begin{eqnarray}
\exp \left( -\frac{w(y,\Omega ,\mu ^{\ast })}{T^{\ast }}\right) &=&\exp
\left( -\frac{1}{T^{\ast }}\frac{1}{y^{12}}\right) \exp \left( \frac{2}{%
T^{\ast }}\frac{1}{y^{6}}+\frac{1}{T^{\ast }}\frac{\mu ^{\ast 2}}{z^{3}}%
g(\Omega )\right)  \notag \\
&=&\exp \left( -\frac{1}{T^{\ast }}\frac{1}{y^{12}}\right) \overset{\infty }{%
\underset{n=0}{\sum }}\frac{1}{n!}\overset{n}{\underset{m=0}{\sum }}\binom{n%
}{m}  \notag \\
&&\times \left( \frac{2}{T^{\ast }}\frac{1}{y^{6}}\right) ^{n-m}\left( \frac{%
1}{T^{\ast }}\frac{\mu ^{\ast 2}}{z^{3}}g(\Omega )\right) ^{m}
\end{eqnarray}
Introduciendo la variable adicional $t=1/T^{\ast }y^{12}$ podemos rescribir
la expresi\'{o}n anterior como 
\begin{eqnarray}
\exp \left( -\frac{w(y,\Omega ,\mu ^{\ast })}{T^{\ast }}\right) &=&e^{-t}%
\overset{\infty }{\underset{n=0}{\sum }}\overset{n}{\underset{m=0}{\sum }}%
\frac{1}{n!}\binom{n}{m}2^{n-m}\left( \frac{1}{T^{\ast }}\right) ^{\frac{n-m%
}{2}}t^{\frac{n-m}{2}}  \notag \\
&&\times \left( \frac{1}{T^{\ast }}\frac{\mu ^{\ast 2}}{z^{3}}g(\Omega
)\right) ^{m}
\end{eqnarray}
Sustituyendo esto en (\ref{BreducidoMSM2}), despu\'{e}s de desarrollar y
simplificar, la integral de $B^{\ast }\left( T^{\ast }\right) $ queda como 
\begin{eqnarray}
B^{\ast }\left( T^{\ast }\right) &=&-\frac{1}{8\pi }\overset{\infty }{%
\underset{n=0}{\sum }}\overset{n}{\underset{m=0}{\sum }}\frac{2^{n-m}}{n!}%
\left( \frac{1}{T^{\ast }}\right) ^{\frac{n+m}{2}}\binom{n}{m}\mu ^{\ast 2m}
\notag \\
&&\times \underset{T}{\int }d\Omega g^{m}\left( \Omega \right) \left(
cy+a\right) ^{3}\frac{\partial }{\partial t}\left\{ e^{-t}t^{\frac{n-m}{2}}%
\frac{1}{z^{3m}}\right\} dt\text{.}
\end{eqnarray}
Para poder integrar debemos completar el cambio de variable en $z$.
Recordando que , $1/z^{3}=1/\left( cy+a\right) ^{3}=1/\left( c/(tT^{\ast
})^{1/12}+a\right) ^{3}$, la expresi\'{o}n anterior queda como 
\begin{eqnarray}
B^{\ast }\left( T^{\ast }\right) &=&-\frac{1}{8\pi }\overset{\infty }{%
\underset{n=0}{\sum }}\overset{n}{\underset{m=0}{\sum }}\frac{2^{n-m}}{n!}%
\left( \frac{1}{T^{\ast }}\right) ^{\frac{n+m}{2}}\binom{n}{m}\mu ^{\ast 2m}%
\underset{T}{\int }d\Omega g^{m}\left( \Omega \right)  \notag \\
&&\times \left( cy+a\right) ^{3}\frac{\partial }{\partial t}\left\{ e^{-t}t^{%
\frac{n-m}{2}}\frac{1}{\left( c/(tT^{\ast })^{1/12}+a\right) ^{3m}}\right\}
dt\text{.}  \label{VirialConDenom}
\end{eqnarray}
Este c\'{a}lculo es una generalizaci\'{o}n del SM por lo que esperamos que
las funciones involucradas en el desarrollo sean similares. Para garantizar
que la integral para $B^{\ast }\left( T^{\ast }\right) $ est\'{e} escrita en
t\'{e}rminos de funciones gamma, debemos tener en el integrando t\'{e}rminos
del tipo, $t^{\alpha -1}e^{-t}$. Haciendo el cambio de variable, $x=\frac{a}{%
c}\left( tT^{\ast }\right) ^{1/12}$, podemos escribir el factor $1/\left(
c/(tT^{\ast })^{1/12}+a\right) ^{3m}$\ que aparece en integrando de la
expresi\'{o}n anterior como 
\begin{eqnarray}
\frac{1}{\left( c/(tT^{\ast })^{1/12}+a\right) ^{3m}} &=&\frac{\left(
tT^{\ast }\right) ^{m/4}}{c^{3m}}\frac{1}{\left( 1+x\right) ^{3m}}  \notag \\
&=&\frac{\left( tT^{\ast }\right) ^{m/4}}{c^{3m}}\overset{\infty }{\underset{%
l=0}{\sum }}\QATOPD\{ \} {3m}{l}\left( \frac{a}{c}\left( tT^{\ast }\right)
^{1/12}\right) ^{l}\text{.}  \label{denominador}
\end{eqnarray}

El significado de $\QATOPD\{ \} {3m}{l}$ en (\ref{denominador}) es como
sigue. Notemos que 
\begin{equation*}
\frac{1}{\left( 1+x\right) ^{3m}}=\left( 1+x\right) ^{-3m}\overset{\infty }{=%
\underset{n=0}{\sum }}\binom{-3m}{l}x^{l}\text{.}
\end{equation*}
Debido a que $m\geqslant 0$, el coeficiente binomial $\binom{-3m}{l}$ no est%
\'{a} definido cuando la primera entrada, $-3m$, es negativa. Hay que
calcularlo espec\'{\i}ficamente. Para $\beta >0$ tenemos que 
\begin{equation*}
\binom{\beta }{1}=\beta \text{; \ \ \ }\binom{\beta }{2}=\frac{\beta \left(
\beta -1\right) }{2!}\text{; \ \ \ }\binom{\beta }{3}=\frac{\beta \left(
\beta -1\right) \left( \beta -2\right) }{3!}\text{.}
\end{equation*}
Entonces 
\begin{equation*}
\binom{\beta }{l}=\frac{\beta \left( \beta -1\right) \left( \beta -2\right)
\cdots \left( \beta -\left( l-1\right) \right) }{l!}\text{,}
\end{equation*}
de manera que 
\begin{eqnarray*}
\binom{-\beta }{l} &=&\frac{-\beta \left( -\beta -1\right) \left( -\beta
-2\right) \cdots \left( -\beta -\left( l-1\right) \right) }{l!} \\
&=&\frac{\left( -1\right) ^{l}}{l!}\beta \left( \beta +1\right) \left( \beta
+2\right) \cdots \left( \beta +l-1\right) \\
&=&\left( -1\right) ^{l}\frac{\left( \beta +l-1\right) !}{l!\left( \beta
-1\right) !}\text{. }
\end{eqnarray*}
Entonces 
\begin{equation*}
\binom{-3m}{l}=\left( -1\right) ^{l}\frac{\left( 3m+l-1\right) !}{l!\left(
3m-1\right) !}\text{.}
\end{equation*}
Y as\'{\i} 
\begin{eqnarray*}
\frac{1}{\left( 1+x\right) ^{3m}} &=&\sum_{l=0}^{\infty }\left\{ \left(
-1\right) ^{l}\frac{\left( 3m+l-1\right) !}{l!\left( 3m-1\right) !}\right\}
x^{l} \\
&=&\sum_{l=0}^{\infty }\QATOPD\{ \} {3m}{l}x^{l}\text{.}
\end{eqnarray*}
Es decir, estamos definiendo la operaci\'{o}n 
\begin{equation}
\QATOPD\{ \} {3m}{l}=\left( -1\right) ^{l}\frac{\left( 3m+l-1\right) !}{%
l!\left( 3m-1\right) !}\text{.}
\end{equation}

Usando (\ref{denominador}) la integral (\ref{VirialConDenom}) queda como 
\begin{eqnarray}
B^{\ast }\left( T^{\ast }\right) &=&-\frac{1}{8\pi }\overset{\infty }{%
\underset{n=0}{\sum }}\overset{n}{\underset{m=0}{\sum }}\overset{\infty }{%
\underset{n=0}{\sum }}\frac{2^{n-m}}{n!}\frac{a^{l}}{\left( 1-a\right)
^{3m+l}}\left( \frac{1}{T^{\ast }}\right) ^{\frac{6n+3m-l}{12}}  \notag \\
&&\times \binom{n}{m}\QATOPD\{ \} {3m}{l}\mu ^{\ast 2m}  \notag \\
&&\underset{T}{\times \int }d\Omega g^{m}\left( \Omega \right) \left(
cy+a\right) ^{3}\frac{\partial }{\partial t}\left\{ e^{-t}t^{\frac{6n+3m-l}{%
12}}\right\} dt\text{.}  \label{BMSMantes de integrar}
\end{eqnarray}
La parte radial en la integral de esta expresi\'{o}n, que denominaremos $I$,
la desarrollamos en varios t\'{e}rminos 
\begin{eqnarray}
I &=&\underset{0}{\overset{\infty }{\int }}dt\left( cy+a\right) ^{3}\frac{%
\partial }{\partial t}\left\{ e^{-t}t^{\frac{6n+3m-l}{12}}\right\}  \notag \\
&=&\underset{0}{\overset{\infty }{\int }}dt\left(
c^{3}y^{3}+3c^{2}ay^{2}+3ca^{2}y+a^{3}\right) \frac{\partial }{\partial t}%
\left\{ e^{-t}t^{\frac{6n+3m-l}{12}}\right\}  \notag \\
&=&c^{3}I_{1}+3c^{2}aI_{2}+3ca^{2}I_{3}+a^{3}I_{4}
\label{Integral I por Cuatro}
\end{eqnarray}
Cada uno de los t\'{e}rminos $I_{i}$ en (\ref{Integral I por Cuatro}) est%
\'{a} asociado con una integral que evaluamos separadamente como sigue. 
\begin{equation}
I_{1}=\underset{0}{\overset{\infty }{\int }}dty^{3}\frac{\partial }{\partial
t}\left\{ e^{-t}t^{\frac{6n+3m-l}{12}}\right\} =\underset{0}{\overset{\infty 
}{\int }}dt\left( \frac{1}{T^{\ast }}\right) ^{\frac{1}{4}}t^{-\frac{1}{4}}%
\frac{\partial }{\partial t}\left\{ e^{-t}t^{\frac{6n+3m-l}{12}}\right\}
\end{equation}
\begin{equation}
I_{2}=\underset{0}{\overset{\infty }{\int }}dty^{2}\frac{\partial }{\partial
t}\left\{ e^{-t}t^{\frac{6n+3m-l}{12}}\right\} =\underset{0}{\overset{\infty 
}{\int }}dt\left( \frac{1}{T^{\ast }}\right) ^{\frac{1}{6}}t^{-\frac{1}{6}}%
\frac{\partial }{\partial t}\left\{ e^{-t}t^{\frac{6n+3m-l}{12}}\right\}
\end{equation}
\begin{equation}
I_{3}=\underset{0}{\overset{\infty }{\int }}dty\frac{\partial }{\partial t}%
\left\{ e^{-t}t^{\frac{6n+3m-l}{12}}\right\} =\underset{0}{\overset{\infty }{%
\int }}dt\left( \frac{1}{T^{\ast }}\right) ^{\frac{1}{12}}t^{-\frac{1}{12}}%
\frac{\partial }{\partial t}\left\{ e^{-t}t^{\frac{6n+3m-l}{12}}\right\}
\end{equation}
\begin{equation}
I_{4}=\underset{0}{\overset{\infty }{\int }}dt\frac{\partial }{\partial t}%
\left\{ e^{-t}t^{\frac{6n+3m-l}{12}}\right\}
\end{equation}
Desarrollamos cada integral para escribirlas en t\'{e}rminos de funciones de
la forma $t^{\alpha -1}e^{-t}$ que sabemos que se integran como $%
\int_{0}^{\infty }$ $t^{\alpha -1}e^{-t}dt=\Gamma (\alpha )$. Despu\'{e}s de
simplificar obtenemos 
\begin{equation}
I_{1}=\frac{1}{4}\left( \frac{1}{T^{\ast }}\right) ^{\frac{1}{4}}\Gamma
\left( \frac{6n-3m+l-3}{12}\right)
\end{equation}
\begin{equation}
I_{2}=\frac{1}{6}\left( \frac{1}{T^{\ast }}\right) ^{\frac{1}{6}}\Gamma
\left( \frac{6n-3m+l-2}{12}\right)
\end{equation}
\begin{equation}
I_{3}=\frac{1}{12}\left( \frac{1}{T^{\ast }}\right) ^{\frac{1}{12}}\Gamma
\left( \frac{6n-3m+l-1}{12}\right)
\end{equation}
\begin{equation}
I_{4}=0
\end{equation}
Sustituimos los resultados de las integrales $I_{i}$ en (\ref{Integral I por
Cuatro}), que es la parte radial en la integral de $B^{\ast }\left( T^{\ast
}\right) $, de manera que (\ref{BMSMantes de integrar}) queda como 
\begin{eqnarray}
B^{\ast }\left( T^{\ast }\right) &=&-\frac{1}{8\pi }\overset{\infty }{%
\underset{n=0}{\sum }}\overset{n}{\underset{m=0}{\sum }}\overset{\infty }{%
\underset{n=0}{\sum }}\frac{2^{n-m}}{n!}\frac{a^{l}}{\left( 1-a\right)
^{3m+l}}\left( \frac{1}{T^{\ast }}\right) ^{\frac{6n+3m-l}{12}}  \notag \\
&&\times \binom{n}{m}\QATOPD\{ \} {3m}{l}\mu ^{\ast 2m}\underset{\Omega }{%
\int }d\Omega g^{m}\left( \Omega \right)  \notag \\
&&\times \left\{ \left( 1-a\right) ^{3}\frac{1}{4}\left( \frac{1}{T^{\ast }}%
\right) ^{\frac{1}{4}}\Gamma \left( \frac{6n-3m+l-3}{12}\right) \right. 
\notag \\
&&+3\left( 1-a\right) ^{2}a\frac{1}{6}\left( \frac{1}{T^{\ast }}\right) ^{%
\frac{1}{6}}\Gamma \left( \frac{6n-3m+l-2}{12}\right)  \notag \\
&&\left. +3\left( 1-a\right) a^{2}\frac{1}{12}\left( \frac{1}{T^{\ast }}%
\right) ^{\frac{1}{12}}\Gamma \left( \frac{6n-3m+l-1}{12}\right) \right\} 
\text{.}  \label{B MSM integrado1}
\end{eqnarray}
Este resultado se puede simplificar recordando que los t\'{e}rminos con $m$
impares son cero. Hacemos $m=2k$, de manera que cuando $m=n$, $2k=n$ y
entonces $k=n/2$. Similarmente cuando $m=0$, $k=0$. Considerando este cambio
de variable y recordando que $G_{k}=(1/8\pi )\int d\Omega \ g^{2k}(\Omega )$
se pueden calcular anal\'{\i}ticamente para toda $k$ finita, podemos
finalmente escribir (\ref{B MSM integrado1}) como 
\begin{eqnarray}
B^{\ast }\left( T^{\ast }\right) &=&-\overset{\infty }{\underset{n=0}{\sum }}%
\overset{k\leq \frac{n}{2}}{\underset{k=0}{\sum }}\overset{\infty }{\underset%
{n=0}{\sum }}\frac{2^{n-2k}}{n!}\frac{a^{l}}{\left( 1-a\right) ^{6k+l}}%
\left( \frac{1}{T^{\ast }}\right) ^{\frac{6n+6k-l}{12}}  \notag \\
&&\times \binom{n}{2k}\QATOPD\{ \} {6k}{l}\mu ^{\ast 4k}G_{k}  \notag \\
&&\times \left\{ \left( 1-a\right) ^{3}\frac{1}{4}\left( \frac{1}{T^{\ast }}%
\right) ^{\frac{1}{4}}\Gamma \left( \frac{6n-6k+l-3}{12}\right) \right. 
\notag \\
&&+\left( 1-a\right) ^{2}a\frac{1}{2}\left( \frac{1}{T^{\ast }}\right) ^{%
\frac{1}{6}}\Gamma \left( \frac{6n-6k+l-2}{12}\right)  \notag \\
&&\left. +\left( 1-a\right) a^{2}\frac{1}{4}\left( \frac{1}{T^{\ast }}%
\right) ^{\frac{1}{12}}\Gamma \left( \frac{6n-6k+l-1}{12}\right) \right\} 
\text{.}  \label{B MSM integrado2}
\end{eqnarray}
N\'{o}tese que este resultado se reduce al de SM cuando $a=0$.

\chapter{\textbf{Funciones }$f_{x}$}

\section{Profundidad y posici\'{o}n del m\'{\i}nimo}

\begin{figure}[h]
        \begin{center}
        \includegraphics[width=.8\hsize]{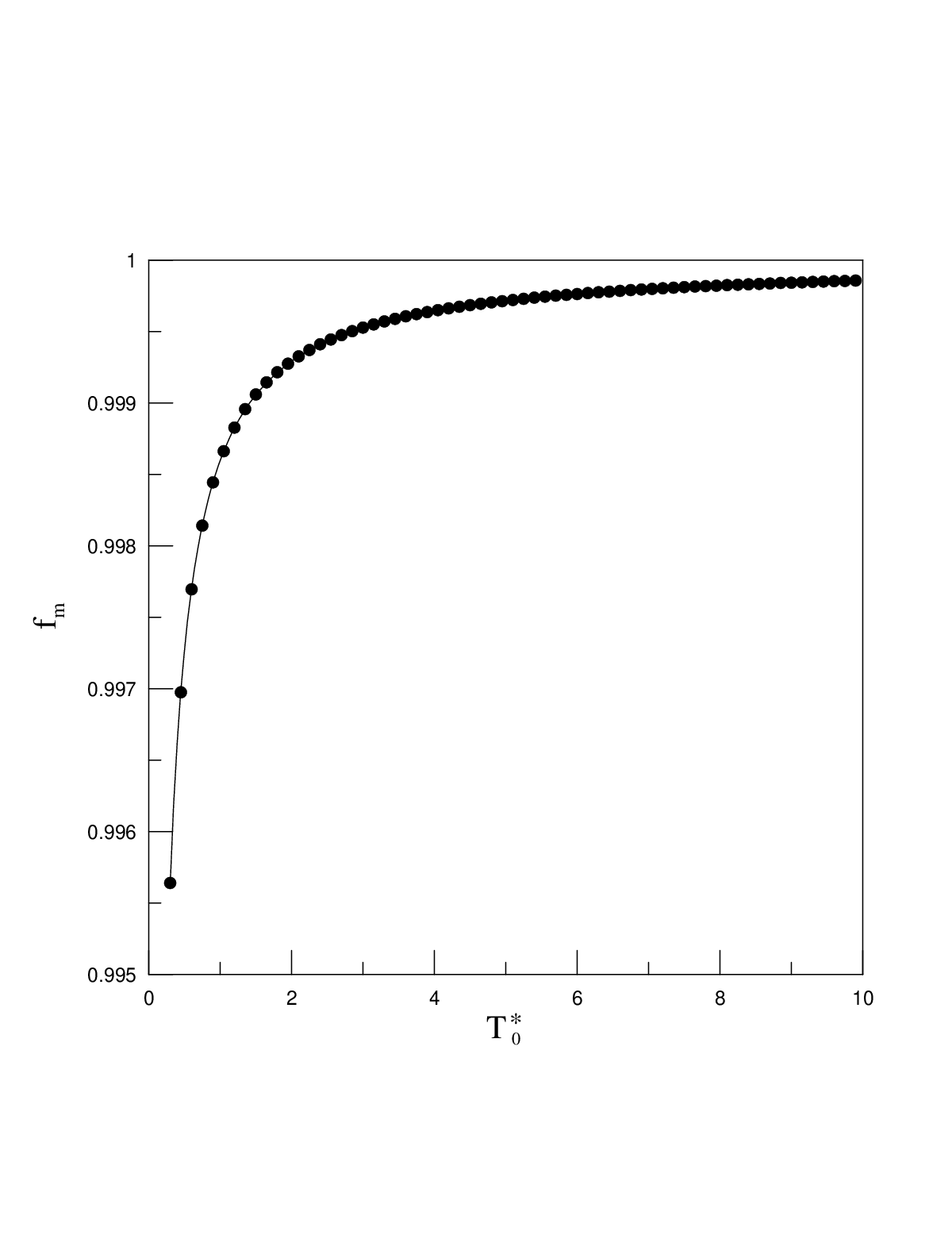}
        \end{center}
        \caption{%
Variaci\'{o}n del di\'{a}metro efectivo, $f_{\text{r}}$, con la
temperatura. Esta cantidad la obtenemos por diferenciaci\'{o}n del potencial
GSM esfericalizado a lo largo de un intervalo de 194 valores de temperatura.
La l\'{\i}nea continua es el ajuste $f_{\text{r}}=1+B_{3}/T_{0}^{\ast
}+B_{4}/T_{0}^{\ast 2}$.
                     }%
       \label{FIGpotancdipminimosm05s1}
\end{figure}

Aqu\'{\i} damos los detalles del c\'{a}lculo de las funciones $f_{x}$. En
primer lugar nos referimos al c\'{a}lculo de los coeficientes que definen
las funciones $f_{\epsilon }$ y $f_{r}$. Como mencionamos en la secci\'{o}n
donde calculamos estas funciones, el procedimiento es como sigue: definimos
un modelo con una pareja de valores $(\mu _{0}^{\ast },s_{0})$ y para cada
temperatura dentro del intervalo 0.3$\leq T_{0}^{\ast }<$10.0 localizamos su
m\'{\i}nimo por diferenciaci\'{o}n y registramos los valores de $\epsilon $
y $r_{\text{m}}$. La variaci\'{o}n de $\epsilon $ y $r_{\text{m}}$ con la
temperatura para cada modelo, la podemos representar como, 
\begin{eqnarray}
f_{\epsilon } &=&1+B_{1}/T_{0}^{\ast }+B_{2}/T_{0}^{\ast 2} \\
f_{r} &=&1+B_{3}/T_{0}^{\ast }+B_{4}/T_{0}^{\ast 2}
\end{eqnarray}
En la Fig. (\ref{FIGpotancdipminimosm05s1}) se muestra el caso con $s_{0}=$1
y $\mu _{0}^{\ast }=$0.5 para $f_{r}$.
%
%
En el
Cuadro \ref{fx 1}\emph{\ }se muestran los coeficientes $B_{i}$ que obtuvimos
para los distintos modelos que usamos. 
\begin{table}[tbp] \centering%
\begin{tabular}{|rrrrrr|}
\hline
$s_{0}$ & $\mu _{0}^{\ast }$ & $B_{1}$ & $B_{2}$ & $B_{3}$ & $B_{4}$ \\ 
\hline
0.5 & 0.25 & 0.00130304 & -0.00000092 & -0.00002221 & 0.00000004 \\ 
& 0.5 & 0.02105990 & -0.00022144 & -0.00036163 & 0.00000992 \\ 
& 0.75 & 0.10973600 & -0.00456762 & -0.00186656 & 0.00017062 \\ 
& 0.9 & 0.23178200 & -0.01566220 & -0.00377248 & 0.00049651 \\ 
& 1.0 & 0.35607800 & -0.03030970 & -0.00551482 & 0.00085130 \\ 
0.7 & 0.25 & 0.00130304 & -0.00000083 & -0.00004354 & 0.00000009 \\ 
& 0.5 & 0.02106400 & -0.00020361 & -0.00070878 & 0.00001995 \\ 
& 0.75 & 0.11006400 & -0.00432076 & -0.00365351 & 0.00034308 \\ 
& 0.9 & 0.23358800 & -0.01513230 & -0.00735742 & 0.00099361 \\ 
& 1.0 & 0.36049800 & -0.02969080 & -0.01071120 & 0.00169438 \\ 
0.8 & 0.25 & 0.00130304 & -0.00000078 & -0.00005687 & 0.00000011 \\ 
& 0.5 & 0.02106660 & -0.00019238 & -0.00092585 & 0.00002609 \\ 
& 0.75 & 0.11026900 & -0.00416281 & -0.00477578 & 0.00045117 \\ 
& 0.9 & 0.23473200 & -0.01478940 & -0.00961823 & 0.00130858 \\ 
& 1.0 & 0.36331900 & -0.02928690 & -0.01399630 & 0.00223124 \\ 
0.9 & 0.25 & 0.00130305 & -0.00000073 & -0.00007197 & 0.00000014 \\ 
& 0.5 & 0.02106940 & -0.00017942 & -0.00117200 & 0.00003282 \\ 
& 0.75 & 0.11050000 & -0.00397900 & -0.00605531 & 0.00057280 \\ 
& 0.9 & 0.23604200 & -0.01438390 & -0.01221200 & 0.00166828 \\ 
& 1.0 & 0.36658300 & -0.02880190 & -0.01778220 & 0.00284930 \\ 
1.0 & 0.25 & 0.00130305 & -0.00000066 & -0.00008885 & 0.00000017 \\ 
& 0.5 & 0.02107240 & -0.00016511 & -0.00144729 & 0.00003997 \\ 
& 0.75 & 0.11075700 & -0.00376635 & -0.00749693 & 0.00070673 \\ 
& 0.9 & 0.23752600 & -0.01390410 & -0.01516070 & 0.00207277 \\ 
& 1.0 & 0.37032500 & -0.02821490 & -0.02211550 & 0.00355266 \\ \hline
\end{tabular}
\caption{Coeficientes $B_{i}$} \label{fx 1} 
\end{table}%

Para cada $s_{0}$ ajustamos la variaci\'{o}n con $\mu _{0}^{\ast }$ de los
coeficientes $B_{i}$ que resulta ser como $B_{1,3}=b_{1,3}\mu _{0}^{\ast 4}$
y $B_{2,4}=b_{2,4}\mu _{0}^{\ast 8}$ y en el Cuadro \ref{fx 2} mostramos los
coeficientes $b_{i,j}$ que corresponden a cada valor de $s_{0}$. 
\begin{table}[tbp] \centering%
\begin{tabular}{|ccccc|}
\hline
$s_{0}$ & $b_{1}$ & $b_{2}$ & $b_{3}$ & $b_{4}$ \\ \hline
0.5 & 0.35464 & -0.03138 & -0.00561 & 0.00091 \\ 
0.7 & 0.35836 & -0.03065 & -0.01091 & 0.00180 \\ 
0.8 & 0.36073 & -0.03018 & -0.01426 & 0.00238 \\ 
0.9 & 0.36346 & -0.02961 & -0.01811 & 0.00303 \\ 
1.0 & 0.36659 & -0.02893 & -0.02250 & 0.00378 \\ \hline
\end{tabular}
\caption{Coeficientes $b_{i}$} \label{fx 2} 
\end{table}%

El lector puede comprobar que los coeficientes $b_{i}$ que aparecen en el
Cuadro \ref{fx 2} admiten un ajuste lineal con $s_{0}$. De manera que las
funciones $f_{\epsilon }$ y $f_{r}$ --Ecs. (\ref{EfeEps}) y(\ref{EfeRm})--
quedan definidas por los coeficientes $c_{ij}$ que resultan de esas
relaciones lineales (ver Cuadro \ref{fx 3}). 
\begin{table}[tbp] \centering%
\begin{tabular}{|c|cccc|}
\hline
$j\backslash i$ & $1$ & $2$ & $3$ & $4$ \\ \hline
$0$ & 0.011898 & -0.002073 & 0.3422 & -0.03393 \\ 
$1$ & -0.03355 & 0.005708 & 0.02373 & 0.004851 \\ \hline
\end{tabular}
\caption{Coeficientes $c_{ij}$} \label{fx 3} 
\end{table}%

\section{Suavidad: aproximaci\'{o}n 2-$s$}

\begin{figure}[h]
        \begin{center}
        \includegraphics[width=.8\hsize]{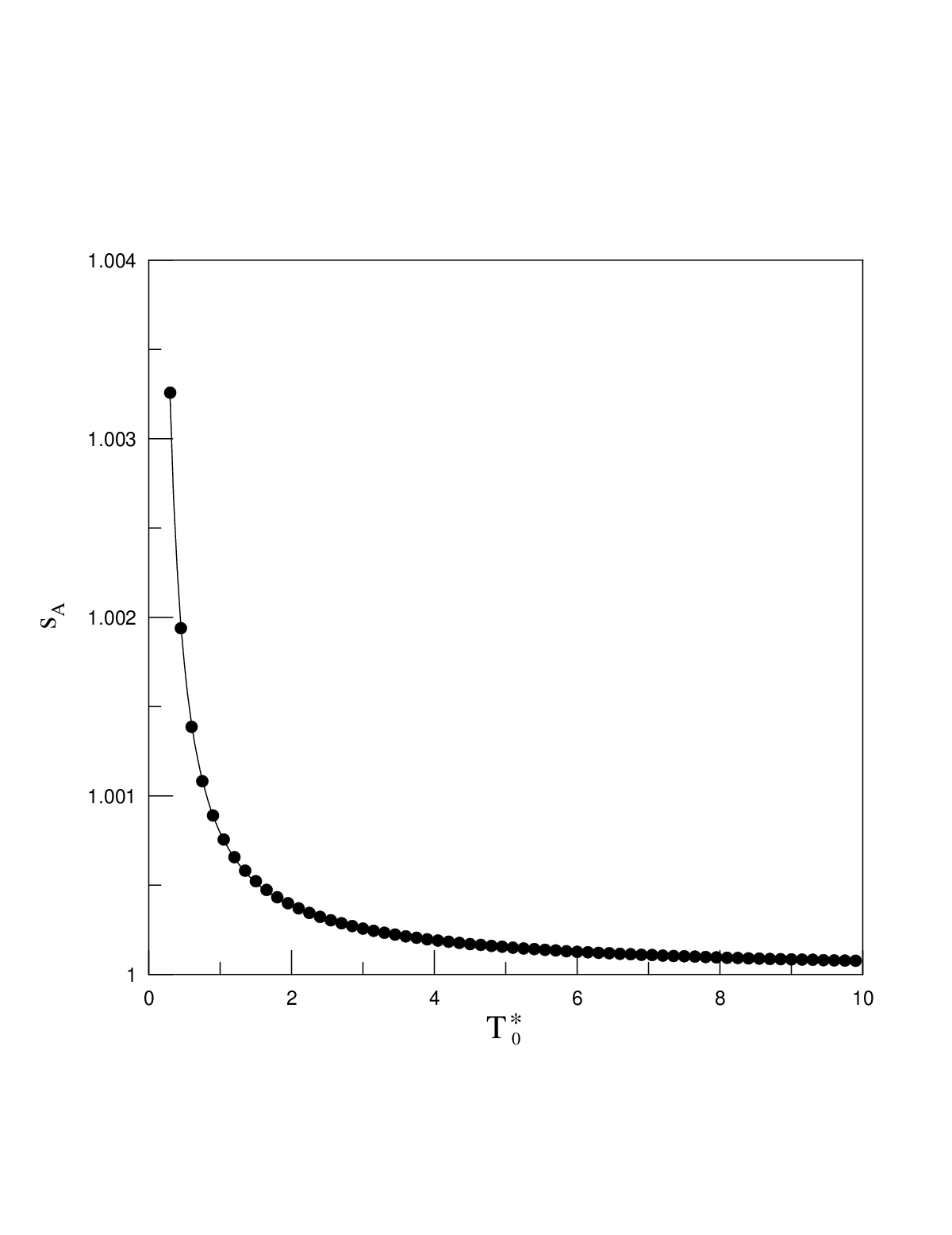}
        \end{center}
        \caption{%
Variaci\'{o}n de la suavidad atractiva, $s_{\text{A}}$, con la
temperatura. Para calcular cada punto evaluamos dos integrales num\'{e}%
ricas. Por cada uno de los 194 puntos que se muestran, evaluamos dos veces
la Ec. (\protect\ref{lambdacalc}). La l\'{\i}nea continua es el ajuste $s_{%
\text{A}}=s_{0}+a_{1}/T_{0}^{\ast }+a_{2}/T_{0}^{\ast 2}$.
                     }%
       \label{FIGsas1mu05}
\end{figure}

Para calcular las funciones $f_{s_{\text{A}}}$ y $f_{s_{\text{R}}}$%
recordamos que cada modelo $\psi (z_{\text{ef}})$ definido por una pareja $%
(s_{0},\mu _{0}^{\ast })$ es \textit{conformal} a $\varphi _{\text{ANC}}(z_{%
\text{ef}},s_{\text{A}},s_{\text{R}})$, por tanto --respectivamente para la
rama atractiva y la repulsiva-- se cumplen las relaciones lineales (\ref{be}%
) y (\ref{lambda}), en donde $s_{\text{R}}$ y $s_{\text{A}}$ son funciones
de $T_{0}^{\ast }$ definidas en intervalo 0.3$\leq T_{0}^{\ast }<$10.0. Para
extraer $s_{\text{A}}(T_{0}^{\ast })$ y $s_{\text{R}}(T_{0}^{\ast })$
integramos num\'{e}ricamente las Ecs. (\ref{lambdacalc}) y (\ref{becalc}) a
lo largo del intervalo $0<z<100$ y usando un paso de temperatura de 0.05; es
decir, realizamos $194$ integrales por cada rama y para cada modelo. La
variaci\'{o}n de $s_{\text{A}}$ y $s_{\text{R}}$ con la temperatura para
cada modelo, la podemos representar como, 
\begin{eqnarray}
s_{\text{A}} &=&s_{0}+a_{1}/T_{0}^{\ast }+a_{2}/T_{0}^{\ast 2} \\
s_{\text{R}} &=&s_{0}+a_{3}/T_{0}^{\ast }+a_{4}/T_{0}^{\ast
2}+a_{5}/T_{0}^{\ast 3}+a_{6}/T_{0}^{\ast 4}
\end{eqnarray}
En la Fig. \ref{FIGsas1mu05} se muestra el caso con $s_{0}=$1 y $\mu
_{0}^{\ast }=$0.5 para $s_{\text{A}}$. 
%
En los Cuadros \ref%
{fss} y \ref{fss Cont}\ se muestran los coeficientes $a_{i}$ que obtuvimos
para los distintos modelos que usamos. 
\begin{table}[tbp] \centering%
\begin{tabular}{|rrrrr|}
\hline
$s_{0}$ & $\mu _{0}^{\ast }$ & $a_{1}$ & $a_{2}$ & $a_{3}$ \\ \hline
0.5 & 0.25 & 4.0800$\times 10^{-4}$ & -3.5385$\times 10^{-7}$ & 8.4638$%
\times 10^{-5}$ \\ 
& 0.5 & 6.4940$\times 10^{-3}$ & -8.2806$\times 10^{-5}$ & 1.3563$\times
10^{-3}$ \\ 
& 0.75 & 0.0318 & -1.5073$\times 10^{-3}$ & 6.7328$\times 10^{-3}$ \\ 
& 0.9 & 0.0622 & -4.4652$\times 10^{-3}$ & 0.0122 \\ 
& 1.0 & 0.0894 & -7.6263$\times 10^{-3}$ & 0.0168 \\ 
0.7 & 0.25 & 2.0607$\times 10^{-4}$ & 3.5763$\times 10^{-8}$ & 1.2045$\times
10^{-4}$ \\ 
& 0.5 & 3.2543$\times 10^{-3}$ & 7.2992$\times 10^{-7}$ & 1.9305$\times
10^{-3}$ \\ 
& 0.75 & 0.0159 & 4.3905$\times 10^{-5}$ & 9.5902$\times 10^{-3}$ \\ 
& 0.9 & 0.0320 & -2.8990$\times 10^{-4}$ & 0.01707 \\ 
& 1.0 & 0.0482 & -1.3902$\times 10^{-3}$ & 0.01969 \\ 
0.8 & 0.25 & 1.3340$\times 10^{-4}$ & 1.6927$\times 10^{-7}$ & 1.4826$\times
10^{-4}$ \\ 
& 0.5 & 2.0869$\times 10^{-3}$ & 3.8581$\times 10^{-5}$ & 2.3763$\times
10^{-3}$ \\ 
& 0.75 & 0.0101 & 5.9608$\times 10^{-4}$ & 0.0118 \\ 
& 0.9 & 0.0210 & 1.1543$\times 10^{-3}$ & 0.0212 \\ 
& 1.0 & 0.0332 & 5.829$\times 10^{-4}$ & 0.0244 \\ 
0.9 & 0.25 & 8.0278$\times 10^{-5}$ & 2.6535$\times 10^{-7}$ & 1.8567$\times
10^{-4}$ \\ 
& 0.5 & 1.2320$\times 10^{-3}$ & 6.1484$\times 10^{-5}$ & 2.9757$\times
10^{-3}$ \\ 
& 0.75 & 5.8142$\times 10^{-3}$ & 1.0063$\times 10^{-3}$ & 0.0149 \\ 
& 0.9 & 0.0129 & 2.1640$\times 10^{-3}$ & 0.0269 \\ 
& 1.0 & 0.0224 & 1.7384$\times 10^{-3}$ & 0.0313 \\ 
1.0 & 0.25 & 4.7891$\times 10^{-5}$ & 3.2693$\times 10^{-7}$ & 2.2870$\times
10^{-4}$ \\ 
& 0.5 & 7.0855$\times 10^{-4}$ & 7.6500$\times 10^{-5}$ & 3.6654$\times
10^{-3}$ \\ 
& 0.75 & 3.1700$\times 10^{-3}$ & 1.2775$\times 10^{-3}$ & 0.0184 \\ 
& 0.9 & 7.9769$\times 10^{-3}$ & 2.7253$\times 10^{-3}$ & 0.0335 \\ 
& 1.0 & 0.0160 & 2.0110$\times 10^{-3}$ & 0.0461 \\ \hline
\end{tabular}
\caption{Coeficientes $a_{1}$, $a_{2}$, y $a_{3}$} \label{fss} 
\end{table}
\begin{table}[tbp] \centering%
\begin{tabular}{|rrrrr|}
\hline
$s_{0}$ & $\mu _{0}^{\ast }$ & $a_{4}$ & $a_{5}$ & $a_{6}$ \\ \hline
0.5 & 0.25 & -6.5364$\times 10^{-7}$ & 2.4427$\times 10^{-7}$ & -1.0913$%
\times 10^{-9}$ \\ 
& 0.5 & -1.7244$\times 10^{-4}$ & 7.1391$\times 10^{-5}$ & -3.6764$\times
10^{-6}$ \\ 
& 0.75 & -4.0038$\times 10^{-3}$ & 1.9211$\times 10^{-3}$ & -2.2434$\times
10^{-4}$ \\ 
& 0.9 & -0.0131 & 6.4622$\times 10^{-3}$ & -8.9663$\times 10^{-4}$ \\ 
& 1.0 & -0.0282 & 0.0146 & -2.3777$\times 10^{-3}$ \\ 
0.7 & 0.25 & -1.0266$\times 10^{-6}$ & 3.6426$\times 10^{-7}$ & -1.5956$%
\times 10^{-9}$ \\ 
& 0.5 & -2.7102$\times 10^{-4}$ & 1.0762$\times 10^{-4}$ & -5.4891$\times
10^{-6}$ \\ 
& 0.75 & -6.3631$\times 10^{-3}$ & 3.0110$\times 10^{-3}$ & -3.5129$\times
10^{-4}$ \\ 
& 0.9 & -0.0208 & 0.0103 & -1.4356$\times 10^{-3}$ \\ 
& 1.0 & -0.0366 & 0.0182 & -2.7492$\times 10^{-3}$ \\ 
0.8 & 0.25 & -1.2625$\times 10^{-6}$ & 4.2745$\times 10^{-7}$ & -1.8129$%
\times 10^{-9}$ \\ 
& 0.5 & -3.3328$\times 10^{-4}$ & 1.2715$\times 10^{-4}$ & -6.3705$\times
10^{-6}$ \\ 
& 0.75 & -7.9160$\times 10^{-3}$ & 3.6703$\times 10^{-3}$ & -4.2603$\times
10^{-4}$ \\ 
& 0.9 & -0.0261 & 0.0128 & -1.7792$\times 10^{-3}$ \\ 
& 1.0 & -0.0457 & 0.0227 & -3.4258$\times 10^{-3}$ \\ 
0.9 & 0.25 & -1.5352$\times 10^{-6}$ & 4.9242$\times 10^{-7}$ & -1.9776$%
\times 10^{-9}$ \\ 
& 0.5 & -4.0528$\times 10^{-3}$ & 1.4755$\times 10^{-4}$ & -7.1842$\times
10^{-6}$ \\ 
& 0.75 & -9.7704$\times 10^{-3}$ & 4.4231$\times 10^{-3}$ & -5.1004$\times
10^{-4}$ \\ 
& 0.9 & -0.0325 & 0.0158 & -2.1871$\times 10^{-3}$ \\ 
& 1.0 & -0.0569 & 0.0282 & -4.2530$\times 10^{-3}$ \\ 
1.0 & 0.25 & -1.8377$\times 10^{-6}$ & 5.6949$\times 10^{-7}$ & -2.1733$%
\times 10^{-9}$ \\ 
& 0.5 & -4.8567$\times 10^{-4}$ & 1.7177$\times 10^{-4}$ & -8.1654$\times
10^{-6}$ \\ 
& 0.75 & -0.0119 & 5.3139$\times 10^{-3}$ & -6.1199$\times 10^{-4}$ \\ 
& 0.9 & -0.0397 & 0.0192 & -2.6774$\times 10^{-3}$ \\ 
& 1.0 & -0.0841 & 0.0444 & -7.3193$\times 10^{-3}$ \\ \hline
\end{tabular}
\caption{Coeficientes $a_{4}$, $a_{5}$, y $a_{6}$} \label{fss Cont} 
\end{table}%

Para obtener $f_{s\text{A}}$ y $f_{s\text{R}}$ ajustamos los coeficientes $%
a_{i}$ de acuerdo a las formas funcionales que se propusieron en las Ecs. (%
\ref{SA}) y (\ref{SR}), en donde los polinomios $p_{i}$ y $q_{i}$ est\'{a}n
dados por las siguientes funciones, 
\begin{eqnarray}
p_{1}(s_{0}) &=&\left. \left( \text{0.2734}-\text{0.4677}s_{0}+\text{0.2087}%
s_{0}^{^{2}}\right) \right/ s_{0} \\
p_{2}(s_{0}) &=&-\left. \left( \text{0.06087}-\text{0.1582}s_{0}+\text{0.1042%
}s_{0}^{^{2}}\right) \right/ s_{0} \\
q_{1}(s_{0}) &=&\left. (\text{0.05736}-\text{0.14846}s_{0}+\text{0.13886}%
s_{0}^{^{2}})\right/ s_{0} \\
q_{2}(s_{0}) &=&-\left. (\text{0.08549}-\text{0.2253}s_{0}+\text{0.2198}%
s_{0}^{^{2}})\right/ s_{0} \\
q_{3}(s_{0}) &=&\left. (\text{0.05724}-\text{0.1546}s_{0}+\text{0.1407}%
s_{0}^{^{2}})\right/ s_{0} \\
q_{4}(s_{0}) &=&-\left. (\text{0.01218}-\text{0.03365}s_{0}+\text{0.02856}%
s_{0}^{^{2}})\right/ s_{0}
\end{eqnarray}

\section{Suavidad: aproximaci\'{o}n $1$-$s$ anal\'{\i}tica}

En esta secci\'{o}n damos los detalles de la aproximaci\'{o}n 1-$s$ anal%
\'{\i}tica. Resumiendo el procedimiento: usamos la aproximaci\'{o}n de $2 $%
-s para calcular $B(T)$, para cada modelo GSM definido con $(s_{0},\mu
_{0}^{\ast })$, dentro del intervalo 0.5$\leq s_{0}<$1.1315\ y 0.0$\leq \mu
_{0}^{\ast }<$1.2, calculamos $T_{0B}^{\ast }$. Representando este valor en
el espacio efectivo como $T_{0B\text{ef}}^{\ast }=T_{0B}^{\ast }/f_{\epsilon
}$, podemos calcular la suavidad efectiva correspondiente (v\'{e}ase la Ref. 
\cite{del Rio 1998 III}). Encontramos que $s_{\text{ef}}(s_{0},\mu
_{0}^{\ast },T_{0}^{\ast })$ es una funci\'{o}n que se puede aproximar
bastante bien como una relaci\'{o}n lineal con $\left. \mu _{0}^{\ast
4}\right/ T_{0}^{\ast }$, tal que la podemos escribir como 
\begin{equation}
s_{\text{ef}}=s_{\text{0}}f_{s}(s_{0},\mu _{0}^{\ast },T_{0}^{\ast
})=s_{0}(1+A(s_{0})\left. \mu _{0}^{\ast 4}\right/ T_{0}^{\ast })
\end{equation}
donde los coeficientes $A$ relativos a los distintos valores de $s_{0}$, los
condensamos en el Cuadro \ref{fs analitic}. Estos coeficientes $A$ los
ajustamos a una funci\'{o}n cuadr\'{a}tica; de manera que, 
\begin{equation}
A(s_{0})=\text{0.44784}-\text{0.88078}s_{0}+\text{0.44504}s_{0}^{2}\text{.}
\end{equation}
\begin{table}[tbp] \centering%
\begin{tabular}{|cc|}
\hline
$s_{0}$ & $A$ \\ \hline
1.1315 & 0.018525 \\ 
1.0 & 0.016152 \\ 
0.8 & 0.029384 \\ 
0.7 & 0.047031 \\ 
0.6 & 0.075901 \\ 
0.5 & 0.121883 \\ \hline
\end{tabular}
\caption{Coeficientes A} \label{fs analitic} 
\end{table}%

\section{Suavidad: aproximaci\'{o}n $1$-$s$ num\'{e}rica}

Aqu\'{\i} damos los detalles de la aproximaci\'{o}n 1-$s$ num\'{e}rica.
Asumimos que la dependencia de $s$ con la temperatura es similar a la de $s_{%
\text{A}}$ \'{o} $s_{\text{R}}$, \textit{i.e}., 
\begin{equation}
s_{\text{ef}}(s_{0},\mu _{0}^{\ast },T_{0}^{\ast })=s_{0}(1+J/T_{0}^{\ast
}+K/T_{0}^{\ast 2})
\end{equation}
Ajustamos los coeficientes $(J,K)$ por inversi\'{o}n de datos de $B(T)$ al
modelo ANC de una suavidad con par\'{a}metros efectivos $\epsilon $ y $r_{%
\text{m}}$. Los datos que usamos en la inversi\'{o}n los calculamos
integrando num\'{e}ricamente distintos modelos del potencial GSM con $%
(s_{0},\mu _{0}^{\ast })$ dentro del intervalo 0.5$\leq s_{0}<$1.1315\ y 0.0$%
\leq \mu _{0}^{\ast }<$1.0.\footnote{%
Para mantener peque\~{n}o el error ($\sigma ^{2}\lesssim 0.001$) en la
regresi\'{o}n cuando $s_{0}\geqslant 1$ y $\mu _{0}^{\ast }\geqslant 0.9$,
la temperatura m\'{\i}nima a partir de la cual empezamos a invertir los
datos de $B(T)$ la escogemos seg\'{u}n el criterio: $T_{0\min }^{\ast
}\geqslant T_{0\text{B}}^{\ast }/10$. Siendo $T_{0\text{B}}^{\ast }$ la
temperatura de Boyle. Siguiendo este criterio la $T_{0\min }^{\ast }$ a la
que calculamos $J$ y $K$ es $0.3$ en la mayor\'{\i}a de los casos; salvo en $%
s_{0}=1,$ $\mu _{0}^{\ast }=0.9,1.0$, en donde usamos $T_{0\min }^{\ast
}=0.4 $ y en $s_{0}=1.1315,$ $\mu _{0}^{\ast }=0.9,1.0$, en donde usamos $%
T_{0\min }^{\ast }=0.45$.} En el Cuadro \ref{fs numeric 1} mostramos los
coeficientes $(J,K)$ que calculamos para los distintos modelos considerados. 
\begin{table}[tbp] \centering%
\begin{tabular}{|rrrr|}
\hline
$s_{0}$ & $\mu _{0}^{\ast }$ & $J$ & $K$ \\ \hline
0.5 & 0.25 & 0.00081 & 0.00014 \\ 
& 0.5 & 0.01391 & 0.00176 \\ 
& 0.75 & 0.08659 & 0.00284 \\ 
& 0.9 & 0.21368 & -0.00497 \\ 
& 1.0 & 0.38625 & -0.02608 \\ 
0.7 & 0.25 & 0.00047 & 0.00016 \\ 
& 0.5 & 0.00847 & 0.00228 \\ 
& 0.75 & 0.05997 & 0.00613 \\ 
& 0.9 & 0.16508 & 0.00047 \\ 
& 1.0 & 0.33508 & -0.02640 \\ 
0.8 & 0.25 & 0.00038 & 0.00017 \\ 
& 0.5 & 0.00705 & 0.00237 \\ 
& 0.75 & 0.05283 & 0.00699 \\ 
& 0.9 & 0.15288 & 0.00171 \\ 
& 1.0 & 0.32834 & -0.02897 \\ 
1.0 & 0.25 & -0.00041 & 0.00039 \\ 
& 0.5 & 0.00495 & 0.00260 \\ 
& 0.75 & 0.04457 & 0.00819 \\ 
& 0.9 & 0.03718 & 0.04688 \\ 
& 1.0 & 0.07825 & 0.06699 \\ 
1.1315 & 0.25 & -0.00220 & 0.00096 \\ 
& 0.5 & 0.00304 & 0.00307 \\ 
& 0.75 & 0.01814 & 0.01874 \\ 
& 0.9 & 0.01147 & 0.06036 \\ 
& 1.0 & 0.03518 & 0.08986 \\ \hline
\end{tabular}
\caption{Coeficientes $J$, $K$} \label{fs numeric 1} 
\end{table}%
Los coeficientes $J$ y $K$ siguen una tendencia con $\mu _{0}^{\ast }$ que,
respectivamente, es de la forma, $J=j(s_{0})\mu _{0}^{\ast 4}$ y $%
K=k(s_{0})\mu _{0}^{\ast 6}+l(s_{0})\mu _{0}^{\ast 8}$. En la Tabla \ref{fs
numeric 2} est\'{a}n condensados los coeficientes $j$, $k$ y $l$ para cada $%
s_{0}$. 
\begin{table}[tbp] \centering%
\begin{tabular}{|cccc|}
\hline
$s_{0}$ & $j$ & $k$ & $l$ \\ \hline
0.5 & 0.36149 & 0.06610 & -0.09230 \\ 
0.7 & 0.30166 & 0.11622 & -0.14261 \\ 
0.8 & 0.29052 & 0.13571 & -0.16456 \\ 
1.0 & 0.07626 & 0.11439 & -0.04568 \\ 
1.1315 & 0.03168 & 0.17869 & -0.08786 \\ \hline
\end{tabular}
\caption{Coeficientes $j$, $k$, $l$} \label{fs numeric 2} 
\end{table}
Finalmente, parametrizamos los coeficientes $j$, $k$ y $l$ como polinomios
cuadr\'{a}ticos en $s_{0}$ que definen la aproximaci\'{o}n (\ref{s numerica}%
),

\begin{eqnarray}
j(s_{0}) &=&\text{0.31411}+\text{0.40327}s_{0}-\text{0.59423}s_{0}^{2}\text{,%
} \\
k(s_{0}) &=&-\text{0.03549}+\text{0.25021}s_{0}-\text{0.06699}s_{0}^{2}\text{%
,} \\
l(s_{0}) &=&\text{0.16070}-\text{0.78800}s_{0}+\text{0.52430}s_{0}^{2}\text{.%
}
\end{eqnarray}

\chapter{\textbf{Funciones }$g_{x}$}

Aqu\'{\i} se encuentran expl\'{\i}citamente los coeficientes que definen las
funciones $g_{x}$. Estos coeficientes son funciones polinomiales que toman
en cuenta los efectos de la elongaci\'{o}n sobre los par\'{a}metros
moleculares. Para obtener estas funciones invertimos el segundo coeficiente
virial de las mol\'{e}culas diat\'{o}micas 2-CLJ\ sin momento dipolar y con
varias elongaciones, al modelo ANC 2-$s$ con par\'{a}metros efectivos. El
lector puede consultar los resultados de la inversi\'{o}n en el Cuadro \ref%
{gx}. En este c\'{a}lculo consideramos que ambas suavidades se corrigen con
el mismo factor, i.e., $g_{S_{\text{A}}}=g_{S_{\text{R}}}=g_{s}$. Realizamos
la inversi\'{o}n dentro de los intervalos de temperatura y elongaci\'{o}n,
0.35$\lesssim T^{\ast }\lesssim $10 y 0.2$\leq \widetilde{L}\leq $ 1.0,
respectivamente. En todos los casos cuando $\widetilde{L}=0$, inducimos los
valores de $g_{r}$, $g_{\epsilon }$ y $s_{0}g_{s}$ a que tomen
respectivamente los valores 1, 1 y 1.1315. 
\begin{table}[tbp] \centering%
\begin{tabular}{|lllll|}
\hline
$\mu ^{\ast }$ & $\widetilde{L}$ & $g_{r}$ & $g_{\epsilon }$ & $s_{0}g_{s}$
\\ \hline
0.0 & 0.0 & 1.0 & 1.0 & 1.1315 \\ 
& 0.1 & 1.0216 & 0.9437 & 1.1257 \\ 
& 0.2 & 1.0483 & 0.8597 & 1.0890 \\ 
& 0.3 & 1.0796 & 0.7786 & 1.0398 \\ 
& 0.4 & 1.1113 & 0.7120 & 0.9884 \\ 
& 0.5 & 1.1430 & 0.6580 & 0.9403 \\ 
& 0.6 & 1.1734 & 0.6150 & 0.8970 \\ 
& 0.7 & 1.2057 & 0.5762 & 0.8623 \\ 
& 0.8 & 1.2354 & 0.5458 & 0.8317 \\ 
& 0.9 & 1.2637 & 0.5206 & 0.8059 \\ 
& 1.0 & 1.2905 & 0.4997 & 0.7844 \\ \hline
\end{tabular}
\caption{Resultados de la inversi\'on para los par\'ametros $g_{x}$} \label%
{gx} 
\end{table}
Ajustamos funciones sencillas de $\widetilde{L}$ a los datos del Cuadro \ref%
{gx} y obtenemos las funciones $g_{r}(\widetilde{L})$, $g_{\epsilon }(%
\widetilde{L})$ y $g_{s}(\widetilde{L})$ que definen las Ecs. (\ref{epsmol}%
), (\ref{rmmol}) y (\ref{smol}). 
\begin{eqnarray}
g_{\epsilon }(\widetilde{L}) &=&1-\text{0.363708}\widetilde{L}-\text{2.880128%
}\widetilde{L}^{2}+\text{7.444486}\widetilde{L}^{3}  \notag \\
&&-\text{7.206996}\widetilde{L}^{4}+\text{2.506727}\widetilde{L}^{5}\text{,}
\\
g_{r}(\widetilde{L}) &=&1+\text{0.213242}\widetilde{L}+\text{0.205587}%
\widetilde{L}^{2}-\text{0.129050}\widetilde{L}^{3}, \\
g_{s}(\widetilde{L}) &=&(\text{1.1315}+\text{0.159028}\widetilde{L}-\text{%
2.628141}\widetilde{L}^{2}+\text{4.525942}\widetilde{L}^{3}  \notag \\
&&-\text{3.351117}\widetilde{L}^{4}+\text{0.947182}\widetilde{L}^{5})/\text{%
1.1315}.
\end{eqnarray}

\chapter{\textbf{Funciones }$h_{x}$}

Aqu\'{\i} se encuentran expl\'{\i}citamente los coeficientes que definen las
funciones $h_{x}$. Estos coeficientes son funciones polinomiales que
calculamos por inversi\'{o}n de datos $B^{\ast }(T^{\ast })$ de los modelos
2-CLJ definidos para distintos valores de la inclinaci\'{o}n del momento
dipolar, $\alpha $, respecto al eje de simetr\'{\i}a molecular en el
intervalo, $0\leq \alpha \leq \pi /2$. El m\'{e}todo de inversi\'{o}n viene
descrito en la secci\'{o}n: \textquotedblleft Diat\'{o}micas con momento
dipolar inclinado\textquotedblright . Realizamos la inversi\'{o}n en los
intervalos, 0.2$\leq \widetilde{L}\leq $ 1.0, y 0.420$\leq \mu _{\text{at}%
}^{\ast }\leq $ 0.841.

La forma de la dependencia funcional de $h_{x}$ con $\widetilde{L}$ la
conjeturamos de manera similar que aquella de los par\'{a}metros $f_{x}$. Es
decir, gr\'{a}ficamente notamos que el mejor ajuste de $h_{\epsilon }$ y $%
h_{s}$ es cuadr\'{a}tico y el de $h_{s}$ resulta ser c\'{u}bico.

\section{Correlaci\'{o}n $h_{\protect\epsilon }$}

La funci\'{o}n $h_{\epsilon }$ la consideramos de la forma, $h_{\epsilon
}=1+d_{11}\widetilde{L}+d_{12}\widetilde{L}^{3}$. Para valores fijos de $\mu
_{0}^{\ast }$ y $\alpha $ ajustamos los coeficientes $d_{11}$ y $d_{12}$
(ver Cuadro \ref{d11,12}). Para cualquier valor de $\mu _{\text{at}}^{\ast }$%
, el lector puede comprobar que la dependencia en $\alpha $ de los
coeficientes es senoidal --de la forma $d_{11}=d_{11}^{0}(\mu _{\text{at}%
}^{\ast })\sin 3\alpha $ y $d_{12}=d_{12}^{0}(\mu _{\text{at}}^{\ast })\sin
\alpha $. Para cada valor fijo de $\mu _{\text{at}}^{\ast }$ ajustamos las
amplitudes $d_{1i}^{0}$ a funciones de orden $4$ en su dependencia con $\mu
_{\text{at}}^{\ast }$ (ver Cuadro \ref{d01i}) para finalmente obtener: 
\begin{eqnarray}
d_{11}(\mu _{\text{at}}^{\ast },\alpha ) &=&-\text{0.54889}\smallskip \mu _{%
\text{at}}^{\ast 4}\sin (3\alpha )\text{,} \\
d_{12}(\mu _{\text{at}}^{\ast },\alpha ) &=&\text{1.58245}\smallskip \mu _{%
\text{at}}^{\ast 4}\sin (\alpha )\text{.}
\end{eqnarray}
\begin{table}[tbp] \centering%
\begin{tabular}{|c|c|c|c|}
\hline
$\mu _{\text{at}}^{\ast }$ & $\alpha $ & $d_{11}$ & $d_{12}$ \\ \hline
0.420 & $0^{\text{o}}$ & 0 & 0 \\ \hline
& $30^{\text{o}}$ & -0.0384 & 0.0480 \\ \hline
& $60^{\text{o}}$ & -0.0137 & 0.0676 \\ \hline
& $90^{\text{o}}$ & 0.01047 & 0.0596 \\ \hline
0.595 & $0^{\text{o}}$ & 0 & 0 \\ \hline
& $30^{\text{o}}$ & -0.1234 & 0.1666 \\ \hline
& $60^{\text{o}}$ & -0.0240 & 0.2223 \\ \hline
& $90^{\text{o}}$ & 0.0408 & 0.2423 \\ \hline
0.728 & $0^{\text{o}}$ & 0 & 0 \\ \hline
& $30^{\text{o}}$ & -0.2008 & 0.3015 \\ \hline
& $60^{\text{o}}$ & 0.0068 & 0.4101 \\ \hline
& $90^{\text{o}}$ & 0.1271 & 0.4991 \\ \hline
0.841 & $0^{\text{o}}$ & 0 & 0 \\ \hline
& $30^{\text{o}}$ & -0.2465 & 0.4323 \\ \hline
& $60^{\text{o}}$ & 0.1165 & 0.5533 \\ \hline
& $90^{\text{o}}$ & 0.2843 & 0.7897 \\ \hline
\end{tabular}
\caption{Coeficientes $d_{11}$ y $d_{12}$} \label{d11,12} 
\end{table}
\begin{table}[tbp] \centering%
\begin{tabular}{|c|c|c|}
\hline
$\mu _{\text{at}}^{\ast }$ & $d_{11}^{0}$ & $d_{12}^{0}$ \\ \hline
0.000 & 0 & 0 \\ \hline
0.420 & -0.0244 & 0.0710 \\ \hline
0.595 & -0.0821 & 0.2590 \\ \hline
0.728 & -0.1639 & 0.5025 \\ \hline
0.841 & -0.2653 & 0.7425 \\ \hline
\end{tabular}
\caption{Coeficientes $d_{1i}^0$} \label{d01i} 
\end{table}%

\section{Correlaci\'{o}n $h_{r}$}

La funci\'{o}n $h_{r}$ la consideramos de la forma, $h_{\epsilon }=1+d_{21}%
\widetilde{L}+d_{22}\widetilde{L}^{2}+d_{23}\widetilde{L}^{3}$. Para valores
fijos de $\mu _{\text{at}}^{\ast }$ y $\alpha $ ajustamos los coeficientes $%
d_{21}$, $d_{22}$ y $d_{23}$ (ver Cuadro \ref{d21,22,23}). Para cualquier
valor de $\mu _{\text{at}}^{\ast }$, el lector puede comprobar que la
dependencia en $\alpha $ de estos coeficientes es cosenoidal --de la forma $%
d_{2i}=d_{2i}^{0}(\mu _{\text{at}}^{\ast })(\cos \alpha -1)$. Para cada
valor fijo de $\mu _{\text{at}}^{\ast }$ ajustamos las amplitudes $%
d_{2i}^{0} $a funciones de orden $4$ en su dependencia con $\mu _{\text{at}%
}^{\ast }$ ( ver Cuadro \ref{d02i}); para finalmente obtener, 
\begin{eqnarray}
d_{21}(\mu _{\text{at}}^{\ast },\alpha ) &=&\text{0.42369}\smallskip \mu _{%
\text{at}}^{\ast 4}\left[ \cos (\alpha )-1\right] \text{,} \\
d_{22}(\mu _{\text{at}}^{\ast },\alpha ) &=&-\text{0.58675}\smallskip \mu _{%
\text{at}}^{\ast 4}\left[ \cos (\alpha )-1\right] \text{,} \\
d_{23}(\mu _{\text{at}}^{\ast },\alpha ) &=&\text{0.51064}\smallskip \mu _{%
\text{at}}^{\ast 4}\left[ \cos (\alpha )-1\right] \text{.}
\end{eqnarray}
\begin{table}[tbp] \centering%
\begin{tabular}{|c|c|c|c|c|}
\hline
$\mu _{\text{at}}^{\ast }$ & $\alpha $ & $d_{21}$ & $d_{22}$ & $d_{23}$ \\ 
\hline
0.420 & $0^{\text{o}}$ & 0 & 0 & 0 \\ \hline
& $30^{\text{o}}$ & 0.0043 & -0.0022 & -0.0041 \\ \hline
& $60^{\text{o}}$ & -0.0022 & -0.0033 & -0.0040 \\ \hline
& $90^{\text{o}}$ & -0.0045 & -0.0042 & -0.0037 \\ \hline
0.595 & $0^{\text{o}}$ & 0 & 0 & 0 \\ \hline
& $30^{\text{o}}$ & 0.0277 & -0.0530 & 0.0174 \\ \hline
& $60^{\text{o}}$ & -0.0127 & -0.0095 & -0.0077 \\ \hline
& $90^{\text{o}}$ & -0.0454 & 0.0407 & -0.0360 \\ \hline
0.728 & $0^{\text{o}}$ & 0 & 0 & 0 \\ \hline
& $30^{\text{o}}$ & 0.03445 & -0.0675 & 0.0147 \\ \hline
& $60^{\text{o}}$ & -0.0559 & 0.0505 & -0.0542 \\ \hline
& $90^{\text{o}}$ & -0.1102 & 0.1173 & -0.0896 \\ \hline
0.841 & $0^{\text{o}}$ & 0 & 0 & 0 \\ \hline
& $30^{\text{o}}$ & 0.0206 & -0.0391 & -0.0149 \\ \hline
& $60^{\text{o}}$ & -0.1235 & 0.1711 & -0.1527 \\ \hline
& $90^{\text{o}}$ & -0.2252 & 0.3522 & -0.2951 \\ \hline
\end{tabular}
\caption{Coeficientes $d_{21}$, $d_{22}$ y $d_{23}$} \label{d21,22,23} 
\end{table}
\begin{table}[tbp] \centering%
\begin{tabular}{|c|c|c|c|}
\hline
$\mu _{\text{at}}^{\ast }$ & $d_{21}^{0}$ & $d_{22}^{0}$ & $d_{23}^{0}$ \\ 
\hline
0.000 & 0 & 0 & 0 \\ \hline
0.420 & 0.0039 & 0.0047 & 0.0048 \\ \hline
0.595 & 0.0379 & -0.0227 & 0.0296 \\ \hline
0.728 & 0.1053 & -0.1052 & 0.0905 \\ \hline
0.841 & 0.2241 & -0.3411 & 0.2945 \\ \hline
\end{tabular}
\caption{Coeficientes $d_{2i}^0$} 
\end{table}%
\label{d02i}

\section{Correlaci\'{o}n $h_{s}$}

La funci\'{o}n $h_{s}$ la consideramos de la forma, $h_{s}(\widetilde{L}%
)=1+d_{31}\widetilde{L}+d_{32}\widetilde{L}^{2}$. Para valores fijos de $\mu
_{\text{at}}^{\ast }$ y $\alpha $ ajustamos los coeficientes $d_{31}$ y $%
d_{32}$ (ver Cuadro \ref{d31, 32}). Para cualquier valor de $\mu _{\text{at}%
}^{\ast }$, el lector puede comprobar que la dependencia en $\alpha $ de los
coeficientes es senoidal --de la forma $d_{31}=d_{11}^{0}(\mu _{\text{at}%
}^{\ast })\sin 3\alpha $ y $d_{32}=d_{12}^{0}(\mu _{\text{at}}^{\ast })\sin
\alpha $. Para cada valor fijo de $\mu _{\text{at}}^{\ast }$ ajustamos las
amplitudes $d_{3i}^{0}$ a funciones de orden $4$ en su dependencia con $\mu
_{\text{at}}^{\ast }$ (ver Cuadro \ref{d03i}) para finalmente obtener: 
\begin{eqnarray}
d_{31}(\mu _{\text{at}}^{\ast },\alpha ) &=&\text{0.31578}\smallskip \mu _{%
\text{at}}^{\ast 4}\sin (3\alpha )\text{,} \\
d_{32}(\mu _{\text{at}}^{\ast },\alpha ) &=&-\text{0.94301}\smallskip \mu _{%
\text{at}}^{\ast 4}\sin (\alpha )\text{.}
\end{eqnarray}
\begin{table}[tbp] \centering%
\begin{tabular}{|c|c|c|c|}
\hline
$\mu _{\text{at}}^{\ast }$ & $\alpha $ & $d_{31}$ & $d_{32}$ \\ \hline
0.420 & $0^{\text{o}}$ & 0 & 0 \\ \hline
& $30^{\text{o}}$ & 0.0263 & -0.0343 \\ \hline
& $60^{\text{o}}$ & 0.0121 & -0.0516 \\ \hline
& $90^{\text{o}}$ & -0.0023 & -0.0485 \\ \hline
0.595 & $0^{\text{o}}$ & 0 & 0 \\ \hline
& $30^{\text{o}}$ & 0.0812 & -0.1136 \\ \hline
& $60^{\text{o}}$ & 0.0191 & -0.1522 \\ \hline
& $90^{\text{o}}$ & -0.0225 & -0.1598 \\ \hline
0.728 & $0^{\text{o}}$ & 0 & 0 \\ \hline
& $30^{\text{o}}$ & 0.1294 & -0.2016 \\ \hline
& $60^{\text{o}}$ & -0.0094 & -0.2434 \\ \hline
& $90^{\text{o}}$ & -0.0914 & -0.2582 \\ \hline
0.841 & $0^{\text{o}}$ & 0 & 0 \\ \hline
& $30^{\text{o}}$ & 0.1500 & -0.2764 \\ \hline
& $60^{\text{o}}$ & -0.0400 & -0.3659 \\ \hline
& $90^{\text{o}}$ & -0.1350 & -0.4347 \\ \hline
\end{tabular}
\caption{Coeficientes $d_{31}$ y $d_{32}$} \label{d31, 32} 
\end{table}
\begin{table}[tbp] \centering%
\begin{tabular}{|c|c|c|}
\hline
$\mu _{\text{at}}^{\ast }$ & $d_{31}^{0}$ & $d_{32}^{0}$ \\ \hline
0.000 & 0 & 0 \\ \hline
0.420 & 0.0143 & -0.0552 \\ \hline
0.595 & 0.0519 & -0.1741 \\ \hline
0.728 & 0.1104 & -0.2849 \\ \hline
0.841 & 0.1425 & -0.4449 \\ \hline
\end{tabular}
\caption{Coeficientes $d_{3i}^0$} \label{d03i} 
\end{table}%

\end{document}